\documentclass[english]{iopart}

\usepackage{xcolor}

\usepackage{graphicx}
\usepackage{url}
\usepackage{iopams} 

\expandafter\let\csname equation*\endcsname\relax
\expandafter\let\csname endequation*\endcsname\relax

\usepackage{amssymb}
\usepackage[latin1]{inputenc}
\usepackage{amsmath}
\usepackage{epsfig}
\newcounter{fig}

\begin{document}

\title[Heun functions and diagonals  of rational functions.]
{\Large  Heun functions and diagonals of rational functions (unabridged version).}
\vskip .3cm 

\author{Y. Abdelaziz$^\dag$,  S. Boukraa$^\pounds$, C. Koutschan$^\delta$ 
J-M. Maillard$^\dag$}

\address{$^\dag$ LPTMC, UMR 7600 CNRS, 
Universit\'e Pierre et Marie Curie, Sorbonne Universit\'e, 
Tour 23, 5\`eme \'etage, case 121, 
 4 Place Jussieu, 75252 Paris Cedex 05, France} 

\address{$^\pounds$  \ LPTHIRM and IAESB,
 Universit\'e de Blida, Algeria}

\address{$^\delta$  Johann Radon Institute for Computational and Applied Mathematics (RICAM), 
Altenberger Strasse 69,  A-4040 Linz, Austria  }

\vskip .2cm 

\begin{abstract}
We provide a set of diagonals of simple rational functions
of four variables that are seen to be  squares of Heun
functions. Each time, these Heun
functions, obtained by creative telescoping,
turn out to be pullbacked $\, _2F_1$ hypergeometric functions
and in fact classical modular forms. We even obtained
Heun functions that are automorphic forms
associated with Shimura curves as solutions
of telescopers of rational functions. 
  
\end{abstract}

\vskip .1cm


\vskip .4cm

\noindent {\bf PACS}: 05.50.+q, 05.10.-a, 02.30.Hq, 02.30.Gp, 02.40.Xx

\noindent {\bf AMS Classification scheme numbers}: 34M55, 
47E05, 81Qxx, 32G34, 34Lxx, 34Mxx, 14Kxx 

\vskip .2cm

\vskip .2cm

{\bf Key-words}: Diagonals of rational functions,  Heun functions, creative telescoping,
extremal rational elliptic surfaces,
pullbacked hypergeometric functions, Schwarz map, Schwarz function, 
modular forms, modular equations, Hauptmoduls,  Belyi coverings, Shimura curves,  automorphic forms.

\vskip .1cm

\section{Introduction}
\label{Introduction}

\vskip .1cm

Diagonals of rational functions naturally emerge in lattice statistical mechanics, enumerative combinatorics
or more generally for $\, n$-fold integrals of theoretical physics~\cite{Short,Big}.
In previous papers~\cite{DiagSelected,unabri,notunabri} we have seen\footnote[1]{These calculations
were performed using the creative telescoping program of C. Koutschan~\cite{Koutschan}.} that many diagonals of
simple rational functions were pullbacked $\, _2F_1$ hypergeometric functions that turn out to be
related to classical modular forms. Sticking with diagonals of
simple rational functions that are solutions of linear differential operators of {\em order two}, it is natural to
study diagonals of simple rational functions that are Heun functions.

 Heun functions emerge in different areas of physics~\cite{Short,Takemura,Valent,Smirnov}
 (see also page 60 of~\cite{Big}) and enumerative combinatorics: the simple cubic lattice Green
 functions~\cite{GlasserGuttmann} can be written as a Heun function.
 Experimentally the Heun functions emerging in physics often\footnote[1]{This is not the case for the Heun functions in~\cite{Smirnov}
   which do not correspond to globally bounded series.} correspond to {\em globally bounded series}~\cite{Short,Big},
 i.e. series that can be recast, after some rescaling, into series with {\em integer coefficients}. Most of the time they turn out
 to be pullbacked $\, _2F_1$ hypergeometric functions  and in fact {\em classical modular forms}. This suggests
 to study the class of Heun functions that are diagonals of rational
 functions\footnote[9]{In~\cite{DiagSelected} we found that diagonals of
   simple rational functions yield quite systematically classical modular forms when the corresponding
 telescoper is of order-two. Diagonals of rational functions are necessarily globally bounded~\cite{Short,Big}.},
 and thus, {\em globally bounded series}~\cite{Big}.
 We will discard the case where the Heun functions are almost trivial, their order-two linear differential operators
 factorising into two order-one linear differential operators. Such rather trivial cases are recalled in~\ref{trivia}. 
 In this paper we examine non trivial Heun functions, which happen to be {\em diagonals of simple rational functions}~\cite{Big}
 of (mostly) four variables or solutions of telescopers of rational functions of (mostly) four variables. 
 We see that they happen to fall into one of three categories:
\begin{itemize}
\item Heun functions that are diagonals of rational functions,  having {\em globally bounded} series expansions,
  and  can be rewritten as pullbacked hypergeometric functions that are {\em classical modular forms}.
\item Heun functions that are diagonals of rational functions, having globally bounded series expansions,
  and can be rewritten as pullbacked hypergeometric functions that are {\em derivatives} of classical modular forms.
\item Heun functions that are solutions of {\em telescopers}\footnote[8]{By ``telescoper''
   of a rational function, say $\, R(x,y,z)$, we here refer to the output of the creative telescoping
   program~\cite{Koutschan}, applied to the {\em transformed} rational function $\, \tilde{R} = R(x/y,y/z,z)/(yz)$.
   Such a telescoper is a differential operator~$T$ in $x,D_x$ such that $T+D_y\cdot U+D_z\cdot V$ annihilates
   $\tilde{R}$, where $U,V$ are rational functions in $x,y,z$. In other words, the telescoper~$T$ represents a
   linear ODE that is satisfied by $ \, Diag(R)$.} of rational functions that have series expansions
that are {\em not} globally bounded\footnote[2]{And hence cannot be diagonals of rational functions.}. They will be seen
to correspond to {\em Shimura automorphic forms} or derivatives of automorphic forms.
\end{itemize} 

The Heun function $\, Heun(a,q,\alpha,\beta,\gamma,\delta, \, \, x)$ is solution of
the order-two Heun linear differential operator with four singularities ($D_x$ denotes $\, d/dx$)
\begin{eqnarray}
  \label{Heun}
  \hspace{-0.98in}&& \quad \quad \, \, \,  
  H_2  \,\, = \, \, \, D_x^2 \,\,\, 
                     + \, \,  \Bigl({{ \gamma} \over { x}} \, + {{ \delta} \over { x \, -1}}
                     \, + {{ \epsilon} \over { x \, -a}}\Bigr) \cdot \, D_x \,\, \,
+ \,  {{ \alpha\, \beta \,x \, -q } \over { x \cdot \, (x-1) \cdot \,  (x \, -a)}}, 
\end{eqnarray}               
where one  has the Fuchsian constraint
$\,  \,  \,  \epsilon   \, = \, \, \alpha \, + \, \beta \, -\gamma \, - \, \delta \, + \, 1$,
where $\alpha,\beta,\gamma,\delta$ need to be rational numbers, and $\, a$ is an algebraic number. The
parameter $\, q$ is called the {\em accessory parameter} and the ratio $\, q/\alpha/\beta\, $
is called the {\em normalised  accessory parameter}.

In the first two sections, we examine the Heun functions
{\em emerging from diagonals of simple rational functions}
that fall into the first or second category above,
and show how they happen to be related to {\em classical modular forms},
or  {\em derivatives of classical modular forms}, corresponding
to pullbacked $\, _2F_1$ hypergeometric functions. These Heun functions have {\em integer} coefficient series,
(or can be recast as series with {\em integer coefficients}~\cite{Big} after a rescaling of the variable),
and are solutions of globally nilpotent~\cite{GloballyNilpotent} 
linear differential operators: the critical exponents of all the singularities are {\em rational numbers}. This
leads  to define a criterion in \ref{notmodular}, that allows to draw up a
list of parameters of the Gauss hypergeometric function $\, _2F_1([a,b],[c],x)$, for which it corresponds
to a {\em classical modular form} (see section \ref{diagfour}). Furthermore, we will find
that some of these Heun functions turned out to be periods of
{\em extremal rational surfaces} (see section \ref{subthree2}).  We do this while
avoiding trivial cases\footnote[5]{(See \ref{trivia}) corresponding to factorizations of the order-two linear
  differential operator of the Heun function (see \ref{facto}), or corresponding to situations where the fourth
  singularity is in fact an apparent singularity (see \ref{apparent}), a situation which often corresponds to
  the previous factorization of the order-two linear differential operator.}.

In the third section, we examine first the solutions of the telescoper of a rational function, corresponding to
a Heun function with a {\em non globally bounded} series expansion, and we show that this Heun function is related
to a specific {\em Shimura curve}~\cite{ElkiesRank19,Voight,Voight2,Shaska,Hallouin,Kurihara}. We then examine
a larger class of Heun functions listed in~\cite{HoeijVidunas},
and show that they are linked to {\em Shimura curves}. We are able to show the link between these Heun functions
and {\em Shimura curves} thanks to a result by  K.Takeuchi~\cite{Takeuchi}.

 \vskip .1cm

\subsection{ Recalls on lattice Green functions as diagonals of rational functions}
\label{subIntroduction}

 The simple cubic lattice Green function~\cite{Joyce}
\begin{eqnarray}
\label{SC}
  \hspace{-0.98in}&& \quad  \quad \quad \quad \quad 
 {{1} \over { (2\, \pi)^3 }} \cdot \,  \int_0^{2\, \pi}\int_0^{2\, \pi}\int_0^{2\, \pi} \, {{ d\theta_1 \,  d\theta_2 \,  d\theta_3 } \over {
         1 \,  \, \,  - \, x \cdot \, (\cos(\theta_1) \, \, +  \cos(\theta_2) \, \, +   \cos(\theta_3)) }},
\end{eqnarray}
is nothing but\footnote[9]{Cooking recipe: change $\, \cos(\theta_i) \, = \,  (1 \, +z_i^2)/2/z_i$
(i.e.  $\, z_i \, = \, \exp(i \, \theta_i)$)
{\em and} $\, x \, \rightarrow \, \, x \cdot \, z_1 \, z_2 \, z_3$.}
the diagonal of the rational function in four
variables $\, x, \, z_1, \, z_2, \, z_3$:
\begin{eqnarray}
\label{DiagSC}
  \hspace{-0.98in}&& \quad\quad   \, {{ 1 } \over {
      1 \,  \,  \, \,  
      - \, x  \cdot \, z_1 \, z_2 \, z_3 \cdot \,
  ((1 \, +z_1^2)/z_1/2 \, \, \,  +   (1 \, +z_2^2)/z_2/2  \,\,  \, +   (1 \, +z_3^2)/z_3/2) }}
                     \nonumber \\
  \hspace{-0.98in}&&   \quad \quad \quad \,
   \, \,  \, \, = \, \,   \,  \,
  {{ 2 } \over { 2 \,  \,  \, \,
      - \, x  \cdot \, z_1 \, z_2 \, z_3 \cdot \,
  (z_1 \, +  1/z_1  \,\, +  \, z_2 \, +  1/z_2 \, \, +  \, z_3 \, +  1/z_3) }}.
\end{eqnarray}
The linear differential operator annihilating the diagonal of
this rational function in {\em four} variables has order three
and is the {\em symmetric square} of a linear differential operator of order two
($\theta$ is the homogeneous derivative $\, x \cdot \, d/dx$):
\begin{eqnarray}
\label{DiagSCL2homog}
\hspace{-0.98in}&& \quad \quad  \quad  \quad   \, 
 9 \, x^4 \cdot \,(2\, \theta \, +3) \cdot \,(2\, \theta \, +1)\, \, \, \, 
 -4\, x^2 \cdot \, (10 \,\theta^2 \, +10\,\theta \, +3) \, \,  \,   +4\, \theta^2, 
\end{eqnarray}
having a Heun function as a solution.
Consequently,
the simple cubic lattice Green function (\ref{SC}), or equivalently the
diagonal of (\ref{DiagSC}) reads:
\begin{eqnarray}
\label{SolSC}
\hspace{-0.98in}&& \quad \quad \quad 
Heun\Bigl({{1} \over {9}}, \, {{1} \over {12}}, \,
{{1} \over {4}}, \,  {{3} \over {4}}, \,1, \,  {{1} \over {2}}, \, \, x^2     \Bigr)^2
\, \, = \, \,   \,  \, Heun\Bigl(9, \,  {{3} \over {4}}, \,   {{1} \over {4}}, \,
 {{3} \over {4}}, \, 1, \,   {{1} \over {2}},   \, \,  9 \, x^2   \Bigr)^2.
\end{eqnarray}
The Heun function on the RHS of (\ref{SolSC}) happens to be a period of an {\em extremal rational curve} as can be seen
in the work of Doran and Malmendier~\cite{Malmendier4}. The diagonal (\ref{SolSC}) can also be written as
a Hadamard product\footnote[1]{Denoted here by a star (*).} of a simple algebraic function and a Heun function:
\begin{eqnarray}
\label{SolSCHadamard}
\hspace{-0.99in}&& \quad \quad  \quad     \quad     \quad    
Heun\Bigl({{1} \over {9}}, \, {{1} \over {12}}, \,
{{1} \over {4}}, \,  {{3} \over {4}}, \,1, \,  {{1} \over {2}}, \, \, x^2     \Bigr)^2
\nonumber \\
  \hspace{-0.99in}&&   \quad  \quad   \quad     \quad     \quad     \quad     \quad     \quad
 = \, \,   \,      (1\, -4 \, x^2)^{-1/2}  \star \,
 Heun\Bigl( {{1} \over {9}},\, {{1} \over {3}}, \, 1, \, 1, \, 1, \, 1, \,  {{x^2} \over {4}}  \Bigr).
\end{eqnarray}
Similarly, considering {\em pencils of K3-surfaces}, Peters and Stienstra introduced~\cite{Stienstra}
the integral\footnote[3]{In section 3 of~\cite{Stienstra},
  the variable $\, x$ is denoted $\, t \, = \, 1/s$. The correspondence between this $\, x$ and the $\, x$ in
  the lattice Green function (\ref{SC}) which corresponds to $\, I(x)/x$, is $\, x \, \rightarrow \, x/2$. Thus 
(\ref{SolSC}) becomes, once divided by $\, x$, the square of the Heun function (\ref{SolSCalsolu}).  } 
\begin{eqnarray}
\label{Stienstra1}
\hspace{-0.98in}&& \quad  \quad  \quad \quad  \, \,   \, 
 I(x) \, \, = \, \, -\, \Bigl({{1} \over {2\, i \, \pi}}\Bigr)^3 \cdot \, \int_{|z_1|=1}\int_{|z_2|=1}\int_{|z_3|=1}
\, {{dz_1\, dz_2\, dz_3} \over {z_1\, z_2\, z_3}}
\nonumber \\
\hspace{-0.98in}&& \quad  \quad  \quad  \quad  \quad \quad  \quad  \quad  \, 
 \times \,  {{ x } \over {
1 \, \,\,
- \,  x \cdot \, (z_1 \, + \, 1/z_1  \,\, +  \, z_2 \, + \, 1/z_2 \, \, +  \, z_3 \, + \, 1/z_3) }}, 
\end{eqnarray}
is annihilated by the linear differential operator of order three that is the
symmetric square of the order-two linear differential operator:
\begin{eqnarray}
\label{Stienstra3}
  \hspace{-0.98in}&& \quad  \quad \quad
 L_2 \, \,\,\, = \, \,\, \,\,
576 \cdot \, x^4 \cdot \, \theta \cdot \, (\theta \, +1)
\,  \,\,\, -8 \cdot \, x^2 \cdot \, (20 \, \theta^2 \, +1) \,\,  \, + \,  (2\, \theta \, -1)^2, 
\end{eqnarray}
where $\, \theta$ denotes the homogeneous derivative $\,\, \theta \, = \, x \cdot \, d/dx$.
Its solution, analytic at $\, x\, = \, 0$, reads:
\begin{eqnarray}
\label{SolSCalsolu}
\hspace{-0.98in}&& \quad  \quad  \quad  \quad  \quad  \quad  \quad  \quad   
       x^{1/2} \cdot \,   Heun\Bigl({{1} \over {9}}, \, {{1} \over {12}}, \,
{{1} \over {4}}, \,  {{3} \over {4}}, \,1, \,  {{1} \over {2}}, \, \, 4 \, x^2  \Bigr),
\end{eqnarray}
the other solution having a formal series expansion with a logarithm.
Note that this square of a Heun function can be recast into a series with {\em integer coefficients}:
\begin{eqnarray}
\label{SolSCser}
  \hspace{-0.98in}&& \quad \quad  \, \,
Heun\Bigl(9, \,  {{3} \over {4}}, \,   {{1} \over {4}}, \,
 {{3} \over {4}}, \, 1, \,   {{1} \over {2}},   \, \,   36  \, z   \Bigr)^2
\, \, \, = \, \, \,  \, \, \, 
 Heun\Bigl({{1} \over {9}}, \,  {{1} \over {12}}, \,   {{1} \over {4}}, \,
                     {{3} \over {4}}, \, 1, \,   {{1} \over {2}},   \, \,   4  \, z   \Bigr)^2
\nonumber \\
  \hspace{-0.98in}&& \quad \quad  \quad \quad \, \,      
\, \, \, = \, \,\, \, 1 \, \, \,  +6\, z \, \, +90\, z^2  \, \, +1860\, z^3 \, \,
+44730 \, z^4 \, +1172556 \, z^5 \, \, +32496156\, z^6 
\nonumber \\
\hspace{-0.98in}&& \quad \quad \quad  \quad \quad  \quad \quad  \quad \, 
 \,+936369720\, z^7 \, \, \, + \, \, 27770358330\, z^8  \, \, \, \, + \, \, \, \cdots 
\end{eqnarray}
Alternative forms of Heun functions like
\begin{eqnarray}
\label{Alter}
  \hspace{-0.98in}&&
Heun\Bigl(9, \, 3, \, 1, \, 1, \, 1, \, 1, \, \, x\Bigr), \, \, \, \, 
Heun\Bigl({{1} \over {9}}, \, {{1} \over {3}}, \, 1, \, 1, \, 1, \, 1, \, \, x\Bigr),  \, \, \, \, 
Heun\Bigl(4, \, {{1} \over {2}}, \, {{1} \over {2}}, \, {{1} \over {2}}, \, 1, \,{{1} \over {2}}, \, \, x  \Bigr),
\nonumber 
\end{eqnarray}
can be introduced for the simple cubic lattice Green function. They are displayed in \ref{Alternative}.
They all reduce to pullbacked $\, _2F_1$ hypergeometric
functions\footnote[2]{See also \ref{pulback2F1represe} below for (\ref{SolSCalsolu}) or (\ref{SolSCser}).}
which turn out to correspond to {\em classical modular forms}\footnote[1]{The emergence, for this
 fibration into K3 surfaces, of modular functions, cusp forms of weight two, via Dedekind's
$\, \eta$-functions, can be found in section 4 of~\cite{Stienstra}.}.

\vskip .2cm 

 \section{Diagonals  of rational functions of three and four variables yielding Heun functions corresponding to classical modular forms}
 \label{diagfour}

We are going to provide a set of exact expressions for diagonals~\cite{JAWexamples}
of simple rational functions
of three and four variables yielding Heun functions. These exact expressions are obtained
using the creative telescoping approch and, more specifically, the program of C. Koutschan~\cite{Koutschan},
these diagonals being analytic at $\, x=0$ globally bounded series, solutions of a telescoper
obtained with this creative telescoping program\footnote[5]{This program also
provides other expressions called the ``certificates'' that we do not use here.}. 

\vskip .1cm

 \subsection{Diagonals  of rational functions of four variables yielding Heun functions}
\label{subfour}

\vskip .1cm

$\bullet$ Example 1. The diagonal of the  rational function
\begin{eqnarray}
\label{Ratfoncfour_7}
  \hspace{-0.98in}&& \quad  \quad 
\, \,  
R(x, \, y, \, z, \, w)  \, \, \,  = \, \,  \quad 
{{1} \over { 1 \, \, \, \,    - \,
  (w \, x \, y \, +w \, x \, z \, +w \, y \, z \, +x \, y \, z \, \,  +w \, x \, +y \, z) }},
\end{eqnarray}
reads:
\begin{eqnarray}
\label{SS7}
  \hspace{-0.98in}&& \quad    
Diag\Bigl(R(x, \, y, \, z, \, w)\Bigr)  \, \, 
\, \, = \, \,\,  \,\,  1 \, \, \,+2\, x \,\,  +18\, x^2 \,\,  +164\, x^3 \,\, +1810\, x^4 
\, \, \,\,  + \, \, \, \cdots 
\end{eqnarray}
A creative telescoping program~\cite{Koutschan} gives the order-three linear differential
operator annihilating the diagonal (\ref{SS7}) of the previous rational function (\ref{Ratfoncfour_7}):
\begin{eqnarray}
\label{L3SS7}
  \hspace{-0.98in}&& \quad  \quad   
 L_3 \, \, \,  = \, \, \,   \,   2 \, +60\, x \,  \, \, \,\,  - (1 \, -40x \, -444 \, x^2) \cdot \,  D_x
  \, \, \,  \, - 3\,  x \cdot \, (1 \, -18 \, x \, -128 \, x^2)\cdot D_x^2
\nonumber \\
 \hspace{-0.98in}&& \quad \quad \quad \quad \quad \quad    \quad \quad   \,
  -x^2 \cdot \, (1 \,+ 4\, x) \cdot \, (1\, -\, 16\, x) \cdot \, D_x^3.
\end {eqnarray}
This  order-three linear differential operator corresponds to the {\em symmetric square} of the order-two
linear differential operator:
\begin{eqnarray}
\label{L2SS7}
\hspace{-0.98in}&& \quad  \quad   \quad   \quad
 L_2 \, \,  = \, \, \,   \, \,    x^2 \cdot \,   (8\, \theta \, +5) \cdot \, (8\, \theta \, +3)
 \, \,  \,   + x \cdot \, (12 \, \theta^2+6\, \theta \, +1) \,  \,  \, \,    -\theta^2.                
\end {eqnarray}
Thus the solution corresponding to the diagonal of (\ref{Ratfoncfour_7}) is given by the square of a Heun function:
\begin{eqnarray}
\label{H1solhyp}
\hspace{-0.98in}&& \quad \quad  
Heun\Bigl(- \, {{1} \over {4}}, \, {{1} \over {16}}, \, {{3} \over {8}},
  \,  {{5} \over {8}}, \, 1, \,  {{1} \over {2}}, \, -4 \, x\Bigr)^2
\nonumber \\
\hspace{-0.98in}&& \quad \quad  \quad \quad \quad
\, \, = \, \, \, \,  \,   1 \, \, \, \,   +2\, x \,\, \,    +18\, x^2 \,\, \,    +164\, x^3 \,\, \,  
+1810\, x^4 \,\, \,    +21252\, x^5 \,\,  \,   +263844\, x^6 
\nonumber \\
\hspace{-0.98in}&& \quad \quad \quad  \quad \quad \quad \quad \quad \quad \quad
\,  +3395016\, x^7\, +44916498\, x^8  \,  \,  \, \, \, + \, \, \, \cdots 
\end{eqnarray}
This Heun function can be written as a pullbacked $\, _2F_1$
hypergeometric function :
\begin{eqnarray}
\label{H1solhypmany}
\hspace{-0.98in}&& \quad \quad \quad  \, \, \, 
Heun\Bigl(- \, {{1} \over {4}}, \, {{1} \over {16}}, \, {{3} \over {8}},
  \,  {{5} \over {8}}, \, 1, \,  {{1} \over {2}}, \, -4 \, x\Bigr)
\, \, \,  = \, \, \, \,  {\cal A} \cdot \, 
 _2F_1\Bigl([{{1} \over {8}}, \, {{3} \over {8}}], \, [1], \, {\cal H} \Bigr),  
\end{eqnarray}
where $\, {\cal A}$ and the Hauptmodul $\, {\cal H}$
are algebraic functions expressed with square roots:
\begin{eqnarray}
\label{H1solhypm1}
\hspace{-0.98in}&& \quad \quad
{\cal H}_{\pm} \, = \, \, -\, 128 \,  x \cdot \, 
 {{1 \, -20\, x \, +50\, x^2 \,+400\, x^3 \, -224\, x^4 \,   -512\, x^5} \over {
 (1 \, -88\, x \, -\, 112\, x^2 \, -256\, x^3)^2}} \,
 \\
\hspace{-0.98in}&& \quad \quad \quad \quad \quad  \quad 
\, \pm \, 128 \,  x \cdot \,   {{(1\, + 2\, x) \, (1\, -12\, x) \,\, (1\, -4\, x) \cdot \,
( 1\, +4\, x)^{1/2} \cdot \,   ( 1\, -16 \, x)^{1/2}  } \over {
(1 \, -88\, x \, -\, 112\, x^2 \, -256\, x^3)^2 }}.
\nonumber
\end{eqnarray}
These Hauptmoduls (\ref{H1solhypm1}) are also given by the {\em genus-zero} quadratic relation
\begin{eqnarray}
\label{H1solhypm1sergenuszero}
  \hspace{-0.98in}&& \quad \quad \quad
(256\,x^3 \, +112\, x^2+88\,x \, -1)^2 \cdot \, {\cal H}_{\pm}^2  \,
 \nonumber          \\
\hspace{-0.98in}&& \quad \quad \quad  \quad \quad \quad
 - \, 256 \cdot \, x  \cdot \,
 (512\,x^5 +224\,x^4\, -400\,x^3 -50\,x^2 \, +20\,x \, -1) \cdot \, {\cal H}_{\pm}
 \nonumber \\
\hspace{-0.98in}&& \quad \quad \quad \quad \quad \quad \quad \quad \quad \quad
 \, \,  +65536 \, x^6 \, \, \, =    \, \ \, \, 0,  
\end{eqnarray}
and have the series expansions:
\begin{eqnarray}
\label{H1solhypm1ser}
  \hspace{-0.98in}&& \quad \quad
{\cal H}_{-} \, = \, \, \, \,
 -256\, x \,  \, \, -39936\, x^2 \,  \, \, -5116416\, x^3 \, \,  \, -595357696\, x^4
 \,  \,\,  -65525931776\, x^5
 \nonumber \\
  \hspace{-0.98in}&& \quad \quad \quad \quad  \quad \quad
 \, -6954923846656\, x^6 \, \,  -719583708750336\, x^7 \, \, \, \, + \, \, \, \cdots 
\nonumber \\
\hspace{-0.98in}&& \quad \quad
{\cal H}_{+} \, = \, \, \, \,
-256\, x^5 \, \, \, -5120\, x^6 \, \, -89600\, x^7 \, \, -1433600\, x^8 \, \, -22201600\, x^9
\nonumber \\
  \hspace{-0.98in}&& \quad \quad \quad \quad \quad  \quad \quad
\, -337755136\, x^{10}
\,\, -5094679040\, x^{11} \, \, \, \, + \, \, \, \cdots 
\end{eqnarray}
The relation between these two Hauptmoduls corresponds to a genus-zero
$\, q \, \leftrightarrow \,  \, q^5$ {\em modular equation} ($q$
denotes the nome of the order-two operator).

\vskip .2cm

This Heun function can also be written alternatively as:
\begin{eqnarray}
\label{H1solhypmanyother}
\hspace{-0.98in}&& \quad \quad \quad  \, \, \, 
Heun\Bigl(- \, {{1} \over {4}}, \, {{1} \over {16}}, \, {{3} \over {8}},
  \,  {{5} \over {8}}, \, 1, \,  {{1} \over {2}}, \, -4 \, x\Bigr)
\, \, \, = \, \, \, \, {\cal A}_1 \cdot \, 
 _2F_1\Bigl([{{1} \over {12}}, \, {{5} \over {12}}], \, [1], \, {\cal H} \Bigr),  
\end{eqnarray}
using the identity
\begin{eqnarray}
\label{H1solhypmanyotherident}
  \hspace{-0.98in}&& \quad \quad \quad  \quad \quad  \, \, \,
_2F_1\Bigl([{{1} \over {8}}, \, {{3} \over {8}}], \, [1], \, x\Bigr)
\,   \, = \, \,   \,  {\cal A}_2 \cdot \,
 _2F_1\Bigl([{{1} \over {12}}, \, {{5} \over {12}}], \, [1], \,  \, H   \Bigr), 
\end{eqnarray}
where $\,  {\cal A}_1$,  $\,  {\cal A}_2$ denote some algebraic functions and where:
\begin{eqnarray}
\label{H1solhwhere}
  \hspace{-0.98in}&&  \, 
 H \,  \,\, = \, \,\, \,
 {{ 27 \cdot \, (27\,x^2\, -414\, x \, +512)\cdot \, x} \over {(9\,x \, +16)^3 }}
\, \,\, \,  - \, \,  {{ 54 \cdot \, (81 \, x \, -256)\cdot \, x} \over {(9\,x \, +16)^3  }}
\cdot \, (1\, -x)^{1/2}. 
\end{eqnarray}

\vskip .2cm 

$\bullet$  Example 2.  Let us consider the rational function in four variables:
\begin{eqnarray}
\label{Ratfoncfour8}
  \hspace{-0.98in}&& \quad  \quad 
\, \,  
R(x, \, y, \, z, \, w)  \, \, \,  = \, \,  \,    \,  
 {{1} \over { 1 \, \, \, - \,
  (w\, x\, y \, +w\, x\, z \, +w\, y\, z \, +x\, y\, +x\, z \, +y \, +z) }}.
\end{eqnarray}
The diagonal of this rational function (\ref{Ratfoncfour8}) reads:
\begin{eqnarray}
\label{SS8}
  \hspace{-0.98in}&&   \quad \quad 
  Diag\Bigl(R(x, \, y, \, z, \, w)\Bigr)  \, \, \,
\, \, = \, \, \,\, 1 \, \, \,  +4\, x \,\,  +48\, x^2 \, \, +760\, x^3
\, \, +13840\, x^4 \,\,  +273504\, x^5
\nonumber \\
 \hspace{-0.98in}&&   \quad \quad \quad \quad  \quad \quad  \quad  \quad  \quad   \quad 
\, +5703096\, x^6 \, \,  \, +123519792\, x^7 \, \, \, + \, \, \, \cdots 
\end{eqnarray}
The linear differential operator annihilating the diagonal of this  rational function
is the third order linear differential operator:
\begin{eqnarray}
\label{S8}
  \hspace{-0.98in}&& \quad  \quad \quad \quad \quad
 x^2 \cdot \, (1 \, +\, x) \cdot \, (1\, -27 \, x) \cdot \,  D_x^3
\, \, \,  \,  \,  +3\, x \cdot \, (1 \, -39\,x \,-54\, x^2) \cdot \, D_x^2
   \nonumber \\
\hspace{-0.98in}&& \quad \quad \quad \quad \quad \quad \quad \quad \, \, \, 
  +(1 \, -86\, x \, -186\, x^2) \cdot \, D_x   \, \,  \,\,  - 4\cdot \, (1 \, +6\, x)
\end{eqnarray}
This third order linear differential operator (\ref{S8}) is the
{\em symmetric square} of an order-two linear differential operator,
having as solution a  (square of a) Heun
function given as series expansion with {\em integer coefficients}:
\begin{eqnarray}
  \hspace{-0.98in}&& \quad  \, \, \,\,
 Heun\Bigl(- \, {{1} \over {27}}, \, {{2} \over {27}}, \, {{1} \over {3}},
  \,  {{2} \over {3}}, \, 1, \,  {{1} \over {2}}, \, -\, x\Bigr)^2
 \, \, \, = \, \, \,\, \,
  1 \,  \, \, +4\, x \, \, +48\, x^2 \,  \,+760\, x^3 \, \, +13840\, x^4 
 \nonumber \\
 \hspace{-0.98in}&&   \quad \quad \quad \quad \quad  \quad \quad  \, \, \, 
 \, \, +273504\, x^5 \,  \,+5703096\, x^6
 \, \, + \, 123519792\, x^7 \, \,  \, +  \, \, \, \cdots       
\end{eqnarray}
We also have the following series expansion with  integer coefficients:
\begin{eqnarray}
\label{alsoHeun}
\hspace{-0.98in}&& \quad  \quad  
Heun\Bigl(- \, {{1} \over {27}}, \, {{2} \over {27}}, \, {{1} \over {3}},
  \,  {{2} \over {3}}, \, 1, \,  {{1} \over {2}}, \, -\, x\Bigr)
 \, \, \, = \, \, \, \,\,
1 \, \,\, +2 \, x \,\, +22 \, x^2\,\, +336 \, x^3\, \,+6006 \, x^4
\nonumber \\
  \hspace{-0.98in}&& \quad  \quad \quad \quad \quad \quad \quad
\,\, +117348 \, x^5 \,
\, +2428272 \, x^6\, \,+52303680 \, x^7  \, \,\,\, + \, \, \cdots 
\end{eqnarray}
This Heun function (\ref{alsoHeun}) can be written as a pullbacked $\, _2F_1$ hypergeometric function
\begin{eqnarray}
\label{AhAh}
\hspace{-0.98in}&& \quad   
Heun\Bigl(- \, {{1} \over {27}}, \, {{2} \over {27}}, \, {{1} \over {3}},
  \,  {{2} \over {3}}, \, 1, \,  {{1} \over {2}}, \, -\, x\Bigr) \, \, = \, \, \,
\\
  \hspace{-0.98in}&& \quad  \quad \, \, 
 \Bigl(25 \, -80 \, x \, \, \, -24 \cdot \, (1\, +\, x)^{1/2} \cdot \, (1\,  \, -27\, x)^{1/2}\Bigr)^{-1/4}
 \cdot \,\,
_2F_1\Bigl([{{1} \over {12}}, \, {{5} \over {12}}], \, [1], \, {\cal H}_{+}\Bigr), \nonumber 
\end{eqnarray}
where the Hauptmodul $\, {\cal H}$ reads:
\begin{eqnarray}
\label{AhAhHaupt}
\hspace{-0.98in}&&  \quad  \quad  \,    
{\cal H}_{\pm} \, \, = \, \, \,
864 \cdot \, x \cdot \,
 {{ (1\,-21\, x \,  +8\, x^2) \,
  \cdot \, (1 \, -42\, x \, +454\, x^2-1008\, x^3 \, -1280\, x^4)  } \over {
 (1 \, +224\, x \, +448\, x^2)^3 }}
 \nonumber \\  
  \hspace{-0.98in}&& \quad   \quad  \quad  \quad \quad  \,      \, \,
 \, \pm \, 864 \cdot \, x \cdot \, (1 \, -8\, x) \cdot \, (1 \, -2\, x)
  \cdot \, (1\, -24\, x) \cdot \,(1 \, -16\, x \, -8\, x^2)
  \nonumber \\
  \hspace{-0.98in}&& \quad   \quad  \quad \quad  \quad \quad  \quad  \quad       \quad     \, \,
  \times \, {{(1\, + \, x)^{1/2} \cdot \, (1 \, -27\,x)^{1/2} } \over {
  (1 \, +224\, x \, +448\, x^2)^3}}. 
\end{eqnarray}
The series expansions of these two  Hauptmoduls (\ref{AhAhHaupt}) read respectively
\begin{eqnarray}
\label{AhAhHauptser1}
\hspace{-0.98in}&&   \quad     
 {\cal H}_{+} \, \, = \, \, \,\,
 1728 \, x \,\, -1270080\, x^2 \,\, +593381376\, x^3 \,\, -226343666304\, x^4
 \nonumber \\
 \hspace{-0.98in}&&  \quad  \quad  \quad   
 \, +76907095308288\, x^5 \, \, -24246668175851520\, x^6\, 
\, +7253781581324351808\, x^7  \nonumber \\
  \hspace{-0.98in}&&  \quad  \quad  \quad \quad
\, -2087529169324932180288\, x^8 \, \,\,\, + \, \, \, \cdots 
\end{eqnarray}
and:
\begin{eqnarray}
\label{AhAhHauptser2}
\hspace{-0.98in}&&    \quad    
 {\cal H}_{-} \, \, = \, \, \,
1728\, x^7 \, +108864\, x^8 \, +4536000\, x^9 \, +158251968\, x^{10}
 \\
\hspace{-0.98in}&&  \quad  \quad  \quad  \quad  \quad  
 \, +5017070016\, x^{11} \,  +150134378688\, x^{12}
 \,  +4328271255168\, x^{13} \,\, \, + \, \, \cdots \nonumber 
\end{eqnarray}
These two Hauptmoduls are the two solutions of the {\em quadratic} genus-zero relation:
\begin{eqnarray}
\label{AhAhHauptquadra}
\hspace{-0.98in}&&     
 1728^2 \cdot \, x^8 \, \,\,
 +1728 \cdot \, (1 \, -21\, x \, +8\, x^2)
 \, (1280\, x^4+1008\, x^3-454\, x^2+42\, x-1) \cdot \, x \cdot \, {\cal H}_{\pm}
\nonumber \\
\hspace{-0.98in}&&  \quad  \quad  \quad  \quad  \quad \quad  \quad
  \, +(1\, + \, 224\, x \, + \, 448\, x^2)^3\cdot  {\cal H}_{\pm}^2
 \, \, \,\, \, = \,\, \,\,  \, 0, 
\end{eqnarray}
and the two $\, j$-invariants (${\cal H}_{\pm}= \, 1728/j_{\pm}$)
are solution of the quadratic relation:
\begin{eqnarray}
\label{AhAhHauptquadrajinv}
\hspace{-0.98in}&&  \quad  \quad  
 x^8 \cdot \, \, j_{\pm}^2  \, \,\,\,\,
+(1 \, -21\, x \, +8\, x^2)\, (1280\, x^4+1008\, x^3-454\, x^2+42\, x-1) \cdot \, x \cdot \, j_{\pm}
\nonumber \\
\hspace{-0.98in}&&  \quad  \quad  \quad  \quad  \quad \quad  \quad
  \,
 +(1\, + \, 224\, x \, + \, 448\, x^2)^3 \, \,\,  = \, \, \, \, 0. 
\end{eqnarray}

Denoting
$\, A \, = \,{\cal H}_{+}$ and $\, B\, = \, {\cal H}_{-}$ and  considering the two (identical)
quadratic relations (\ref{AhAhHauptquadra}) $\, Q(x, \, A) \, = \, \, 0$ and
$\, Q(x, \, B) \, = \, \, 0$, one easily gets by elimination of $\, x$
(performing the resultant between $\, Q(x,A) \, = \, \, 0$ and
$\, Q(x,B) \, = \, \, 0$ in $\, x$),
the {\em modular equation} $\, P(A, \, B) \, = \, \, 0$. One
gets a quite large {\em modular equation}
(corresponding to $\, q \leftrightarrow \,  q^7$ in the nome $\, q$, see (\ref{AhAhHauptser1})
and (\ref{AhAhHauptser2})):
\begin{eqnarray}
\label{7tau}
\hspace{-0.98in}&&  \quad   
81600^9 \cdot \, A^6 \, B^6 \cdot \, (343\, A^2 \, +286\, A\, B \, +343\, B^2)
\, \, + \,\, \cdots \, \, \, \,  -2^{36} \, 3^{18} \cdot \, A \, B \, \, \, = \, \,\,  \, 0.
\end{eqnarray}
Note that this (symmetric) algebraic curve is a {\em genus-zero} curve. 

Also note that the previous Heun function can be written alternatively with another algebraic
Hauptmodul $\, H$ (and another algebraic function  $\, {\cal A}$)
\begin{eqnarray}
\label{33}
\hspace{-0.98in}&& \quad  \quad  \quad   
Heun\Bigl(- \, {{1} \over {27}}, \, {{2} \over {27}}, \, {{1} \over {3}},
 \,  {{2} \over {3}}, \, 1, \,  {{1} \over {2}}, \, -\, x\Bigr)
 \, \, \, = \, \, \,\,
 {\cal A}  \cdot \,
_2F_1\Bigl([{{1} \over {12}}, \, {{5} \over {12}}], \, [1], \,  H\Bigr), 
\end{eqnarray}
where this alternative Hauptmodul is solution of a degree six equation
\begin{eqnarray}
\label{deg6}
\hspace{-0.98in}&& \quad   
 \, p_6(x)^3   \cdot \, (1\, -2\,x)^6 \cdot \, H^6
 \, \,\,     + \,  3 \cdot \, 1728 \cdot \, x^4  \cdot \, p_{20}(x)
 \cdot \, (1\, -2\,x)^3 \cdot \, H^5  \, \,
 \nonumber \\
  \hspace{-0.98in}&& \quad \, \,
 - \,  1728^2 \cdot \, x  \cdot \, p_{23}(x) \cdot \, H^4      \, \,
   \, + \,  1728^3 \cdot \, x^3  \cdot \, p_{21}(x) \cdot \, H^3      \, \,
  \,  + \, 1728^4  \cdot \, x^{8}  \cdot \, p_{16}(x) \cdot \, H^2
    \nonumber \\
  \hspace{-0.98in}&& \quad   \,  \, \,\, \quad   \quad \,
  - \, 1728^5 \cdot \, x^{10} \cdot \, p_{14}(x)  \cdot \, H
 \, \, \, + 1728^6 \cdot \, x^{24} \, \, \, = \, \,  \, \,\, 0,
\end{eqnarray}
where the polynomials $\, p_{6}(x)$,  $\, p_{14}(x)$, $\, p_{16}(x)$,
$\, p_{20}(x)$, $\, p_{21}(x)$,  $\, p_{23}(x)$
are given in \ref{Append1}.
Note that the curve (\ref{deg6}) is a {\em genus-one} curve.
This degree six polynomial equation (\ref{deg6}) in  $\, H$, gives
Hauptmoduls having the following series expansions:
\begin{eqnarray}
 \label{ABQ6}
  \hspace{-0.98in}&&  \quad \quad 
1728\, x^2 \,\,  +31104\, x^3 \, \, -689472\, x^4\, -34193664\, x^5\,  \,-431329536\, x^6
\nonumber \\
\hspace{-0.98in}&& \quad \quad \quad \quad 
\, +4925546496\, x^7 \,\,  +262313555328\, x^8 \, \, +3508587850752\, x^9 \, \, \, \, + \, \, \cdots                 
\end{eqnarray}
and
\begin{eqnarray}
 \label{ABQ6other}
  \hspace{-0.98in}&& \, \, \quad \quad 
  1728\,x^{14} \,\,  +217728\, x^{15} \,\,  +15930432\, x^{16} \,\,  +888039936\, x^{17} \,
 +41880888000\, x^{18} 
 \nonumber \\
 \hspace{-0.98in}&& \quad \quad \quad   \quad \quad 
 +1763242411392\, x^{19}  \,  +68405965290432\, x^{20} \, \,  \,  \, + \, \, \cdots 
\end{eqnarray}
corresponding to  $\,  q \, \, \leftrightarrow \, q^7 \,\, $ in the nome $\, q$.

Denoting $\, A $ and $\, B$ two Hauptmoduls solutions of the two identical degree
six relations (\ref{deg6}), $\, Q_6(x,A) \, = \, \, 0$ and $\, Q_6(x,B) \, = \, \, 0$,
one easily gets\footnote[1]{By elimination of $\, x$ performing a resultant of  $\, Q_6(x,A)$ and
 $\, Q_6(x,B)$ in $\, x$.} the modular equation $\, P(A, \, B) \, = \, \, 0$.
This {\em modular curve} is also a {\em genus-one} curve.

\vskip .3cm 

$\bullet$  Example 3. The rational function in four variables
\begin{eqnarray}
\label{Ratfoncfour9}
  \hspace{-0.98in}&& \quad  \,  \quad  \quad \quad 
\, \,  
R(x, \, y, \, z, \, w)  \, \, \,  = \, \,  \,  \,   \,  
 {{1} \over { 1 \,\, \,  \, - \, 
  (y \, +z \, \, +w\, z \, +x\, y \, +x\,z \, +\,  w\, x\, y) }},
\end{eqnarray}
 has a diagonal whose series expansion with integer ocoefficients reads:
\begin{eqnarray}
\label{SS9}
  \hspace{-0.98in}&& \quad  
\, \,  
Diag\Bigl(R(x, \, y, \, z, \, w)\Bigr)  \, \, \,  
\, \, = \, \, \,\, 1 \, \,\,  +4\, x \, \, +60\, x^2 \, \,
+1120\, x^3 \, \, +24220\, x^4 \, \, +567504\, x^5
\nonumber \\
\hspace{-0.98in}&& \quad  \quad    \quad  \quad  \quad  \quad  \quad  \quad 
\, +14030016\, x^6 \, \, +360222720\, x^7 \,\,  \, \, + \, \, \, \cdots 
\end{eqnarray}
The linear differential operator annihilating the diagonal of this
rational function (\ref{Ratfoncfour9}) has order three:
\begin{eqnarray}
  \label{S9}
  \hspace{-0.98in}&& \quad   \quad   \,  
  4 \, +96\cdot x \,\,\,\,\,  - (1 \, -92 \cdot x \, -864 \cdot x^2) \cdot \,  D_x
  \,\,\, \, - 3 \, x \cdot \, (1 \, - 42\cdot x \, - 256 \cdot\,  x^2)\cdot D_x^2
   \nonumber \\
  \hspace{-0.98in}&& \quad   \quad  \quad  \quad \quad  \quad    \quad   
  \, - x^2 \cdot \, (1 \, +4 \, x) \cdot \,  (1 \, -32 \, x) \cdot \, D_x^3.
\end{eqnarray}
This order-three  linear differential operator is the {\em symmetric square}
of an order-two linear differential operator:
\begin{eqnarray}
  \label{L2S9}
  \hspace{-0.98in}&& \quad   \quad    \quad    \quad   \,
 L_2 \, \, = \, \, \,\,
 8 \, x^2 \cdot \, (4\, \theta \, +3)  \cdot \,  (4\, \theta \, +1)
 \,  \, \, + 2 \cdot \, (14 \, \theta^2 \, +7 \, \theta \, +1) \, \, \, \, -\theta^2.
\end{eqnarray}
The solution of the linear differential operator (\ref{S9}), analytic at $\, x\, = \, 0$,
is thus given by the square of a Heun function which has
a series expansion with integer coefficients:
\begin{eqnarray}
  \label{HeunS9}
  \hspace{-0.98in}&& \quad    \quad   
 Heun\Bigl(-{{1} \over {8}}, \, {{1} \over {16}}, \, {{1} \over {4}},
 \, {{3} \over {4}}, \, 1,\, {{1} \over {2}},\,\,  -4  \, x\Bigr)^2
 \, \, = \, \, \,\,\,  \,  1 \,\,\,  +4\, x \,\,  +60\, x^2\,  \, +1120\, x^3
\nonumber \\
 \hspace{-0.98in}&& \quad  \quad \quad  \quad \quad \quad    \quad  \quad  \quad 
\, +24220\, \, x^4 \,  \,+567504\, \, x^5 \, \,  \, \, + \, \, \, \cdots
\end{eqnarray}
The linear differential operator operator (\ref{S9}) is the
{\em symmetric square} of a linear differential operator of order two,
such that one of its solutions can be written as a pullbacked $\, _2F_1$ hypergeometric function:
\begin{eqnarray}
\label{Heunxxx}
\hspace{-0.98in}&& \quad    
Heun\Bigl(-{{1} \over {8}}, \, {{1} \over {16}}, \, {{1} \over {4}},
 \, {{3} \over {4}}, \, 1,\, {{1} \over {2}},\, - \, 4 \, x\Bigr)
\, \, \, = \, \, \,\,  \, 1 \, \,  \, + \, 2\, x \,\, \,
 +28\, x^2 \,\,  +504\, x^3 \,\,  +10710\, x^4
\nonumber \\
 \hspace{-0.98in}&& \quad    \quad   \quad  
\,  \, +248220\, x^5 \,\,  +6091680\, x^6
\,\,  +155580000\, x^7 \,\,  +4092325500\, x^8 \,\,  \, + \, \, \, \cdots
\nonumber \\
\hspace{-0.98in}&& \quad  \quad   
\, \, = \, \, \, \,  {\cal A}^{(1)}_{\pm}  \cdot \,
_2F_1\Bigl([{{1} \over {6}}, \,  {{2} \over {3}}], \, [1], \, {\cal H}^{(1)}_{\pm}  \Bigr)
\, \, = \, \, \, \,  {\cal A}^{(2)}_{\pm}  \cdot \,
 _2F_1\Bigl([{{1} \over {8}}, \,  {{5} \over {8}}], \, [1], \, {\cal H}^{(2)}_{\pm}  \Bigr), 
\end{eqnarray}
where $\, {\cal A}^{(1)}_{\pm}$, $\,  {\cal A}^{(2)}_{\pm} \, $ and the two Hauptmoduls
$\,\, {\cal H}^{(1)}_{\pm}\, $ are square root algebraic functions:
\begin{eqnarray}
\label{Heunxxxyyy}
  \hspace{-0.98in}&& \quad \quad \quad \quad \quad 
{\cal H}^{(1)}_{\pm} \, \, = \, \, \,
 -54 \, x \cdot \, {{1 \, -19\, x \, -200\, x^2} \over {
   (1 \, +4\, x)\cdot \, (1 \, -50\, x)^2}}
\nonumber \\
  \hspace{-0.98in}&& \quad \quad \quad \quad \quad \quad \quad \quad \quad 
  \, \, \, \pm \, 54 \cdot \, x \cdot \,  (1\, -32\, x)^{1/2}
   \cdot \, {{ 1\, -5\, x} \over { (1 \, +4\, x)\cdot \, (1 \, -50\, x)^2}}. 
\end{eqnarray}
The two Hauptmoduls $\, {\cal H}^{(1)}_{\pm}$ are solutions of the quadratic relation:
\begin{eqnarray}
\label{HeunxxxyyyTT}
  \hspace{-0.98in}&& \quad  \quad \,
(1 \, +4\, x) \cdot \, (1\, -50\, x)^2 \cdot \, ({\cal H}^{(1)}_{\pm})^2
\, \,\,\,
 -108 \,x \cdot \, (200\, x^2 \, +19\, x \, -1) \cdot \,  {\cal H}^{(1)}_{\pm}
\nonumber   \\
\hspace{-0.98in}&& \quad \quad \quad \quad \quad \quad  \quad \quad  \quad 
\, \, +11664\,x^3   \,\, \, = \, \, \,\, 0.
\end{eqnarray}
The two Hauptmoduls $\, {\cal H}^{(2)}_{\pm}$ in (\ref{Heunxxx})
are also square root algebraic functions:
\begin{eqnarray}
\label{HeunxxxyyyTT}
  \hspace{-0.98in}&&   \quad \quad \quad 
{\cal H}^{(2)}_{\pm}  \, \, = \, \, \,
-28 \cdot \, x \cdot \, {{1\, -30 \, x \, +64 \, x^2 } \over { (1 \, -96 \, x)^2}}  \,
\nonumber \\
  \hspace{-0.98in}&&  \quad \quad \quad \quad \quad \quad \quad 
 \,  \pm \, 28 \cdot \, x \cdot \,  (1\, -16 \, x) \cdot \, {{ (1 \, +4\,x)^{1/2}
 \cdot \, (1 \, -32\, x)^{1/2}} \over { (1 \, -96 \, x)^2}}, 
\end{eqnarray}
solutions of the quadratic relation
\begin{eqnarray}
\label{HeunxxxyyyTT}
  \hspace{-0.98in}&& \quad \quad \quad \quad
(1 \, -96\, x)^2 \cdot \,  ({\cal H}^{(2)}_{\pm})^2
\, \, \,  +256 \,x \cdot \, (64\, x^2 \, -30\, x \, +1) \cdot \,  {\cal H}^{(2)}_{\pm}
 \nonumber \\
  \hspace{-0.98in}&& \quad \quad \quad \quad \quad \quad \quad \quad
\, \, +65536 \, x^4 \, \,\, = \, \,\, \, 0,
\end{eqnarray}
the algebraic function  $\, {\cal A}^{(1)}_{\pm}$ being solution of
\begin{eqnarray}
\label{Heunxxxyyzz}
  \hspace{-0.98in}&&\,  \,  \quad \quad \quad 
 512 \, \, \, \, -27 \cdot \, (1\, -20\, x) \cdot \,
 (19 \, -312\, x \, -6000\, x^2 \, -80000\, x^3)\cdot \, Y \, \,
 \nonumber \\
\hspace{-0.98in}&& \quad \quad \quad \quad  \quad  \quad \quad \quad \quad \quad 
+(1\, +4\, x)^3 \cdot \,(1\, -50\,x)^6 \cdot \,Y^2
\, \, \, \, = \, \, \, \, \, 0,         
\end{eqnarray}
where $\, \, \, Y \, = \, \, ({\cal A}^{(2)}_{\pm})^{18}$,
the algebraic function  $\, {\cal A}^{(2)}_{\pm}$ being solution of
\begin{eqnarray}
\label{Heunxxxyyzztt}
  \hspace{-0.98in}&&\quad  \quad\quad 
1 \, \,\,  + \, 2  \cdot \, q_8(x) \cdot Y \, \, \,
 + \, 3^{32} \cdot \, (1\, - 96 \, x)^{16} \cdot \, Y^2 \,\,  \, \, = \,\,  \, \, 0,
\quad   \quad \quad  \quad  \quad    \hbox{where:}
 \\
  \hspace{-0.98in}&&
q_8(x) \, \, = \, \, \,92393273930231100473344\, x^8 \,
-182396792383587915661312\, x^7
\nonumber \\
 \hspace{-0.98in}&&\quad \, \, 
\, +7442201965961886564352\, x^6 \, +10564527655702470066176\, x^5
\nonumber \\
\hspace{-0.98in}&&\quad \, \, 
\, -1994146206485388984320\, x^4 \, +154408466296830427136\, x^3
 \\
\hspace{-0.98in}&&\quad \, \, 
 \, -6048257896412868608\, x^2 \, +118593292086518528\, x \,  \, -926510094425921,
 \nonumber           
\end{eqnarray}
where $\, \, \, Y \, = \, \, ({\cal A}^{(2)}_{\pm})^{64}$.
The series expansions of the Hauptmoduls $\, {\cal H}^{(1)}_{\pm}$
read:
\begin{eqnarray}
\label{H11}
  \hspace{-0.98in}&&\quad
{\cal H}^{(1)}_{-} \, = \, \,\,
-108\, x \,\,  -8208\, x^2\, \, -547776\, x^3\, \, -34193664\, x^4
 \,\,  -2048523264\, x^5\,
 \nonumber \\
  \hspace{-0.98in}&& \quad \quad \quad \, \,  -119335292928\, x^6
  \, -6811411267584\, x^7\, -382782182326272\, x^8  \,  \, +  \,  \, \cdots               
\end{eqnarray}
and:
\begin{eqnarray}
\label{H12}
\hspace{-0.98in}&&\quad 
{\cal H}^{(1)}_{+} \, = \, \,
 -108\, x^2\,\,  -2160\, x^3\,\,  -56592\, x^4\, \, -1475712 \, x^5\,\,  -39711168\, x^6\,
\nonumber \\
  \hspace{-0.98in}&& \quad \quad \quad  \quad  \quad 
-1088716032\, x^7\, -30317739264\, x^8\, -854924599296\, x^9
 \,  \, \,  \, +  \,  \, \cdots              
\end{eqnarray}
The relation between these two Hauptmoduls corresponds to the genus-zero {\em modular equation}:
\begin{eqnarray}
\label{modeqH12}
  \hspace{-0.98in}&& \quad  \quad \quad  
 625\,\, A^3\, B^3\,\,\,
  -525\, A^2\, B^2 \cdot \, (A+B)\, \,\,  -96\, \, A\,B \, \cdot \,  (A^2+B^2) \,\,\,
 -3\, A^2\, B^2  
 \nonumber \\
   \hspace{-0.98in}&&   \quad   \quad  \quad   \quad  \quad  
\, \,\, \, -4 \cdot \, (A^3+B^3) \, \, \,
+ 528 \cdot \, \, A\,B \cdot \,  (A+B) \, \,\, \, -432 \cdot \, A\, B
 \, \,\, = \, \, \, \,  \, 0,              
\end{eqnarray}
which can (for instance)  be rationally parametrised as follows:
\begin{eqnarray}
\label{modeqH12param}
\hspace{-0.98in}&& \,  \quad   \,   
A(v) \, \, = \, \, \, \,
{\frac {108 \cdot \, v \cdot \, (1 \, +v)^{2}}{ (16 \, +15\,v)  \cdot \, (2 \, +3\,v)^{2}}},
 \quad \, \,
B(v)  \, \, = \, \, \, \,
- \,{\frac {108 \cdot \, (1 \, +v) \cdot \,  {v}^{2}}{(4 \, +3\,v)  \cdot \, (32 \, +33\,v)^{2}}}, 
\end{eqnarray}
where $\, A(v)$ and $\, B(v)$ are related by an involution: 
\begin{eqnarray}
\label{modeqH12param2}
 \hspace{-0.98in}&& \, 
 \quad  \quad  \quad    B(v)  \, \,
\, = \, \, \, \, A\Bigl( -\,{\frac {64 \cdot \, (1+v)}{63\,v+64}}\Bigr),
 \quad  \quad  \quad A(v)  \, \,
 \, = \, \, \, \, B\Bigl( -\,{\frac {64 \cdot \, (1+v)}{63\,v+64}}\Bigr).
\end{eqnarray}
The series expansions of the Hauptmoduls $\, {\cal H}^{(2)}_{\pm}$
read:
\begin{eqnarray}
\label{H21}
  \hspace{-0.98in}&&\quad \quad 
{\cal H}^{(2)}_{-} \, = \, \, \,\,
-56\,x \, \, \, \, -9072\,{x}^{2} \,\, \, -1229256\,{x}^{3}
\,\, \, -152418672\,{x}^{4} \,\, \, -17935321320\,{x}^{5}
\nonumber \\
  \hspace{-0.98in}&& \, \, \quad \quad \quad \quad \quad  \, \,
 -2038883437584\,{x}^{6} \,\,  -226173478925520\,{x}^{7} \,
 \,  \,\,\, + \, \, \, \cdots 
\end{eqnarray}
and 
\begin{eqnarray}
\label{H22}
  \hspace{-0.98in}&& \quad \quad 
{\cal H}^{(2)}_{+} \, = \, \,  \,\,
 -56\,{x}^{3} \,  \,\, -1680\,{x}^{4} \, \, -46872\,{x}^{5} \, \, \,-1291248\,{x}^{6} \,\,  \, -35752752\,{x}^{7}
\nonumber \\
  \hspace{-0.98in}&& \quad \quad  \quad    \quad  \quad  \,
 -998627616\,{x}^{8} \, \,  -28151491032\,{x}^{9} \, \,  -800518405680\,{x}^{10}
 \,\,\,\,   + \, \, \, \cdots 
\end{eqnarray}
The relation between these last two Hauptmoduls $\, {\cal H}^{(2)}_{\pm}$
corresponds to the (genus-zero) {\em modular equation}:
\begin{eqnarray}
\label{modeqH22}
\hspace{-0.99in}&& \quad \quad  \quad \quad 
640000 \cdot \, A^2\, B^2 \cdot \, (9\,A^2 \, +14\, A\, B \, +9\, B^2)
\nonumber \\ \quad 
\hspace{-0.98in}&&  \quad  \quad  \quad\quad \quad 
\, \,   +4800\, A \, B \cdot \, (A+B) \cdot \, (A^2 \, -1954\, A\, B \, +B^2)
\nonumber \\ \quad 
  \hspace{-0.98in}&&  \quad   \quad  \quad \quad  \quad\quad \quad 
\, +A^4+B^4 \,  \, -56196\, A\, B\cdot \, (A^2+B^2) \,  \,  +3512070\, A^2\, B^2
\nonumber \\
 \hspace{-0.98in}&& \quad  \quad  \quad  \quad   \quad   \quad \quad   \quad  \quad
+ 116736 \cdot \, A \, B \cdot \, (A+B) \,  \, \,   -65536 \cdot \, A \, B
\, \,\,\, = \, \,\,\, \, 0.
\end{eqnarray}

\vskip .2cm

$\bullet$  Example 4.  The rational function in four variables
\begin{eqnarray}
\label{Ratfoncfour12}
  \hspace{-0.98in}&& \quad  \quad  \,  \quad 
\, \,  
R(x, \, y, \, z, \, w)  \, \, \,  = \, \,  \,    \,  
 {{1} \over { 1 \,\,\, \,  - \,
  (w\, x\, z \, +w\, y \, +w\, z \, +x\, y \, +x\, z \, +y \, +z) }},
\end{eqnarray}
has a diagonal that reads:
\begin{eqnarray}
\label{SS12}
  \hspace{-0.98in}&&  
\, \,  
Diag\Bigl(R(x, \, y, \, z, \, w)\Bigr)  \, \, \,
\, \, = \, \, \,\,\,   1 \,\,\, +6\, x \,\,\, +114\, x^2
\,\, +2940\, x^3 \,\, +87570\, x^4 \, \, \, + \, \, \, \cdots 
\end{eqnarray}
The telescoper\footnote[1]{By abuse of terminology we will call, everywhere in this paper,
  ``telescoper'' of a rational function $\, R(x,y,z)$
  the output of the creative telescoping program~\cite{Koutschan}. For instance,
  for a rational function of three variables $\, R(x,y,z)$, we will call  ``telescoper''
  of a rational function $\, R(x,y,z)$, what is stricto sensu,
  the telescoper  of the rational function $\, R(x/y,y/z,z)/(yz)$.}
of the diagonal (\ref{SS12}) of this  rational function
of four variables (\ref{Ratfoncfour12}) reads:
\begin{eqnarray}
  \label{S12}
  \hspace{-0.98in}&& \quad  \quad  \, \,\, 
  6 \, +12\cdot x \, \,\,  \, - (1 \, -144\cdot x \, -108 \cdot x^2) \cdot\,  D_x \,\,\, 
  - x \cdot \, (3 \, -198\cdot x \, -96 \cdot x^2) \cdot \,  D_x^2
\nonumber \\
  \hspace{-0.98in}&& \quad  \quad \quad    \quad  \quad     \quad  \quad          \,
  - x^2 \cdot \, (1 \, -44 \cdot x \, -16 \cdot x^2) \cdot \,  D_x^3,
\end{eqnarray}
It is the {\em symmetric square} of an order-two linear differential operator
which has a Heun solution analytic at $\, x \, = \, 0$. Consequently
the order-three telescoper (\ref{S12}) 
has a square of a Heun solution. It has a series expansion
with {\em integer coefficients}:
\begin{eqnarray}
  \label{H12}
  \hspace{-0.98in}&& \quad  
  Heun\Bigl(-{{123} \over {2}} \, +{{55} \over {2}} \cdot \, 5^{1/2}, \,
  -{{33} \over {8}} \, +{{15} \over {8}} \cdot\, 5^{1/2},
  \, {{1} \over {4}}, \,  {{3} \over {4}},\, 1, \,   {{1} \over {2}}, \,  \,
  2 \, \cdot (11 \, -5 \cdot\,  5^{1/2}) \cdot \, x\Bigr)^2
  \nonumber \\
  \hspace{-0.98in}&& \quad   \quad   
  \, \, = \, \, \,\,\,
  1 \,\,\, +6\, x \,\,\, +114\, x^2 \,\, +2940\, x^3 \,\, +87570\, x^4
  \, \,+2835756\, x^5 \,\, +96982116\, x^6
  \nonumber \\
  \hspace{-0.98in}&& \quad \quad    \quad \quad  \quad  \quad   \quad       \quad    
  \, +3446781624\, x^7 \,
  \, +126047377170\, x^8 \,\, \, \, + \, \, \cdots
\end{eqnarray}
\vskip .3cm

The  square of the Heun funcion (\ref{H12}), solution of (\ref{S12}), can be rewritten as a pullbacked
$\, _2F_1$ hypergeometric function:
\begin{eqnarray}
\label{H122F1}
\hspace{-0.98in}&& \quad  \quad \quad   \quad   \quad   \quad \quad \quad \quad \quad \quad
{\cal A} \cdot \,
 _2F_1\Bigl([{{1} \over {12}}, \, {{7} \over {12}}], \, [1], \, {\cal H} \Bigr)^2
\end{eqnarray}
where $ \, {\cal A}$ is an algebraic function and where 
the Hauptmodul $\,  {\cal H}$
reads:
\begin{eqnarray}
\label{H122F1}
\hspace{-0.98in}&& \quad   \,   \,   \,   
{\cal H}  \, \, = \, \, \,
- \, 864 \cdot \,
{{7776 \, x^4 \, -12600\, x^3 \, +1890\, x^2 \, -80\, x \, +1} \over {
    (6480\,x^2 \, +540\,x\, -1)^2 }} \cdot \, x
  \\
  \hspace{-0.98in}&& \,   \quad   \quad  \quad  \quad   \,  \,   
 \, + \, 864 \cdot \, (1\, -4\, x) \cdot \,  (1\, -18\, x)
\cdot \,  (1\, -36\, x) \cdot \,
      {{(1\, -44\, x \, -16 \, x^2)^{1/2} } \over {
          (1 \, -540\, x \, -6480\, x^2)^2  }}  \cdot \, x
\nonumber \\
\hspace{-0.98in}&& \, \,  \, \,    
\, = \, \, \,
-1728 \, x^5 \, -138240\, x^6 \, -7793280\, x^7
\, -383961600\, x^8 \, -17716017600\, x^9
\,\, \,   + \, \, \cdots
\nonumber 
\end{eqnarray}
The pullback ${\cal H}$ is solution of the {\em genus-zero} quadratic relation:
\begin{eqnarray}
\label{H122F1altAA}
\hspace{-0.98in}&& \quad   \quad  \quad  \quad
(6480\, x^2+540\,x-1)^2\cdot \, {\cal H}^2
\nonumber \\
  \hspace{-0.98in}&& \quad   \quad  \quad  \quad \quad  \quad \, \,
+1728\cdot \, (7776\, x^4-12600\,x^3+1890\,x^2-80\,x+1) \cdot \, x \cdot \, {\cal H}
\nonumber \\
 \hspace{-0.98in}&& \quad   \quad  \quad \quad  \quad  \quad \quad  \quad \quad  \quad \, \,
    \, \, +2985984\, x^6 \, \,   \, = \, \, \, 0.
\end{eqnarray}
Note that changing the sign of the square root in (\ref{H122F1}) (Galois conjugate)
yields the alternative expansion:
\begin{eqnarray}
\label{H122F1alt}
\hspace{-0.98in}&& \quad   \quad  \quad
-1728\,\, x \,\,  \,-1728000\, x^2 \,\, \, -1388016000\, x^3 \,\, \, -1005452352000\, x^4
\nonumber \\
\hspace{-0.98in}&& \quad  \,  \, \quad \quad \quad  \quad  \quad
 \, -686965980744000\, x^5 \, \, \, -451977565258368000\, x^6
\, \, \, \, + \, \, \cdots 
\end{eqnarray}
These two Hauptmoduls series (\ref{H122F1})  and (\ref{H122F1alt}) are related
by the {\em genus-zero} modular equation:
\begin{eqnarray}
\label{H122F1altMod}
  \hspace{-0.98in}&& \quad  
 383093207587837762627239936 \cdot \, A^4 \, B^4 \cdot \, (25\, A^2+14\, A\, B+25\, B^2)
 \nonumber \\
  \hspace{-0.98in}&& 
\, \, -331453065290799513600 \cdot \,
A^3\, B^3 \cdot \, (A+B) \cdot \, (15047\, A^2+31514658\, A\, B+15047\, B^2)
\nonumber \\
  \hspace{-0.98in}&& 
\, \, 
+4480842240\, A^2\, B^2 \cdot \,
\Bigl(144903770079 \cdot \, (A^4+B^4)
\nonumber \\
  \hspace{-0.98in}&& \quad  \quad \quad  \quad
 \, -7730345599747300 \cdot \, A\, B \cdot \, (A^2\,  +\, B^2)
 +401951713284567050 \cdot \, A^2\, B^2\Bigr)
\nonumber
\end {eqnarray}
\begin{eqnarray}
\label{H122F1altMod2}
  \hspace{-0.98in}&& \quad  
\, \, +3386880 \cdot \, A\, B \cdot \, (A+B) \cdot\, \Bigl(15047  \cdot \, (A^4+B^4)
\nonumber \\
  \hspace{-0.98in}&& \quad  \quad \quad  \quad
 \, -419175723722072 \cdot \, A \, B \cdot \, (A^2 + \, B^2)
\, -4206296569303686878 \cdot  \, A^2\, B^2 \Bigr) \, \,
\nonumber \\
  \hspace{-0.98in}&& 
+A^6 +B^6 \, \,\, \,
 -25\, A\, B \cdot \,  \Bigl(72243325686 \cdot\, (A^4+\, B^4)
\nonumber \\
  \hspace{-0.98in}&& \quad  \, 
\, -38887039753371909735 \cdot\, A \, B \cdot \, (A^2 + \, B^2)
 +17585442099134941585204 \cdot \, A^2\, B^2  \Bigr)
\nonumber \\
\hspace{-0.98in}&& 
\, \, +1382400 \cdot \, A \, B \cdot \, (A+B) \cdot \,
\Bigl(8142703\, (A^2+B^2) \, -149242947792862 \cdot \, A\, B \Bigr)
\nonumber 
\end {eqnarray}
\begin{eqnarray}
\label{H122F1altMod3}
\hspace{-0.98in}&& 
\, \, 
-1373552640\, A\, B \cdot \, \Bigl(18909\, (A^2+B^2) \,  -3621715210 \cdot \, A\, B \Bigr)
 \nonumber \\
  \hspace{-0.98in}&&  \quad 
\, \, 
+25386119331840 \cdot \, A\, B \cdot \, (A+B)
\, \, \,  \, 
-8916100448256 \cdot \, A\, B \, \,\, \,   = \, \, \,  \, 0.
\end{eqnarray}

The  square of the Heun solution (\ref{H12}) can also be rewritten as a pullbacked
$\, _2F_1$ hypergeometric function:
\begin{eqnarray}
\label{H122F1bis}
\hspace{-0.98in}&& \quad  \quad \quad \quad \quad \quad \quad \quad \quad \quad
{\cal A}_{5/12}(x) \cdot \,
 _2F_1\Bigl([{{1} \over {12}}, \, {{5} \over {12}}], \, [1], \, \,{\cal H}\Bigr)^2, 
\end{eqnarray}
where $\, {\cal A}_{5/12}(x)$ is an algebraic function
and where the Hauptmodul  $\,{\cal H}$ in (\ref{H122F1bis}) 
is solution of the quadratic relation:
\begin{eqnarray}
\label{H122F1bisbis}
  \hspace{-0.98in}&& \quad  \quad  \quad  
 (144\, x^2 +216\, x+1)^3\cdot \, {\cal H}^2\, \,
 \nonumber \\
\hspace{-0.98in}&& \quad  \quad \quad  \quad \quad \quad   
  -1728\, x\cdot \, (3456\,x^5+7776\,x^4-12600\,x^3+1890\,x^2-80\,x \, +1)\cdot \, {\cal H}
\nonumber \\
\hspace{-0.98in}&& \quad \quad \quad   \quad \quad \quad \quad \quad   \quad \quad
\, \, +2985984\, x^6  \, \, = \, \, \, 0.               
\end{eqnarray}
The two  Hauptmoduls  read
\begin{eqnarray}
\label{H122F1bisread}
  \hspace{-0.98in}&&   \quad \quad  \quad  
{\cal H}_{\pm} \, \, = \, \, \,
\,{\frac {864\, x \cdot \,  (3456\,{x}^{5}+7776\,{x}^{4}-12600\,{x}^{3}+1890\,{x}^{2}-80\,x+1) }{ (144\,{x}^{2}+216\,x+1)^{3}}}
\nonumber \\
 \hspace{-0.98in}&&  \quad \quad \quad \quad
  \quad \pm \,    \, {\frac { 864\, (1 -36\,x)  \cdot \, (1 -18\,x)  \left( 1-4\,x \right) x}{ (144\,{x}^{2}+216\,x+1)^{3}}} \cdot \, (1 -44\,x-16\,{x}^{2})^{1/2}, 
\end{eqnarray}
which expands respectively as:
\begin{eqnarray}
\label{H122F1bisexp}
  \hspace{-0.98in}&&  \quad \quad
{\cal H}_{+} \, \, = \, \, \,
1728\,x \,  \,-1257984\,{x}^{2} \, +575828352\,{x}^{3} \, -214274336256\,{x}^{4}  \, \, \, \, + \, \, \cdots
     \nonumber \\
  \hspace{-0.98in}&&   \quad \quad
{\cal H}_{-} \, \, = \, \, \,
 1728\,{x}^{5} \, +138240\,{x}^{6} \, +7793280\,{x}^{7} \, +383961600\,{x}^{8} \,  \, \,\,  + \, \, \cdots                
\end{eqnarray}
These two Hauptmoduls series (\ref{H122F1bisread})  are related by a {\em genus-zero} modular equation
which can be parametrized rationally\footnote[2]{It corresponds to $\, N=5$ in Table 4 and Table 5 of~\cite{Maier1}.} as:  
\begin{eqnarray}
\label{H122F1bisexpq5}
  \hspace{-0.98in}&&  \quad \quad \quad \quad 
{\cal H}_{+} \, \, = \, \, \,\,{\frac { 1728\, z}{ ( {z}^{2} \, +10\,z \, +5)^{3}}}, \quad \quad \quad 
{\cal H}_{-} \, \, = \, \, \, \,{\frac {728 \, {z}^{5}}{ ({z}^{2}+250\,z+3125)^{3}}}. 
\end{eqnarray}

\vskip .3cm

{\bf Note:}  The  Heun function (\ref{H12}) resembles
the  Heun function associated with
{\em extreme rational surfaces}~\cite{Malmendier4,HigherMalmendier} 
($\phi$ denotes the golden number $\, (1\, +\sqrt{5})/2$):
\begin{eqnarray}
\label{extreme}
\hspace{-0.98in}&& \,  \quad
Heun\Bigl( {{8 \, - 5\, \phi} \over {3 \, + 5\, \phi}}, \,
    {{816 \, +165\, \phi} \over {(3 \, + 5\, \phi)^3}},
    \, 1, \,   1, \,  1, \,   1, \,  \,  t\Bigr)
 \\
\hspace{-0.98in}&& \quad \,  \quad \quad
      \, \, = \, \, \,
Heun\Bigl(-{{123} \over {2}} \, +{{55} \over {2}} \cdot \, 5^{1/2}, \,
{{3} \over {2}} \cdot \, (145185 \cdot \, 5^{1/2} \, -324643 ),
 \, 1, \,   1, \,  1, \,   1, \,  \,  t\Bigr).
\nonumber 
\end{eqnarray}
This Heun function (\ref{H12}) is also {\em reminiscent} of the Heun
function solution of the  Ap\'ery operator~\cite{Short,Big}:
\begin{eqnarray}
\label{Apery}
\hspace{-0.98in}&& \, 
 Heun\Bigl(-{{123} \over {2}} \, +{{55} \over {2}} \cdot \, 5^{1/2}, \,
  -{{33} \over {2}} \, +{{15} \over {2}} \cdot\, 5^{1/2},
  \, 1, \, 1,\, 1, \,  1, \,  \, \,
     {{1} \over {2}} \, \cdot (11 \, -5 \cdot\,  5^{1/2}) \cdot \, x\Bigr)
\nonumber \\
\hspace{-0.98in}&& \, \, \,
\, \, = \, \, \,\, \,1 \,\, \,+ 3 \, x \, \,+ \, 19 \, x^2 \,\,
+ \, 147 \, x^3 \,\, + \, 1251 \, x ^4 \,\, + \, 11253 \, x^5
\,\, + \, 104959 \, x^6 \, \,\, + \, \, \cdots
\end{eqnarray}
This Heun function (\ref{Apery}) can be rewritten
as a pullbacked hypergeometric function:
\begin{eqnarray}
\label{Apery2F1}
\hspace{-0.98in}&& \quad \quad  \quad \quad
{{1} \over {(1 \,-12\,x \, +14\,x^2\, +12 \,x^3 +x^4)^{1/4} }} 
\nonumber \\
\hspace{-0.98in}&& \quad \quad \quad \quad \quad \quad
\times \,
 _2F_1\Bigl([{{1} \over {12}}, \, {{5} \over {12}}], \, [1], \,
 \, {{ 1728 \cdot \, x^5 \cdot  \, (1 \, -11 \, x -x^2)} \over {
     (1 \,-12\,x \, +14\,x^2\, +12 \,x^3 +x^4)^3 }} \Bigr),
\end{eqnarray}
or
\begin{eqnarray}
\label{Apery2F1other}
\hspace{-0.98in}&& \quad \quad \quad
{{1} \over {(1 \, +228\,x \, +494\,x^2\, -228 \,x^3 +x^4)^{1/4} }} 
\nonumber \\
\hspace{-0.98in}&& \quad \quad \quad \quad \quad
\times \,
 _2F_1\Bigl([{{1} \over {12}}, \, {{5} \over {12}}], \, [1], \,
 \, {{ 1728 \cdot \, x \cdot  \, (1 \, -11 \, x -x^2)^5} \over {
     (1 \, +228\,x \, +494\,x^2\, -228 \,x^3 +x^4)^3 }} \Bigr),
\end{eqnarray}
where the Hauptmodul in (\ref{Apery2F1}) can be written
\begin{eqnarray}
\label{HauptApery2F1}
  \hspace{-0.98in}&& \quad \quad \quad \quad \quad 
{{1728 \, x} \over {(x^2 \, + \, 10\, x \, \, +5)^3 }}
 \, \, \circ \, \,   {{   1 \, -11 \, x -x^2} \over { x}}
\nonumber \\
 \hspace{-0.98in}&& \quad \quad \quad \quad \quad \quad \quad \quad \quad \quad
\, \, = \, \,   \, \,
{{1728 \, x^5} \over {(x^2 \, + \, 250\, x \, \, +3125)^3 }}
\, \, \circ \, \,   {{ 125 \, x } \over {  1 \, -11 \, x -x^2 }},    
\end{eqnarray}
when the  Hauptmodul in (\ref{Apery2F1other}) can be written:
\begin{eqnarray}
\label{HauptApery2F1}
  \hspace{-0.98in}&& \quad \quad \quad \quad \quad 
{{1728 \, x^5} \over {(x^2 \, + \, 250\, x \, \, +3125)^3 }}
\, \, \circ \, \,   {{   1 \, -11 \, x -x^2} \over { x}}
\nonumber \\
 \hspace{-0.98in}&& \quad \quad \quad \quad \quad \quad \quad \quad \quad  \quad \quad
\, \,  \, \,= \, \,  \,  \,    {{1728 \, x} \over {(x^2 \, + \, 10\, x \, \, +5)^3 }}
 \, \, \circ \, \, {{ 125 \, x } \over {  1 \, -11 \, x -x^2 }}.   
\end{eqnarray}

$\bullet$  Example 5.  The rational function in four variables
\begin{eqnarray}
\label{Ratfoncfour10}
  \hspace{-0.98in}&& \quad  \quad   
\, \,  
R(x, \, y, \, z, \, w)  \, \, \,  = \, \,  \,    \,  
 {{1} \over { 1 \,\, \, \, - \,\,
  (y \, +z \,  \, +w\, z \, +x\, y \, +x\, z \, \, +w\, x\, y   \, +  w\, x\, y\, z ) }},
\end{eqnarray}
has a diagonal that reads:
\begin{eqnarray}
\label{SS10}
  \hspace{-0.98in}&& \quad 
\, \,  
Diag\Bigl(R(x, \, y, \, z, \, w)\Bigr)  \, \, 
\, \, = \, \, \,\, 1 \, \, \, +5\,x \, \, \,+73\, x^2 \,\,
+1445\, x^3 \,\,  +33001\, x^4 \, \, + \, \, \, \cdots 
\end{eqnarray}
The telescoper of the diagonal of this  rational function (\ref{Ratfoncfour10}) of four variables
reads:
\begin{eqnarray}
\label{S10}
  \hspace{-0.98in}&& \quad  \quad  \quad  \quad 
 L_3 \, \, = \, \, \, x^2 \cdot \, (1\, -34 \, x\,  +x^2) \cdot \,  D_x^3
   \, \, \,\,\,   + 3 \, x \cdot \, (1 \, -51 \, x \, +2 \, x^2) \cdot \,  D_x^2
\nonumber \\
 \hspace{-0.98in}&& \quad   \quad \quad  \quad  \quad  \quad   \quad \quad  \quad  \quad   \quad   
  +(1 \, -112 \, x \, +7 \, x^2) \cdot \,  D_x \, \,\,\,  \,   + x \, -5. 
\end{eqnarray}
It is the {\em symmetric square} of an order-two linear differential operator
with a Heun solution, analytic at $\, x\, = \, 0$. Consequently the diagonal of
(\ref{Ratfoncfour10}), solution of (\ref{S10}), 
can be written in terms of the square of two (Galois conjugate) Heun functions
which have a series expansion with {\em integer coefficients}:
\begin{eqnarray}
\label{H10}
\hspace{-0.98in}&& \quad
(1 \, -34\,\, x \, +x^2)  \,\,  \times
  \nonumber \\
 \hspace{-0.98in}&& \quad \quad \,\,
 Heun\Bigl(577 \, +408 \cdot \, 2^{1/2}, \,  \,  {{663} \over {2}} \, +234 \cdot \, 2^{1/2}, \, \, 
 {{3} \over {2}}, \,  \, {{3} \over {2}},\, \,   1,\, {{3} \over {2}}, \, 
 \,\,  \, (17\,+12\cdot  \,2^{1/2}) \cdot \, x \Bigr)^2
  \nonumber \\
  \hspace{-0.98in}&&
\, = \, \, \,\, \, (1 \, -34\,\, x \, +x^2)  \,\,  \times
  \nonumber \\
 \hspace{-0.98in}&& \quad \quad \,\,
 Heun\Bigl(577 \, -408 \cdot \, 2^{1/2}, \,  \,  {{663} \over {2}} \, -234 \cdot \, 2^{1/2}, \, \, 
 {{3} \over {2}}, \,  \, {{3} \over {2}},\, \,   1,\, {{3} \over {2}}, \, 
 \,\,  \, (17\, -12\cdot  \,2^{1/2}) \cdot \, x \Bigr)^2
  \nonumber \\
  \hspace{-0.98in}&& \quad  \,\,    \quad
\, \, \, = \, \, \,\, \, 1 \, \, \, +5\,x \, \, \,+73\, x^2 \,\,  +1445\, x^3 \,\,  +33001\, x^4
 \,\,  +819005\, x^5 \, \, +21460825\, x^6
 \nonumber \\
 \hspace{-0.98in}&& \quad \quad \quad \quad \quad \quad \quad  \quad \quad \quad
 \, +584307365\, x^7
 \, \,  \, + \, \, \, \cdots 
\end{eqnarray}
It can also be written as a pullbacked  $\, _2F_1$ hypergeometric function
\begin{eqnarray}
\label{H102F1}
\hspace{-0.98in}&& \quad \quad \quad \quad \quad\quad \quad \quad \quad \quad
{\cal A}_{-} \cdot \,
 _2F_1\Bigl([{{1} \over {3}}, \, {{2} \over {3}}], \, [1], \,\, {\cal H}_{-} \Bigr)^2, 
\end{eqnarray}
where the Hauptmodul $\,  {\cal H}_{\pm}$ reads
\begin{eqnarray}
\label{H102F1bis}
\hspace{-0.98in}&& \quad  \quad
{\cal H}_{\pm}  \, \, = \, \, \,\,\,
{{ 1\, -24\, x \, +30\, x^2 \,  + x^3 } \over { 2 \cdot \, (1\, +x)^3 }} \, \,\, \,\, \,\,
 \pm \, \,  {{ 1\,-7\, x \,  +x^2} \over {
  2 \cdot \, (1\, +x)^3 }} \cdot \, (1\,-34\, x \,  +x^2)^{1/2},  
\end{eqnarray}
with the expansions:
\begin{eqnarray}
\label{pas1moinsH}
\hspace{-0.98in}&& \quad 
{\cal H}_{-}  \, \, = \, \, \,\,\,
 27\,{x}^{2}+648\,{x}^{3}+15471\,{x}^{4}+389016\,{x}^{5}+10234107\,{x}^{6}+278861616\,{x}^{7}
\nonumber \\
\hspace{-0.98in}&& \quad \quad \quad   \quad  \quad  \quad  \quad  \quad   \quad  
\,+7808397759\,{x}^{8} \, +223397228880\,{x}^{9}
\, \,\,\,  + \, \, \, \cdots 
\end{eqnarray}
\begin{eqnarray}
\label{1moinsH}
\hspace{-0.98in}&& \quad  
1\, -{\cal H}_{+} \, \, = \, \, \,\, \, 27\,\, x \, \,  \,\,-81\, x^2 \,\, +891\, x^3\,\,
+15039\, x^4 \,\, +389691\, x^5 
\, \,\, + \, \, \, \cdots 
\end{eqnarray}
and where the algebraic factor $\, {\cal A}_{-}$ reads:
\begin{eqnarray}
\label{A102F1bis}
\hspace{-0.98in}&& \quad  \quad \quad 
{\cal A}_{-} \, \, \, = \, \, \,\,\,\,\,
{{ 3} \over { 2}} \cdot \, {{ 1 \, -x  } \over {  (1 \, + \, x)^2}}
                   \, \,\,\,\,  - \,   {{ (1 \, -34 \, x \,+ \, x^2)^{1/2}} \over { 2 \cdot \,(1 \, + \, x)^2 }}
  \\
\hspace{-0.98in}&& \quad  \quad \quad \quad \quad 
\, \, = \, \, \,   1 \, +5\,x \, +61\,{x}^{2} \, +1097\,{x}^{3}
\, +23737\,{x}^{4} \, +569549\,{x}^{5} \, \, + \, \, \cdots     \nonumber           
\end{eqnarray}
The two series (\ref{pas1moinsH}) and  (\ref{1moinsH}) are related by the (symmetric) modular equation
\begin{eqnarray}
\label{Mod1moinsH}
  \hspace{-0.98in}&& \quad  \quad \quad \quad
 8\,{A}^{3}{B}^{3}\,  \,\,  \,\, -12\,{B}^{2}{A}^{2} \cdot \, (A +B) \, \, \,\, 
  +3\,AB \cdot \, (2\,{A}^{2}+13\,AB+2\,{B}^{2})
 \nonumber \\
 \hspace{-0.98in}&& \quad  \quad \quad  \quad \quad \quad
\, \,  - \, (A+B) \cdot \,  ({A}^{2}+29\,AB+{B}^{2}) \,\,  \,\,   +27\,AB
 \, \, = \, \, \, 0.       
\end{eqnarray}

\vskip .1cm

\vskip .2cm

$\, \bullet$  Example 6. Let us consider the following rational function
in four variables$\, x$, $\, y$, $\, z$ and $\, w$
\begin{eqnarray}
\label{Ratfoncfour11first}
  \hspace{-0.98in}&& \quad  \quad  \quad  \quad 
\, \,  
R(x, \, y, \, z, \, w)  \, \, \,  = \, \,  \quad 
{{1} \over { 1 \, \,\,  -(y +z\, \, +w\,y \, +x\,z \,\, +w\,x\,y \, +w\,x\, z)}}, 
\end{eqnarray}
or the rational function:
\begin{eqnarray}
\label{Ratfoncfour20}
  \hspace{-0.98in}&& \quad  \quad  \quad  \quad 
\, \,  
R(x, \, y, \, z, \, w)  \, \, \,  = \, \,  \quad 
{{1} \over { 1 \, \, \, +x\,y \, +y\,z \, +z\,w \, +w\,x \, +y\,w \, +x\,z}}.
\end{eqnarray}
The diagonals of these two rational functions (\ref{Ratfoncfour11first}), (\ref{Ratfoncfour20})
give the {\em same} series
expansion with {\em integer coefficients}:
\begin{eqnarray}
\label{Ratfoncfour11ser}
  \hspace{-0.98in}&& \quad  
 Diag\Bigl(R(x, \, y, \, z, \, w)\Bigr)  \, \, \,  = \, \,  \,  \,    \,
 1 \, \,\,  +6\, x \, \,\,  +90\, x^2 \,\, \,  +1860\, x^3
\, \, \, +44730\, x^4 \,\, \,  +1172556\, x^5
\nonumber \\
\hspace{-0.98in}&& \quad  \quad  \quad  \quad   \quad    \quad    \quad  
\, +32496156\, x^6 \, \,\, +936369720\, x^7   \, \,  \,  + \, \, \, \cdots              
\end{eqnarray} 
The order-three linear differential operator annihilating this series (\ref{Ratfoncfour11ser})
 is the {\em symmetric square} of the linear differential operator of order two, and is given by:
\begin{eqnarray}
\label{Ratfoncfour11L3}
\hspace{-0.98in}&&  \quad \quad
L_3 \,  \, = \, \, \, \,  \, x^2 \cdot \, (1\, -36\,x) \cdot \, (1\, -4\, x) \cdot \, D_x^3
 \,  \, \, + 3\, x\cdot \, (1\, -60\, x \, +\, 288\, x^2) \cdot \, D_x^2
\nonumber \\
\hspace{-0.98in}&& \quad \quad \quad \quad \quad \quad  \quad 
\, \,\, +\, (1\, -132\, x \, + 972\, x^2)  \cdot \, D_x
\,  \,  \, \, \,  - \,  6 \cdot \, (1 -\, 18\, x) . 
\end{eqnarray}
The solution of this order-three telescoper  $\, L_3$ reads:
\begin{eqnarray}
\label{Ratfoncfour11}
\hspace{-0.98in}&& \quad  \quad  
Heun\Bigl({{1} \over {9}},\,  {{1} \over {12}}, \,   {{1} \over {4}},
  \,  {{3} \over {4}},\,  1,\,  {{1} \over {2}},  \, \, 4\, x\Bigr)^2
\, \, = \, \, \,   (1 \, -4\,x) \cdot \, Heun\Bigl({{1} \over {9}}, \, {{5} \over {36}},\,
 {{3} \over {4}},\, {{5} \over {4}},\, 1,\, {{3} \over {2}}, \,\, 4\,x\Bigr)^2
 \nonumber \\            
 \hspace{-0.98in}&& \quad  \quad \, \,  \quad 
\, \,  = \, \, \,\,\, 1 \,\, \, +6\, x \,\, \, +90\, x^2 \,\, \, +1860\, x^3
\, +44730\, x^4 \,\,  +1172556\, x^5
\, \,\, \,  + \, \, \, \cdots 
\end{eqnarray}
This series expansion (\ref{Ratfoncfour11}) already occurred in (\ref{SolSCser})
for the simple cubic lattice Green function. 
This Heun function (\ref{Ratfoncfour11}) is quite simply related\footnote[2]{This corresponds
  to the fact that the order-three linear differential operator of example 3 by (\ref{S9})  is equal to
 $\, (1\, +4\,x)^{-1/2} \cdot  \, pullback(L_3, x/(1\, +4\,x)) \cdot \, (1\, +4\,x)^{1/2}$
  where  $\, L_3$ is the order-three linear differential operator given by (\ref{Ratfoncfour11L3}).}
to the  Heun function of example 3:
\begin{eqnarray}
\label{Ratfoncfour11simply}
\hspace{-0.98in}&& \quad  \quad \quad
Heun\Bigl(-{{1} \over {8}},\,  {{1} \over {16}}, \,   {{1} \over {4}},
  \,  {{3} \over {4}},\,  1,\,  {{1} \over {2}},  \, \, -4\, x\Bigr)^2
\nonumber \\
\hspace{-0.98in}&& \quad \quad \quad \quad  \quad \quad \quad
 \, \, = \, \, \,       (1\, + 4 \, x)^{-1/2} \cdot \,
Heun\Bigl({{1} \over {9}},\,  {{1} \over {12}}, \,   {{1} \over {4}},
 \,  {{3} \over {4}},\,  1,\,  {{1} \over {2}},  \, \, {{4\, x} \over { 1\, + \, 4 \, x}}\Bigr)^2.
 \end{eqnarray}
 This Heun function  in (\ref{Ratfoncfour11}) can also be written 
 as a pullbacked  $\, _2F_1$ hypergeometric function
\begin{eqnarray}
\label{Ratfoncfour11hyp}
\hspace{-0.98in}&& \quad   \quad 
 Heun\Bigl({{1} \over {9}},\,  {{1} \over {12}},\,   {{1} \over {4}},
  \,  {{3} \over {4}},\,  1,\,  {{1} \over {2}},  \, \, 4\, x\Bigr)
  \\
  \hspace{-0.98in}&& \quad     \quad  \quad 
\, \, = \, \,  1 \, +3\,x \, +{\frac{81}{2}}{x}^{2} \, +{\frac{1617}{2}}{x}^{3} \, +{\frac{152955}{8}}{x}^{4}
\, +{\frac{3969405}{8}}{x}^{5} \, \,  \, + \, \, \cdots 
 \nonumber \\
\hspace{-0.98in}&& \quad    \quad  \quad  
\, \, = \, \, \,\,  {\cal A}^{(1)}_{\pm} \cdot \,
_2F_1\Bigl([{{1} \over {6}}, \, {{2} \over {3}}],\, [1], \, {\cal H}^{(1)}_{\pm} \Bigr)
\, \,\, = \, \, \, \, {\cal A}^{(2)}_{\pm} \cdot \,
_2F_1\Bigl([{{1} \over {8}}, \, {{5} \over {8}}],\, [1], \,   {\cal H}^{(2)}_{\pm}\Bigr),
\end{eqnarray}
where $\,\, {\cal H}^{(1)}_{\pm}$, $\,\, {\cal H}^{(2)}_{\pm}\,$ are
algebraic functions expressed in terms of square roots.
The detailed calculations which are similar to the ones of example 3 are given in
\ref{pulback2F1represe}. As a consequence of identity (\ref{Ratfoncfour11simply}),
there is a close relation between
example 3 and example 6 and of course the simple cubic lattice Green functions (see (\ref{SolSCser})).

\subsection{A few comments on example 6}
\label{subthree2}

Considering the second form for the rational function of example 6, namely (\ref{Ratfoncfour20}),
it is straightforward to see that the {\em one-parameter family} of rational functions
\begin{eqnarray}
  \label{Ratfoncfour20lambda}
  \hspace{-0.98in}&& \quad  \quad  \quad  \quad 
\, \,  
R_{\lambda}(x, \, y, \, z, \, w)  \, \, \,  = \, \,  \quad 
{{ 1} \over { 1 \, \, \, +\lambda \cdot \, (x\,y \, +y\,z \, +z\,w \, +w\,x \, +y\,w \, +x\,z)}},
\end{eqnarray}
has a diagonal\footnote[1]{Deduced, without any calculation, for the scaling
  transformation $ \, (x, \, y, \, z, \, u) \, \rightarrow \, \,
  (\lambda^{1/2} \cdot \, x, \, \lambda^{1/2} \cdot \, y, \, \lambda^{1/2} \cdot \, z, \, \lambda^{1/2} \cdot \, u)$.}
deduced from (\ref{Ratfoncfour11})
\begin{eqnarray}
\label{Ratfoncfour11lambdaHeun}
\hspace{-0.98in}&& \quad 
Heun\Bigl({{1} \over {9}},\,  {{1} \over {12}}, \,   {{1} \over {4}},
  \,  {{3} \over {4}},\,  1,\,  {{1} \over {2}},  \, \, 4\,\,\, \lambda^2  \,\,  x\Bigr)^2
\, \,  = \, \, \,\,\, 1 \,\, \, +6\,\lambda^2 \, x \,\, \, +90\, \lambda^4 \, x^2 
\, \,\, \,  + \, \, \, \cdots, 
\end{eqnarray}
the telescoper of (\ref{Ratfoncfour20lambda}) being the pullback of the order-three
linear differential operator (\ref{Ratfoncfour11L3})
by $\, \, \, x \, \rightarrow \, \lambda^2 \cdot \,  x$.

Let us now consider the rational function
\begin{eqnarray}
  \label{Ratfoncfour20xyzu}
  \hspace{-0.98in}&& \quad  \quad 
\, \,  
R(x, \, y, \, z, \, w)  \, \, \,  = \, \,  \quad 
{{ 1} \over { 1 \, \, \, + x\, y\, z\, w \cdot \, (x\,y \, +y\,z \, +z\,w \, +w\,x \, +y\,w \, +x\,z)}},
\end{eqnarray}
deduced from (\ref{Ratfoncfour20}) by the following monomial\footnote[2]{Note a typo in
  the footnote of section 4.2 in~\cite{unabri,notunabri}. A determinant equal $\, \pm \, 1$ condition
  is missing to have a birational transformation. A monomial transformation
like (\ref{Monomial}) is not birational (the determinant is $\, -3$ here).} transformation:  
\begin{eqnarray}
\label{Monomial}
  \hspace{-0.98in}&& \quad  \quad \quad \quad  \quad       (x, \, y, \, z, \, w)
 \, \, \,  \,  \, \quad \longrightarrow \, \quad  \quad
 \, (y z w, \,  \, \,\, x z w, \,\, \,  x y w, \,\,  \,  x y z).
\end{eqnarray}
It is straightforward to see, from its definition,
that the diagonal of the rational function (\ref{Ratfoncfour20xyzu}), actually corresponds to 
the diagonal of the rational function (\ref{Ratfoncfour20lambda}) where the parameter $\, \lambda$ is
taken to be equal to the product $\, \lambda\,=\, x y z u$,  
thus reading the formula (\ref{Ratfoncfour11lambdaHeun}) with $\, \lambda \, = \, x$:
\begin{eqnarray}
\label{Ratfoncfour11x3}
\hspace{-0.98in}&& 
Heun\Bigl({{1} \over {9}},\,  {{1} \over {12}}, \,   {{1} \over {4}},
  \,  {{3} \over {4}},\,  1,\,  {{1} \over {2}},  \, \, 4 \,\,  x^3\Bigr)^2
\, \,  = \, \, \,\,\, 1 \,\, \, +6\, \, x^3 \,\, \, +90\,  \, x^6 \, +1860 \, x^9
\, \,\, \,  + \, \, \, \cdots 
\end{eqnarray}
One verifies that the  telescoper of the rational function (\ref{Ratfoncfour20xyzu})
is actually the pullback\footnote[5]{Performing the same monomial change of variable (\ref{Monomial}) on 
 the rational function (\ref{Ratfoncfour11first}) instead of (\ref{Ratfoncfour20}), one gets, as it should,
the same telescoper pullback of the order-three linear differential operator (\ref{Ratfoncfour11L3}) 
by $\, \, \, x \, \rightarrow \,  \,  x^3$, with the same diagonal series (\ref{Ratfoncfour11x3}).} of the order-three
linear differential operator (\ref{Ratfoncfour11L3})
by $\, \, \, x \, \rightarrow \,  \,  x^3$. 
In contrast let us now consider the rational function
\begin{eqnarray}
\label{Ratfoncfour20xyzubirat}
\hspace{-0.98in}&& \quad \quad \quad \, \, 
R(x, \, y, \, z, \, w)  \, \, \,  = \, \, \, \, {\frac {x}{w{x}^{3}y \,\, +{x}^{3}yz \, \,+wxz \,\, +{x}^{2}y \,\, +x \,\,+z \,\, +w}}, 
\end{eqnarray}
deduced from (\ref{Ratfoncfour20}) by the following (involutive) {\em birational} monomial transformation:  
\begin{eqnarray}
\label{biratMonomial}
  \hspace{-0.98in}&& \quad \, \, \,  \quad \quad \quad    \quad       (x, \, y, \, z, \, w)
 \,  \quad \longrightarrow \, \quad  \quad
 \, \Bigl({{1} \over {x}}, \,\,  \, x^2 \, y, \, \, z, \,\,  w\Bigr).
\end{eqnarray}
The telescoper of (\ref{Ratfoncfour20xyzubirat}) is actually {\em the same telescoper
as the one for} (\ref{Ratfoncfour20}), namely the order-three linear differential operator
$\, L_3$ given by  (\ref{Ratfoncfour11L3}). This telescoper has, thus, the Heun function (\ref{Ratfoncfour11})
as a solution. However (\ref{Ratfoncfour11}) does not coincide with the
diagonal\footnote[9]{Which is not well-defined (depending on the ordering of the four variables).}
of the rational function (\ref{Ratfoncfour20xyzubirat}). The solutions of the telescoper
of a rational function and the diagonal of a rational function
{\em are two concepts which do not necessarily coincide}. We will underline
this point several times in this paper (see section \ref{vanHoeijVidunas},
\ref{Duality1} and \ref{Duality2}). 

\vskip .1cm

\subsection{Comment on the simplicity of the rational functions yielding Heun functions}
\label{comment}

Let us consider the rational function (\ref{Ratfoncfour9}) of example 3, 
and perform the simple transformation
$\, (x, \, y, \, z, \, w) \, \rightarrow \, \, (1\, +x, \, 1\, +y, \, 1\, +z, \, 1\, +w)$
on (\ref{Ratfoncfour9}). One obtains that way the following new rational function:
\begin{eqnarray}
\label{Ratfoncfour9shift}
  \hspace{-0.98in}&& \, \,  \quad  \quad 
-\, {{1 } \over { wxy \, +wx \, +wy \, +wz \, +2\,xy \, +xz \, +2\,w \, +3\,x+3\,y+3\,z \, +5}}.
\end{eqnarray}
The telescoper of this  quite simple rational function (\ref{Ratfoncfour9shift}) is a {\em very large}
order-nine linear differential operator of degree $\, 48$ in $\, x$.
Along this line let us consider the rational function
\begin{eqnarray}
\label{Ratfoncfour9screw}
  \hspace{-0.98in}&& \quad  \,  \, \,   \quad \quad 
\, \,  
R(x, \, y, \, z, \, w)  \, \, \,  = \, \,  \,  \,   \,  
 {{1} \over { 1 \,\, \,  \, - \, 
  (y \, +2\, z \, \, +w\, z \, +x\, y \, +x\,z \, +\,  w\, x\, y) }},
\end{eqnarray}
which corresponds to a very simple modification\footnote[1]{We have just changed one coefficient
  in the denominator of (\ref{Ratfoncfour9}): the $\, z$ term becomes a  $\,2\, z$ term.} of the
rational function (\ref{Ratfoncfour9}) of example 3. The telescoper of this new simple rational
function (\ref{Ratfoncfour9screw}) is an irreducible order-five linear differential
operator\footnote[5]{This  order-five linear differential
operator is homomorphic to its adjoint, its differential
Galois group being $\, SO(5, \,\mathbb{C})$.}. Again we are far from having diagonals
of rational functions and solutions of telescopers that can be simply expressed
as Heun functions (or even solutions of order-two linear differential operators).
Considering simple rational functions of the form $\, 1/P(x,y,z,w)$ with a polynomial
$\, P(x,y,z,w)$ of degree at most one in $ \, x, \, y, \, z, \, w$ {\em is far from being sufficient
to get Heun functions}.
Conversely let us consider the slightly more involved rational function
\begin{eqnarray}
\label{Ratfoncfour9ppreserving}
  \hspace{-0.98in}&& \quad  \,  \, \,   \quad 
\, \,  
R(x, \, y, \, z, \, w)  \, \, \,  = \, \,  \,  \,   \,  \\
\hspace{-0.98in}&& \quad  \,  \, \,   \quad \, \,   \quad  \quad 
{\frac {1\,+x}{1 \,\,+x-2\,y-2\,z\,  -3\,xy-3\,xz  \,  \, -uxy \, -uxz \, -{x}^{2}y \, -{x}^{2}z  \,   -u{x}^{2}y }}.
\nonumber 
\end{eqnarray}                  
The diagonal of the rational function (\ref{Ratfoncfour9ppreserving}) reads:
\begin{eqnarray}
\label{Ratfoncfour9ppreservingDiag}
  \hspace{-0.98in}&& \,  \, \quad  \,
_2F_1\Bigl([{{1} \over {2}}, \,  {{1} \over {2}}], \, [1], \, 32 \, x  \Bigr) \, \, \, =  \, \, \,  1 \, +8\,x \, +144\,{x}^{2}
                     \, +3200\,{x}^{3} \, +78400\,{x}^{4}
                     \, \, \, + \, \, \, \cdots 
\end{eqnarray}        
The telescoper of  the rational function (\ref{Ratfoncfour9ppreserving}) is an order-six
linear differential operator which is the {\em direct sum} of two order-three linear differential operators
$\, L_6 \, = \, \, L_3 \oplus \, M_3$,
where $\, L_3$ factorises into the product of an order-one and an order-two  linear differential operator
$\, L_3 \, = \, \, L_1 \cdot \, L_2$,  and 
where $\, M_3$ is {\em exactly}\footnote[2]{It is thus is  
the symmetric square of the  order-two  linear differential operator (\ref{L2S9}) which 
has the  Heun solution (\ref{Heunxxx}).} the  order-three  linear differential operator (\ref{S9}) which 
has the  Heun solution (\ref{HeunS9}), namely:
\begin{eqnarray}
\label{Ratfoncfour9ppreservingHeun}
 \hspace{-0.98in}&& \, \, \,  \,  
 Heun\Bigl(-{{1} \over {8}}, \, {{1} \over {16}}, \, {{1} \over {4}},
                     \, {{3} \over {4}}, \, 1,\, {{1} \over {2}},\, - \, 4 \, x\Bigr)^2 \,  \,  = \, \, \, \, 
Heun\Bigl(-8, \, {{1} \over {2}}, \, {{1} \over {4}},
                     \, {{3} \over {4}}, \, 1,\, {{1} \over {2}},\, - \, 32 \, x\Bigr)^2.
\end{eqnarray}        
The diagonal of the rational function (\ref{Ratfoncfour9ppreserving}) is, in fact, solution of the
order-two  linear differential operator $\, L_2$, which is thus, the miminal order
linear differential operator operator annihilating the diagonal (\ref{Ratfoncfour9ppreservingDiag}). The
creative telescoping method yields
a higher order telescoper which provides a ``companion'' order-three linear differential operator  $\, M_3$
with the same square of Heun solution\footnote[9]{Which corresponds to a ``Period'' over
another cycle than the evanescent cycle of the diagonal.} (\ref{Ratfoncfour9ppreservingHeun}) as example 3. 

\vskip .1cm

\subsection{Periods of extremal rational surfaces}
\label{subthree2}
Let us now introduce the rational function in just {\em three} variables:
\begin{eqnarray}
\label{Ratfonc4}
  \hspace{-0.98in}&& \quad  \quad \quad \quad \quad 
\, \, 
R(x, \, y, \, z)  \, \, \,  = \, \,  \quad 
 {{1} \over { 1 \, \,  \,\,  + x \, + y \, + z  \, \, \,+ x \,y \, + y \, z \,  \,  \, - x^3 \,y \,z }}.
\end{eqnarray}
The diagonal of this rational function (\ref{Ratfonc4}) has the following series expansion:
\begin{eqnarray}
\label{Ratfonc4diag}
  \hspace{-0.98in}&& \quad \quad
\, \, 
 Diag\Bigl( R(x, \, y, \, z)\Bigr)   \, \, \,  = \, \, \, \, \, 
 1 \,  \,  \, -2\,x \,  \, +6\,{x}^{2} \, \,  -11\,{x}^{3} \, \, \,  -10\,{x}^{4} \, \,  +273\,{x}^{5}
\, -1875\,{x}^{6}
\nonumber \\
 \hspace{-0.98in}&& \quad  \, \quad  \quad   \quad     \quad      \quad    \quad  
\, +9210\,{x}^{7} \,  \, -34218\,{x}^{8} \, \,  \,  +78721\,{x}^{9} \, \,  +108581\,{x}^{10}
 \, \, \, \, \,  + \,  \, \, \cdots 
\end{eqnarray}

\vskip .2cm

In order to find the diagonal of this rational function of three variables,
one gets the telescoper annihilating this  diagonal using creative telescoping~\cite{Koutschan}.
This telescoper is actually an order-four linear differential operator $\, L_4$
which, not only factorizes into two order-two  linear differential operators,
but is actually the {\em direct sum} (LCLM) of two\footnote[1]{
  These two order-two  linear differential operators $\, L_2$ and $\, M_2$ are not homomorphic.}
order-two  linear differential operators
$\, L_4 \, = \, \, L_2 \oplus \, M_2$. These two order-two
linear differential operators read respectively
\begin{eqnarray}
\label{L2andM2}
\hspace{-0.98in}&&   \quad  \quad
L_2 \, \, = \,  \, \,\,
27 \, x^2 \cdot \, (\theta \, +1)^2 \,\, + 3 \, x \cdot \, (3\, \theta^2 \, + 3 \, \theta \, +1)
\, \, + \, \theta^2
 \\
\hspace{-0.98in}&& \quad  \quad  \quad
                   \, \,\,\,  = \,  \, \, \,
(1 \,+9\,x \,  +27\,x^2) \cdot \, x^2 \cdot \,D_x^2
 \, \,\,  +(1 \, +9\, x)^2 \cdot \, x \cdot \, D_x \, \, \, +3 \, x \cdot  \, (1 \, +9\, x),
\nonumber  
\end{eqnarray}
and:
\begin{eqnarray}
\label{M2andL2}
\hspace{-0.98in}&& \quad  \quad  \quad \quad 
 M_2  \, \, = \,  \, \, \,
(1 \,+9\,x \,  +27\,x^2) \cdot \, (5\, +18\, x) \cdot \, (1\, -2\, x) \cdot \, x^2 \cdot \,D_x^2
\nonumber \\
\hspace{-0.98in}&& \quad  \quad \quad  \quad\quad \quad 
 \, \,  +(5 \, +70\, x \, +261\, x^2 \, -756\, x^3 \, -2916\, x^4) \cdot \, x \cdot \, D_x
 \nonumber \\
\hspace{-0.98in}&& \quad  \quad \quad  \quad \quad  \quad \quad \quad 
\, \, + \, x \cdot  \, (1 \, -9\, x) \cdot  \, (5\, + \, 60 \, x \, + 108 \, x^2).        
\end{eqnarray}
Note that 
$\, L_2$ and $\, M_2$ {\em share exactly the same singularities} $\, x= \, 0, \, \infty$ and
$\, 1 \,+9\,x \,  +27\,x^2 \, = \, \, 0$.
In contrast, the factor $ \, (5\, +18\, x)$  in (\ref{M2andL2}) corresponds to an {\em apparent} singularity,
when the factor $\, (1\, -2\, x)$ corresponds to a true singularity.
One can get rid
of the  $ \, 5\, +18\, x \, = \, 0$ apparent singularity
performing the following desingularization $\, L_2 \rightarrow \, L_3 \, = \, \, L_1 \cdot \, L_2$,
changing the order-two operator $\, L_2$ into an order-three
linear differential operator $\, L_3$, the order-one operator $\, L_1$ reading:
\begin{eqnarray}
\label{desingextrem}
\hspace{-0.98in}&& \quad \quad \quad
L_1 \, \,  = \, \, \, D_x \,\,  \,  + \, {{ A'(x)} \over {A(x)}}
 \quad \quad \quad  \quad \quad  \quad  \quad  \hbox{where:} 
  \\
  \hspace{-0.98in}&& \quad \quad \quad
 A(x) \, \,   = \, \,  \,  (5\, +18\, x) \cdot \, (1\, -2 \, x) \cdot \, (1\, +9\, x \, +27 \, x^2) \cdot \, x^{2/7}.
                     \nonumber 
\end{eqnarray}
The solution of the order-two linear differential operator $\, L_2 \,$ has the
following Heun function\footnote[2]{This Heun function $\, Heun(a, q, \alpha, \beta, \, \gamma, \, \delta, \, \rho \, x) \, $ is such that
$\, q \, = \, a/(1+a)$, $\, q/\rho= \, -1/9$, $\, a/\rho^2 = \, 1/27$, $\, \, 1/\rho \, $ and $\, a/\rho \, $ being complex conjugate.

} solution, analytic at $\, x \, = \, 0$:
\begin{eqnarray}
\label{Ratfonc4HeunSol}
\hspace{-0.98in}&& \quad  \quad
{\cal S}_1 \, \, = \, \, \, Heun \Bigl({{1} \over {2}}\, -{{i \sqrt{3}} \over {2}}, \,
{{1} \over {2}}\, -{{i \sqrt{3}} \over {6}}
, \, 1, \, 1, \, 1, \,1, \,\,\,
{{3} \over {2}} \cdot \, \Bigl(-3 \, + \, i \sqrt{3}  \Bigr) \cdot \,x \Bigr)
\\
\hspace{-0.98in}&& \quad   \quad \quad   \quad 
\, = \, \, \, \, \,  1 \,\,\,  \,  -3\,x \,\,\,\,   +9\, x^2 \,\, \,
-21\, x^3 \, \, \, +9\, x^4\,\,\,   +297\, x^5
\,\, \,  -2421\, x^6 \,\, \,  +12933\, x^7
\nonumber \\
\hspace{-0.98in}&& \quad \quad    \quad    \quad  \quad \quad  \quad \quad  \,
-52407\, x^8 \,  \, +145293\, x^9 \, -35091\, x^{10} \,
- 2954097 \, x^{11} \, \,\, \,  + \, \, \cdots 
\end{eqnarray}
This Heun function (\ref{Ratfonc4HeunSol}) can also be written alternatively
in terms of other $\, _2F_1$ hypergeometric functions:
\begin{eqnarray}
\label{Ratfonc4HeunSol2}
\hspace{-0.98in}&& \quad    \ 
Heun \Bigl({{1} \over {2}}\, -{{i \sqrt{3}} \over {2}}, \,
{{1} \over {2}}\, -{{i \sqrt{3}} \over {6}}
, \, 1, \, 1, \, 1, \,1, \,\,\,
{{3} \over {2}} \cdot \, \Bigl(-3 \, + \, i \sqrt{3}  \Bigr) \cdot \,x \Bigr)
\nonumber \\
\hspace{-0.98in}&& \quad  \quad \quad \quad \quad \quad
\, = \, \,\,  {{1} \over {1 \, + \, 3 \, x}}  \cdot \,
_2F_1\Bigl([{{1} \over {3}}, {{2} \over {3}}], \, [1], \, \,
   {{27 \cdot \, x^3} \over {(1 \, + \, 3 \, x)^3 }})
                   \nonumber
\end{eqnarray}         
\begin{eqnarray}
\label{Ratfonc4HeunSol2.2}
\hspace{-0.98in}&& \quad   
\,\, = \, \, \Bigl({{1} \over {1\, +9\, x \, +27\, x^2 \, -27\, x^3}}\Bigr)^{1/3}
\cdot \,
_2F_1\Bigl([{{1} \over {6}}, {{2} \over {3}}], \, [1], \, \,
-\, {{108 \cdot \, x^3 \cdot \,(1 \, +9\, x \, + 27\, x^2) } \over {
    (1\, +9\, x \, +27\, x^2 \, -27\, x^3)^2 }}\Bigr)
                   \nonumber
\end{eqnarray}         
\begin{eqnarray}
\label{Ratfonc4HeunSol2.3}
\hspace{-0.98in}&& \quad  
\,\, = \, \,  \Bigl({{1} \over { 1\, +3 \, x }}\Bigr)^{1/4}
\cdot \, \Bigl({{1} \over { 1 \, +9\, x \, +27\, x^2 \, +3\, x^3 }}\Bigr)^{1/4}
\\
\hspace{-0.98in}&& \quad \quad \quad \quad \quad \quad \quad  \quad \quad 
\times \,\,
_2F_1\Bigl([{{1} \over {12}}, {{5} \over { 12 }}], \, [1], \, \,
 {{1728 \cdot \, x^9 \cdot \, (1 \, + 9 \, x \, + 27 \, x^2) } \over {
 (1 \, +3 \, x)^3 \cdot \,   (1 \, +9\, x \, +27\, x^2 \, +3\, x^3)^3 }} \Bigr)
                   \nonumber
\end{eqnarray}         
\begin{eqnarray}
\label{Ratfonc4HeunSol2.4}
\hspace{-0.98in}&& \quad          
                     \,\, = \, \,
  (1\, +9\, x)^{-1/4} \cdot \,      (1\, +3\, x)^{-1/4} \cdot \, (1\, +27 \, x^2)^{-1/4}
                  \nonumber     \\
\hspace{-0.98in}&& \quad \quad \quad \quad \quad \quad \quad  \quad \quad 
\times \,\,
  _2F_1\Bigl([{{1} \over {12}}, {{5} \over { 12 }}], \, [1], \, \,
 {{1728 \cdot \, x^3 \cdot \, (1 \, + 9 \, x \, + 27 \, x^2)^3 } \over {
  (1 \, +3 \, x)^3 \cdot \,   (1 \, +9\, x)^3  \cdot \, (1\, +27 \, x^2)^3 }} \Bigr),
\nonumber 
\end{eqnarray}
\begin{eqnarray}
\label{Ratfonc4HeunSol2.5}
\hspace{-0.98in}&& \quad          
 \,\, = \, \,
  (1\, +9\, x)^{-1/4} \cdot \,      (1\, +243\,x \, +2187\,{x}^{2} \, +6561\,{x}^{3})^{-1/4}
\nonumber     \\
\hspace{-0.98in}&&   \quad \quad \quad  \quad \quad 
\times \,\,
  _2F_1\Bigl([{{1} \over {12}}, {{5} \over { 12 }}], \, [1], \, \,
 {{1728 \cdot \, x \cdot \, (1 \, + 9 \, x \, + 27 \, x^2) } \over {
 (1 \, +9\, x)^3  \cdot \, (  1\, +243\,x \, +2187\,{x}^{2} \, +6561\,{x}^{3}  )^3 }}
  \Bigr).
\nonumber 
\end{eqnarray}
Note that the Hauptmoduls in (\ref{Ratfonc4HeunSol2}) can be rewritten
as the composition of two pullbacks:
\begin{eqnarray}
\label{Hauptrewrt}
  \hspace{-0.98in}&& \quad \quad \quad \quad \quad \quad
 {{1728 \cdot \, x^9 \cdot \, (1 \, + 9 \, x \, + 27 \, x^2) } \over {
 (1 \, +3 \, x)^3 \cdot \,   (1 \, +9\, x \, +27\, x^2 \, +3\, x^3)^3 }}
 \nonumber \\
 \hspace{-0.98in}&& \quad \quad \quad \quad \quad \quad \quad \quad \quad
 \, \, = \, \, \,
\Bigl(  {\frac {1728 {z}^{3}}{ (z \, +27)  \, (z+243)^{3}}}\Bigr)
      \, \circ \, \Bigl( {\frac { 729 \,{x}^{3}}{1 \,+9\,x + 27\,{x}^{2}}}\Bigr),
\end{eqnarray}
\begin{eqnarray}
\label{Hauptrewrt2}
  \hspace{-0.98in}&& \quad \quad \quad \quad \quad \quad
 {{1728 \cdot \, x^3 \cdot \, (1 \, + 9 \, x \, + 27 \, x^2)^3 } \over {
 (1 \, +3 \, x)^3 \cdot \,   (1 \, +9\, x)^3  \cdot \, (1\, +27 \, x^2)^3 }}
 \nonumber \\
 \hspace{-0.98in}&& \quad \quad \quad \quad \quad \quad \quad \quad \quad
 \, \, = \, \, \,
\Bigl(  {\frac {1728 {z}}{ (z \, +27)  \, (z \, +3)^{3}}}\Bigr)
      \, \circ \, \Bigl( {\frac { 729 \,{x}^{3}}{1 \,+9\,x + 27\,{x}^{2}}}\Bigr),
\end{eqnarray}
\begin{eqnarray}
\label{Hauptrewrt3}
  \hspace{-0.98in}&& \quad \quad \quad \quad \quad \quad
  {{1728 \cdot \, x \cdot \, (1 \, + 9 \, x \, + 27 \, x^2) } \over {
            (1 \, +9\, x)^3  \cdot \, (  1\, +243\,x \, +2187\,{x}^{2} \, +6561\,{x}^{3}  )^3 }}                    
 \nonumber \\
 \hspace{-0.98in}&& \quad \quad \quad  \quad \quad \quad \quad \quad
 \, \, = \, \, \,
\Bigl(  {\frac {1728 {z}}{ (z \, +27)  \, (z \, +3)^{3}}}\Bigr)
\, \circ \, \Bigl( 729 \,  \, x \cdot \, (1 \,+9\,x + 27\,{x}^{2})
 \Bigr).
\end{eqnarray}
The modular equation relating the Hauptmodul (\ref{Hauptrewrt})
with the Hauptmodul (\ref{Hauptrewrt3}) 
corresponds to $\, \,\, q \, \leftrightarrow \, q^9 \,\, $
in the nome $\, q$ (see also Table 4 and Table 5 in~\cite{Maier1}).   

This Heun function (\ref{Ratfonc4HeunSol2})
is in fact the {\em period} of an {\em extremal rational surface}~\cite{Malmendier4},
and was shown to be related\footnote[5]{Change $\, x \, \rightarrow \, \, x/27$ to match
  $\,{\cal S}_1$,  given by (\ref{Ratfonc4HeunSol}),
with (\ref{selectedHeun}).} to {\em classical modular forms} in table 15 in~\cite{Maier1} for $\, N=\, 9$:  
\begin{eqnarray}
\label{selectedHeun}
 \hspace{-0.98in}&& \quad  \quad    \quad  
 Heun\Bigl({{-9 \mp 3\, \sqrt{3} \, i} \over {-9 \pm 3\, \sqrt{3} \, i}},
 \,  {{9 \pm \, 3\, \sqrt{3} \,  i} \over {18}},  \,  1, \,
 1, \, 1, \,  1,   \, \,   \, {{2\, \, x} \over {-9 \pm  3\, \sqrt{3} \, \, i }}   \Bigr)
  \nonumber  \\
  \hspace{-0.98in}&& \quad  \quad  \quad \quad \quad \, \, = \, \,\,
Heun\Bigl( {{1 \pm \, \sqrt{3} \, i} \over {2}}, \,  \, {{3 \pm \, \, \sqrt{3} \,  i} \over {6}}, \,  1, \,
        1, \, 1, \,  1,   \, \,   \, {{-3 \mp  \, \sqrt{3} \, \, i } \over { 18 }} \cdot \, x   \Bigr).
\end{eqnarray}               

The other order-two linear differential operator $\, M_2\, $ has the following
(classical modular form, see \ref{extremparam}) pullbacked $\, _2F_1$ hypergeometric
solution\footnote[2]{It seems that this  pullbacked
  $\, _2F_1$ hypergeometric (\ref{spurious}) cannot be seen as a (simple)
  Heun function: see \ref{extremparam}.}
analytic at $\, x=\, 0$:
\begin{eqnarray}
\label{spurious}
\hspace{-0.98in}&& \quad  \quad 
{\cal S}_2 \,\,  = \, \,  \,
{{1} \over { (1 \, + 4\, x \,-2\, x^2 \,-36\, x^3 \, + 81\, x^4 )^{1/4}}} \cdot \,
\nonumber \\
\hspace{-0.98in}&& \quad  \quad \quad \quad  \quad   \quad        \times \, 
   \, _2F_1\Bigl([{{1} \over {12}}, {{5} \over { 12 }}], \, [1], \, \,
 {{1728 \cdot \, x^5 \, \cdot \, (1 +9 \, x \, +27 \, x^2)  \cdot \,(1 \, -2\,x)^2  } \over {
(1 \, + 4\, x \,-2\, x^2 \,-36\, x^3 \, + 81\, x^4)^3}}\Bigr)
\nonumber \\
\hspace{-0.98in}&& \quad  \quad         
\, \, = \, \,  \, \,\,
1 \,\, \, \, -x\, \,\, \,  +3\,{x}^{2}\, \,\, -{x}^{3}\,\,\,  -29\,{x}^{4}
\, \, \,+249\,{x}^{5}\,\,\, -1329\,{x}^{6}\,\,\,
+5487\,{x}^{7}\, \,-16029\,{x}^{8}
\nonumber \\
  \hspace{-0.98in}&& \quad  \quad \quad  \quad \quad  \quad \quad  \quad
 \, +12149\,{x}^{9} \,\,  \, +252253\,{x}^{10}
 \, \,\,  \, + \, \, \, \cdots               
\end{eqnarray} 
This second order-two linear differential operator $\, M_2$ is {\em not homomorphic}
to the previous one\footnote[9]{They cannot be  homomorphic: they do not have exactly the same singularities.
The order-two linear differential operator $\, M_2$ has the extra $\, x \, = \, 1/2$ singularity.
}.
The Hauptmoduls $\,  {\cal H}_1 \, $  for $\, L_2$  (see (\ref{Ratfonc4HeunSol2.4}))
and $\,  {\cal H}_2 \, $  for $\, M_2$ (see (\ref{spurious}))
\begin{eqnarray}
\label{calH1}
\hspace{-0.98in}&& \quad  \quad \quad   \quad \quad   \quad
 {\cal H}_1 \, \, = \, \, {{1728 \cdot \, x \cdot \, (1 \, + 9 \, x \, + 27 \, x^2) } \over {
            (1 \, +9\, x)^3  \cdot \, (  1\, +243\,x \, +2187\,{x}^{2} \, +6561\,{x}^{3}  )^3 }} , 
\end{eqnarray}
\begin{eqnarray}
\label{calH2}
  \hspace{-0.98in}&& \quad  \quad  \quad \quad   \quad \quad   \quad
{\cal H}_2 \, \, = \, \,
{{1728 \cdot \, x^5 \, \cdot \, (1 +9 \, x \, +27 \, x^2)  \cdot \,(1 \, -2\,x)^2  } \over {
(1 \, + 4\, x \,-2\, x^2 \,-36\, x^3 \, + 81\, x^4)^3}}, 
\end{eqnarray} 
are  {\em not simply related}\footnote[1]{Their associated nomes
  are not simply related. However, one can imagine that these Hauptmoduls (\ref{calH1}),  (\ref{calH2}) are Igusa invariants
  of an algebraic  surface (for instance a split Jacobian of a genus-two algebraic curve).}. They just
both vanish at $\, \, 1 \, + 9 \, x \, + 27 \, x^2 \, = \, \, 0$.
A rational parametrisation is introduced in \ref{extremparam} for these two order-two
linear differential operators making clear the differences and similarities
of these two linear differential operators. 

\vskip .2cm 

One finds that the diagonal of  (\ref{Ratfonc4}) {\em is actually the half-sum
of the two series} (\ref{Ratfonc4HeunSol}) and (\ref{spurious}): 
\begin{eqnarray}
\label{spurioushalf}
  \hspace{-0.98in}&& \quad  \quad \quad  \quad \quad \quad  \quad \quad \quad 
Diag\Bigl( R(x, \, y, \, z)\Bigr)          \,   \, \, = \, \,  \, \,\,\,
 {{ {\cal S}_1 \,  + {\cal S}_2} \over { 2}}. 
\end{eqnarray}
The order-four linear differential operator $\,\, L_4 \, = \, \, L_2 \oplus \, M_2\,\,$
is thus the {\em minimal order telescoper}. The diagonal of the rational function (\ref{Ratfonc4})
is the {\em sum of two classical modular forms}.

\vskip .2cm

{\bf Remark 1:} The previous results can also be understood as follows. The telescoper of the rational function
\begin{eqnarray}
\label{Ratfonc4other}
  \hspace{-0.98in}&& \quad  \quad \quad \quad \quad \quad
\, \, 
R(x, \, y, \, z)  \, \, \,  = \, \,  \quad 
 {{x} \over { 1 \, \,\,  \, + x \, + y \, + z  \, \, \,+ x \,y \, + y \, z \, \, \,  - x^3 \,y \,z }},
\end{eqnarray}
similar to (\ref{Ratfonc4}) (where the numerator of the rational function
has been changed from $\, 1$ to $\, x$)
is actually the {\em same as the one for} (\ref{Ratfonc4}), namely $\, L_4 \, = \, L_2 \oplus M_2$. 
The diagonal of (\ref{Ratfonc4other}) reads:
\begin{eqnarray}
\label{Ratfonc4other}
  \hspace{-0.98in}&& \,  \quad 
Diag\Bigl( R(x, \, y, \, z)\Bigr)             \, \, = \, \,  \, \,
    {{ {\cal S}_2 \, - \,  {\cal S}_1 } \over { 2}}  \, \, = \, \,  \, \,\,
 x \,  \, \, \, -3\,{x}^{2} \, \,   +10\,{x}^{3} \,  \,  -19\,{x}^{4} \,  \,  -24\,{x}^{5} 
 \nonumber \\
  \hspace{-0.98in}&& \quad  \,   \quad  \quad  
\, \,   +546\,{x}^{6}\, \,  -3723\,{x}^{7}\, \, +18189 \,{x}^{8}
\,\,  -66572\,{x}^{9} \,\,  +143672\,{x}^{10} \, \, \,   + \, \, \, \cdots 
\end{eqnarray}
The  telescoper of the rational function of three variables
\begin{eqnarray}
\label{Ratfonc4other1}
  \hspace{-0.98in}&& \quad  \quad \quad \quad \quad \quad
\, \, 
R(x, \, y, \, z)  \, \, \,  = \, \,  \quad 
 {{1 \, -x} \over { 1 \, \,  \, \,  + x \, + y \, + z  \, \, \,+ x \,y \, + y \, z \,  \, - x^3 \,y \,z }}, 
\end{eqnarray}
is the order-two  linear differential operator $\, L_2$ with the hypergeometric solution $\,{\cal S}_1 $. 
The  telescoper of the rational function of three variables
\begin{eqnarray}
\label{Ratfonc4other2}
  \hspace{-0.98in}&& \quad  \quad \quad \quad \quad \quad
\, \, 
R(x, \, y, \, z)  \, \, \,  = \, \,  \quad 
 {{1\, +x} \over { 1 \, \, \,  \, + x \, + y \, + z  \, \, \,+ x \,y \, + y \, z \,  \,  \,  - x^3 \,y \,z }}, 
\end{eqnarray}
is the order-two  linear differential operator $\, M_2\,$ with the hypergeometric solution $\,{\cal S}_2 $. 
Note however that the  telescoper of the rational function
\begin{eqnarray}
\label{Ratfonc4other4}
  \hspace{-0.98in}&& \quad  \quad \quad \quad \quad \quad
\, \, 
R(x, \, y, \, z)  \, \, \,  = \, \,  \quad 
 {{x \, y} \over { 1 \, \,  \, \,  + x \, + y \, + z  \, \, \,+ x \,y \, + y \, z \, \,  - x^3 \,y \,z }},
\end{eqnarray}
is an order-five linear differential operator $\, L_5$ which is the
direct sum of the order-two operator $\, L_2$ and an order-three linear differential 
operator $\, L_3 \, = \, N_2 \cdot \, D_x$, namely
$\, L_5 \, = \, \,  L_2 \, \oplus \, (N_2 \cdot \, D_x)\,$ 
where $\, N_2$ {\em is non-trivially homomorphic to} $\, M_2$
(with new apparent singularities $\, \, 36\,x^2 \, -6\, x\, +1 \, = \,\,  0$
and $\,(1\, +3\, x)  \, = \,\,  0$). 
The series expansion of the diagonal of this last rational function
is non trivial and reads:
\begin{eqnarray}
\label{Ratfonc4otherser}
  \hspace{-0.98in}&& \quad   \quad    \quad    \quad  
Diag\Bigl( R(x, \, y, \, z)\Bigr)             \, \, = \, \,  \, \,
 -x \,  \, \,  \, \, +3\,{x}^{2} \, \, \,  \,  -10\,{x}^{3} \,  \,  \,
 +23\,{x}^{4} \, \,  \,  -6\,{x}^{5} \,  \,  \, -378\,{x}^{6}
 \nonumber \\
 \hspace{-0.98in}&& \quad  \quad  \quad   \quad \quad \quad \quad
\, +3009\,{x}^{7}  \,  \, -15993\,{x}^{8}  \, \, +64394\,{x}^{9}
\, \,  -175102\,{x}^{10}  \,\, \, \,    \, + \, \, \cdots 
\end{eqnarray}

\vskip .2cm

{\bf Remark 2:} All the previous Heun functions occurring as diagonals of simple rational functions
can all be rewritten in terms of pullbacked $\, _2F_1$ hypergeometric functions
which turn out to correspond to {\em classical modular curves}.
These  $\, _2F_1$ hypergeometric
functions are not arbitrary, they are ``special'' $\, _2F_1$'s corresponding to
selected parameters, namely  $\, _2F_1$'s related to {\em classical modular curves}.
\ref{notmodular} gives a simple condition {\em on the nome} of these $\, _2F_1$'s
to be related to {\em classical modular curves}.
In \ref{Special} we give the exhaustive list of these $\, 28$ hypergeometric $\, _2F_1$'s
related to {\em classical modular curves}.

\vskip .2cm

 \subsection{Derivatives of classical modular forms}
\label{Derivclassforms}

Let us recall example 6, and let us consider, instead of the rational function (\ref{Ratfoncfour20}),
its {\em homomogeous partial derivative} with respect to one of its four variables: 
\begin{eqnarray}
\label{Ratfoncfour20deriv}
  \hspace{-0.98in}&& \quad 
\, \,  \, \,\, \,    
 x \cdot \, {{\partial R(x, \, y, \, z, \, w)} \over {\partial x}}   \, \, \,  = \, \,  \quad 
{{x \cdot \, (y\, +z\, +w)} \over { (1 \, \, \, +x\,y \, +y\,z \, +z\,w \, +w\,x \, +y\,w \, +x\,z)^2 }}.
\end{eqnarray}
The telescoper of this rational function (\ref{Ratfoncfour20deriv}) is an order-three
linear differential operator $\, M_3$ which is homomorphic to the order-three
operator $\, L_3$ given by (\ref{Ratfoncfour11L3}) which was the telescoper of
the rational function (\ref{Ratfoncfour20}). This homomorphism reads:
\begin{eqnarray}
\label{Ratfoncfour20derivHomo}
  \hspace{-0.98in}&& \quad  \quad   \quad \, \,
M_3  \cdot \, \theta  \, \, = \, \, \,  L_1 \cdot \, L_3
 \, \, \quad  \quad \hbox{where:}  \quad  \quad \, \, \, \,
L_1 \, \, = \, \, \, (1\, -18\, x) \cdot \, \theta \,  \,  \,  + \, 18\, x, 
\end{eqnarray}
where  $\, \theta$ is the {\em homogeneous derivative} $\, \theta \, = \, x \cdot \, D_x$.
Consequently the solutions of the order-three linear differential operator
$\, M_3$ are simply obtained by taking the
homogeneous derivative $\, \theta \, = \, x \cdot \, D_x$ of the solutions
 of the order-three linear differential operator  $\, L_3$.
In particular, the diagonal of the  rational function (\ref{Ratfoncfour20deriv})
is the {\em homogeneous derivative} of  the diagonal of the
rational function (\ref{Ratfoncfour20}):
\begin{eqnarray}
\label{Ratfoncfour20derivgen}
  \hspace{-0.98in}&& \quad   \quad  \quad 
\, \,  \, \,
 Diag\Big( x \cdot \, {{\partial R(x, \, y, \, z, \, w)} \over {\partial x}} \Bigr) 
\, \, \,  = \, \,  \,  \,
x \cdot \, {{ d} \over {dx}} \Big( Diag\Big(  R(x, \, y, \, z, \, w) \Bigr) \Bigr), 
\end{eqnarray}
The diagonal of (\ref{Ratfoncfour20deriv}) will thus be the  {\em homogeneous derivative of
the classical modular form} (\ref{Ratfoncfour11}). This is a general result on diagonals of rational functions.
We have the following identity valid for any order-$\, N$ linear differential operator $\, L$
\begin{eqnarray}
\label{Ratfoncfour20derividentity}
  \hspace{-0.98in}&& \quad \quad \quad
\, \,  
 Diag\Big( {\cal L} \Big( R(x, \, y, \, z, \, w) \Bigr) \Bigr) 
 \, \, \,  = \, \,  \,  \,  \,  L \Big( Diag\Big(  R(x, \, y, \, z, \, w) \Bigr) \Bigr), 
  \\
 \label{Ratfoncfour20derividentityPol} 
 \hspace{-0.98in}&&   \quad \quad \hbox{where:}  \quad  \quad \quad 
 L \, \, = \, \,     \sum_{n=0}^{N} \, P_n(x) \cdot \, \theta^n,  \quad \quad
 {\cal L}  \, \, = \, \,   \sum_{n=0}^{N} \, P_n(x\, y\, z\, w) \cdot \, \Theta^n, 
  \\
 \hspace{-0.98in}&& \quad \quad  \hbox{with:}  \quad  \quad   \quad  \quad \quad 
 \theta \, \, = \, \, \,    x \cdot \, {{ d } \over {d x}}, 
 \quad   \quad  \cdots   \quad \quad  \quad 
 \Theta \, \, = \, \, \,    w \cdot \, {{ \partial } \over {\partial w}},
\end{eqnarray}
where the $\, P_n$'s  are polynomials. This identity can, of course be generalised to
the diagonal of rational functions of an arbitrary number of variables.
For any  Heun function or classical modular form of this paper, obtained as a diagonal
of a rational function, we can use these identities (\ref{Ratfoncfour20derivgen}),
(\ref{Ratfoncfour20derividentity}) to get other rational functions that will not  be 
Heun functions or classical modular forms, but
{\em derivatives of Heun functions or classical modular forms}.

\vskip .2cm

Note that the derivative of a classical modular form, or more generally an order-one linear
differential operator like (\ref{Ratfoncfour20derividentityPol}) acting on a classical
modular form, is {\em no longer a  classical modular form}.
With this example we see that a Heun function which has a series expansion
with  {\em integer coefficients} (or more generally is globally bounded  series),
is  {\em not necessarily a  classical modular form}, but can be 
{\em an order-one linear differential operator acting on a classical modular form}.

\vskip .2cm

A simple example of diagonals of rational functions of three variables,
corresponding to derivatives of Heun functions,  is given in \ref{subthree}. Along
this line see also \ref{negative}.

\vskip .2cm

\section{Heun function solutions of telescopers of rational functions related to Shimura curves}
\label{vanHoeijVidunas}

The rational function of four variables
\begin{eqnarray}
\label{Ratfonc4gg0}
  \hspace{-0.98in}&& 
\, \,  \, 
R(x, \, y, \, z, \, u)  \, \, \,  = \, \,  \, 
  {{x\, y \, z} \over { 1 \, \,  \,  \, - x \,  y \,  z \, u \, \, 
   \, +\, x \,  y \,  z   \cdot \, (x \, +y\, +z) \,   \, \, + \, x \,y  + \, y \,z \, + \, x \, z}},
\end{eqnarray}
has a telescoper that is a linear differential operator of order three:
\begin{eqnarray}
\label{telesHeun}
  \hspace{-0.98in}&&  \quad   \quad   \quad  \quad 
 L_3 \,\, \, = \, \, \, \, 8 \,  x \cdot \, (1\, -x) \cdot \, (1\, -4\, x) \cdot \,  D_x^3 \, \, \, \, 
+ 12 \cdot \, (1\, -10\, x \,+12 \, x^2)  \cdot \,  D_x^2
\nonumber \\
  \hspace{-0.98in}&& \quad  \quad \quad   \quad   \quad  \quad  \quad  \quad  \quad  \quad 
  - 6 \cdot \, (7 \, - 17 \cdot \,  x) \cdot  \, D_x
\, \, \,    \,  +3,  
\end{eqnarray}
which corresponds to the {\em symmetric square} of an order-two linear differential operator
reading in terms of the homogeneous derivative $\, \theta$:
\begin{eqnarray}
\label{telesHeunL2}
\hspace{-0.98in}&&  \quad   \quad  \quad \,  
 \,   \,   x^2 \cdot \, (8\, \theta \, +3) \cdot \,  (8\, \theta \, +1) \, 
 \,  \,  \, \,   \,   - \, x  \cdot \, (80 \, \theta^2 \, +1)  \, \, \, \, +8 \cdot \, \theta  \cdot \, (2\, \theta \, -1).           
\end{eqnarray}
The solutions of order-three linear differential operator $\, L_3$
are, thus, expressed in terms of the following Heun functions
\begin{eqnarray}
 \label{HeunShimura}
  \hspace{-0.98in}&& \quad  \, \,  \,   \,  
 Heun\Bigl( {{1} \over {4}}, \, {{1} \over {64}}, \, {{1} \over {8}},
  \, {{3} \over {8}}, \, {{1} \over {2}}, \, {{1} \over {2}}, \, \, x \Bigr)^2, \quad \,  \,  \,  \, 
 x \cdot \, Heun\Bigl( {{1} \over {4}}, \, {{21} \over {64}}, \, {{5} \over {8}},
  \, {{7} \over {8}}, \, {{3} \over {2}}, \, {{1} \over {2}}, \, \, x \Bigr)^2, 
\end{eqnarray}
or:
\begin{eqnarray}
 \label{HeunShimura3}
  \hspace{-0.98in}&& \quad  \quad  \,
 x^{1/2} \cdot \, Heun\Bigl( {{1} \over {4}}, \, {{1} \over {64}}, \, {{1} \over {8}},
     \, {{3} \over {8}}, \, {{1} \over {2}}, \, {{1} \over {2}}, \, \, x \Bigr)
 \cdot \, Heun\Bigl( {{1} \over {4}}, \, {{21} \over {64}}, \, {{5} \over {8}},
  \, {{7} \over {8}}, \, {{3} \over {2}}, \, {{1} \over {2}}, \, \, x \Bigr). 
\end{eqnarray}
The series expansion of the first expression in (\ref{HeunShimura}) reads:
\begin{eqnarray}
 \label{HeunShimuraexp}
  \hspace{-0.98in}&&  \quad \, \,
1 \,\,  \; +{\frac{1}{4}}\,x \, \, \,\, +{\frac{5}{16}}\,{x}^{2}\, \, +{\frac{5}{8}}\,{x}^{3} \,
\, +{\frac{2795}{1792}}\,{x}^{4}\, \, +{\frac{15691}{3584}}\,{x}^{5}
\, \,+{\frac{1039363}{78848}}\,{x}^{6}
  \, \,+{\frac{1872975}{45056}}\,{x}^{7}
 \nonumber \\
  \hspace{-0.98in}&& \quad  \quad \quad  \, \, \,
+{\frac{4786080975}{35323904}}\,{x}^{8} \, \, +{\frac{1142244025}{2523136}}{x}^{9} \,
\, +{\frac{1182929670845}{767033344}}\,{x}^{10}
 \, \, \, \, \, +    \,    \,\,   \cdots   
\end{eqnarray}
Do note that the diagonal of this rational function (\ref{Ratfonc4gg0}) {\em is actually equal to zero}:
{\em it is different from this solution of the telescoper} (\ref{telesHeun}). The two concepts, namely
being the diagonal of a rational function and being the solution  of the telescoper of
that rational function {\em do not necessarily identify}. The solutions of the telescoper
are  $\,n$-fold integrals of that  rational function integrand  {\em over all possible cycles}:
a solution like (\ref{HeunShimura}) is thus a ``Period'' of an algebraic variety
corresponding to a particular {\em non-evanescent} cycle. It is different from
the  diagonal of that  rational function which is a ``Period''  over  {\em evanescent cycles}.

In contrast with all the other (square of) Heun functions of this paper,
which are associated with {\em classical modular forms}, the series expansion (\ref{HeunShimuraexp})
is {\em not globally bounded: it cannot be recast\footnote[5]{After a rescaling of the variable.}
  into a series with integer coefficients}. 
{\em It cannot be a diagonal of a rational function: it is only
a solution of the telescoper of a rational function}.
The order-three linear differential operator (\ref{telesHeun}), is
the {\em symmetric square} of the linear differential operator of order two $\, L_2$:
\begin{eqnarray}
\label{L2Shimura}
\hspace{-0.98in}&&   \,   \,  
L_2 \, \,\, = \, \,\, \,
D_x^2  \,\, \,
+  \frac{ 1 -10\,x +12 \,x^2}{ 2 \, x \cdot \, (1\, -4\,x )\, \cdot (1\, -x)} \cdot  \, D_x
 \,\, \,\, -\frac{1\, -3 \,x}{16 \cdot x \cdot (1\, -4 \, x) \cdot \, (1\, -x)}, 
\end{eqnarray}
whose (formal) series expansions at $ \, 0$, $\,1$, and $\,\infty$ 
do not contain\footnote[2]{We have three elliptic points.} {\em logarithms}. This
order-two linear differential operator
$\, L_2 \, $ admits the solutions: 
\begin{eqnarray}
\label{Ratfonc4gg2hyp}
 \hspace{-0.98in}  \quad   \,  \quad \,  \,  \quad   \quad   \, \,
  x^{1/2} \cdot \, (1\, -x)^{-7/8} \cdot \,
  _2F_1\Bigl([{{ 7} \over { 24}}, \, {{ 11} \over { 24}}], \, [{{ 5} \over {4}}],
  \,\,  \, {{27} \over { 4}} \cdot \, {{x^2} \over {(1\, -x)^3 }}\Bigr), 
  \\
 \hspace{-0.98in} \quad \,  \quad  \,\, \quad    \quad   \, \,
  (1\, -x)^{-1/8} \cdot \,   _2F_1\Bigl([{{ 1} \over { 24}}, \, {{ 5} \over { 24}}], \,
  [{{ 3} \over {4}}] ,\,\,  \, {{27} \over { 4}} \cdot \, {{x^2} \over {(1\, -x)^3 }}\Bigr).
  \nonumber
\end{eqnarray}
The precise correspondence with the Heun functions in (\ref{HeunShimura}) reads:
\begin{eqnarray}
\label{corresp}
 \hspace{-0.98in}  \quad   \,  \quad \,  \quad    \quad  
Heun\Bigl( {{1} \over {4}}, \, {{1} \over {64}}, \, {{1} \over {8}},
  \, {{3} \over {8}}, \, {{1} \over {2}}, \, {{1} \over {2}}, \, \, x \Bigr)
  \nonumber \\
\hspace{-0.98in}  \quad   \,   \,   \, \, \quad  \quad   \quad   \quad  \quad   \quad  
  \, \, = \, \, \,
  (1\, -x)^{-1/8} \cdot \,   _2F_1\Bigl([{{ 1} \over { 24}}, \, {{ 5} \over { 24}}], \,
  [{{ 3} \over {4}}] ,\,\,  \, {{27} \over { 4}} \cdot \, {{x^2} \over {(1\, -x)^3 }}\Bigr),
  \\
 \label{corresp2} 
  \hspace{-0.98in}  \quad   \,  \quad \quad  \,  \quad  
 Heun\Bigl( {{1} \over {4}}, \, {{21} \over {64}}, \, {{5} \over {8}},
  \, {{7} \over {8}}, \, {{3} \over {2}}, \, {{1} \over {2}}, \, \, x \Bigr)
  \nonumber \\
\hspace{-0.98in}  \quad   \,   \,    \, \,\quad  \quad   \quad   \quad  \quad   \quad  
  \, \, = \, \, \,  (1\, -x)^{-7/8} \cdot \,
  _2F_1\Bigl([{{ 7} \over { 24}}, \, {{ 11} \over { 24}}], \, [{{ 5} \over {4}}],
  \,\,  \, {{27} \over { 4}} \cdot \, {{x^2} \over {(1\, -x)^3 }}\Bigr).
\end{eqnarray}
The pullbacks in all the $\, _2F_1$ hypergeometric functions of this paper are
special rational (or algebraic) functions: they correspond to the concept of
{\em Belyi maps}\footnote[9]{An important area where Belyi functions~\cite{HoeijVidunas,Belyi0,Belyi} appear
is {\em precisely Shimura curves}.
Any Belyi covering gives a modular curve with respect to some (not necessarily congruence) subgroup.}.
In this case which {\em does not correspond to a classical modular form} but a (Shimura)
{\em automorphic form}\footnote[1]{ They lie at the crossroads of many areas of mathematics.
 They have played an important role in the proof of Fermat's last theorem. A Shimura curve is simply
 a Riemann surface which is uniformized by an arithmetic Fuchsian group.},
the  pullback $\,  {{27} \over { 4}} \cdot \, {{x^2} \over {(1\, -x)^3 }}\, $
in (\ref{Ratfonc4gg2hyp}) being also ``special'' (see \ref{Kilianapp}). 

The two solutions of the linear differential operator (\ref{L2Shimura}) can be used to construct a basis
for space of {\em automorphic forms}, which can then be used to construct Hecke operators relative
to this basis\footnote[5]{See example 9 in~\cite{YifanHecke} for more details.}. The
second solution in (\ref{Ratfonc4gg2hyp}) corresponds to an
{\em automorphic form associated with  a Shimura curve}
with signature $\, (0,\, 4, \, 2, \, 6)$ which appears 
in Table $ \,1$ in~\cite{Takeuchi}. More details on Heun, or  $\, _2F_1$,
{\em automorphic forms} associated with
{\em Shimura curves}~\cite{FangTingTu,FangTingTuYifang,Filipuk} are given in \ref{Shimura}. Note in particular
the fact that there exists {\em an algebraic series} $\, y(x)$ corresponding to a 
{\em  modular equation}, such that the two  $\, _2F_1$ hypergeometric functions
(\ref{corresp}),  (\ref{corresp2}) {\em actually verify the  following identity/symmetry}:
\begin{eqnarray}
  \label{ModularinftyL2inftysolident2first}
  \hspace{-0.98in}&&  \quad   \quad   \quad  \quad \, \,\,
 w^{3/8}   \cdot \,     \rho \cdot \, y'(x)^{1/2}   \cdot \,     x^{3/8} \cdot \, (1-x)^{1/4} \cdot \,
 _2F_1\Bigl( [{{ 1} \over {24 }}, \,  {{5 } \over {24 }}], \, [{{ 3} \over {4 }} ], \, x\Bigr)
 \nonumber \\
  \hspace{-0.98in}&&  \quad   \quad   \quad  \quad  \quad \quad   \quad 
  \, \, = \,  \,   \,  \,  y(x)^{3/8} \cdot \, (1\, -y(x))^{1/4} \cdot \,
 _2F_1\Bigl( [{{ 1} \over {24 }}, \,  {{5 } \over {24 }}], \, [{{ 3} \over {4 }} ], \, y(x)\Big),
\end{eqnarray}
where the two complex numbers  $\, w$ and $\, \rho \, $ {\em are on the unit circle}. More
details are given in \ref{Shimura}. 
Such an identity is reminiscent of the hypergeometric
identities we studied in~\cite{Youssef} for classical modular forms.

This ``Shimura''  $\, _2F_1$ hypergeometric function
can be seen to correspond to other Heun functions than
(\ref{corresp}) or (\ref{corresp2}). Using the general identity~\cite{Valent,ReduceMaier}
\begin{eqnarray}
\label{general}
\hspace{-0.98in}&& \quad \quad \quad  \quad \, \,
Heun(2, \, 16\,a\,b, \, 4\,a, \, 4\, b, \, a+b \,+1/2, \, 2\, (a+b), \, \, x)
\nonumber \\
\hspace{-0.98in}&& \quad \quad \quad \quad \quad  \quad  \, \,
 \, \, = \, \, \,
 _2F_1\Bigl([a, \, b], \, [a+b \, +{{1} \over {2}}], \, \, 4 \cdot \, x \cdot \, (2\, -x) \cdot \, (1\, -x)^2\Bigr), 
\end{eqnarray}
one deduces the identity
\begin{eqnarray}
\label{generaldeduce}
\hspace{-0.98in}&&  \quad   \quad   \quad \quad  \quad  \quad 
Heun\Bigl(2, \, {{5} \over {36}}, \, \, {{1} \over {6}}, \, {{5} \over {6}}, \, {{3} \over {4}}, \,  {{1} \over {2}}, \, \, x\Bigr)
\nonumber \\
\hspace{-0.98in}&& \quad \quad \quad \quad \quad  \quad   \quad  \quad          
\, \, = \, \, \, _2F_1\Bigl([{{1} \over {24}}, \, {{5} \over {24}}], \, [{{3} \over {4}}],
 \, \, 4 \cdot \, x \cdot \, (2\, -x) \cdot \, (1\, -x)^2\Bigr), 
\end{eqnarray}
as well as the  identity:
\begin{eqnarray}
\label{generaldeduce2}
\hspace{-0.98in}&&  \quad   \quad   \quad \quad \quad  \quad 
Heun\Bigl(2, \, {{77} \over {36}}, \, \, {{7} \over {6}}, \, {{11} \over {6}}, \, {{5} \over {4}}, \,  {{3} \over {2}}, \, \, x\Bigr)
\nonumber \\
\hspace{-0.98in}&& \quad \quad \quad \quad \quad  \quad   \quad  \quad         
\, \, = \, \, \, _2F_1\Bigl([{{7} \over {24}}, \, {{11} \over {24}}], \, [{{5} \over {4}}],
\, \, 4 \cdot \, x \cdot \, (2\, -x) \cdot \, (1\, -x)^2\Bigr). 
\end{eqnarray}

More generally, Heun functions, related to Shimura curves, often emerge in the context
of {\em Belyi maps} where Heun functions with four singularities, are expressed
as pullbacked $\, _2F_1$ hypergeometric functions.
For example Table 3.4.4 of~\cite{HoeijVidunas} (see also~\cite{FSU}),
corresponds to the $\, _2F_1$ hypergeometric function
$ _2F_1([{{1} \over {3}}, \, {{1} \over {12}}], \, [{{3} \over {4}}],  \,  \,  x) \, $
with three different pullbacks
\begin{eqnarray}
\label{Table344B21}
 \hspace{-0.98in}&& \quad \quad \quad  \quad \quad
 _2F_1\Bigl([{{1} \over {3}}, \, {{1} \over {12}}], \, [{{3} \over {4}}],
 \, \,  {\frac {  {x}^{4}\cdot \, ({x}^{2}-3) }{ 1\, -3\,{x}^{2}}} \Bigr) 
\nonumber \\
\hspace{-0.98in}&& \, \,\quad \quad \quad \quad \quad \quad \, \,  \quad
\, \, = \, \,  \, \,
(1\, -3 \, x^2)^{1/3} \cdot \, Heun\Bigl({{1} \over {9}},\, {{1} \over {6}},\, {{1} \over {2}},
  \, 1,\, {{1} \over {2}},\, {{3} \over {4}}, \, \,  {{x^2} \over {3}} \Bigr), 
\end{eqnarray}
namely to the pullback B3 in the Table 3.4.4 in~\cite{HoeijVidunas}. 
In the paper~\cite{HoeijVidunas} most of the  $\, _2F_1$ hypergeometric functions are in fact associated with
{\em Shimura curves}: all tables\footnote[2]{In~\cite{HoeijVidunas} the table number, e.g. 3.4.4, means
that the elliptic points are 3, 4 and 4. For an example on how to obtain the hypergeometric function
 associated to these elliptic points, see for example, paragraph 2.5 in \cite{FangTingTu}.}
except 2.3.13,  2.3.14, 2.5.7, correspond to Heun functions
corresponding to pullbacked $\, _2F_1$  hypergeometric functions, that are
{\em automorphic forms associated with Shimura curves}\footnote[1]{Because they do not appear 
  in Takeuchi's~\cite{Takeuchi} table 1, which gives a complete list of
  hypergeometric functions that are associated with Shimura curves.}.

On a related issue
we found the transformation\footnote[8]{Which might be
  in the literature, yet we have not seen it.}:
\begin{eqnarray}
\label{Table344}
  \hspace{-0.98in}&& \, \, \, \, \, \, \,  \, \, 
_2F_1\Bigl([{{1} \over {3}}, \, {{1} \over {12}}], \, [{{3} \over {4}}],  \,  x  \Bigr)
 \, \, \, = \, \, \,\,
(1\, -x)^{-1/12}  \cdot \,
 _2F_1\Bigl([{{1} \over {24}}, \, {{5} \over {24}}], \, [{{3} \over {4}}],
 \,  \,  -\, {{ 4 \, x } \over { (1\, -x)^2 }}  \Bigr). 
\end{eqnarray}
Relations (\ref{Table344B21}) and (\ref{Table344}) show that there is a relation between
several Heun functions corresponding to {\em automorphic forms associated with Shimura curves}, namely 
$\, Heun\Bigl({{1} \over {9}},\, {{1} \over {6}},\, {{1} \over {2}},  \, 1,\, {{1} \over {2}},\, {{3} \over {4}}, \, \,  x) \, $
in  (\ref{Table344B21}), and the  Heun functions (\ref{HeunShimura}) which emerged as
solutions of telescoper of the rational function (\ref{Ratfonc4gg0}). We actually have the relation
\begin{eqnarray}
  \label{relation}
 \hspace{-0.98in}&&    \quad \quad  \quad  \, \, 
(1\, -3 \, x^2)^{3/8} \cdot \, \Bigl(1\, -\, {{x^2}  \over {3}}\Bigr)^{1/8}  \cdot \,
Heun\Bigl({{1} \over {9}},\, {{1} \over {6}},\, {{1} \over {2}},\, 1,\, {{1} \over {2}},\, {{3} \over {4}}, \, \,  {{x^2} \over {3}} \Bigr)
\nonumber \\
\hspace{-0.98in}&&    \quad  \quad  \quad  \quad  \quad  \quad  \quad  \, \, 
\, \, = \, \, \,
Heun\Bigl( {{1} \over {4}}, \, {{1} \over {64}}, \, {{1} \over {8}},
  \, {{3} \over {8}}, \, {{1} \over {2}}, \, {{1} \over {2}}, \, \, {{ 4\, x^2} \over {(1\, -3 \, x^2) \cdot \, (x^2\, -3) }}  \Bigr),              
\end{eqnarray}
which is a consequence of the identity on Belyi maps:
\begin{eqnarray}
  \label{Belyirelation}
  \hspace{-0.98in}&&    \quad \,\, \, \,
 {{-4\, x} \over { (1\, -x)^2}} \, \circ \, \,    {\frac {  {x}^{4}\cdot \, ({x}^{2} \, -3) }{ 1\, -3\,{x}^{2}}}
\, \,\,\,  = \, \, \,  \, \,
{{27} \over { 4}} \cdot \, {{x^2} \over {(1\, -x)^3 }}  \, \circ \, \,   {{ 4\, x^2} \over {(1\, -3 \, x^2) \cdot \, (x^2\, -3) }}. 
\end{eqnarray}

\vskip .1cm

\subsection{Other Heun functions solutions of telescopers of rational functions related to Shimura curves}
\label{othervanHoeijVidunas}

The rational function of four variables
\begin{eqnarray}
\label{Ratfonc4gg0OTHER}
  \hspace{-0.98in}&& \quad  \, \,  
\, \, 
 R(x, \, y, \, z, \, u)  \, \, \,  = \, \,  \quad
\nonumber \\
 \hspace{-0.98in}&&  \quad  \quad   \quad  \, \,    \quad  
  {{x\, y \, z \, u} \over {
u{x}^{2}{y}^{2}{z}^{2} \,  \, +u{x}^{2}yz \, +ux{y}^{2}z \, +uxy{z}^{2} \,  \, +uxy \, +uxz \, +uyz \, -xyz }}, 
\end{eqnarray}
has a telescoper that is a linear differential operator of order three:
\begin{eqnarray}
\label{telesHeunOTHER}
  \hspace{-0.98in}&&  \quad  \, \, \, \, \, 
 M_3 \, \, = \, \, \, -3 \, \, \,     +6 \cdot \,(1\,-7\,x\, +4\,{x}^{2}) \cdot \,x \cdot \, D_x
 \,\, \,  \, +12 \cdot \, (4\,-10\,x \, +3\,{x}^{2}) \cdot \, x^2 \, \cdot \, D_x^2
\nonumber \\
 \hspace{-0.98in}&&  \quad \quad  \quad    \quad  \quad  \quad \,\,\,
\, \, +8\, \cdot \, (x \, -1)  \cdot \, (x \, -4) \cdot \, x^3 \, \cdot \, D_x^3, 
\end{eqnarray}
which actually corresponds to the {\em symmetric square} of an order-two linear differential operator
readings in terms of the homogeneous derivative $\, \theta$:  
\begin{eqnarray}
\label{telesHeunL2OTHER}
\hspace{-0.98in}&&  \quad   \quad  
M_2   \,\, \, = \, \, \,   \,  \,    x^2 \cdot \, \theta^2
\,  \,  \, \, - \, 2 \, x  \cdot \, (14 \, \theta^2 \,  -7 \, \theta \, +1)
\, \, \, \, \, -8 \cdot \, (4\, \theta \, -1) \cdot \, (4\, \theta \, -3).           
\end{eqnarray}
The solutions of order three operator $\, M_3 \, \, $ are, thus,
expressed in terms of the following (square and product of) Heun functions:
\begin{eqnarray}
 \label{HeunShimuraOTHER}
  \hspace{-0.98in}&&  \,  \,  \quad  \, \, 
 x^{1/4} \cdot \, Heun\Bigl( 4, \, {{9} \over {64}}, \, {{1} \over {8}},
  \, {{5} \over {8}}, \, {{3} \over {4}}, \, {{1} \over {2}}, \, \, x \Bigr)^2,  \, \quad   \, \,  \,
  x^{3/4} \cdot \, Heun\Bigl( 4, \,  {{49} \over {64}}, \, {{3} \over {8}},
 \, {{7} \over {8}}, \, {{5} \over {4}}, \, {{1} \over {2}}, \, \,  x \Bigr)^2,
 \nonumber 
\end{eqnarray}
and
\begin{eqnarray}
 \label{HeunShimuraOTHER2}
  \hspace{-0.98in}&&   \,   \quad \quad \,  \,  \,
x^{1/2} \cdot \,
Heun\Bigl( 4, \, {{9} \over {64}}, \, {{1} \over {8}},\, {{5} \over {8}}, \, {{3} \over {4}}, \, {{1} \over {2}}, \, \, x \Bigr)
\cdot \,
Heun\Bigl( 4, \,  {{49} \over {64}}, \, {{3} \over {8}}, \, {{7} \over {8}}, \, {{5} \over {4}}, \, {{1} \over {2}}, \, \,  x \Bigr).
\end{eqnarray}
This order-two linear differential operator
$\, M_2 \, $ has the pullbacked $\, _2F_1$ solutions: 
\begin{eqnarray}
\label{Ratfonc4gg2hypOTHER}
 \hspace{-0.98in}  \quad   \,  \quad \,  \,  \quad   \quad   \, \,
  x^{3/8} \cdot \, (1\, -x)^{-7/8} \cdot \,
  _2F_1\Bigl([{{ 7} \over { 24}}, \, {{ 11} \over { 24}}], \, [{{ 5} \over {4}}],
  \,\,  \, {{- 27} \over { 4}} \cdot \, {{x} \over {(1\, -x)^3 }}\Bigr), 
  \\
 \hspace{-0.98in} \quad \,  \quad  \,\, \quad    \quad   \, \,
  x^{1/8} \cdot \,  (1\, -x)^{-1/8} \cdot \,   _2F_1\Bigl([{{ 1} \over { 24}}, \, {{ 5} \over { 24}}], \,
  [{{ 3} \over {4}}] ,\,\,  \,  {{- 27} \over { 4}} \cdot \, {{x} \over {(1\, -x)^3 }}\Bigr).
  \nonumber
\end{eqnarray}
reminiscent of (\ref{Ratfonc4gg2hyp}). One recovers the same (Shimura) $\, _2F_1$ hypergeometric function
as the one in  (\ref{Ratfonc4gg2hyp}), {\em but with another selected pullback}.
Similarly to the pullback in  (\ref{Ratfonc4gg2hyp}),
this last pullback  $\, \,  {{- 27} \over { 4}} \cdot \, {{x} \over {(1\, -x)^3 }}\, $ is also ``special'' as can be seen
in \ref{Kilianapp} with equations (\ref{suchthatxsplit}) and (\ref{suchthatxsplitLH}). 

These $\, _2F_1$ solutions (\ref{Ratfonc4gg2hypOTHER}) can also be rewritten\footnote[1]{All
  these $\, _2F_1$ hypergeometric functions are Shimura hypergeometric functions~\cite{Voight,Takeuchi}
  corresponding to $ \, (4, \, 4, \, 3)$ and  $ \, (6, \, 6, \, 2)$
  difference of exponents in Table (1) of~\cite{Voight,Takeuchi}.}
as 
\begin{eqnarray}
\label{Ratfonc4gg2hypOTHERrewritten}
  \hspace{-0.98in}  \,    \, 
  {\cal A}_1(x) \cdot \,
 _2F_1\Bigl([{{ 1} \over { 3}}, \, {{ 2} \over { 3}}], \, [{{ 5} \over {4}}],  \,\,  \, {\cal H}_1(x)\Bigr),  \quad \quad   
  {\cal A}_2(x) \cdot \,
  _2F_1\Bigl([{{ 1} \over { 12}}, \, {{ 5} \over { 12}}], \, [{{ 3} \over {4}}],  \,\,  \, {\cal H}_2(x)\Bigr), 
\end{eqnarray}
or
\begin{eqnarray}
\label{Ratfonc4gg2hypOTHERrewritten2}
  \hspace{-0.98in}  \,   \, 
  {\cal A}_3(x) \cdot \,
 _2F_1\Bigl([{{ 1} \over { 12}}, \, {{ 1} \over { 4}}], \, [{{ 1} \over {2}}],  \,\,  \, {\cal H}_3(x)\Bigr),  \quad \quad   
  {\cal A}_4(x) \cdot \,
 _2F_1\Bigl([{{ 7} \over { 12}}, \, {{ 3} \over { 4}}], \, [{{ 3} \over {2}}],  \,\,  \, {\cal H}_4(x)\Bigr), 
\end{eqnarray}
where the $\,  {\cal A}_i(x)$'s  are algebraic functions, and where the $\,  {\cal H}_i(x)$'s are algebraic functions
that can be simply expressed with square roots.

As far as, not Heun, but  $\, _2F_1$ hypergeometric functions related with {\em Shimura curves}
are concerned, several identities also appear in the litterature (see also \ref{identities}). Note
that the set of Gauss hypergeometric functions, or Heun functions, 
that are associated with Shimura curves {\em is a finite set}~\cite{Voight,Takeuchi}.

\vskip .2cm

{\bf Remark 1:}  There exist ``true'' Heun functions
(that {\em cannot be reduced to pullbacked $\, _2F_1$ hypergeometric
functions}) which correspond to {\em automorphic forms associated with a Shimura curve}.
One example comes from the order-two linear differential operator 
\begin{eqnarray}
  \hspace{-0.98in}&& \quad    \quad   \quad \,   \, \quad   
L_2 \, \,\,  = \, \,  \,\,  D_x^2
\, \, \, +   \,{\frac {12\,{x}^{4} \, -238\,{x}^{3} \, +3157\,{x}^{2} \, -3648\,x \, +2592}{
16 \cdot \, {x}^{2} \cdot \, (x-1)^{2} \cdot \, (2\,x-27)^{2}}}, 
\end{eqnarray}
which has the two  Heun solutions
\begin{eqnarray}
\label{trueHeun}
\hspace{-0.98in}&&
\quad    \quad \,  \, 
x^{1/3} \cdot \,  (1 \,\, -\, x)^{1/4} \cdot \,  (27 \,\, -2\, x)^{1/4} \cdot \,
Heun\Bigl({{27} \over {2}},\, {{7} \over {36}}, \, {{1} \over {12}}, \,  {{7} \over {12}}, \,  {{2} \over {3}}, \, {{1} \over {2}}, \, \, x\Bigr),
\nonumber \\
\hspace{-0.98in}&&
\quad    \quad \,  \, 
x^{2/3} \cdot \,  (1 \,\, -\, x)^{1/4} \cdot \,  (27 \,\, -2\, x)^{1/4} \cdot \,
Heun\Bigl({{27} \over {2}},\, {{47} \over {18}}, \, {{5} \over {12}}, \,  {{11} \over {12}}, \,  {{4} \over {3}}, \, {{1} \over {2}}, \, \, x\Bigr),
\end{eqnarray}
{\em We have not (yet ...) been able to see such ``true'' Heun functions as solutions of telescopers of rational functions}.
This Heun example (\ref{trueHeun}) has a  genus-zero {\em level three modular equation}
given in~\cite{ElkiesComput,YifanHecke} and in \ref{HeunTRUE}.

\vskip .3cm

{\bf Remark 2:} The Heun functions we found as diagonals of rational functions
or solutions of telescopers of  rational functions, 
were pullbacked $\, _2F_1$ hypergeometric functions which turn out to correspond to
{\em classical modular forms or (Shimura) automorphic forms}. In both cases
this means that the Heun function can be rewritten as a
$\, _2F_1$ hypergeometric function with, {\em not just one pullback},
but an {\em infinite number of pullbacks} (generated by the {\em modular
equations}, see for instance (\ref{ModularinftyL2inftysolident2first})). 
Is it possible for a  Heun function to correspond to a
{\em globally bounded series} and to reduce to
pullbacked $\, _2F_1$ hypergeometric function
with a rational or algebraic pullback, without being automatically
a  classical modular forms  ? 
In \ref{ReduceMaierHeun}  we show that a Heun function can actually correspond
to a globally bounded series, being reducible to a pullbacked $\, _2F_1$ hypergeometric function,
without necessarily corresponding to a classical modular form.
We have not yet been able to find such Heun functions as diagonal
of rational functions, or even, as solutions of telescopers of rational functions.

\vskip .2cm

{\bf Remark 3:} More  rational functions yielding Heun functions for their diagonals can be obtained
using\footnote[2]{Or, more generally monomial transformations. } the
$ \, (x, \, y, z, \, u) \, \, \rightarrow \, \, (x^n, \, y^n, \, z^n, \, u^n)$
transformation for {\em positive} integers $\, n$. The case where the integer $\, n$ is negative,
in particular $\, n=-1$, is different and sketched in  \ref{Duality}.

\vskip .2cm 
\vskip .2cm 

\section{Conclusion}
\label{Conclusion}

The examples of diagonals  of  rational functions in three or four variables,
that we presented here, illustrate cases where the
diagonal of the rational functions are given by  Heun functions with
{\em integer coefficients series}, and can be expressed either in terms of  pullbacked
hypergeometric functions that are {\em classical modular forms}, or  {\em  derivatives} of
classical  modular forms. Furthermore, we
constructed in subsection \ref{subthree2}, a rational function whose diagonal is given by a
Heun function that has already been
identified as a ``Period'' of an {\em extremal rational elliptic surface}~\cite{Malmendier4}, and
that has also emerged in the context of pullbacked $\, _2F_1$ hypergeometric functions~\cite{Maier1}.
The emergence of squares of Heun functions for most of the diagonals of rational functions of
this paper, suggests a
{\em ``Period'' of algebraic surfaces (possibly product of elliptic curves) interpretation}.
The exact expressions of the diagonal of rational functions in this paper, or in previous
papers~\cite{DiagSelected,unabri,notunabri},  are always obtained using the creative telescoping
approach, being globally bounded series, solutions of the telescopers of these rational functions. 
Finally we have also seen a case where the rational function has a telescoper with  Heun function
solutions, that can be expressed as  pullbacked $\, _2F_1$ hypergeometric functions that are
{\em not} globally bounded, and happen to be associated with one of the 77 cases of
{\em Shimura curves}~\cite{Takeuchi}. Such  remarkable  $\, _2F_1$ hypergeometric functions 
solutions of a telescoper of a rational function are {\em not} diagonals of that rational function
(the series are not  globally bounded). They can be interpreted as ``Periods''~\cite{KontZagier,Zagier}
of an algebraic variety over some non-evanescent\footnote[1]{Diagonals are periods over evanescent cycles.} cycles.
With these  $\, _2F_1$ Shimura examples one sees clearly that 
solutions, analytic at $\, x \, = \, 0$, of telescopers of  rational functions  are {\em not}
diagonals of these rational functions.

All these examples seem to suggest an {\em algebraic geometrical} link between the diagonals/solutions
of the telescopers, and the original rational functions, and this link should be investigated. This
study should help shed light on the geometrical nature of the algebraic varieties associated with
the denominators of the rational functions (K3, Calabi-Yau threefolds, extremal rational elliptic
surfaces, Shimura varieties).
In a forthcoming paper which is a work in progress at the current stage, we intend to introduce
an {\em algebraic geometry approach} that proves to be efficient in explaining this link, in the cases
where the  order-two linear differential telescopers of the rational functions
or the diagonals of rational functions
are related to {\em classical modular forms}.

\vskip .2cm
\vskip .3cm 

{\bf Acknowledgments.}
Y. A. and J-M. M. would like to thank the Stat. and Math. Department of Melbourne University for hospitality,
where part of this paper was written. Y. A. and J-M. M. would like to thank A. J. Guttmann for numerous discussions
on lattice Green functions. Y. A. would like to thank A. Malmendier for several enlightening discussions
on the periods of rational extremal surfaces. Y. A. would like to thank Dahlia Abdelaziz, Rashida Abdelaziz and
Siwa Abdelaziz for the great time spent together while working on this paper, and for being the joyful and loving cousins they are.
Y. A. would like to thank Mark van Hoeij for the explanations he provided him on Heun to Hypergeometric functions pullbacks.
Y.~A. would like to thank John Voight for enlightening explanations on Shimura curves 
and for pointing out the reference \cite{Takeuchi}.
Y. A. would like to thank Yifang Yang for several very explanatory emails on automorphic functions and Shimura curves.
Y. A. would like to thank Wadim Zudilin for an enlightening exchange on the modular parametrization of hypergeometric functions.
J-M. M. would like to thank Pierre Charollois for many enlightening discussions on modularity and automorphic forms.
J-M. M. would like to thank Raschel Kilian for enlightening discussions on walks on the quarter of a plane.
S. B. would like to thank the LPTMC and the CNRS for kind support. We thank the Research Institute for Symbolic Computation,
for access to the RISC software packages. We thank M. Quaggetto for technical support.


\appendix

\section{Trivialization cases of Heun functions}
\label{trivia}

We already encountered in~\cite{Youssef} (see section 3 equations (47), (48) and (C.12) in~\cite{Youssef})
an interesting example of Heun function solution of an order-two linear differential operator
which {\em factorises into two order-one operators}. The Heun function 
\begin{eqnarray}
\label{HeunYoussef}
\hspace{-0.98in}&& \quad \quad  \quad \quad \quad \quad \,\,
\Phi(x) \, \, = \, \,  \,  \,  x^{1/2} \cdot \,
Heun\Bigl(M, \, {{M+1} \over {4}}, \,   {{1} \over {2}}, \,  1,  \,   {{3} \over {2}}, \,     {{1} \over {2}}, \, \,   x\Big),                   
\end{eqnarray}
satisfies the identity:
\begin{eqnarray}
\label{HeunYoussef2}
\hspace{-0.98in}&& \quad \quad \quad \quad \quad \quad  \,\,
\Phi\Bigl( {{4 \cdot \, x \cdot \, (1\, -x)  \cdot \, (1 \, -x/M)} \over { (1 \, -x^2/M)^2}}    \Bigr)
 \, \,\, = \, \, \, \,  2 \cdot \, \Phi(x).  
\end{eqnarray}
For $\, M\, = 2, \, 1/2, \, -1$,  the Heun function (\ref{HeunYoussef})
can be written as pullbacked $\, _2F_1$ hypergeometric functions.
Let us recall, in the next subsections, the more general results of Ronveaux~\cite{Ronveaux,RonveauxLame}.

\subsection{Factorization cases of the  order-two operator for Heun functions}
\label{facto}

Table 1 page 181 in~\cite{Ronveaux} gives a set of six cases
for which the order-two linear differential operator factors into the product
of two order-one linear differential operators.
Let us recall the conditions of  Table 1 in~\cite{Ronveaux}   to get a factorization of the
order-two Heun linear differential operator. The order-two Heun linear differential operator reads: 
\begin{eqnarray}
  \label{Heunapp}
  \hspace{-0.98in}&& \quad \quad \,\,
  H_2  \,\, = \, \, \, \, \,D_x^2 \,\,\, 
+   \Bigl({{ \gamma} \over { x}}
\, + {{ \delta} \over { x \, -1}} \, + {{ \epsilon} \over { x \, -a}}\Bigr) \cdot \, D_x \,\, \, 
+ \,  {{ \alpha\, \beta \,x \, -q } \over { x \cdot \, (x-1)  \cdot \, (x \, -a)}}. 
\end{eqnarray}               
where one  has the Fuchsian constraint
$\, \,  \epsilon   \, = \, \, \alpha \, + \, \beta \, -\gamma \, - \, \delta \, + \, 1$.
One can easily verify that the order-two Heun linear differential operator (\ref{Heunapp})
factorizes into two order-one linear differential operators when 
\begin{eqnarray}
\label{VI}
 \hspace{-0.98in}&& \quad \quad \quad \quad \quad \quad \quad \quad\quad 
 q \, \,= \, \, \, \, \, a \, \delta \cdot \, (\gamma \, -1) \,\,\, +\epsilon \, +\gamma \, \, -2
\nonumber \\
  \hspace{-0.98in}&& \quad \quad \quad \hbox{and:} \quad  \quad \quad \quad \quad \quad \quad \quad 
(\gamma \, +\epsilon \, -2) \cdot \, (\delta+1) \,\, \,   = \,\,  \,\, \alpha \, \beta
\end{eqnarray}
i.e. after using the  Fuchsian constraint:
\begin{eqnarray}
\label{VIbis}
 \hspace{-0.98in}&& \quad \quad \quad \quad  \quad \quad \quad \quad \quad 
 q   \,\, = \, \,\, \,\,   a \, \delta \cdot \, (\gamma \, -1) \,\,\, +\alpha \, + \beta \, -\delta \,\,  -1,
\nonumber \\
\hspace{-0.98in}&& \quad \quad \quad  \hbox{and:}  \quad \quad \quad \quad  \quad  \quad  \quad
(\alpha \, +\beta -\delta \, -1) \cdot \, (\delta \, +1) \,  \, \,  = \, \,  \,\,  \alpha \, \beta.
\end{eqnarray}
One has the following factorization $\,  \,  H_2  \, = \, \, L_1 \cdot \, M_1\,  \, $ where:
\begin{eqnarray}
\label{factoVIbis}
 \hspace{-0.98in}&& \quad \quad \quad \quad 
L_1 \, \, = \, \, \, \, (x\,-1)\cdot \, D_x \, \, \, +\delta, \quad \, \,
\nonumber \\
\hspace{-0.98in}&& \quad \quad \quad \quad 
M_1 \, = \, \, \, x \cdot \, (x\, -a)\cdot \, D_x \, \, 
\, +(\gamma \, +\epsilon \, -2) \cdot \, x \,\, \,  +a \cdot \, (1-\gamma) 
 \\
\hspace{-0.98in}&&\quad \quad  \quad \quad \quad \quad \quad  
\, \, = \, \,  \,  x \cdot \, (x-a)\cdot \, D_x
\, \, \,\, +\,   (\alpha \, +\beta \, -\delta \, -1)  \cdot \, x
\, \,  \, + \, a \cdot \, (\gamma \, -1).
\nonumber 
\end{eqnarray}
The case VI in Table 1 of Ronveaux~\cite{Ronveaux},  $\, \,  \alpha \, = \, \, \delta \, +1 $, and thus
$\, \,  \beta \, = \, \, \gamma \, +\epsilon \, -2$ (since one  has the Fuchsian constraint
$\,  \,  \epsilon   \, = \, \, \alpha \, + \, \beta \, -\gamma \, - \, \delta \, + \, 1$),
corresponds to this case.
Along this line, let us consider the Heun function\footnote[1]{For
  $\, a=\, 9$, $\, \beta \, =\,\gamma \, =\,   \delta  \, =\,  1$,
the Heun function  (\ref{VIbisHeun})
is the simple rational function
$\,\, Heun(9,1,2,1,1,1,\, 27\, x) \, = \, \,  1/(1\, -3\, x)$.}:
\begin{eqnarray}
\label{VIbisHeun}
  \hspace{-0.98in}&& \quad \quad \quad \quad \quad \quad
Heun\Bigl(a, \,\,  a  \, \, \delta \cdot \, (\gamma \, -1) \, \, + \beta, \,\,
  \delta \, +1, \,\, \beta, \, \, \gamma, \,\, \delta, \, \, \, \, \, x \Bigr).          
\end{eqnarray}
The previous factorisation (\ref{factoVIbis}) yields
\begin{eqnarray}
\label{VIbisHeunyields}
  \hspace{-0.98in}&& \, \, \, \,  \, \, 
M_1\Bigl[ Heun\Bigl(a, \,\,  a  \, \, \delta \cdot \, (\gamma \, -1) \, \, + \beta, \,\,
\delta \, +1, \,\, \beta, \, \, \gamma, \,\, \delta, \, \, \, \, \, x \Bigr)\Bigr]
\, \, \, = \, \, \,  \,  \lambda \cdot \,   (1\, -x)^{-\delta},       
\end{eqnarray}
the RHS of (\ref{VIbisHeunyields}) being an algebraic function when the parameter $\, \delta$ is a rational number.
The LHS of (\ref{VIbisHeunyields}) {\em can thus be written as the diagonal of a rational function}~\cite{Fu,Denef}.
The Heun function (\ref{VIbisHeun}) is not necessarily the diagonal of a rational  function
but the order-one operator $\, M_1$ acting on that Heun function is a diagonal\footnote[5]{We have,
  of course, a similar result for the previous Heun function (\ref{HeunYoussef}).} of a
rational function~\cite{Fu,Denef}. This Heun function (\ref{VIbisHeun})  is {\em locally bounded}. 
For instance $\,\, Heun(9,92,6,2,3,5, \, 9\, x)\,$ is {\em not globally bounded}, but the series expansion of
$\, (\theta \, +2)[Heun(9,92,6,2,3,5, \, 9\, x)]$ is actullay globally bounded (it is a series with
{\em integer coefficients} corresponding to the expansion of an algebraic function):
\begin{eqnarray}
\label{VIbisHeunyieldslocally}
\hspace{-0.98in}&& 
x \cdot \, {{d   Heun(9,92,6,2,3,5, \, 9\, x)} \over { dx}}
\, + 2 \cdot \,  Heun(9,92,6,2,3,5, \, 9\, x)  \, \, = \, \, \, 2 \, \, +92\,x \, +2522\,{x}^{2}
\nonumber \\
\hspace{-0.98in}&& \quad   \quad   \quad   \quad   \quad  \, \, \, \, \,
  \, +53552\,{x}^{3} \,  \, +972092\,{x}^{4} \,  \, +15852440\,{x}^{5} \, 
 \, +239057660\,{x}^{6}  \,  \, \,\, + \, \, \, \cdots
 \nonumber \\
\hspace{-0.98in}&& \quad \quad \quad  \quad   \quad  \quad  \quad   \quad  
\, \, = \, \, \, \, {{2} \over {(1\, -x) \cdot \, (1\, -9 \, x)^5}}.
\end{eqnarray}

\vskip .2cm

$\, \bullet$  For  the parameters $\, a \, = \, 9$,
$\, q\, = \, 3$, $\, \alpha \, =1/2$, $\, \beta \, = \, 1$,
$\gamma$ and $\, \delta$ being deduced from (\ref{Fuchscondi}), (\ref{apparentcondi2}), 
one finds that
\begin{eqnarray}
\label{Letsconsider}
\hspace{-0.98in}&&  \,  \,
 Heun\Bigl(9, \, 3, \, 1/2, \, 1, \, 40, \, - \, {{73}   \over {2}}, \,\, x\Bigr)
 \,  \, = \, \,  \,\, 1 \, \,\,\,  +  {{1} \over {120}} \cdot \, x  \,\,\, 
 + \,{{1} \over {6560}}  \cdot \, x^2  \, \, \,   -{{1} \over {1353984}} \cdot \, x^4 
\nonumber \\
\hspace{-0.98in}&& \quad \quad \quad \,  \,
 \, -{{3} \over {19858432}}   \cdot \, x^5 \,\, \, -{{1} \over {36106240}} \cdot \, x^6
\, \, \,- \, {{13} \over {2491330560}} \cdot \, x^7 \, \, \, \, \, + \, \, \cdots 
\end{eqnarray}
is an {\em algebraic function}. Let us consider the Laurent series expansion
of the algebraic function $\,  {\cal A}(x)$
\begin{eqnarray}
\label{Letsconsider}
  \hspace{-0.98in}&&  \, \quad
{\cal A}(x) \,\, = \, \, \, \, 
 {{2^{73}} \over {729027183996402643275 }} \cdot \,  {{ (1\, -x)^{75/2} \cdot \, (x \, -12) } \over {x^{39}}}
  \nonumber \\
\hspace{-0.98in}&&  \, \quad  \quad   \quad  
 \,  \, = \, \,  \,  \, - \, {{ 2^{75} } \over { 243009061332134214425 \, \, x^{39} }}
 \,\,  \, \,\, + \, \cdots \, \, \, \,\,
-{{292448} \over {5 \, x^3}} \, \, +{{9880} \over {3\, x^2}} \, \, \, -{{104} \over {x}}
 \nonumber \\
\hspace{-0.98in}&&  \, \quad  \quad  \quad \quad  \quad  \quad
\, +1 \,\, \, +{{1} \over {120}} \cdot \, x  \, \,
+ \, {{1} \over {6560}}  \cdot \, x^2  \,  \, \,    -{{1} \over {1353984}} \cdot \, x^4
 \, \, \, \,  \,\,  + \, \, \, \cdots  
\nonumber \\
  \hspace{-0.98in}&&  \, \quad  \quad  \quad \
\,  \,      \, = \, \,  \,  \,  \,
PP({\cal A}(x)) \, \, \, + \, Heun\Bigl(9, \, 3, \, 1/2, \, 1, \, 40, \, -{{73} \over {2}}, \,\,  x\Bigr),
\end{eqnarray}
where $\, PP({\cal A}(x))\, $ denotes the principal part (negative powers) of the
Laurent expansion of the algebraic function $\,  {\cal A}(x)$.
Note that the Heun function $\, Heun\Bigl(9, \, 3, \, 1/2, \, 1, \, 40, \, -73/2, \, x\Bigr)$
is solution of an order-two  linear differential operator
$\, L_2$, which is the direct-sum (LCLM)
$\, \, L_2 \, = \, \, L_1 \oplus \, M_1\,\, $ of two order-one linear differential 
operators. One order-one linear differential operator 
\begin{eqnarray}
\label{L1M1}
\hspace{-0.98in}&&  \, \quad \quad \quad \quad \quad \quad \quad
L_1 \,\, = \, \, \, \,  D_x \,\, \,\, 
+ \, {{ 3} \over {2}} \, {{ x-26} \over {(x-1) \cdot \, x}}
\,  \, \, \, \,  - \, \, {{ 1} \over {x\, -\, 12}}, 
\end{eqnarray}
has the algebraic function $\, {\cal A}(x)$ as solution,  
and the other one $\, M_1$ is a quite large  order-one linear differential
operator having the rational function $\,  PP({\cal A}(x)) \, = \, P(x)/x^{39}\, \, $
given in (\ref{Letsconsider}) as solution.

\vskip .2cm

$\, \bullet$  For other values of the parameters $\, a \, = \, 9$,
$\, q\, = \, 3$, $\, \alpha \, = \, 1/2$, $\, \beta \, = \, 1$,
$\gamma$ and $\, \delta$ being deduced from (\ref{Fuchscondi}), (\ref{apparentcondi2}), 
one finds that the Heun function
\begin{eqnarray}
\label{L1L2Heunbisrelation}
\hspace{-0.98in}&&  \,   \,   \,   \,   \,   \,  \,  
Heun(9, \,  3, \,  1, \,  1,  \, 37,  \, -33, \, \, x) \,  \, \, = \, \,  \,  \,  \, \,
  1 \,  \, \,\,  + \, {{ x} \over {111}} \,\, \,\, - \,{{ x^2} \over {2109 }} \, \, \,  \,
 - \,{{ x^3} \over {9139 }}  \, \, \, \, -\,{{ x^4 } \over {54834}}
\nonumber \\
 \hspace{-0.98in}&&  \,   \quad   \,  \quad  \quad  \quad \, \, 
  \, - \,{{ 7\, \, x^5 } \over {2248194}} \, \, \,
 - \, {{3 \, \, x^6} \over {5245786  }} \,\, \,  - \,  {{ 11 \, \, x^7} \over { 96672342}}
\,  \, \, - \,  {{x^8} \over {40899837 }} \,  \, \, \, \, \, + \, \, \cdots 
\end{eqnarray}                 
is solution of an order-two linear differential operator $\, L_2$ which 
factorises  in the product of two order-one operators
$\,\, L_2 \, = \, \, N_1 \cdot \, P_1\,\, $ where:
\begin{eqnarray}
\label{L1L2Heunfact}
\hspace{-0.98in}&& \quad \quad \quad \quad \quad \,  \,
N_1 \, \,\, = \, \, \, \,   (x-9) \cdot  \, D_x \, \,  \, \, - \, 1,
\nonumber \\
\hspace{-0.98in}&& \quad \quad \quad \quad \quad \,  \,
P_1  \, \,\, = \, \, \, \,  \, 
 x \cdot \, \, (x\, -1) \cdot \, (x\, -15)  \cdot \, D_x \, \,  \, \, + x^2 \, -65\, x \, +540.
\end{eqnarray}
where the order-one linear differential operator $\, P_1$ has the rational function solution: 
\begin{eqnarray}
\label{L1L2Heunbis}
\hspace{-0.98in}&& \quad  \quad  \quad \quad  \quad  \quad  \quad \,  \,
{\cal R}(x) \, \, = \, \, \, {{ (x \, -15) \cdot \, (x \, -1)^{34} } \over {x^{36}}}.           
\end{eqnarray}
One deduces:
\begin{eqnarray}
\label{thereforeL1L2Heunbis}
\hspace{-0.98in}&& \quad \quad   \,  \,
 x \cdot \, \, (x\, -1) \cdot \, (x\, -15)
 \cdot \, {{d} \over {dx}}\Bigl(Heun(9,\, 3,\, 1,\, 1,\, 37,\, -33, \,\, x) \Bigr)
\nonumber \\
\hspace{-0.98in}&& \quad \quad  \quad \quad   \,  \,
 \, \, + (x^2 \, -65\, x \, +540) \cdot \,Heun(9, \,  3, \,1,\, 1,\, 37, \,-33,\, \, x)
 \, \, \, +  60 \cdot \, (x-9)
\nonumber \\
\hspace{-0.98in}&& \quad \quad \quad  \,  \,\, \, = \, \, \, \,
P_1\Bigl(Heun(9, \,  3, \,  1, \,  1,  \, 37,  \, -33, \, \,  x)\Bigr)
\, \, \,  \,   +  60 \cdot \, (x-9)    
  \, \, \,\,  = \, \, \, \, \, 0. 
\end{eqnarray}
Note that the series (\ref{L1L2Heunbisrelation}) is {\em not globally bounded}
but  $\, P_1(Heun(9, 3, 1, 1, 37, -33, x))$ is
{\em globally bounded}\footnote[9]{A situation we already encountered~\cite{Big}.}.

\vskip .1cm

$\, \bullet$  For other values of the parameters $\, a \, = \, 10$,
$\, q\, = \, 3$, $\, \alpha \, = \, 1/2$, $\, \beta \, = \, 3/2$,
$\gamma$ and $\, \delta$ being deduced from (\ref{Fuchscondi}), (\ref{apparentcondi2}), 
one finds that the corresponding Heun function
$\, Heun(10,3,1/2,3/2,81/2,-73/2,x)$ is solution of  an order-two linear differential operator
\begin{eqnarray}
\label{Bigfacto}
\hspace{-0.98in}&& \quad \quad \quad \quad \, \,  \quad \,  \,
L_2 \, \, \, = \, \, \,   \,   4\, \, x \cdot \, (x-1)\cdot \, (x-10) \cdot \, D_x^2
\nonumber \\
\hspace{-0.98in}&& \quad \quad \quad  \quad  \quad \quad \, \,   \quad  \quad \quad \,  \,
\, +6\cdot \, (2\, x^2 -53\, x \, +270) \cdot \, D_x
\,  \, \, \, +3 \cdot \, (x-4), 
\end{eqnarray}
which {\em factorises}\footnote[1]{The fact that the second order linear differential operator
associated with the Heun function factorises is noticed in the Remark of page 15 of~\cite{MaierHeunG}.}
in the product of two order-one  linear differential operators
$\,\, L_2 \, = \, \, N_1 \cdot \, P_1$,  
where the order-one linear differential operator  $\, P_1$ has an algebraic solution of the form
$\, x^{-79/2} \cdot \, P_{38}(x)$,  where $\,  P_{38}(x)$ is a polynomial of degree $\, 38$,
when the order-one  linear differential operator  $\, N_1$ has an algebraic solution of the form
$\, (1\, -x)^{73/2} \cdot \, (x\, -10)/x/P_{38}(x)$, with the {\em same} polynomial $\,  P_{38}(x)$.
This polynomial of degree $\, 38$ is solution of the non-linear ODE:
\begin{eqnarray}
\label{nonlin}
  \hspace{-0.98in}&& \quad  \quad  \quad 
2  \cdot \,    (152\,{x}^{2}-1579\,x+770) \cdot \, {{P'(x)} \over {P(x)}}
\, \,\,\, \,\,  -  \, 4 \cdot \, x \cdot \, (x \, -1)
\cdot \,  (x \, -10)   \cdot \,    {{P''(x)} \over {P(x)}}
\nonumber \\
\hspace{-0.98in}&& \quad \quad \quad \quad \quad \quad \quad \quad
\, \, \, = \, \, \, \, \,  24 \cdot \, (247\, x \, -2410).
\end{eqnarray}
Besides this selected polynomial solution $\, P_{38}(x)$, the  solutions of  (\ref{nonlin})
can be expressed in terms of Heun functions like
$\, Heun(10, 84, 5/2, 7/2, 81/2, -73/2,  \,  \,  x) \, $ or
$\, Heun(10, 14383, -37, -36, -77/2, -73/2,  \,  \,  x)$,  which are
{\em not globally bounded} series\footnote[5]{These Heun functions
 are not diagonals of rational functions, they are just solutions
 of the non-linear ODE (\ref{nonlin}).}.

\vskip .2cm


\vskip .2cm
  
\subsection{Heun functions where the fourth singularity is an apparent singularity}
\label{apparent}

Let us recall some results of~\cite{MaierHeunG}. 
Let us consider a Heun function $\, Heun(a,  \, q, \, \alpha, \, \beta, \, \gamma, \, \delta, \, x) \, $
where\footnote[2]{i.e. $\, \epsilon \, = \, \, -1$
  see the definition of $\,\, \epsilon \, $ in (\ref{Heun}) or (\ref{Heunapp}).}:
\begin{eqnarray}
\label{Fuchscondi}
\hspace{-0.98in}&& \quad \quad \quad \quad \quad \quad  \quad \quad  \quad \quad 
\,  \delta \, \,   \,  = \, \,  \,   \, \alpha \, \,  + \, \beta  \, \, -\gamma \, \, \,  \,  +2.
\end{eqnarray}        
The fourth singularity $\, a$ will be an apparent singularity when
\begin{eqnarray}
  \label{apparentcondi}
  \hspace{-0.98in}&& \quad \quad \quad \quad \quad \quad \quad 
q^2 \, \, \,
 + \, \Bigl( (\gamma\, -1) \, -(2\, \alpha\, \beta \, + \,  \alpha\, +\, \beta) \cdot \, a\Bigr) \cdot \, q
\nonumber \\
  \hspace{-0.98in}&& \quad \quad \quad  \quad \quad  \quad \quad  \quad \quad \quad 
\, + \, \alpha \, \beta \, \, a \cdot \,\Bigl((\alpha \, +1) \cdot \, (\beta\, +1) \cdot \, a \,\,  -\, \gamma  \Bigr)
\, \, \, = \, \,\,  \, 0,
\end{eqnarray}               
or:
\begin{eqnarray}
  \label{apparentcondi2}
  \hspace{-0.98in}&&  \quad  
\gamma \, \, \, = \, \, \,  \, 
  {{ a^2 \, \alpha \, \beta \cdot \, (\alpha \, +1) \cdot \, (\beta \, +1) \,
    \, - \, a \, q \cdot \, (2\, \alpha\, \beta \, + \,  \alpha\, +\, \beta )
\,  \, + \, q \cdot \, (q \, -1)
 } \over { \alpha \, \beta \, a \,\, -q  }}.
\end{eqnarray}
This condition (\ref{apparentcondi}) can also be rationally parametrized as:
\begin{eqnarray}
\label{apparentcondi3}
\hspace{-0.98in}&& \quad \quad \,  \,  \, \quad  \,
a \, \, = \, \, \,
 {{ e \cdot \, (e\, -\gamma \, +1)} \over {  (e\, -\alpha) \, (e\, -\beta)}}, \, \, \quad \quad \quad 
 q \, \, = \, \, \,
   \alpha  \, \beta \cdot \, {{ (e \, +1) \cdot \, (e\, -\gamma \, +1)} \over {  (e\, -\alpha) \, (e\, -\beta)}}. 
\end{eqnarray}
Introducing the order-three linear differential operator
$\, \, L_3 \, = \, \, x^3 \cdot \, (x \, -1) \cdot \, L_1 \cdot \, L_2 \,\,  $ where 
\begin{eqnarray}
  \label{L1L2}
  \hspace{-0.98in}&&     \quad  \quad  \quad  \quad 
L_1  \, \, = \, \,\,\, 
D_x \, \, \, \,  + {{ e+1} \over { x}} \,\, \,   + {{ 1} \over { x \, -1}} \,\,  \,  + {{ 1} \over { x \, -a}},
\end{eqnarray}
and where $\, L_2$ is the previous Heun operator (\ref{Heunapp}) for $\, \epsilon \, = \, -1$
\begin{eqnarray}
  \label{L1L2L2}
  \hspace{-0.98in}&&     \quad  \quad 
L_2  \, \, = \, \, \, \,  D_x^2 \,\, \,  \, 
+  \Bigl({{ \gamma} \over { x}} \, + {{ \delta} \over { x \, -1}}
 \, - {{ 1} \over { x \, -a}}\Bigr) \cdot \, D_x \,\,  \, \, 
+  {{ \alpha\, \beta \,x \, -q } \over { x \cdot \, (x-1) \cdot \, (x \, -a)}},   
\end{eqnarray}
one finds that the order-three  linear differential operator $\, L_3$ reads:
\begin{eqnarray}
  \label{L1L2L3}
  \hspace{-0.98in}&&
\, \, \, \,  L_3 \, \, = \, \, \, \, (x \, -1)  \cdot \,  x^3 \cdot \, D_x^3
 \, \,  \,  \,  \, +\Bigl((\alpha+\beta+e+4) \cdot \, x \, \, \,  -e-\gamma-1\Bigr) \cdot \, x^2 \cdot \, D_x^2
\\
\hspace{-0.98in}&&   \,\, \,  \,   \quad                \,
+\Bigl(\Bigl(\alpha\, \beta \,  \, + (\alpha\,  +\beta \, +1) \cdot (e+2)\Bigr)
 \cdot \, x \, \,  -e\, \gamma \Bigr) \cdot \, x \cdot \,  D_x
\,\,\,  + \alpha \, \beta\, (e+1) \cdot \, x,
\nonumber
\end{eqnarray}
which, in term of the homogeneous derivative $\, \theta \, = \, \, x \, D_x$, can be written in a more compact way:
\begin{eqnarray}
  \label{L1L2L3theta}
  \hspace{-0.98in}&& \, \,  \,
 L_3 \, \, \, = \, \, \, \, 
  (\theta \,+e+1) \cdot \, (\theta \,+\beta) \cdot \, (\theta \,+\alpha) \cdot \, x \,  \,\, \,
  -\theta \cdot \, (\theta \, +\gamma-1)\cdot \, (\theta \, +e-1).
\end{eqnarray}
This last expression (\ref{L1L2L3theta}) shows, very clearly, that $\, L_3$ corresponds
to a $\, _3F_2$ hypergeometric function (see page 16 of~\cite{MaierHeunG}):
\begin{eqnarray}
  \label{3F2}
  \hspace{-0.98in}&& \quad \quad \quad \quad \quad \quad  \quad \quad \quad \,  \,
  _3F_2\Bigl([\alpha, \, \beta, \, e \, +1], \, [\gamma, \, e], \,\, x\Bigr).
\end{eqnarray}
The singularity $\, x \, = \, a\, \, $ in the  Heun linear differential  operator (\ref{L1L2L2})
is an {\em apparent singularity}. The head polynomial of (\ref{L1L2L3})
does not have this apparent singularity $\, x \, = \, a$.
This apparent singularity can thus be removed introducing
a (non minimal) {\em higher order} linear differential  operator $\, L_3$: this is called~\cite{Chen,Barkatou,Barkatou2} the
{\em desingularization}\footnote[1]{Often the desingularization removes the apparent singularities but creates
  unpleasant irregular singularities (for instance at $\, x \, = \, \infty$). Such desingularization destroys the
  Fuchsian character of the original linear differential operators. In physics one is interested in
  {\em desingularization preserving the  Fuchsian character of the linear differential operators}. This is precisely the case here.}
of the linear differential  operator $\, L_2$.
Introducing the order-two linear differential operator $\, M_2$ having the $\, _2F_1$ hypergeometric solution
$\, _2F_1([\alpha, \, \beta], \, [\gamma], \,\, x)$, one finds that  the  Heun linear differential  operator (\ref{L1L2L2}),
and  $\, M_2$ are actually homomorphic:
\begin{eqnarray}
  \label{3F2homo}
  \hspace{-0.98in}&& \quad \quad \quad \quad \, \, 
 L_2 \cdot \, (\theta \, + \, e)    \, \, \, = \, \, \, \,
 \Bigl(\theta \,  \,  \, + e\, +2 \, \, + {{1} \over {x-1}} \, - \, {{a} \over { x\, -a}}  \Bigr)      \cdot \, M_2.       
\end{eqnarray}
Consequently, as far as series expansions at $\, x \, = \, 0$ are concerned, the Heun function\footnote[2]{When $\, \gamma$
  is not an integer, the series expansions
 of these Heun functions (\ref{3F2homomany}) are not generically globally bounded.}
such that its parameters verify (\ref{Fuchscondi}) and (\ref{apparentcondi3}), can be written in several ways: 
\begin{eqnarray}
  \label{3F2homomany}
  \hspace{-0.98in}&& \quad \quad \quad \quad \quad 
 Heun(a,  \, q, \, \alpha, \, \beta, \, \gamma, \, \delta, \, x) \, \, = \, \, \,
 _3F_2\Bigl([\alpha, \, \beta, \, e \, +1], \, [\gamma, \, e], \,\, x\Bigr)
\nonumber \\
  \hspace{-0.98in}&& \quad  \quad \quad \quad \quad \quad \quad \quad  \quad 
\, \, = \, \, \,
 {{1} \over {e}} \cdot \, (\theta \, + \, e) \Bigl[ \,  _2F_1([\alpha, \, \beta], \, [\gamma], \,\, x) \Bigr].               
\end{eqnarray}

\vskip .3cm

\section{Alternative Heun functions for the simple cubic lattice Green function.}
\label{Alternative}

Note that such a Heun function, like (\ref{SolSCser}), can also be written as a
$\, Heun(9, 3, 1, 1, 1, 1, \, {\cal A}(x)) \, $  function, where $\, \,  {\cal A}(x))\, $
is an algebraic function, using the identity~\cite{MaierHeunG}
\begin{eqnarray}
\label{SolSCser1zero}
\hspace{-0.98in}&& \quad 
Heun\Bigl(9, \, 3, \, 1, \, 1, \, 1, \, 1, \, \, x\Bigr) \, \,   = \, \,   \,
\\
 \hspace{-0.98in}&&  \quad  \quad  \quad   \quad  \quad 
\Bigl( 1\, -{{x^2} \over {9}} \Bigr)^{-1/2}
 \,  \cdot \, Heun\Bigl(9, \,  {{3} \over {4}}, \,   {{1} \over {4}}, \,
 {{3} \over {4}}, \, 1, \,   {{1} \over {2}},   \, \,
 {{ 36 \cdot \, x \cdot \, (1\, -x)  \cdot \, (9\, -x) } \over { (9\, -x^2)^2}}  \Bigr),
 \nonumber
\end{eqnarray}
or
\begin{eqnarray}
\label{SolSCser1zero1}
\hspace{-0.98in}&& \quad 
Heun\Bigl({{1} \over {9}}, \, {{1} \over {3}}, \, 1, \, 1, \, 1, \, 1, \, \, x\Bigr) \, \,   = \, \,   \,
\nonumber \\
  \hspace{-0.98in}&&  \quad  \quad  \quad   \quad  \quad 
\Bigl( 1\, - 9 \, x^2 \Bigr)^{-1/2}
 \,  \cdot \, Heun\Bigl({{1} \over {9}}, \,  {{1} \over {12}}, \,   {{1} \over {4}}, \,
 {{3} \over {4}}, \, 1, \,   {{1} \over {2}},   \, \,
 {{ 4 \cdot \, x \cdot \, (1\, -x)  \cdot \, (1\, -9\, x) } \over { (1 \, -9\, x^2)^2}}  \Bigr),
\nonumber
\end{eqnarray}
which is a special case of: 
\begin{eqnarray}
\label{SolSCser1}
\hspace{-0.98in}&& \quad \quad \quad 
Heun\Bigl(a, \,q, \, 1, \, 1, \, 1, \, 1, \, \, x\Bigr) \, \,   = \, \,   \,
  \\
\hspace{-0.98in}&&  \quad  \quad  \quad   \quad  \quad \, \, 
\Bigl( 1\, -{{x^2} \over {a}} \Bigr)^{-1/2}
 \,  \cdot \, Heun\Bigl(a, \,  {{q} \over {4}}, \,   {{1} \over {4}}, \,
 {{3} \over {4}}, \, 1, \,   {{1} \over {2}},   \, \,
{{ 4 \,  \, a \cdot \, x \cdot \, (1\, -x)  \cdot \, (a\, -x) } \over { (a\, -x^2)^2}}  \Bigr).
\nonumber
\end{eqnarray}

\subsection{Other Heun functions for the simple cubic lattice Green function.}
\label{Alternative2}

One has the identity
\begin{eqnarray}
\label{quadraidentity}
  \hspace{-0.98in}&&   \quad \quad \quad 
Heun\Bigl(9, \, {{3} \over {4}}, \, {{1} \over {4}}, \, {{3} \over {4}}, \, 1, \,{{1} \over {2}}, \, \, x\Bigr)
 \\
\hspace{-0.98in}&& \quad  \quad      \quad \,  \,  \,  
\, \, = \, \, \,  {{3} \over {8 \, x}} \cdot \,   \
\Bigl(72 \, -40\, x \, - 72 \cdot \,
\Bigl(1\, -  \, {{x} \over {9}}\Bigr)^{1/2} \cdot \, \Bigl(1\, -  \, x\Bigr)^{1/2}  \Bigr)^{1/2} \nonumber 
 \\
  \hspace{-0.98in}&& \quad  \quad  \quad    \quad \, 
                     \,  \,   \,  \,  \times \,
Heun\Bigl(4, \, {{1} \over {2}}, \, {{1} \over {2}}, \, {{1} \over {2}}, \, 1, \,{{1} \over {2}}, \, \,
{{5} \over {2}} \, - {{9} \over {2 \, x}} \, + \,  {{9} \over {2 \, x}} \cdot \,
 \Bigl(1\, -  \, {{x} \over {9}}\Bigr)^{1/2} \cdot \, (1\, -  \, x)^{1/2}    \Bigr),
 \nonumber     
\end{eqnarray}
which can be written using the parametrization (\ref{productellipticparam}): 
\begin{eqnarray}
\label{quadraidentity}
  \hspace{-0.98in}&& \quad  \quad  \quad  \quad   
Heun\Bigl(9, \, {{3} \over {4}}, \, {{1} \over {4}}, \, {{3} \over {4}}, \, 1, \,{{1} \over {2}},
 \, \,  -\, {{ 9\, y } \over { (y\, -1) \cdot \, ( y\, -4)  }} \Bigr)
\\
\hspace{-0.98in}&& \quad  \quad \quad \quad    \quad  \quad \quad  \quad    \quad  
\, \, = \, \, \, \,
\Bigl( {{y\, -1) \cdot \, (y\, -4)} \over {4}} \Bigr)^{1/4} \cdot  \,
 Heun\Bigl(4, \, {{1} \over {2}}, \, {{1} \over {2}}, \, {{1} \over {2}}, \, 1, \,{{1} \over {2}}, \, \, y   \Bigr).
\nonumber       
\end{eqnarray}
One also has the identity:
\begin{eqnarray}
\label{lastidentity}
\hspace{-0.98in}&& \quad  \,\,   
  Heun\Bigl(9, \, {{3} \over {4}}, \, {{1} \over {4}}, \, {{3} \over {4}}, \, 1, \,{{1} \over {2}}, \, \,
   {{36 \cdot \, u \cdot \, (u\, -1)^2 \cdot \, (u\, -4) \cdot \, (u^2\, -4) } \over {
          (u^2\, -2\, u -2)^2 \cdot \, (u^2\, -2\, u \, +4)^2}} \Bigr) 
              \\
\hspace{-0.98in}&& \quad \,\,  \quad  
\, \,\,\,  = \, \, \,
\Bigl( {{(2 \, +2\, u \,  -u^2) \cdot \, (4 \,-2\, u + \,  u^2)} \over { 8 \cdot \,(1\, -u)^3 }}  \Bigr)^{1/2}  \cdot \,
 _2F_1\Bigl([{{1} \over {2}}, \, {{1} \over {2}}], \, [1],
\,  \,    -\, {{ u \, \cdot \, (4\, -u)^3} \over { 16 \cdot \, (1\, -u)^3}}     \Bigr).
\nonumber 
\end{eqnarray}

\subsection{Simple cubic lattice Green function: focus on
  $\, Heun(4, \, {{1} \over {2}}, \, {{1} \over {2}}, \, {{1} \over {2}}, \, 1, \,{{1} \over {2}}, \, \, x)$}
\label{Alternative3}

Also note that the square of this Heun function can be written, in a different way, as
a {\em product} of pullbacked  $\, _2F_1$ hypergeometric functions. Let us recall the identity~\cite{Joyce}
\begin{eqnarray}
\label{productelliptic}
\hspace{-0.98in}&& \quad    \quad  \quad \quad \quad \quad 
Heun\Bigl(4, \, {{1} \over {2}}, \, {{1} \over {2}}, \, {{1} \over {2}}, \, 1, \,{{1} \over {2}}, \, \, x\Bigr)^2 
\nonumber \\
\hspace{-0.98in}&& \quad  \quad \quad \quad  \quad   \quad   \quad \quad  \quad  
   \, \, = \, \, \,  \,  _2F_1\Bigl([{{1} \over {2}}, \, {{1} \over {2}}], \, [1], \,    {\cal H}_ {+}    \Bigr)
  \cdot \,  _2F_1\Bigl([{{1} \over {2}}, \, {{1} \over {2}}], \, [1], \,   {\cal H}_ {-}\Bigr), 
\end{eqnarray}
where:
\begin{eqnarray}
\label{productellipticHaupt}
  \hspace{-0.98in}&& \quad \quad \quad \, \, \, \,\, \,
{\cal H}_{\pm}  \, \, = \, \,  \, \, \, {{1} \over {2}} \,  \,  \, \,
 \pm \, {{x} \over {2}} \cdot \, \Bigl(1\, -  \, {{x} \over {4}} \Bigr)^{1/2} \,
 \,   -\, {{1} \over {2}}   \cdot \,
\Bigl(1\, -  \, {{x} \over {2}} \Bigr)   \cdot \,   \Bigl(1\, -  \, x\Bigr)^{1/2}.    
\end{eqnarray}
Their series expansion reads:
\begin{eqnarray}
\label{productellipticHauptser}
\hspace{-0.98in}&& \quad \quad \quad  \, \,
 {\cal H}_{+}  \, \, = \, \, \,\,
 x \, \, \,\,  -{{1} \over { 8}}  \, x^2 \, \,\,  \, -{{1} \over {256}}  \, x^3\,
 \, \,  +{{7} \over {2048}} \, x^4 \, \, \,  +{{251} \over {65536}}  \, x^5 \, \, 
 \, +{{1785} \over {524288}}  \, x^6 \,  
 \nonumber \\
 \hspace{-0.98in}&& \quad \quad \quad \quad \quad 
\,   \, +{{24555 } \over {8388608}}   \, x^7  \, +{{168927} \over {67108864}}  \, x^8
\,  \,  \, \, + \, \, \, \cdots 
 \\
\hspace{-0.98in}&& \quad  \quad  \quad \, \,
{\cal H}_{-}  \, \, = \, \, \, \,
{{1} \over {256}} \, x^3 \,\,\,  + {{9} \over {2048}}\, x^4 \, \,\,  +{{261} \over {65536}} \, x^5
 \,\,\,  +{{1799} \over {524288}} \, x^6 \,\,\,  +{{24597} \over {8388608}} \, x^7
\nonumber \\
 \hspace{-0.98in}&& \quad \quad \quad \quad \quad 
\, +{{168993} \over {67108864}} \, x^8 \,\, +{{9372077} \over {4294967296}} \, x^9 \,
\, +{{65602251} \over {34359738368}}\, x^{10} \,\, \, \,  + \, \, \, \cdots \nonumber 
\end{eqnarray}

The relation between $\, x$ and these two Hauptmoduls gives the {\em genus-zero}
quartic relation\footnote[1]{Note that this quartic relation (\ref{xH})
is invariant by $\,\,\, {\cal H}_{\pm} \, \rightarrow \, \, 1\, -\, {\cal H}_{\pm}$.}:
\begin{eqnarray}
\label{xH}
 \hspace{-0.98in}&&  \quad \quad  \quad  \quad  \quad  \, \,
 256 \cdot \, {\cal H}_{\pm}^2 \cdot \, (1\, -  \, {\cal H}_{\pm})^2
  \\
 \hspace{-0.98in}&& \quad \quad  \quad \quad  \quad \quad \quad \quad  \,\,
 -32 \cdot \, x \cdot \, (2\,x^2 \, -9\,x \, +8)
   \cdot \, {\cal H}_{\pm} \cdot \, (1\, -  \, {\cal H}_{\pm}) \, \, \, \,\,  +x^4
\, \,\,\, = \, \,\,\,  \, 0, \nonumber 
\end{eqnarray}
the relation between these two Hauptmoduls reading  the {\em genus-zero modular equation}:
\begin{eqnarray}
\label{xHH}
\hspace{-0.98in}&& \quad  \quad  \, \,
 -256 \cdot \, A^3\,B^3 \,\, \,  +384 \,A^2 \,B^2 \cdot \,(A+B) \, \,\,
+A^4\,+B^4 \, 
 \, \, -132 \cdot \, A \,B \cdot \,(A^2+B^2)
\nonumber \\
\hspace{-0.98in}&& \quad \quad  \quad \quad \quad   \, \,
\, -762 \cdot \, A^2\, B^2 \, \, \, +384 \cdot  \,A \,B \cdot  \,(A+B) \,
\, -256 \cdot \, A\, B  \, \,\,   = \,  \,\,  \,  \, 0,
\end{eqnarray}
corresponding to $\, \, q \,  \, \leftrightarrow \,  \, q^3 \,$ in the nome $\, q$.

One can rewrite the two  $\, _2F_1$  hypergeometric functions in the RHS of (\ref{productelliptic}) 
\begin{eqnarray}
\label{3F2identbisinparticul}
  \hspace{-0.99in}&& 
_2F_1\Bigl([{{1} \over {2}}, \,  {{1} \over {2}}], \, [1], \,  {\cal H}_ {\pm} \Bigr)
\,  = \,   \, _2F_1\Bigl([{{1} \over {4}}, \,  {{1} \over {4}}], \, [1], \,
4 \, {\cal H}_ {\pm} \cdot \, (1\, -{\cal H}_ {\pm})\Bigr)
 \,  = \, \,  _2F_1\Bigl([{{1} \over {4}}, \,  {{1} \over {4}}], \, [1], \, H_{\pm}\Bigr), 
\nonumber 
\end{eqnarray}
where the two pullbacks $\, H_{\pm}$ 
\begin{eqnarray}
\label{3F2identbisinparticulplusmoins}
\hspace{-0.98in}&&  \,
H_{\pm} \,  \, = \, \, \, \, \,  {{1} \over {4}}  \cdot \, (2\, x^2 \, -9\, x \, +8)  \cdot \,x
 \,\, \,\,  \pm \,  (x-2) \cdot \, (1-x)^{1/2} \cdot \, \Bigl(1 \, -{{x} \over {4}}\Bigr)^{1/2} \,  \cdot \, x, 
\end{eqnarray}
are solutions of the genus-zero quadratic relation
\begin{eqnarray}
\label{3F2identbisinparticulquadra}
\hspace{-0.98in}&&  \quad  \quad  \quad \quad  \quad  \quad  \, \, 
 16 \cdot \,  H_{\pm}^2 \,  \, \, \, \, \,   \,
 -8 \cdot \, x \cdot \, (2\, x^2-9\, x+8) \cdot \,  H_{\pm}   \, \, \, \, \,  \, +x^4
\,  \, \, \, = \,  \,  \,\, \, 0, 
\end{eqnarray}
their expansions reading:
\begin{eqnarray}
\label{3F2identbisinparticulmoins}
\hspace{-0.98in}&&    \quad   \quad   \quad 
 H_{-} \,  \, = \, \,  \,   \,   4 \, x \,  \, \,   \, -{{9} \over {2}} \, x^2 \, \,    \,
 +{{63} \over {64}} \, x^3 \, \, \,   - {{9} \over {512}} \, x^4 \, \, \,   -{{261} \over {16384}} \, x^5
\, \, \,   - {{1791} \over {131072}}  \, x^6 \,   \,  \, \, + \, \, \, \cdots,
\nonumber  
\end{eqnarray}
\begin{eqnarray}
\label{3F2identbisinparticulplus}  
\hspace{-0.98in}&&  \quad   \quad   \quad 
H_{+} \,  \, = \, \, \,    \,   {{1} \over {64}} \, x^3 \, \, \,    + {{9} \over {512}} \, x^4
\, + {{261} \over {16384}} \, x^5 \, + {{1791} \over {131072}}\, x^6
\, \,\,  \, + \, \, \, \cdots  
\end{eqnarray}
Thus (\ref{productelliptic}) can also be rewritten as:
\begin{eqnarray}
\label{productellipticbis}
  \hspace{-0.98in}&& \quad \quad  \quad  \quad  \quad  \quad 
Heun\Bigl(4, \, {{1} \over {2}}, \, {{1} \over {2}}, \, {{1} \over {2}}, \, 1, \,{{1} \over {2}}, \, \, x\Bigr)^2
\nonumber  \\
\hspace{-0.98in}&&  \quad \quad \quad \quad  \quad  \quad  \quad  \quad  \quad 
   \,  = \, \, \,   _2F_1\Bigl([{{1} \over {4}}, \, {{1} \over {4}}], \, [1], \,   H_ {+}    \Bigr)
  \cdot \,  _2F_1\Bigl([{{1} \over {4}}, \, {{1} \over {4}}], \, [1], \,  H_ {-}\Bigr). 
\end{eqnarray} 
Besides (\ref{productellipticHaupt}), let us introduce the two other Hauptmoduls:
\begin{eqnarray}
\label{productellipticHauptother}
  \hspace{-0.98in}&& \quad \quad  \quad \, 
{\cal H}^{(+)}_{\pm}  \, \, = \, \, \, \, {{1} \over {2}} \,  \, \, \,
 \pm \, {{x} \over {2}} \cdot \, \Bigl(1\, -  \, {{x} \over {4}} \Bigr)^{1/2} \,
 \,   +\, {{1} \over {2}}   \cdot \,
\Bigl(1\, -  \, {{x} \over {2}} \Bigr)  \cdot \,   \Bigl(1\, -  \, x\Bigr)^{1/2}.    
\end{eqnarray}
One can easily see that the four pullbacked
hypergeometric functions, corresponding to $\, _2F_1([1/2,1/2],[1],{\cal H}^{(+)}_{\pm}) \, $
and  $\, _2F_1([1/2,1/2],[1],{\cal H}_{\pm})$,  are all solutions of the order-four
linear differential operator $\, H_4$:
\begin{eqnarray}
\label{H4}
  \hspace{-0.98in}&& \quad \quad \quad \quad 
H_4 \, \, = \, \, \,  \,  256 \cdot \, \, (x-1)^2 \cdot \, (x-4)^2 \cdot \, x^2 \cdot \, D_x^4
\nonumber \\
\hspace{-0.98in}&& \quad \quad \quad \quad \quad \quad \, \,
+256 \cdot \, (x-1)\, (x-4) \cdot \, (8\, x^2-33\, x+16) \cdot \, x \cdot \, D_x^3
 \\
\hspace{-0.98in}&& \quad \quad \quad \quad \quad \quad \, \,\, \,
+32 \cdot \, (115\, x^4-991\, x^3+2370\, x^2-1696\, x+256) \cdot \, D_x^2
\nonumber \\
\hspace{-0.98in}&& \quad \quad \quad\quad \quad \quad  \quad  \, \, \, \, \,\, \,
+16 \cdot \, (76\, x^3-693\, x^2+1128\, x-448) \cdot \, D_x
\, \, \, \, \, \,  +9 \cdot \, x \cdot \, (x\, -16). \nonumber 
\end{eqnarray}
Note that this order-four linear differential operator is homomorphic to its adjoint
with an order-one linear differential intertwiner
\begin{eqnarray}
\label{adjointH4}
  \hspace{-0.98in}&& \quad \quad \quad \quad \quad
H_4 \cdot \, L_1 \, \, \, = \, \,\,  \, adjoint(L_1) \cdot  \,       adjoint(H_4)
\quad \quad  \quad \quad  \quad  \quad  \hbox{where:} 
\nonumber \\
\hspace{-0.98in}&& \quad \quad \quad \quad 
L_1 \, \, \,  = \, \, \, \, \,  2 \cdot  \,(x-1) \cdot  \, (x-4) \cdot \, x \cdot \,D_x
\, \, \, +5\, x^2 \, -4\, x \, -4,
\end{eqnarray}
and with an order-three intertwiner
\begin{eqnarray}
\label{adjointH4bis}
 \hspace{-0.98in}&& \quad \quad \quad \quad \quad \quad \quad \quad
  adjoint(H_4)  \cdot \, L_3 \, \, \, = \, \,\, \, adjoint(L_3)  \cdot \, H_4,  
\end{eqnarray}
suggesting that the differential Galois group of this order-four linear differential
operator could be $\, SO(3,\, \mathbb{C})$. This is not the case. In fact this order-four
linear differential operator is
{\em not irreducible}\footnote[1]{At first sight using the command DFactor of Maple, one
  could imagine that this linear differential operator is irreducible.}: it has an
  {\em absolute} factorization (see (\ref{quadraidentitypullback}) below).
Introducing the parametrization
\begin{eqnarray}
\label{productellipticparam}
  \hspace{-0.98in}&& \quad \quad \quad \quad \quad \quad \quad \quad \quad
x \, \, = \, \,  -\, {{ u \cdot \, (4\, -u) \cdot \, (4 \, -u^2)} \over { 4 \cdot \, (1\, -u)^2 }},  
\end{eqnarray}
the product identity (\ref{productelliptic}) becomes:
\begin{eqnarray}
\label{productelliptic2}
\hspace{-0.98in}&&  \quad    \quad 
Heun\Bigl(4, \, {{1} \over {2}}, \, {{1} \over {2}}, \, {{1} \over {2}}, \, 1, \,{{1} \over {2}},
\, \,  -\, {{ u \cdot \, (4\, -u) \cdot \, (4 \, -u^2)} \over { 4 \cdot \, (1\, -u)^2 }}\Bigr)^2
 \\
\hspace{-0.98in}&& \quad  \quad    \quad 
  \, \, \, \, = \, \, \,  \,
_2F_1\Bigl([{{1} \over {2}}, \, {{1} \over {2}}], \, [1], \,  \,
-\, {{ u \, \cdot \, (4\, -u)^3} \over { 16 \cdot \, (1\, -u)^3}}     \Bigr)
\cdot \,  _2F_1\Bigl([{{1} \over {2}}, \, {{1} \over {2}}], \, [1],
\, \,   -\, {{ u^3 \, \cdot \, (4\, -u)^3} \over { 16 \cdot \, (1\, -u)^3}}   \Bigr).
 \nonumber 
\end{eqnarray}
Note that these two pullbacked $\, _2F_1$ hypergeometric functions are simply related:
\begin{eqnarray}
\label{simplyrelated}
  \hspace{-0.98in}&& \quad \quad \quad   \quad    \quad  \quad     
_2F_1\Bigl([{{1} \over {2}}, \, {{1} \over {2}}], \, [1], \,  \,
 -\, {{ u \, \cdot \, (4\, -u)^3} \over { 16 \cdot \, (1\, -u)^3}}     \Bigr)
\nonumber \\
  \hspace{-0.98in}&& \quad \quad \quad \quad \quad \quad   \, \,  \quad  \quad
\, \,\, \,  = \, \, \,  \,  (1\, -u) \cdot \, 
_2F_1\Bigl([{{1} \over {2}}, \, {{1} \over {2}}], \, [1],
\, \,   -\, {{ u^3 \, \cdot \, (4\, -u)^3} \over { 16 \cdot \, (1\, -u)^3}}   \Bigr). 
\end{eqnarray}
Therefore the identity (\ref{productelliptic2}), in fact, relates the square of a
Heun function with the square of a pullbacked $\, _2F_1$ hypergeometric function,
or more simply, gives this Heun function as a function of  a
pullbacked $\, _2F_1$ hypergeometric function:
\begin{eqnarray}
\label{productelliptic2bis}
\hspace{-0.98in}&&  \quad    \quad      \quad            \quad      
Heun\Bigl(4, \, {{1} \over {2}}, \, {{1} \over {2}}, \, {{1} \over {2}}, \, 1, \,{{1} \over {2}},
  \, \,  -\, {{ u \cdot \, (4\, -u) \cdot \, (4 \, -u^2)} \over { 4 \cdot \, (1\, -u)^2 }}\Bigr) 
\nonumber \\
\hspace{-0.98in}&& \quad     \quad   \quad \quad  \quad \quad     \quad \quad        
  \, \, = \, \, \,  \,  (1\, -u)^{-1/2}  \cdot \, 
 _2F_1\Bigl([{{1} \over {2}}, \, {{1} \over {2}}], \, [1], \,  \,
 -\, {{ u \, \cdot \, (4\, -u)^3} \over { 16 \cdot \, (1\, -u)^3}}     \Bigr),
 \end{eqnarray}
 or:
\begin{eqnarray}
\label{productelliptic2ter}
\hspace{-0.98in}&&  \quad    \quad      \quad            \quad      
Heun\Bigl(4, \, {{1} \over {2}}, \, {{1} \over {2}}, \, {{1} \over {2}}, \, 1, \,{{1} \over {2}},
\, \,  -\, {{ t \cdot \, (t\, +2) \cdot \,
(3\, t \, +2)\cdot \, (3\, t \, +4)} \over { 4 \cdot \, (t\, +1)^2 }}\Bigr) 
\nonumber \\
\hspace{-0.98in}&& \quad     \quad   \quad\quad  \quad \quad  \quad \quad           
  \, \, = \, \, \,  \,  (1 \, +t)^{-1/2}  \cdot \, 
 _2F_1\Bigl([{{1} \over {2}}, \, {{1} \over {2}}], \, [1],
\,  \,    -\, {{ t \, \cdot \, (3\, t \, +4)^3} \over { 16 \cdot \, (t\, +1) }}     \Bigr).
 \end{eqnarray}
 Let us note that the order-four linear differential operator $\, H_4$, given by (\ref{H4}),
 pullbacked by the $\,\, x \rightarrow \, \, x(u)\, $  pullback (\ref{productellipticparam}),  
 gives an order-four linear differential operator which, not only factorizes, but has a {\em direct-sum}
 factorization into two  order-two linear differential operators that are simply conjugated:
\begin{eqnarray}
\label{quadraidentitypullback}
 \hspace{-0.98in}&& 
 pullback\Bigl(H_4, \, \,  -\, {{ u \cdot \, (4\, -u) \cdot \, (4 \, -u^2)} \over { 4 \cdot \, (1\, -u)^2 }}\Bigr)
\, \, = \, \, \,   L_2 \, \oplus \, M_2, \quad \, \,\,
M_2 \, = u \cdot \, L_2 \cdot \, {{1} \over {u}}. 
\end{eqnarray}
These two linear differential operators $\, L_2$ and $\, M_2 \, $ have respectively
the pullbacked $\, _2F_1$ hypergeometric solutions:
\begin{eqnarray}
\label{solpullback}
 \hspace{-0.98in}&& \,
 _2F_1\Bigl([{{1} \over {2}}, \, {{1} \over {2}}], \, [1],
\, \,   -\, {{ u^3 \, \cdot \, (4\, -u)^3} \over { 16 \cdot \, (1\, -u)^3}}   \Bigr), \,  \, \, \,  \, 
u \cdot \, _2F_1\Bigl([{{1} \over {2}}, \, {{1} \over {2}}], \, [1],
\, \,   -\, {{ u^3 \, \cdot \, (4\, -u)^3} \over { 16 \cdot \, (1\, -u)^3}}   \Bigr).      
\end{eqnarray}
Recalling the identity
\begin{eqnarray}
\label{EllipticK0}
\hspace{-0.98in}&&
\quad  \quad    \quad    \quad   
_2F_1\Bigl([{{1} \over { 2}}, \, {{1} \over { 2}}], \, [1], \, \, x\Bigr)
 \\
\hspace{-0.98in}&&
\quad  \quad \quad  \quad  \quad  \quad  \quad     
\, \,\, \,  = \, \, \,  (1\, -\, x \, +\, x^2)^{-1/4}
\cdot  \, _2F_1\Bigl([{{1} \over { 12}}, \, {{5} \over { 12}}], \, [1], \,
\, {{27 \, x^2 \cdot \, (1\, - \, x)^2} \over {
4 \cdot \, (1\, -\, x \, +\, x^2)^3 }}\Bigr),
\nonumber 
\end{eqnarray}
one finds that the $\, _2F_1$ hypergeometric functions in (\ref{productelliptic})
can be rewritten as:
\begin{eqnarray}
\label{EllipticK02F1}
  \hspace{-0.98in}&& \quad \quad  \quad   
_2F_1\Bigl([{{1} \over {2}}, \, {{1} \over {2}}], \, [1], \,    {\cal H}_ {\pm}    \Bigr)
\nonumber \\
\hspace{-0.98in}&&
\quad  \quad \quad  \quad  \quad  \, \,  
\, \, = \, \, \,     (1\, -\,  {\cal H}_ {\pm} \, +\,  {\cal H}_ {\pm}^2)^{-1/4}
\cdot  \, _2F_1\Bigl([{{1} \over { 12}}, \, {{5} \over { 12}}], \, [1], \,
\, {{27 \cdot \,  {\cal H}_ {\pm}^2 \cdot \, (1\, - \,  {\cal H}_ {\pm})^2} \over {
4 \cdot \, (1\, -\,  {\cal H}_ {\pm} \, +\,  {\cal H}_ {\pm}^2)^3 }}\Bigr)
\nonumber \\
\hspace{-0.98in}&&
\quad  \quad \quad  \quad  \quad   \, \, 
\, \, = \, \, \,     (1\, -\,  {\cal H}_ {\pm} \, +\,  {\cal H}_ {\pm}^2)^{-1/4}
\cdot  \, _2F_1\Bigl([{{1} \over { 12}}, \, {{5} \over { 12}}], \, [1], \,  H_ {\pm} \Bigr),    
\end{eqnarray}
where  $\, H_ {\pm} \, $ is no longer solution of a quartic equation,
but of a {\em quadratic genus-zero} equation: 
\begin{eqnarray}
\label{EllipticK02F1whereH}
  \hspace{-0.98in}&& \quad  \, 
 (x-4)^3 \cdot \, (x^3 \, -60\, x^2 \, +48\, x \, -64)^3  \cdot \,  H_ {\pm}^2
 \nonumber \\
 \hspace{-0.98in}&& \quad  \quad  \,   \,  
 \, + \, \,  216 \cdot \, x^2  \cdot \,
(2\, x^6 \, +15\, x^5+92\, x^4 \, +464\, x^3 \, +3648\, x^2 \, -12288\, x \, +8192)  \cdot \,  H_ {\pm}
 \nonumber \\
 \hspace{-0.98in}&& \quad \quad  \quad \quad  \quad \quad \quad 
 \, \, +11664 \cdot \, x^8 \,  \, \, = \, \, \,  \,  \, 0. 
\end{eqnarray}
The relation between the two Hauptmoduls $\, H_ {\pm} \, $ corresponds to the {\em genus-zero}
modular equation\footnote[1]{Corresponding to $\, q \, \, \leftrightarrow \, \, q^3 \,$ in the nome $\, q$.}
\begin{eqnarray}
\label{modEllipticK02F1whereH}
  \hspace{-0.98in}&& \quad \quad \quad
 262144000000000 \cdot \, A^3 \,B^3 \cdot \, (A+B)
\nonumber \\
 \hspace{-0.98in}&& \quad \quad  \quad  \quad 
 +4096000000 \cdot \, A^2\, B^2 \cdot \, \Bigl(27  \cdot \, (A^2+B^2) \, -45946  \cdot \,  A\, B \Bigr)
\nonumber \\
 \hspace{-0.98in}&& \quad \quad  \quad  \quad 
 +15552000 \cdot \, A \, B \cdot \, (A+B) \cdot \, (A^2+B^2 +241433 \cdot \, A\, B)
\nonumber \\
 \hspace{-0.98in}&& \quad \quad  \quad  \quad 
+729 \cdot \, \Bigl(A^4+B^4 \, \,   - 1069956 \cdot \, A \, B\, (A^2 + \,B^2)
 \, \, +2587918086 \cdot \, A^2\, B^2\Bigr)
\nonumber \\
 \hspace{-0.98in}&& \quad \quad  \quad  \quad 
+2811677184 \cdot \, A\, B \cdot \, (A+B) \, \, 
-2176782336 \cdot \, A\, B  \, \,\,  = \, \, \, \, \, 0,              
\end{eqnarray}
The series expansion of the two Hauptmoduls $\, H_ {\pm} \, $ reads respectively:
\begin{eqnarray}
\label{modEllipticK02F1Hser}
  \hspace{-0.98in}&& \quad \, \quad  
 {{27} \over { 4}} \cdot \, x^2 \, \, \, +{{81} \over { 16}} \cdot \, x^3 \, \, \, - {{8181} \over { 512}} \cdot \, x^4
 \, \,  \,- {{ 27351} \over {1024}} \cdot \,x^5  \, \, \, \, + \, \, \, \cdots,  \quad \, \,  \, \, \quad \hbox{and:}
          \\
  \hspace{-0.98in}&& \quad \, 
 {{27} \over {262144}} \cdot \, x^6  \, \, +{{243} \over {1048576}} \cdot \, x^7 \, +{{11421} \over {33554432}}  \cdot \,x^8
 \, +{{14013} \over {33554432}} \cdot \, x^9 \, \, \, \, + \, \, \, \cdots 
\end{eqnarray}

\vskip .2cm

{\bf Anecdotal remark:} The order-three intertwiner $\, L_3$ in (\ref{adjointH4bis})
is, in fact\footnote[2]{When written in a unitary way $\,\, D_x^3 \, + \,\cdots$}, 
the {\em symmetric square} of an order-two linear differential operator
\begin{eqnarray}
\label{L2Heun}
 \hspace{-0.98in}&& \, \, \,  \, \,   \,
 L_2 \, \, = \, \, \,  D_x^2 \, \, \,
 + \, {\frac {3\,{x}^{2}-10\,x +4}{2 \, x \cdot \, (x-4)  \cdot \, (x-1) }} \cdot \, D_x \,  \, \,
+ \,{\frac { (x+4)  \cdot \, (x \, -2) }{ 32 \cdot \, (x-1) \cdot \, (x-4) \cdot \, {x}^{2}}}, 
\end{eqnarray}
which has the two Heun solutions
$\,\,  S_{\pm} \, \, = \, \, \,  x^{(1 \, \pm  \sqrt{2})/4} \,  \cdot \, h_{\pm} \, \, $
where: 
\begin{eqnarray}
\label{L2Heuntwo}
  \hspace{-0.98in}&& \quad \, 
h_{\pm} \, \, \, =  \,  \, \,\, 
Heun\Bigl(4, \, {{11} \over {8}} \, \pm {{ 7 \cdot \, 2^{1/2}} \over {8}},
\, 1 \,  \pm  {{ 3 \cdot \, 2^{1/2}} \over {8}}, \, 1 \,  \pm  {{ 2^{1/2}} \over {8}},
 \, 1 \,  \pm  {{ 2^{1/2}} \over {2}},  \, \, \, x   \Bigr).  
\end{eqnarray}
It is clear, from the $\,\, x^{(1 \, \pm   2^{1/2})/4}\,\, $ factors,  that the order-two operator $\, L_2$
is  not even\footnote[5]{The intertwiners of Fuchsian (resp. globally nilpotent) linear differential operators
  have no reason to be Fuchsian (resp. globally nilpotent) linear differential operators as well.}
{\em globally nilpotent}~\cite{GloballyNilpotent}. 
Furthermore, the two series $\, h_{+} \, + h_{-} \, \, $ and  $\, 2^{-1/2} \cdot  \, (h_{+} \, - h_{-}) \, \, $
are series with rational number coefficients that are {\em not globally bounded} series. Such Heun
functions (\ref{L2Heuntwo}) cannot be written as pullbacked $\, _2F_1$ hypergeometric functions.

\vskip .2cm

\section{Polynomials of the degree six equation for the Hauptmodul in (\ref{33}).}
\label{Append1}

The polynomials of the degree six equation for the Hauptmodul in (\ref{33}) read respectively:
\begin{eqnarray}
  \label{p6}
  \hspace{-0.98in}&& 
p_{6}(x)   \, = \, \, \, 1 \,  \,\, -53324\, x \, \, \, +3340572\, x^2
\,  \, \,\,  -47158880\, x^3 \,  \,\,  +453452848\, x^4
 \nonumber \\
  \hspace{-0.98in}&& \quad  \quad \quad  \quad  \quad  \quad
                     \, +867240000\, x^5 \, \, +729000000\, x^6, 
\end{eqnarray}
\begin{eqnarray}
  \label{p14}
  \hspace{-0.98in}&& 
  p_{14}(x)  \, = \, \, \, 1\, -126\, x \, +6657\, x^2\, -191100\, x^3 \, +3224004\, x^4 \, -32165952\, x^5
 \nonumber \\
  \hspace{-0.98in}&& \quad \quad \,
+179161346\, x^6  \,-459836304\, x^7 \, +116094384\, x^8 \, +1082203136\, x^9
\nonumber \\
  \hspace{-0.98in}&& \quad \quad \,
-247538592\, x^{10}\, -690095616\, x^{11} \, -102971392\, x^{12}
\nonumber \\
  \hspace{-0.98in}&& \quad \quad \, +15237120\, x^{13} \, +324000\, x^{14}, 
\end{eqnarray}
\begin{eqnarray}
  \label{p16}
  \hspace{-0.98in}&& 
p_{16}(x)   \, = \, \, \,
1 \, -144\, x \, +12624\, x^2 \, +42210112\, x^3 \, +35493701376\, x^4 \, +4373215830144\, x^5
\nonumber \\
  \hspace{-0.98in}&& \quad \quad \,
 +146527536091776\, x^6 +1709973141608448\, x^7+8301405990184512\, x^8
\nonumber \\
  \hspace{-0.98in}&& \quad \quad  \,+19700334651209215\, x^9+25456135068016191\, x^{10}
 +18571208108112576\, x^{11}\nonumber \\
  \hspace{-0.98in}&& \quad \quad \,\,
 +7732095471912574\, x^{12}+1556868770685984\, x^{13}+183059891926656\, x^{14}
 \nonumber \\
  \hspace{-0.98in}&& \quad \quad \, \, +2050894080000\, x^{15}\, +43740000000\, x^{16}, 
\end{eqnarray}
\begin{eqnarray}
  \label{p20}
  \hspace{-0.98in}&& 
  p_{20}(x)   \, = \, \, \, 496 \, +14229477\, x \, +10755437160\, x^2 \, +607313973993\, x^3
  \nonumber \\
  \hspace{-0.98in}&& \quad \,
   +21837165846834\, x^4 \, -8741350651741356\, x^5 \, +602696000526139688\, x^6
   \nonumber \\
  \hspace{-0.98in}&& \quad \,
   -18362650954659075270\, x^7+237729206666512798092\, x^8
  \nonumber \\
  \hspace{-0.98in}&& \quad \,
  -755131861209486545984\, x^9 -4078730236814710350912\, x^{10}
  \nonumber \\
  \hspace{-0.98in}&& \quad \,
 +4455455555487369556416\, x^{11} +8298505398959353031040\, x^{12}
 \nonumber \\
  \hspace{-0.98in}&& \quad \,
 -501211331403375909060096\, x^{13}\,+32930923081507234916352\, x^{14}
 \nonumber \\
  \hspace{-0.98in}&& \quad
 -7365760252808436401159680\, x^{15}-14299198145937514719360000\, x^{16}
 \nonumber \\
  \hspace{-0.98in}&& \quad \,
-5550618706232520960000000\, x^{17}-2323303457201280000000000\, x^{18}
 \nonumber \\
  \hspace{-0.98in}&& \quad \,
 -2534505901920000000000000\, x^{19}+114791256000000000000000\, x^{20},
\end{eqnarray}
\begin{eqnarray}
\label{p21}
\hspace{-0.98in}&& 
p_{21}(x)   \, = \, \, \, 
1\, -189\, x+15939\, x^2-790713\, x^3+25604460\, x^4-567479130\, x^5
 \nonumber \\
  \hspace{-0.98in}&& \quad \,+8729096106\, x^6-129136524678\, x^7-3128791781472\, x^8
  \nonumber \\
  \hspace{-0.98in}&& \quad \,
  -301592422936140\, x^9+9302223231205632\, x^{10}-898829709897155904\, x^{11}
  \nonumber \\
  \hspace{-0.98in}&& \quad \,-3001729628561501184\, x^{12}  \, -14056123657998705984\, x^{13}
   \nonumber \\
  \hspace{-0.98in}&& \quad -75146837553583537200\, x^{14} \,-220865053128551921712\, x^{15}
 \nonumber \\
  \hspace{-0.98in}&& \quad  \, -233016707230759517184\, x^{16} -25485724878879707232\, x^{17}
  \nonumber \\
  \hspace{-0.98in}&& \quad \,+57908320494660830720\, x^{18}+11705232438547200000\, x^{19}
  \nonumber \\
  \hspace{-0.98in}&& \quad \,-65745768960000000\, x^{20}-3149280000000000\, x^{21}, 
\end{eqnarray}
\begin{eqnarray}
  \label{p23}
  \hspace{-0.98in}&& 
 p_{23}(x)   \, = \, \, \, 1 \, -207\, x \, +21552\, x^2+41491618\, x^3 +32829303696\, x^4
 \nonumber \\
  \hspace{-0.98in}&& \quad \,
 +2194878922992\, x^5-81778493396032\, x^6-2027922617204064\, x^7
 \nonumber \\
  \hspace{-0.98in}&& \quad \,
 +50756763414324000\, x^8 +304451170309086240\, x^9
                     -5117266473854922240\, x^{10}
  \nonumber \\
  \hspace{-0.98in}&& \quad
\,-23872757678772284352\, x^{11}+92761784722387529728\, x^{12}
 \nonumber \\
  \hspace{-0.98in}&& \quad \,
 -1131857205540040786944\, x^{13}\, +4168576271341671432192\, x^{14}
 \nonumber \\
  \hspace{-0.98in}&& \quad \,
 -35184687910528656122881\, x^{15}+2169420555967834017888\, x^{16}
 \nonumber \\
  \hspace{-0.98in}&& \quad \,
 -270473856235208160230976\, x^{17} -471011087555724046299136\, x^{18}
\nonumber \\
  \hspace{-0.98in}&& \quad \,
-105170018593490449009152\, x^{19} \,-71294201738328407040000\, x^{20}
\nonumber \\
  \hspace{-0.98in}&& \quad \,
-141314879220788736000000\, x^{21}\, +1238649615360000000000\, x^{22}
\nonumber \\
  \hspace{-0.98in}&& \quad \, \, -127545840000000000000\, x^{23}, 
\end{eqnarray}

\vskip .1cm

{\bf Remark:} Taking the resultant (eliminating $\, x$) between equation (\ref{AhAhHauptquadra}),
in the example 2,  with the following genus-zero curve
\begin{eqnarray}
\label{immediately}
  \hspace{-0.98in}&& \quad  \quad \quad \quad \quad  \, \,
8 \, z^3\, x^3 \, \,  -12\, z^2\, x^2 \cdot \, (x+z) \, \,\, \, \,
+x\, z \cdot \, (6\, x^2-83\, x\, z+6\, z^2)
\nonumber \\
 \hspace{-0.98in}&& \quad  \quad \quad \quad \quad  \, \, \,\quad \quad \quad
\, \,  -(x+z) \cdot \, (x^2+17\, x\, z+z^2)  \, \, \, \, +x\, z
 \,  \, \, = \, \,  \, \, 0, 
\end{eqnarray}
one obtains immediately equation (\ref{deg6}) and polynomials (\ref{p6})-(\ref{p23}) of this
appendix.

\vskip .2cm

\section{Pullbacked $\, _2F_1$ representation of
  $\, Heun({{1} \over {9}}, \, {{1} \over {12}},\, {{1} \over {4}},\, {{3} \over {4}},\, 1,\, {{1} \over {2}}, \,\, 4\,x)$.}
\label{pulback2F1represe}

Let us consider example 6 of section (\ref{subfour}), which diagonal is the square of 
the Heun function  (\ref{Ratfoncfour11hyp}), which can be written as a pullbacked $\, _2F_1$
hypergeometric function
\begin{eqnarray}
\label{Ratfoncfour11hypapp}
\hspace{-0.98in}&& \quad  \quad \quad \quad \quad 
 Heun\Bigl({{1} \over {9}},\,  {{1} \over {12}},\,   {{1} \over {4}},
  \,  {{3} \over {4}},\,  1,\,  {{1} \over {2}},  \, \, 4\, x\Bigr)
  \\
\hspace{-0.98in}&& \quad  \quad   \quad  \quad   \quad \quad 
\, \, = \, \, \,\,  {\cal A}^{(1)}_{\pm} \cdot \,
_2F_1\Bigl([{{1} \over {6}}, \, {{2} \over {3}}],\, [1], \, {\cal H}^{(1)}_{\pm} \Bigr)
\, \,\, = \, \, \, \, {\cal A}^{(2)}_{\pm} \cdot \,
_2F_1\Bigl([{{1} \over {8}}, \, {{5} \over {8}}],\, [1], \,   {\cal H}^{(2)}_{\pm}\Bigr),
\nonumber 
\end{eqnarray}
where the two pullbacks $\,\, {\cal H}^{(1)}_{\pm}$, $\,\, {\cal H}^{(2)}_{\pm}\,$
are square root algebraic functions
\begin{eqnarray}
\label{Ratfoncfour11Haupt}
\hspace{-0.98in}&& \quad    \quad  \quad  
 {\cal H}^{(1)}_{\pm} \, \, = \, \, \,
  -54 \cdot \, x \cdot \, {{1 \, -27\,x \, -108\, x^2 } \over {(1 \, -54\,x)^2 }}
 \nonumber  \\
  \hspace{-0.98in}&& \quad  \quad  \quad  \quad \quad \quad \quad  \quad \quad  \quad  
  \pm  \, 54 \cdot \, x \cdot \, (1 \, -9\, x) \cdot \,
 {{ (1 \, -4\, x)^{1/2} \cdot \, (1 \, -36\, x)^{1/2}    } \over {(1 \, -54\,x)^2 }}, 
  \\           
  \hspace{-0.98in}&& \quad    \quad  \quad   
 {\cal H}^{(2)}_{\pm} \, \, = \, \, \,  -128 \cdot \, x  \cdot \,
 {{ 1\, -38\, x \, +200\, x^2 } \over {(1 \, -100\,x)^2\cdot \, (1 \, -4\, x) }} 
       \nonumber  \\
  \hspace{-0.98in}&& \quad  \quad  \quad  \quad \quad \quad \quad  \quad \quad  \quad  
  \pm  \,   128 \cdot\, x \cdot \, (1\, -120 \, x) \cdot \,
   {{ (1\, -36 \, x)^{1/2} } \over {(1 \, -100\,x)^2 \cdot \, (1 \, -4\, x) }},                
\end{eqnarray}
where   $\,\, Y_{\pm}\, = \, \, ( {\cal A}^{(1)}_{\pm} )^{12} \,\, $  are simple algebraic functions
respectively solutions of 
\begin{eqnarray}
\label{Ratfoncfour11HauptAlg1p}
\hspace{-0.98in}&& \quad \quad  \quad  \quad   \quad 
64 \,\, \,  \,  + \,  p_3(x) \cdot \, Y_{+} \,  
\, \,   + \,  \, (1\, - \, 54\, x)^{4} \cdot \, Y_{+}^2
                   \, \, \, \, \,   = \, \, \,  \,\, 0,
  \\
  \label{Ratfoncfour11HauptAlg1m}
\hspace{-0.98in}&& \quad \quad  \quad  \quad   \quad 
1 \,\, \,  \,  + \,  p_3(x) \cdot \, Y_{-} \,  
\, \,\,  + \,  64  \cdot \, (1\, - \, 54\, x)^{4} \cdot \, Y_{-}^2
\, \, \, \, \,   = \, \, \,  \,\, 0, 
\end{eqnarray}
where
\begin{eqnarray}
\label{Ratfoncfour11HauptAlg1}
\hspace{-0.98in}&& \quad \quad  \quad  \quad \quad 
  p_{3}(x) \, \,= \, \, \,  186624\, x^3\, \, -15552\, x^2 \, \,\,  +2484\, x \, \,\,    -65,
\end{eqnarray}
and  where $\,\, Y_{\pm} \, = \, \, ( {\cal A}^{(2)}_{\pm} )^{8}\,\,  $ are simple
algebraic functions, respectively solutions of:
\begin{eqnarray}
\label{Ratfoncfour11HauptAlg2p}
\hspace{-0.98in}&& 
81  \, -2 \cdot \, (41 \, -900 \, x) \cdot \, (1\, -4\, x) \cdot \, Y_{+}
\, +(1 \, -100\, x)^{2} \cdot \, (1 \, -4\, x)^{2} \cdot \, Y_{+}^2
                 \,  \, \, = \, \,\, \, 0,
\end{eqnarray}
\begin{eqnarray}
\label{Ratfoncfour11HauptAlg2m}
\hspace{-0.98in}&& 
1  \, -2 \cdot \, (41 \, -900 \, x) \cdot \, (1\, -4\, x) \cdot \, Y_{-}
\, + 81 \cdot \,  (1 \, -100\, x)^{2} \cdot \, (1 \, -4\, x)^{2} \cdot \, Y_{-}^2
                \, = \, \, 0.
\end{eqnarray}
The two Hauptmoduls $\, {\cal H}^{(1)}_{\pm}$ have the following series expansions
\begin{eqnarray}
\label{Ratfoncfour11Hauptser1}
\hspace{-0.98in}&& \quad 
 {\cal H}^{(1)}_{-}     \,   \, = \, \,\,   \,
-108\, x \,\, \, -8640\, x^2\,\, \, -615168\, x^3 \,\,\,  -41167872\, x^4\,\, \, -2650337280\, x^5
 \nonumber  \\
\hspace{-0.98in}&& \quad \quad  \, -166137937920\, x^6 \, -10213026103296\, x^7
\, -618505440067584\, x^8 \, \,\, \, + \, \, \, \cdots
\end{eqnarray}
and 
\begin{eqnarray}
\label{Ratfoncfour11Hauptser2}
\hspace{-0.98in}&& \quad 
{\cal H}^{(1)}_{+}     \,   \, = \, \,   \,
-108\, x^2 \,\,\,  -3024\, x^3 \,\, \, -87696\, x^4 \,\, \, -2616192\, x^5\,\,\,  -79800768\, x^6 \, 
\nonumber  \\
\hspace{-0.98in}&& \quad \quad \quad 
 -2477350656\, x^7 \,\,\, 
 -78006945024\, x^8 \,\,\,  -2485113716736\, x^9 \, \,  \,
 \, \, + \, \, \, \cdots
\end{eqnarray}
and are related by the   {\em genus-zero\footnote[1]{Its parametrisation is in
    agreement with the rational parametrisation given below in \ref{ATTENTION}.} modular equation}:
\begin{eqnarray}
\label{mod11Hauptser2}
\hspace{-0.98in}&& \quad  \quad  \quad
625\, A^3\, B^3 \, \, \,  -525\, A^2\, B^2 \cdot \, (A+B)
 \, \, \,  -96\,  A  B\,\cdot \, (A^2+B^2) \, \, \,   -3\, A^2\, B^2
                   \nonumber  \\
  \hspace{-0.98in}&& \quad\quad  \quad \quad \quad
\, \, -4 \cdot \, (A^3+B^3)\,\, \,
+ \, 528 \cdot \, A \, B\, \cdot \, (A+B) \, \,  \,
-432 \cdot \, A\, B  \,\,  \, \, =\, \, \, \, \, 0,        
\end{eqnarray}which is the same {\em modular equation} as (\ref{modeqH12}).
Note that one can get rid of these square root algebraic expressions
for the Hauptmoduls introducing some rational parametrisation
such that all the arguments of the miscellaneous $\, _2F_1$
hypergeometric functions (\ref{Ratfoncfour11}) introduced to rewrite
the Heun solution,  are rational functions.
Such a rational parametrisation is given in \ref{ATTENTION}. 
The previous genus-zero modular equation (\ref{mod11Hauptser2}) has the following
rational parametrisation (see (\ref{Ratfonc3HeunparamSol2}) and  (\ref{Ratfonc3HeunparamSol3})
in \ref{ATTENTION}):
\begin{eqnarray}
\label{mod11Hauptser2param}
  \hspace{-0.98in}&& \quad  \quad  \quad \quad  \quad
 A \, = \, \,-\,  {{108 \, \, z \cdot \, (1\, +\, 3 \, z)
     \cdot \, (1 \, -4\, z \, -12\, z^2)^2 } \over {
 (1 \, -36\, z \, -108\, z^2)^2  }},  
\nonumber \\
\hspace{-0.98in}&& \quad  \quad  \quad \quad  \quad
 B \, = \, \, -\,  {{108 \,\, z^2 \cdot \, (1\, +\, 3 \, z)^2
     \cdot \, (1\, -\, 6 \, z) \cdot \, (1\, +\, 2 \, z) } \over {
 (1\, +\, 6 \, z)^2 \cdot \, (1 \, -6\, z \, -18\, z^2)  }} .
\end{eqnarray}   
The two Hauptmoduls $\, {\cal H}^{(2)}_{\pm} \, $ have the following series expansions:
\begin{eqnarray}
\label{Ratfoncfour11Hauptser1}
\hspace{-0.98in}&& 
{\cal H}^{(2)}_{-}     \,   \, = \, \,   \,
-256\, x \,   \,  \, -42496\, x^2 \,\,  \,    -5955328\, x^3 \, \,  \,  -766211584\, x^4 \,  \,  \,  -93688527616\, x^5
 \nonumber       \\
\hspace{-0.98in}&& \quad  \quad  \quad  \quad 
    -11075543976448\, x^6 \,
-1278221854881280\, x^7 \, \,\,  \,   + \, \, \cdots               
                \\
  \hspace{-0.98in}&&
{\cal H}^{(2)}_{+}     \,   \, = \, \,   \,
 -256\,x^3 \, \,  \,  -10752\,x^4 \,  \, \, -361728\,x^5 \,\,  \,   -11580928\,x^6 \,\,  \,   -367348224\,x^7
 \nonumber \\
  \hspace{-0.98in}&& \quad \quad  \,   \,
-11679151104\,x^8 \, \, -373464444160\,x^9 \,\,  -12021474516480\,x^{10}   \, \,\, \,   + \, \, \, \cdots                 
\end{eqnarray}
and are related by the  {\em genus-zero  modular equation}:
\begin{eqnarray}
\label{mod11Hauptser22}
  \hspace{-0.98in}&& \quad  \quad \quad  \quad
640000 \cdot \, A^2\, B^2 \cdot \, (9\, A^2+14\, A\, B+9\, B^2) \,
\nonumber \\
\hspace{-0.98in}&& \quad  \quad \quad \quad  \quad \quad
+4800\cdot \, A \, B \cdot \, (A+B) \cdot \, (A^2-1954\, A\, B+B^2) \,
 \\
\hspace{-0.98in}&& \quad  \quad \quad \quad  \quad \quad
+A^4+B^4 \,  \,\, \,  
-56196 \cdot \, A\, B\cdot \, (A^2+B^2) \, \,  \,  
+3512070 \cdot \, A^2 \, B^2  \,
\nonumber \\
\hspace{-0.98in}&& \quad  \quad  \quad  \quad  \quad \quad  \quad \quad \quad
+116736 \cdot \, A\, B \cdot \, (A+B) \, \, \,   \, \,  -65536 \cdot \, A\, B
   \, \, \,    \,  \, = \, \, \, \, 0.  \nonumber
\end{eqnarray}
{\bf Remark:} The  modular equation (\ref{mod11Hauptser22}) is actually the same as the
modular equation (\ref{modeqH22}) of example 3. This can be seen
as a consequence of identity (\ref{Ratfoncfour11simply}), showing a relation between
example 3 and example 6.

\section{A rational parametrisation for
  $\, Heun({{1} \over {9}}, \, {{1} \over {12}},\, {{1} \over {4}},\, {{3} \over {4}},\, 1,\, {{1} \over {2}}, \,\, 4\,x)$.}
\label{ATTENTION}
The diagonal of  the following rational function
of the four variables$\, x$, $\, y$, $\, z$ and $\, w$:
\begin{eqnarray}
\label{Ratfoncfour20bis}
  \hspace{-0.98in}&& \quad  \quad  \quad  \quad 
\, \,  
R(x, \, y, \, z)  \, \, \,  = \, \,  \quad 
{{1} \over { 1 \,  \,  \, \, +x\,y \, +y\,z \, +z\,w \, +w\,x \, +y\,w \, +x\,z}}.
\end{eqnarray}
reads:
\begin{eqnarray}
\label{Ratfoncfour20serbis}
\hspace{-0.98in}&&
 \quad  \quad  \quad  \quad  \quad  \quad   \quad  \quad \, \, 
  (1 \, -4\,x) \cdot \, Heun\Bigl({{1} \over {9}}, \, {{5} \over {36}},\,
         {{3} \over {4}},\, {{5} \over {4}},\, 1,\, {{3} \over {2}}, \,\, 4\,x\Bigr)^2.
\end{eqnarray}
This Heun function
can be rewritten as miscellaneous  $\, _2F_1$ with rational pullbacks.
Let us introduce the following rational parametrisation:
\begin{eqnarray}
\label{xz}
\hspace{-0.98in}&& \quad \quad \quad \quad \quad \quad \quad
x \, \, = \, \, \, \,
{{  z \cdot \, (1 \,+2\, z) \cdot \,(1 \, -6\, z) \cdot \,(1 \, +3\, z) } \over {
    (1 \, +6\, z)^2  }}. 
\end{eqnarray}
With this rational parametrisation (\ref{xz}) the diagonal of the rational function (\ref{Ratfoncfour20bis})
can be rewritten in the following ways:
\begin{eqnarray}
\label{Ratfonc3HeunparamSol}
  \hspace{-0.98in}&& \quad  \quad
 Heun\Bigl({{1} \over {9}}, \, {{1} \over {12}},\,
 {{1} \over {4}},\, {{3} \over {4}},\, 1,\, {{1} \over {2}}, \,\,
 {{ 4\,  z \cdot \, (1 \,+2\, z)\cdot \,(1 \, -6\, z)
 \cdot \,(1 \, +3\, z) } \over {(1 \, +6\, z)^2  }}\Bigr)^2
 \nonumber \\
\hspace{-0.98in}&& \quad \quad  \quad 
 \, = \, \,\,\,
 {{(1 \, +4\, z \, +12\, z^2)^2 } \over {(1\, +\, 6 \, z)^2 }} 
\nonumber \\
\hspace{-0.98in}&& \quad \quad \quad \quad \quad \quad
\times \, 
 Heun\Bigl({{1} \over {9}}, \, {{5} \over {36}},\,
 {{3} \over {4}},\, {{5} \over {4}},\, 1,\, {{3} \over {2}}, \,\,
 {{ 4\,  z \cdot \, (1 \,+2\, z)\cdot \,(1 \, -6\, z)
\cdot \,(1 \, +3\, z) } \over {(1 \, +6\, z)^2  }}\Bigr)^2
\nonumber 
\end{eqnarray}
\begin{eqnarray}
\label{Ratfonc3HeunparamSol2}
\hspace{-0.98in}&& \quad \quad \quad 
 \, = \, \,\,\, {{(1\, +\, 6 \, z) } \over {
     (1 \, -36\, z \, -108\, z^2)^{2/3}   }}
  \\
 \hspace{-0.98in}&& \quad \quad  \quad \quad  \quad \quad \quad \quad \times \,
 _2F_1\Bigl([{{1} \over {6}}, \, {{2} \over {3}}], \, [1], \,
 -\,  {{108 \, \, z \cdot \, (1\, +\, 3 \, z)
     \cdot \, (1 \, -4\, z \, -12\, z^2)^2 } \over {
 (1 \, -36\, z \, -108\, z^2)^2  }}  \Bigr)^2
 \nonumber
\end{eqnarray}
\begin{eqnarray}
\label{Ratfonc3HeunparamSol3}
 \hspace{-0.98in}&& \quad  \quad \quad 
 \, = \, \,\,\, {{(1\, +\, 6 \, z)^{1/3} } \over {
     (1 \, -6\, z \, -18\, z^2)^{2/3}  }}
 \\
 \hspace{-0.98in}&& \quad \quad  \quad \quad  \quad \quad \quad \quad \times \,
 _2F_1\Bigl([{{1} \over {6}}, \, {{2} \over {3}}], \, [1], \,
 -\,  {{108 \,\, z^2 \cdot \, (1\, +\, 3 \, z)^2
     \cdot \, (1\, -\, 6 \, z) \cdot \, (1\, +\, 2 \, z) } \over {
 (1\, +\, 6 \, z)^2 \cdot \, (1 \, -6\, z \, -18\, z^2)  }}  \Bigr)^2
 \nonumber 
  \end{eqnarray}
\begin{eqnarray}
\label{Ratfonc3HeunparamSol4}
 \hspace{-0.98in}&& \quad   \quad \quad 
 \, = \, \,\,\, {{(1\, +\, 6 \, z)^{1/2} } \over {
     (1 \, +234\, z \, +972\, z^2 \, + \, 1080 \, z^3)^{1/2}  }}
 \nonumber \\
  \hspace{-0.98in}&& \quad \quad  \quad \quad  \quad \quad \quad
 \times \,
 _2F_1\Bigl([{{1} \over {12}}, \, {{5} \over {12}}], \, [1], \,
 \,  {{1728 \,\, z \cdot \, (1\, +\, 3 \, z)^3
     \cdot \, (1\, -\, 6 \, z)^6 \cdot \, (1\, +\, 2 \, z)^2 } \over {
 (1\, +\, 6 \, z)^3 \cdot \,  (1 \, +234\, z \, +972\, z^2 \, + \, 1080 \, z^3)^3  }}  \Bigr)^2
 \nonumber
\end{eqnarray}
\begin{eqnarray}
\label{Ratfonc3HeunparamSol5}
 \hspace{-0.98in}&&  \quad   \quad \quad 
 \, = \, \,\,\, {{(1\, +\, 6 \, z)^{1/2} } \over {
     (1 \, -6 \, z \, +12\, z^2 \, + \, 120 \, z^3)^{1/2}   }}
 \nonumber \\
 \hspace{-0.98in}&& \quad \quad  \quad \quad   \quad \quad \quad \times \,
 _2F_1\Bigl([{{1} \over {12}}, \, {{5} \over {12}}], \, [1], \,
 \,  {{1728 \,\, z^3 \cdot \, (1\, +\, 3 \, z)
 \cdot \, (1\, -\, 6 \, z)^2 \cdot \, (1\, +\, 2 \, z)^6 } \over {
 (1\, +\, 6 \, z)^3 \cdot \,  (1 \, -6 \, z \, +12\, z^2 \, + \, 120 \, z^3)^3  }}  \Bigr)^2. 
\nonumber
\end{eqnarray}
\begin{eqnarray}
\label{Ratfonc3HeunparamSol6}
 \hspace{-0.98in}&& \, 
 \, = \, \,\,\, 
 \,  1\,\, \,  +6\,z \,\,\,   +12\,{z}^{2}\, +96\,{z}^{3}\, +360\,{z}^{4}\, +2160\,{z}^{5}\,
 +10488\,{z}^{6}\, +58464\,{z}^{7}
 \, \,\, \,  + \, \, \, \cdots 
\nonumber
\end{eqnarray}

\vskip .2cm

\section{A nome necessary condition to be a classical modular form: why $\, _2F_1([1/5,1/5],[1],\, x)$  is not a classical modular form.}
\label{notmodular}

Consider the identity:
\begin{eqnarray}
\label{identite12}
  \hspace{-0.98in}&&  \,   \quad  \quad  \quad   \quad 
_2F_1 \Bigl([{{1} \over {3}}, \, {{2} \over {3}}], \, [1], \,\, \,  x\Bigr)
  \\
  \hspace{-0.98in}&&  \,  \quad  \quad  \quad  \quad    \quad  \quad  \quad
  \, \, = \, \, \,
(1\, + \, 8 \, x)^{-1/4} \cdot \, 
 _2F_1 \Bigl([{{1} \over {12}},\, {{5} \over {12}}], \, [1],
\,\, \,  64 \cdot \,{\frac {x \cdot \, (1 \, -x)^{3}}{(1 \, +8\,x)^{3}}}\Bigr).
 \nonumber 
\end{eqnarray}
The nome associated to the linear differential operator of order two having $\,_2F_1([1/3,\, 2/3], \, [1],\, x) \, $
as a solution is given by:
\begin{eqnarray}
\label{nome1}
\hspace{-0.98in}&& 
 Q(x) \, \, = \, \, \, \, 
 x \,\,  \, +{{5} \over {9}} \,{x}^{2} \, \, +{\frac {31}{81}}\,{x}^{3} \,
+{\frac {5729 }{19683}}\,{x}^{4} \, +{\frac {41518 }{177147}}\,{x}^{5}
\, +{\frac {312302}{1594323}}\,{x}^{6} \, \, \, \, + \, \, \, \cdots 
\end{eqnarray}
and the nome associated to the operator of order two having
$\,\,_2F_1([1/12,5/12], \, [1],\, x)\,$ as a solution expands as follows:
\begin{eqnarray}
\label{nome2}
\hspace{-0.98in}&& 
q(x) \, \, = \, \,\, \, x \,\, +{\frac {31}{72}}\,{x}^{2}
\, +{\frac {20845}{82944}}\,{x}^{3} \, +{\frac {27274051}{161243136}}\,{x}^{4}
\, +{\frac {183775457147 }{1486016741376}}\,{x}^{5}\, \,\,  \, + \, \, \, \cdots 
\end{eqnarray}
The two  $\, _2F_1$ hypergeometric series are {\em globally bounded}, the series of the corresponding
nomes (\ref{nome1}) and  (\ref{nome2}) {\em are also globally bounded}, as one expects
for a classical modular form. 
The identity (\ref{identite12}) on the other solutions of the linear differential
operators annihilating  $\,\,_2F_1([1/3,\, 2/3], \, [1],\, x) \, $
 and  $\,\,_2F_1([1/12,5/12], \, [1],\, p(x))$,  gives the following
 identity on their respective ratio $\, \tau$
\begin{eqnarray}
\label{tau12}
  \hspace{-0.98in}&& \quad   \, \,   \quad \quad  \quad
 \tau\Bigl([{{1} \over {3}}, \, {{2} \over {3}}], \, [1],\, x\Bigr)
\, \,\, = \, \, \,\,
\mu \cdot \,    \tau\Bigl([{{1} \over {12}},\, {{5} \over {12}}], \, [1],
 \,  64 \cdot \,{\frac {x \cdot \, (1 \, -x)^{3}}{(1 \, +8\,x)^{3}}} \Bigr),            
\end{eqnarray}
where $\, \mu$ is a constant, which gives after exponentiation:
\begin{eqnarray}
\label{q12}
  \hspace{-0.98in}&& \quad  \quad \quad \quad  \quad  \quad\quad \quad
 64 \cdot \, Q(x) \, \, \, = \, \,  \, \,\,
q\Bigl(64 \cdot \,{\frac {x \cdot \, (1 \, -x)^{3}}{(1 \, +8\,x)^{3}}} \Bigr).               
\end{eqnarray}
Now, the RHS of (\ref{q12}) is {\em necessarily globally bounded}, which agrees with the  globally bounded
character of the nome (\ref{nome1}).

In contrast, let us consider $_2F_1([1/5,1/5],[1],\, x)$. The corresponding series
{\em is globally bounded}\footnote[5]{Any $_2F_1(a, \, b],[c],\, x)$ with $ \, c= \, 1$
  is globally bounded since it is of {\em weight} zero: it is of the form $ \, _nF_{n-1}$,
  and has $ \, c$ given by an integer and not a fractional number.}, however
the corresponding nome which reads
\begin{eqnarray}
\label{nome5}
\hspace{-0.98in}&& \,  \quad   \quad   \quad  
Q_{[1/5,1/5]}(x) \, \, \, = \, \, \, \,\,
x \,  \, \,\, +{\frac {8}{25}}\,{x}^{2} \,  \,\,  +{\frac {102 }{625}}\,{x}^{3} \, \,  \,
+{\frac {4744 }{46875}} \, x^4  \, \, \,  +{\frac {81914}{1171875}} \, x^5
 \nonumber \\
  \hspace{-0.98in}&& \quad  \quad  \quad  \quad  \quad  \quad \quad \quad
\, \,  +{\frac {63094248}{1220703125}} \, x^6  \,
\, \,  +{\frac {11003093386\,{x}^{7}}{274658203125}} \,\, \,\, \, + \, \, \, \cdots 
\end{eqnarray} is {\em not} globally bounded. Therefore, it is not possible to find
any {\em algebraic (or rational) pullback} $\, p(x)$ such that
\begin{eqnarray}
\label{nome57}
  \hspace{-0.98in}&& \quad  \quad \quad \quad \quad \quad \quad \quad \quad
  \mu \cdot \,  Q_{[1/5,1/5]}(x) \,\, \, = \, \, \,\, q\Bigl(p(x)\Bigr), 
\end{eqnarray}
since the RHS of (\ref{nome57}) is {\em necessarily globally bounded} when
$\,\, \mu \cdot \,  Q_{[1/5,1/5]}(x)\,\, $
{\em cannot be globally bounded} regarless of the constant  $\, \mu$.

\vskip .1cm

\subsection{The globally bounded nome  condition to be a classical modular form for Heun functions.}
\label{notmodularsub}

Let us recall the order-two linear differential operator (\ref{Heun}), which has
$\, Heun(a, \, q, \, \alpha, \, \beta, \, \gamma, \, \delta, \, x)$ as a solution.
Since we are interested in diagonals of rational functions, we focus on series expansions at $\, x\, = \, 0$.
To have a nome requires the other solution to have a formal series expansion with a logarithm
which corresponds to impose that the parameter\footnote[1]{The condition to have a logarithm
  for the formal series of  (\ref{Heun}) at  $\, x\, = \, 1$ is that the parameter $\, \delta$  is an integer.
The  condition to have a logarithm at  $\, x\, = \, a$ is
$\, \epsilon \, = \, \, \alpha \, + \, \beta \, -\gamma \, -\delta \, +1$  being an integer.} $\, \gamma$ is an integer.
To have a logarithm is a necessary but not sufficient condition to have a nome which is analytic
at $\, x \, = \, 0$ (mirror map). When\footnote[2]{This actually corresponds to what we had called
``premodular condition'' in~\cite{Youssef}.  The premodular condition
$\, W(x) \, = \, \, -1/2/x^2 \, + \, \, \cdots$
yields exactly $ \,\gamma \, = \, \, 1$. In contrast with $\, _2F_1$ hypergeometric functions, this
premodular condition  is far from imposing that the
corresponding Heun series is globally bounded: $\, Heun(3,5,7,11,1,13, \, x)$,
for instance, is not globally bounded. }  $\, \gamma\, = \, \, 1 \,  \, $  we have a nome $\, Q$,
analytic at $\, x \, = \, 0$, which series expansion reads:
\begin{eqnarray}
\label{nomeHeun}
  \hspace{-0.98in}&& \quad  \, \quad  \quad \quad  \quad \quad  \quad
Q \, \,\,\, = \, \,\,\,\,  x \, \, \,\,\, +   \Bigl( \delta \, -2 \, \, \,
+  {{\alpha \, + \, \beta \, - \, \delta} \over {a}}\Bigr)   \cdot \, x^2
\,\,\,\, \, + \, \, \cdots 
\end{eqnarray}
The next terms become more and more involved rational expressions of the parameters of the Heun function.
For a given Heun function with $\, \gamma\, = \, \, 1$  one can easily use the  globally bounded nome  condition.
In practice this is an efficient way to discard the   $\, \gamma\, = \, \, 1 \, $
Heun functions which are not classical  modular forms.
However, finding exhaustively the Heun  functions with $\, \gamma\, = \, \, 1$ corresponding
to  classical  modular forms, remains a quite involved task (see~\cite{ReduceMaier} and  \ref{ReduceMaierHeun} below).

\vskip .1cm

\section{Periods of an extremal rational surface: a rational parametrisation.}
\label{extremparam}

Let us introduce a rational parametrisation
for the linear differential operator $\, M_2$
and its solution (\ref{spurious}):
\begin{eqnarray}
\label{changeextrem}
  \hspace{-0.98in}&& \quad  \, \quad   \quad \quad   \quad  \quad  \quad  \quad \quad
x \, \, = \, \, \,    {{z^2} \over {(1\, -z) \cdot (2\, +\, z)}}.
\end{eqnarray}
With that rational change of variable (\ref{changeextrem})  the solution (\ref{spurious}) of $\, M_2$
becomes:
\begin{eqnarray}
\label{changeextremsolident}
  \hspace{-0.98in}&& \quad \quad \quad \quad \quad 
 {\cal S}_2 \, \, = \, \, \, {\frac { (z+2)^{2} \cdot (1 \, -z) }{  2 \cdot \, p_{12}(z)^{1/4}}}
 \cdot \, _2F_1\Bigl([{{1} \over {12}}, \, {{5} \over {12}}], \, [1], \, \, H_1 \Bigr)
   \nonumber \\
  \hspace{-0.98in}&& \quad  \quad \quad \quad \quad \quad \quad \quad 
 \, \, = \, \, \, {\frac { (z+2)^{2} \cdot \, (1 \,-z) }{ q_{12}(z)^{1/4}}}  \cdot \, \,
_2F_1\Bigl([{{1} \over {12}}, \, {{5} \over {12}}], \, [1], \, \, H_2 \Bigr)
  \\
  \hspace{-0.98in}&& \quad \quad  \quad  \quad  \quad 
 \, \, = \, \, \,\, \, \,
 1 \, \, \, -{\frac{1}{4}}{z}^{2} \,\, \,  -{\frac{1}{16}}{z}^{5}\,  \,
 +{\frac{1}{8}}{z}^{6} \, \, +{\frac{3}{64}}{z}^{8} \, \, +{\frac{3}{32}}{z}^{9}
 \,\,  +{\frac{45}{512}}{z}^{10}    \, \, \, \, \, \, +   \,  \, \,  \cdots
 \nonumber   
\end{eqnarray}
where
\begin{eqnarray}
\label{changeextremsol}
  \hspace{-0.98in}&& 
 H_1  \, \, = \, \,  
 {\frac {1728  \cdot \, {z}^{5} \cdot \,
 (1\, - z^2)^{4}  \, (1  \, -z^3)^{2}  \, (z+2)^{3} \, ({z}^{2}+4\,z-4)  \, ({z}^{2}-2\,z+4) }{
  p_{12}(z)^{3}}}, 
 \\
  \hspace{-0.98in}&& 
H_2  \, \, = \, \, 
{\frac { 1728 \cdot \,  {z}^{10}  \cdot \,
(1\, -{z}^{3})  \, (1\, -{z}^{2})^{2} \, (z+2)^{6} \, ({z}^{2}+4\,z-4)^{2} \, ({z}^{2}-2\,z+4)^{2}}{
 q_{12}(z)^{3}}},                    
\end{eqnarray}
with:
\begin{eqnarray}
  \label{p12P12}
\hspace{-0.98in}&&\quad  \quad  \quad
  p_{12}(z) \, \, = \, \, \, 
 {z}^{12} \, +8\,{z}^{11} \, +16\,{z}^{10} \, +8\,{z}^{9} \, +48\,{z}^{8} \, +112\,{z}^{7} \, -96\,{z}^{5}
  \nonumber \\
  \hspace{-0.98in}&& \quad \quad  \quad  \quad \quad  \quad  \quad  \quad  \quad
+16\,{z}^{4} \, -16\,{z}^{3} \, -32\,{z}^{2} \, +16,
   \\
  \hspace{-0.98in}&&\quad  \quad  \quad
 q_{12}(z) \, \, = \, \, \, {z}^{12} \, +8\,{z}^{11} \,
  +16\,{z}^{10} \, +8\,{z}^{9} \, +48\,{z}^{8} \, -128\,{z}^{7} \, +384\,{z}^{5}
 \nonumber \\
  \hspace{-0.98in}&& \quad \quad  \quad  \quad \quad  \quad  \quad \quad  \quad
+256\,{z}^{4} \, -256\,{z}^{3} \, -512\,{z}^{2} \, +256.                
\end{eqnarray}
Relation (\ref{changeextremsolident}) makes crystal clear that the solution (\ref{spurious}) of $\, M_2$
corresponds to a classical modular form, associated with the well-known genus-zero
fundamental modular
equation\footnote[1]{Corresponding to $q \,\,  \leftrightarrow \,\,  q^2$
  in the nome, see equation (4) in~\cite{Youssef}.}:
\begin{eqnarray}
\label{p12P12}
\hspace{-0.98in}&& \quad  \quad  \quad   \quad  
1953125 \cdot \, A^3\, B^3 \, \,\, \,  -187500 \cdot  \, A^2\, B^2 \cdot \, (A+B) \, \,\, 
\nonumber \\
  \hspace{-0.98in}&&\quad  \quad  \quad  \quad  \quad \quad \quad \quad 
+375 \cdot  \, A\, B \cdot \, (16\, A^2-4027\, A\, B+16\, B^2)
\\
  \hspace{-0.98in}&&\quad  \quad  \quad \quad \quad
\, \,  -64 \cdot \, (A+B) \cdot  \, (A^2+1487\, A\, B+B^2) \,\, \, \,   +110592 \cdot \, A\, B
\, \, \,  = \, \, \,\,  0.
 \nonumber
\end{eqnarray}
With the same rational change of variable (\ref{changeextrem}), 
the solution (\ref{Ratfonc4HeunSol2}) of the order-two linear differential operator
$\, L_2$ can be written in term of a {\em very simple} Heun function
associated with the following remarkable Heun identity:
\begin{eqnarray}
\label{identitycrazy}
\hspace{-0.98in}&& \, \, 
 {\cal S}_1\, \, = \, \, \,  Heun \Bigl({{1} \over {2}}\, -{{i \sqrt{3}} \over {2}}, \,
{{1} \over {2}}\, -{{i \sqrt{3}} \over {6}}
, \, 1, \, 1, \, 1, \,1, \,\,\,
                   {{3} \over {2}} \cdot \, \Bigl(-3 \, + \, i \sqrt{3}  \Bigr) \cdot \, {{z^2} \over {(1\, -z) \cdot (2\, +\, z)}}\Bigr)
\nonumber \\
 \hspace{-0.98in}&& \quad  \, \quad  \quad \quad
\, \, = \, \, \, \,  {{1} \over {4}} \cdot \, (1\, -z) \cdot \, (2\, +\, z) \cdot \, 
    Heun \Bigl( -\, {{1} \over {8}}, \,  {{1} \over {4}}, \, 1, \, 1, \, 1, \, 1, \, \,  -\, {{z^3 } \over { 8}}\Bigr).                   
\end{eqnarray}
The solution $\, S_1$ can also be written:
\begin{eqnarray}
\label{identitycrazyhyp}
  \hspace{-0.98in}&& \, \, \quad  \quad  \quad  \quad 
{\cal S}_1\, \,\, = \, \, \,\,
{{ (z\, +2)^2 \cdot \, (1 \, -z)} \over { 2 \cdot \, (z^3 \, +2) }} \cdot \,
 _2F_1\Bigl([{{1} \over {3}}, \, {{2} \over {3}} ], \, [1], \, \Bigl({{3 \, z} \over {z^3 \, +2}} \Bigr)^3 \Bigr).              
\end{eqnarray}

\vskip .1cm

\section{Special $\, _2F_1$ hypergeometric functions associated with classical  modular forms }
\label{Special}

The Heun functions of this paper
can all be rewritten in terms of pullbacked $\, _2F_1$ hypergeometric functions
which turn out to correspond to {\em classical modular curves} (with the exception of the ``Shimura'' Heun functions of
section (\ref{vanHoeijVidunas}), see also \ref{Shimura} below). These  $\, _2F_1$ hypergeometric
functions are not arbitrary, they are ``special'' $\, _2F_1$'s corresponding to
selected parameters, namely  $\, _2F_1$'s related to {\em classical modular curves}.  These
various $\, _2F_1$ are often simply related
\begin{eqnarray}
  \label{classicalIdentities}
  \hspace{-0.98in}&& \quad \quad \quad 
  _2F_1\Bigl([{{1} \over {2}}, \,{{1} \over {2}}], \, [1], \, \, x\Bigr)
 \\
 \hspace{-0.98in}&& \quad  \quad  \quad \quad \quad 
 \, \, = \, \, \,\,
(1\, +\omega^2\, x)^{-1/2} \cdot \,  _2F_1\Bigl([{{1} \over {6}}, \,{{1} \over {2}}], \, [1],
\, \, \, {{ 3 \, \cdot \,  \omega \cdot \, (\omega \, -1) \cdot \, x \, \cdot \, (1 \, -x)} \over { (1 \, + \omega^2 \cdot x)^3}}  \Bigr),
 \nonumber
\end{eqnarray}
where $\,\,\, 1\, + \,  \omega \, + \omega^2 \, = \, 0$ ($ \, \omega \, $ is a third root of unity), or:
\begin{eqnarray}
\label{derivative2}
\hspace{-0.98in}&&   \,
_2F_1\Bigl([{{1} \over {4}}, \, {{1} \over {4}}],\, [1], \, 64 \, x\Bigr)
\, \, = \, \, \,
   {{1} \over { (1 \, - \, 64 \, x)^{1/4} }} \cdot \,
   _2F_1\Bigl([{{1} \over {4}}, \, {{3} \over {4}}],\, [1], \,
  - \, {{ 64 \, x} \over {1 \, - \, 64 \, x}} \Bigr)
\nonumber \\
\hspace{-0.98in}&&  \quad  \quad \,\quad 
\, \, = \, \, \,  {{1} \over { (1 \, + \, 64 \, x)^{1/4} }} \cdot \,
   _2F_1\Bigl([{{1} \over {8}}, \, {{5} \over {8}}],\, [1], \,
   \, {{ 256 \, x} \over {(1 \, + \, 64 \, x)^2}} \Bigr)
 \\
\hspace{-0.98in}&&  \quad  \quad \,\quad \quad 
\, \, = \, \, \,
                   {{1} \over { (1 \, - \, 64 \, x)^{1/12} \cdot \, (1 \, +\, 8 \, x )^{1/6} }}
\nonumber  \\
\hspace{-0.98in}&&  \quad  \quad \quad \quad  \quad \quad  \quad \,
                   \times \,
 _2F_1\Bigl([{{1} \over {12}}, \, {{7} \over {12}}],\, [1], \,
 \, - \, {{ 1728 \, x^2} \over { (1 \, - \, 64 \, x) \cdot \, (1 \, +\, 8 \, x )^2}} \Bigr)
\nonumber 
\end{eqnarray}
\begin{eqnarray}
  \label{focus}
  \hspace{-0.98in}&&
  _2F_1\Bigl([{{1} \over {2}}, \,{{5} \over {6 }}], \, [1], \, 144 \, \, x  \Bigr)
  \, \, = \, \, \,  \, 
  (1\, -144 \, x)^{-1/3} \cdot \,
  _2F_1\Bigl([{{1} \over {2}}, \,{{1} \over {6 }}], \, [1], \, 144 \, \, x  \Bigr)
     \nonumber \\
\hspace{-0.98in}&& \quad \quad \quad 
     \, \, = \, \, \, \,  (1\, -144 \, x)^{-5/12} \cdot \,
     _2F_1\Bigl([{{1} \over {6}}, \,{{5} \over {6 }}], \, [1], \,
        {{1} \over {2}} \,
        -\, {{1} \over {2}} \cdot \, {{1 \, - 72\, x } \over { (1\, -144 \, x)^{1/2} }} \Bigr)
        \nonumber \\
\hspace{-0.98in}&& \quad \quad \quad  
  \, \, = \, \, \, \, 
  (1 \, - 72\, x)^{-1/6}   \cdot \, (1\, -144\, x)^{-1/3} \, \cdot \, 
  _2F_1\Bigl([{{1} \over {12}}, \,{{7} \over {12 }}], \, [1], \,
     {{ 5184\, x^2 } \over { (1 \, - 72\, x)^2}}  \Bigr)
           \nonumber \\
  \hspace{-0.98in}&& \quad \quad \quad   \quad
 \, \, = \, \, \, \, 
  (1 \, - 144\, x)^{-5/12}   \cdot \,
  _2F_1\Bigl([{{1} \over {12}}, \,{{5} \over {12 }}], \, [1], \,
     - \, {{ 5184\, x^2 } \over { (1 \, - 144\, x)}}  \Bigr),  
\end{eqnarray}
\begin{eqnarray}
  \label{focus23}
  \hspace{-0.98in}&&  
  _2F_1\Bigl([{{2} \over {3}}, \,{{2} \over {3}}], \, [1], \, 27 \, \, x  \Bigr)
 \nonumber \\
\hspace{-0.98in}&& \quad \quad \quad \quad   \quad  
 \, \, = \, \, \, 
  (1\, - 729\, x^2)^{-1/3} \cdot \,
  _2F_1\Bigl([{{1} \over {6}}, \,{{2} \over {3}}], \, [1], \, {{ 108 \, x} \over { (1\, + \, 27\, x)^2}}  \Bigr)
\end{eqnarray}
 \begin{eqnarray}
  \label{focus23bisx}
  \hspace{-0.98in}&&   \quad \quad \quad 
 \, \, = \, \, \, {{1} \over { (1\, - \, 27\, x)^{5/12} \cdot \, (1\, -3\, x)^{1/4}  }}
\nonumber  \\
\hspace{-0.98in}&&  \quad  \quad\quad  \quad \quad  \quad  \quad \,
\times \,    \,
 _2F_1\Bigl([{{1} \over {12}}, \,{{5} \over {12}}], \, [1],
  \,  \, - \, {{ 1728 \, x^3} \over { (1\, - \, 27\, x) \cdot \, (1\, -3\, x)^3}}  \Bigr)
 \nonumber
\end{eqnarray}                  
\begin{eqnarray}
  \label{focus23bisxbisx}
  \hspace{-0.98in}&&  \quad\quad 
\, \, = \, \, \, {{1} \over { (1\, - \, 27\, x)^{5/12} \cdot \, (1\, -243\, x)^{1/4}  }}
 \nonumber  \\
\hspace{-0.98in}&&  \quad  \quad \quad  \quad \quad  \quad  \quad \,
\times \,  \,
 _2F_1\Bigl([{{1} \over {12}}, \,{{5} \over {12}}], \, [1], \,  \,
  - \, {{ 1728 \, x} \over { (1\, - \, 27\, x) \cdot \, (1\, -243\, x)^3}}  \Bigr).
  \nonumber
\end{eqnarray}
\begin{eqnarray}
  \label{focus2}
  \hspace{-0.98in}&&
  _2F_1\Bigl([{{3} \over {8}}, \,{{7} \over {8}}], \, [1], \, 256 \, \, x  \Bigr)
  \, \, = \, \, \, 
  (1\, -256\, x)^{-1/4} \cdot \,
  _2F_1\Bigl([{{1} \over {8}}, \,{{5} \over {8}}], \, [1], \, 256 \, \, x  \Bigr),
 \end{eqnarray}
\begin{eqnarray}
  \label{focus3}
  \hspace{-0.98in}&& 
  _2F_1\Bigl([{{1} \over {3}}, \,{{5} \over {6}}], \, [1], \, 108 \, \, x  \Bigr)
  \, \, = \, \, \, 
  (1\, -108\, x)^{-1/6} \cdot \,
  _2F_1\Bigl([{{1} \over {6}}, \,{{2} \over {3}}], \, [1], \, 108 \, \, x  \Bigr), 
 \end{eqnarray}
\begin{eqnarray}
  \label{focus4}
  \hspace{-0.98in}&&
  _2F_1\Bigl([{{2} \over {3}}, \,{{5} \over {6}}], \, [1], \, 108 \, \, x  \Bigr)
  \, \, = \, \, \, 
  (1\, -108\, x)^{-2/3} \cdot \,
  _2F_1\Bigl([{{1} \over {6}}, \,{{2} \over {3}}], \, [1], \,
      -\, {{108 \, \, x} \over {1 \, -\,  108 \, \, x  }}  \Bigr), 
\end{eqnarray}
\begin{eqnarray}
  \label{focus5}
  \hspace{-0.98in}&&
  _2F_1\Bigl([{{1} \over {2}}, \,{{3} \over {4}}], \, [1], \, 32 \, \, x  \Bigr)
  \, \, = \, \, \, 
  (1\, -32\, x)^{-1/4} \cdot \,
  _2F_1\Bigl([{{1} \over {4}}, \,{{1} \over {2}}], \, [1], \, 32 \, \, x  \Bigr)
  \\
 \hspace{-0.98in}&& \quad \quad \quad \quad \quad  \, \, = \, \, \, 
  (1\, -32\, x)^{-1/4} \cdot \, (1\, -16\, x)^{-1/4} \cdot \,
 _2F_1\Bigl([{{1} \over {8}}, \,{{5} \over {8}}], \, [1],
 \, {{256\,x^2 } \over {(1\, -16\, x)^2 }} \Bigr),  \nonumber
 \end{eqnarray}
\begin{eqnarray}
\label{focus7}
\hspace{-0.98in}&& 
_2F_1\Bigl([{{5} \over {6}}, \,{{5} \over {6}}], \, [1], \, 432 \, \, x  \Bigr)
 \, \,  \,  = \, \, \, 
 (1\, - 432\, x)^{-5/6} \cdot \,
  _2F_1\Bigl([{{1} \over {6}}, \,{{5} \over {6}}], \, [1],
  \, -\,  {{432 \, \, x} \over { 1\, - 432\, x }}  \Bigr)
  \nonumber \\
 \hspace{-0.98in}&&  \, 
  \, \, = \, \, \,  (1\, - 432\, x)^{-2/3}   \cdot \, (1\, + 432\, x)^{-1/6}   \cdot \,
  _2F_1\Bigl([{{1} \over {12}}, \,{{7} \over {12}}], \, [1], \,
       {{1728\, x } \over { (1\, + 432\, x)^2}}  \Bigr).
\end{eqnarray}
\begin{eqnarray}
  \label{focus8}
  \hspace{-0.98in}&& 
  _2F_1\Bigl([{{3} \over {4}}, \,{{3} \over {4}}], \, [1], \, 64 \, \, x  \Bigr)
  \, \, = \, \, \,  
  (1\, - 64\, x)^{-3/4} \cdot \,
  _2F_1\Bigl([{{1} \over {4}}, \,{{3} \over {4}}], \, [1],
  \, -\,  {{64 \, \, x} \over { 1\, - 64\, x }}  \Bigr)
  \\
   \hspace{-0.98in}&& \quad \quad  \quad  \quad 
 \, \, = \, \, \, 
  (1\, - 64\, x)^{-1/2} \cdot \, (1\, + 64\, x)^{-1/4} \cdot \,
  _2F_1\Bigl([{{1} \over {8}}, \,{{5} \over {8}}], \, [1],
 \, \,  {{256 \, \, x} \over {  (1\, + 64\, x)^2 }}  \Bigr),  \nonumber 
 \end{eqnarray}
\begin{eqnarray}
  \label{focus9}
  \hspace{-0.98in}&&
  _2F_1\Bigl([{{1} \over {2}}, \,{{2} \over {3}}], \, [1], \, 36 \, \, x  \Bigr)
  \, \,\, = \, \,\, \,  
  (1\, - 36\, x)^{-1/6} \cdot \,
  _2F_1\Bigl([{{1} \over {3}}, \,{{1} \over {2}}], \, [1],
  \, 36 \, \, x  \Bigr)
  \\
  \hspace{-0.98in}&& \quad \quad \quad \quad  
  \, \, = \, \, \, 
  (1\, - 36\, x)^{-1/6} \cdot \, (1\, -18\, x)^{-1/3} \cdot \,
  _2F_1\Bigl([{{1} \over {6}}, \,{{2} \over {3}}], \, [1],
 \, \,  {{324 \, \, x^2} \over {  (1\, - 18 \, x)^2 }}  \Bigr), \nonumber 
\end{eqnarray}
\begin{eqnarray}
  \label{classicalIdentities}
  \hspace{-0.98in}&& 
  \label{ratpullback}         
  _2F_1\Bigl([{{1} \over {3}}, \,{{2} \over {3}}], \, [1], \, \, x\Bigr)
  \, \, = \, \, \,\,
(1\, + 8\, x)^{-1/4} \cdot \,  _2F_1\Bigl([{{1} \over {12}}, \,{{5} \over {12}}], \, [1],
 \, \, \, {{ 64 \cdot \, x  \cdot \, (1\, -x)^3} \over { (1 \, + \, 8\, x)^3}}  \Bigr),
 \end{eqnarray}
\begin{eqnarray}
  \label{ratpullback2}
  \hspace{-0.98in}&&     
  _2F_1\Bigl([{{1} \over {4}}, \,{{3} \over {4}}], \, [1], \, \, x\Bigr)
  \, \, = \, \, \,\,
(1\, + 3\, x)^{-1/4} \cdot \,  _2F_1\Bigl([{{1} \over {12}}, \,{{5} \over {12}}], \, [1],
\, \, \, {{ 27 \cdot \, x  \cdot \, (1\, -x)^2} \over { (1 \, + \, 3\, x)^3}}  \Bigr). 
\end{eqnarray}
In fact, all these ``special'' $\, _2F_1$'s hypergeometric functions correspond to
{\em classical modular forms} because they can be rewritten~\cite{Youssef} as  
 $ \, {\cal A} \cdot \, _2F_1([1/12,5/12],[1], p(x))$ where the pullback $\, p(x)$ 
 is in general more involved than simple rational pullbacks like (\ref{ratpullback})
 or (\ref{ratpullback2}), being often {\em algebraic} functions.
 For instance one has the following identities
\begin{eqnarray}
 \label{forinstance}
 \hspace{-0.98in}&&  
_2F_1\Bigl([{{1} \over {8}}, \,{{5} \over {8}}], \, [1],   \, \, x\Bigr) \, \, = \, \, \,
  \\
\hspace{-0.98in}&& \quad \, \, = \, \, \,
\Bigl({{8} \over {(5 \cdot \, (1\, -x)^{1/2}\, +3)^{1/4} }} \Bigr)^{1/4}
\cdot \, _2F_1\Bigl([{{1} \over {12}}, \,{{5} \over {12}}], \, [1],  \, \,
 27 \cdot \, x \cdot \, {{1 \, - (1\, -x)^{1/2} } \over { 3 \, +5 \cdot \,  (1\, -x)^{1/2} }} \Bigr)
 \nonumber \\
  \hspace{-0.98in}&&  \quad  \, \, = \, \, \,
\Bigl({{2} \over {(5 \cdot \, (1\, -x)^{1/2} \, -3 )^{1/4} }} \Bigr)^{1/4}
\cdot \, _2F_1\Bigl([{{1} \over {12}}, \,{{5} \over {12}}], \, [1],  \, \,
27 \cdot \, x \cdot \, {{1 \, + (1\, -x)^{1/2} } \over { 3 \, -5 \cdot \,  (1\, -x)^{1/2} }} \Bigr),
\nonumber
\end{eqnarray}
or
\begin{eqnarray}
 \label{forinstance2}
 \hspace{-0.98in}&&   \quad \quad  \quad  \quad 
_2F_1\Bigl([{{1} \over {6}}, \,{{2} \over {3}}], \, [1],   \, \, x\Bigr) \, \,  \, \, = \, \, \, \, {\cal A}(x)
 \cdot \, _2F_1\Bigl([{{1} \over {12}}, \,{{5} \over {12}}], \, [1],  \, \, {\cal H}(x) 
 \Bigr), 
\end{eqnarray}
where $\, {\cal H}(x)$ reads
\begin{eqnarray}
 \label{forinstance2}
  \hspace{-0.98in}&&
 {\cal H}(x) \, \, = \, \, \,  4 \cdot \, x \cdot \, {\frac { 1458 \, -1215\,x \, + 125\,{x}^{2} }{( 25\,x \, -9)^{3}}}
 \,  \,  \, +8 \cdot \, x \cdot \,{\frac {(27-11\,x)  \cdot \, (27-25\,x) }{\sqrt {1-x} \cdot \, (9-25\,x)^{3}}}, 
\end{eqnarray}          
and
where $\, {\cal A}(x)$ reads:
\begin{eqnarray}
 \label{forinstance3}
\hspace{-0.98in}&& \quad \quad \quad 
{\cal A}(x)   \, \, = \, \, \,
(1\, -x)^{-1/24} \cdot \,  \Bigl( {{ 81} \over { 9 \, -25 \, x}}  \Bigr)^{1/8}
 \cdot \,  \Bigl( {{ 5 \cdot \, (1\, -x)^{1/2} \, -4 } \over {  5 \cdot \, (1\, -x)^{1/2} \, +4}}   \Bigr)^{1/8},                 
\end{eqnarray}          
which is solution of
\begin{eqnarray}
 \label{forinstance3}
\hspace{-0.98in}&&
\, \,   531441 \, \, -7290\, \cdot \, (x-1)  \cdot \,  (25\,x-73) \cdot \,  Z
\, \, + (x-1)  \cdot \,  (25\,x-9)^{3} \cdot \, {Z}^{2} \, \, = \,\, \, 0,            
\end{eqnarray}
where $\, Z \, = \, \,  {\cal A}(x)^{12}$.

Generalizing the globally bounded nome condition approach of the previous \ref{notmodular},  we
looked for all possible $\, _2F_1$ hypergeometric functions related\footnote[3]{See ~\cite{Short,Big},
  and the hypergeometric functions in the previous
  sections in this paper.}  to pullbacked $_2F_1([1/12,5/12],[1],x)$,
{\em looking at the globally bounded condition of their nome} (see (\ref{nome57})).  We
give here a {\em finite list} of only $\, 28$ hypergeometric functions that have
{\em integer coefficient series}, that are related\footnote[1]{By related to classical modular forms,
we mean, from now, that any of the hypergeometric functions below can be rewritten as a pullbacked
  $ \,  \, _2F_1([1/12,5/12],[1],x) \, $ function, and hence is necessarily a  classical modular form.}
to {\em modular forms}. 
\begin{eqnarray}
  \label{Specialliste}
  \hspace{-0.98in}&& \quad \quad  \quad 
 _2F_1\Bigl([{{1} \over {2}}, \,{{1} \over {2 }}], \, [1], \, 16 \, \, x  \Bigr), \, \,  \, \,  \, 
 _2F_1\Bigl([{{1} \over {2}}, \,{{1} \over {3 }}], \, [1], \, 36 \, \, x  \Bigr), \, \,  \, \,  \, 
 _2F_1\Bigl([{{1} \over {3}}, \,{{1} \over {3 }}], \, [1], \, 27 \, \, x  \Bigr),
  \nonumber \\
    \hspace{-0.98in}&& \quad \quad  \quad 
 _2F_1\Bigl([{{1} \over {3}}, \,{{2} \over {3 }}], \, [1], \, 27 \, \, x  \Bigr), \, \,  \, \,  \, 
  _2F_1\Bigl([{{1} \over {6}}, \,{{1} \over {2 }}], \, [1], \, 108 \, \, x  \Bigr), \, \,  \, \,  \, 
    _2F_1\Bigl([{{1} \over {6}}, \,{{1} \over {3 }}], \, [1], \, 108 \, \, x  \Bigr),
    \nonumber \\
    \hspace{-0.98in}&& \quad \quad  \quad 
 _2F_1\Bigl([{{1} \over {6}}, \,{{2} \over {3 }}], \, [1], \, 432 \, \, x  \Bigr), \, \,  \, \,  \, 
 _2F_1\Bigl([{{1} \over {6}}, \,{{1} \over {6 }}], \, [1], \, 432 \, \, x  \Bigr), \, \,  \, \,  \, 
    _2F_1\Bigl([{{1} \over {6}}, \,{{5} \over {6 }}], \, [1], \, 432 \, \, x  \Bigr),
    \nonumber \\
    \hspace{-0.98in}&& \quad \quad  \quad 
 _2F_1\Bigl([{{1} \over {4}}, \,{{1} \over {4 }}], \, [1], \, 64 \, \, x  \Bigr), \, \,  \, \,  \, 
 _2F_1\Bigl([{{1} \over {4}}, \,{{1} \over {2 }}], \, [1], \, 32 \, \, x  \Bigr), \, \,  \, \,  \, 
    _2F_1\Bigl([{{1} \over {4}}, \,{{3} \over {4 }}], \, [1], \, 64 \, \, x  \Bigr),
    \nonumber \\
    \hspace{-0.98in}&& \quad \quad  \quad 
 _2F_1\Bigl([{{1} \over {8}}, \,{{3} \over {8 }}], \, [1], \, 256 \, \, x  \Bigr), \, \,  \,  
_2F_1\Bigl([{{1} \over {8}}, \,{{5} \over {8 }}], \, [1], \, 256 \, \, x  \Bigr), \, \,  \,
 _2F_1\Bigl([{{3} \over {8}}, \,{{7} \over {8 }}], \, [1], \, 256 \, \, x  \Bigr), 
   \nonumber 
\end{eqnarray}
\begin{eqnarray}
  \label{Specialliste2}
  \hspace{-0.98in}&& \quad \quad  \quad
 _2F_1\Bigl([{{2} \over {3}}, \,{{5} \over {6 }}], \, [1], \, 108 \, \, x  \Bigr), \, \,  \, \, 
 _2F_1\Bigl([{{1} \over {3}}, \,{{5} \over {6 }}], \, [1], \, 108 \, \, x  \Bigr),\, \,  \, \,
 _2F_1\Bigl([{{1} \over {2}}, \,{{3} \over {4 }}], \, [1], \, 32 \, \, x  \Bigr), 
 \nonumber \\
 \hspace{-0.98in}&& \quad \quad  \quad
 _2F_1\Bigl([{{3} \over {4}}, \,{{3} \over {4 }}], \, [1], \, 64 \, \, x  \Bigr), \, \,  \, \, 
 _2F_1\Bigl([{{5} \over {8}}, \,{{7} \over {8 }}], \, [1], \, 256 \, \, x  \Bigr), \, \,  \, \, 
    _2F_1\Bigl([{{2} \over {3}}, \,{{2} \over {3 }}], \, [1], \, 27 \, \, x  \Bigr),
 \nonumber \\
 \hspace{-0.98in}&&  \quad \quad  \quad
 _2F_1\Bigl([{{5} \over {6}}, \,{{5} \over {6 }}], \, [1], \, 432 \, \, x  \Bigr), \, \,  \, \, 
  _2F_1\Bigl([{{1} \over {2}}, \,{{5} \over {6 }}], \, [1], \, 144 \, \, x  \Bigr),
\, \, \, _2F_1\Bigl([{{1} \over {2}}, \,{{2} \over {3 }}], \, [1], \, 36 \, \, x  \Bigr)
\nonumber \\
  \hspace{-0.98in}&&  \quad \quad  \quad
  _2F_1\Bigl([{{1} \over {12}}, \,{{7} \over {12 }}], \, [1], \, 1728 \, \, x  \Bigr),
 \, \,  \,  \quad  \quad 
 _2F_1\Bigl([{{1} \over {12}}, \,{{5} \over {12 }}], \, [1], \, 1728 \, \, x  \Bigr), \, \,  \,
 \\
  \hspace{-0.98in}&&  \quad \quad \quad
 _2F_1\Bigl([{{5} \over {12}}, \,{{11} \over {12 }}], \, [1], \, 1728 \, \, x  \Bigr), \quad  \quad 
 \, \,  \,   _2F_1\Bigl([{{7} \over {12}}, \,{{11} \over {12 }}], \, [1], \, 1728 \, \, x  \Bigr)               .
 \nonumber 
\end{eqnarray}

\vskip .2cm 

Using this globally bounded condition of the nome criterion, we wrote a program that went
through all the values of $ \, a$ and $ \, b$ in $ \,[-1, \, 1] \,$  (with small increments like $ \,1/200$),
with $c= \, 1$, singling out the $\, _2F_1$ hypergeometric functions that have integer coefficients
(or more generally globally bounded)  series expansions, both for the $\, _2F_1$  hypergeometric functions,
and {\em for the nome}. Running this program returned to us exactly the $\, _2F_1$  hypergeometric functions
in the above list (\ref{Specialliste2}), and {\em only this list of twenty eight hypergeometric functions}.

\vskip .1cm 

\subsection{Derivatives of classical modular forms}
\label{derivclassimod}

Recalling identities (\ref{3F2homomany}),
where the parameters in (\ref{3F2homomany}) verify (\ref{Fuchscondi}) and (\ref{apparentcondi3}),
one can also deduce, for instance for $\, e \, = \,1/2$,  some identities
on some (homogeneous) derivatives of the classical modular form
$\, _2F_1([1/6,1/6],\, [1], \, x)$ of the previous list (\ref{Specialliste2}): 
\begin{eqnarray}
\label{anecdotalHeun2_1}
\hspace{-0.98in}&& \quad \quad \quad  \,   
 Heun\Bigl({{9} \over {4}}, \, {{3} \over {16}}, \, {{1 } \over {6}}, \,{{1 } \over {6}}, \, \, 1, \,  {{4} \over {3}}, \, \,    x \Bigr)
\,  \, \, \, = \, \, \,   \,
(2\, \theta \, +1) \Bigl[  \,  _2F_1([{{1 } \over {6}}, \,{{1} \over {6}}], \, [1], \, \, x\Bigr)   \Bigr]
\nonumber \\
 \hspace{-0.98in}&& \quad  \quad \quad \quad \quad  \quad   \quad                   \,  \, = \, \,   \, \,
_3F_2([{{1 } \over {6}}, \,{{1} \over {6}}, \, \, {{3 } \over {2}}], \, [{{1 } \over {2}}, \, 1], \, \, x\Bigr). 
\end{eqnarray}        
Such homogeneous derivatives of classical modular forms are not classical modular forms. We have however seen, 
in section (\ref{Derivclassforms}), that when a diagonal of a rational function
can be expressed as a classical modular form,  the homogeneous derivatives of that classical modular form
are also diagonals of other rational functions simply deduced from the first rational function.

\vskip .1cm

\subsection{Hypergeometric functions with negative values related to classical modular forms}
\label{negative}

One remarks that {\em no negative values} of  $ \, a$ and $ \, b \, $
occur in the previous list of $_2F_1$ classical modular forms (\ref{Specialliste2}).
But what about negative values of the parameters of the $_2F_1$ functions ? Can we have $\, _2F_1$ hypergeometric
functions with  negative values of the parameters  $ \, a$ and $ \, b \, $ such that it is not a
classical modular form, but a derivative of a classical modular form ?

Denoting $\, \theta \, = \, x \cdot \, D_x \, \, $ the homogeneous derivative, one has the following identities:
\begin{eqnarray}
  \label{relatedtomodforms}
  \hspace{-0.98in}&&  \quad \,  \,  \, 
_2F_1\Bigl([-{{1} \over {6}}, \, {{5} \over {6 }}], \, [1], \, \, x  \Bigr) \, \, = \,   \, \,
(1 \, -x)^{1/3} \cdot \, (1 \, +6\cdot \, \theta)\,
 \Bigl[ \, _2F_1\Bigl([{{1} \over {6}}, \,{{1} \over {6 }}], \, [1], \, \, x  \Bigr) \Bigr],      
\nonumber \\
 \hspace{-0.98in}&&   \quad \,  \,  \, 
_2F_1\Bigl([{{1} \over {6}}, \, {{1} \over {6 }}], \, [1], \, \, x  \Bigr) \, \, = \,   \, \,
 (1 \, -x)^{2/3} \cdot \, (1 \, -6\cdot \, \theta)\,
 \Bigl[ \, _2F_1\Bigl([{{-1} \over {6}}, \,{{5} \over {6 }}], \, [1], \, \, x  \Bigr) \Bigr], 
\end{eqnarray}
and thus, by elimination,  one finds that the order-two operators annihilating respectively
$\,_2F_1([{{-1} \over {6}}, \,{{5} \over {6 }}], \, [1], \, \, x  ) \, \,  $ and
$\,\, _2F_1([{{1} \over {6}}, \,{{1} \over {6 }}], \, [1], \, \, x) \, \, $ read:
\begin{eqnarray}
\label{pseudofacto}
\hspace{-0.98in}&& \quad  \quad  \quad  \quad 
 L_2 \, \, = \, \,\,\, 
 (1\, -x)^{1/3} \cdot \, (1 \, +6\cdot \, \theta) \cdot \, (1\, -x)^{2/3} \cdot \, (1 \, -6\cdot \, \theta) \,  \,  \, \, - \, 1
 \\
 \hspace{-0.98in}&&  \quad  \quad \quad  \quad  \quad  \quad   \quad  \quad  \quad   \quad  \quad 
 \, \, = \, \,\, x \cdot \, (6\, \theta \, +5) \cdot \,  (6\, \theta \, -1)  \,\,\,\,  - \, 36 \, \theta^2, 
\nonumber \\
 \hspace{-0.98in}&&  \quad  \quad  \quad  \quad 
M_2 \, \, = \, \,\,\,   (1\, -x)^{2/3} \cdot \, (1 \, -6\cdot \, \theta)
\cdot \, (1 \, -x)^{1/3} \cdot \, (1 \, +6\cdot \, \theta) \,   \, \, - \, 1
 \\
 \hspace{-0.98in}&&  \quad  \quad  \quad   \quad  \quad  \quad  \quad  \quad  \quad  \quad  \quad 
  \, \, = \, \,\, x \cdot \, (6\, \theta \, +1) \cdot \,  (6\, \theta \, +1)  \,\,\,\,  - \, 36 \, \theta^2.
  \nonumber       
\end{eqnarray}
These two order-two linear differential operators $\, L_2$ and $\, M_2\, $ are
{\em equivalent}\footnote[1]{Use the command ``equiv'' of Mark van Hoeij and not the command  ``Homomorphisms''
  of DEtools in Maple. This command is the algebraic extension of the command  ``Homomorphisms''.}
with order-one intertwiners (with algebraic coefficients):
\begin{eqnarray}
\label{pseudofactointertwin}
  \hspace{-0.98in}&& \quad  \quad  \quad  \quad
 M_2 \cdot \,  (1-x)^{2/3} \cdot \,  (6 \, \theta \, -1)
 \, \,\,\,   = \, \, \, \,  (1\, -x)^{2/3} \cdot \,  (6 \, \theta \, -1)  \cdot \,   L_2, 
\end{eqnarray}
and:
\begin{eqnarray}
\label{pseudofactointertwin2}
  \hspace{-0.98in}&& \quad  \quad  \quad  \quad
 L_2 \cdot \,  (1-x)^{1/3} \cdot \,  (6 \, \theta \, +1)
 \, \, \,\,  = \, \, \,\,   (1\, -x)^{1/3} \cdot \,  (6 \, \theta \, +1)  \cdot \,   M_2, 
\end{eqnarray}
Furthermore we also have the identities
\begin{eqnarray}
\label{furthermore}
  \hspace{-0.98in}&&   \quad  \quad \,
 _2F_1\Bigl([{{5} \over {6}}, \, {{7} \over {6 }}], \, [1], \, \, x  \Bigr) \, \,\,  = \,   \, \,\, 
 (1\, -x)^{-5/6}    \cdot \,
_2F_1\Bigl([-{{1} \over {6}}, \, {{5} \over {6 }}], \, [1], \, \, -\, {{x} \over {1\, -x}}  \Bigr),        
\end{eqnarray}
or:
\begin{eqnarray}
\label{furthermore2}
  \hspace{-0.98in}&&   \quad \quad \,
 _2F_1\Bigl([-{{1} \over {6}}, \, {{5} \over {6 }}], \, [1], \, \, x  \Bigr)\, \, \, = \,   \, \,\, 
  (1\, -x)^{-5/6}    \cdot \, _2F_1\Bigl([{{5} \over {6}}, \, {{7} \over {6 }}],
   \, [1], \, \, -\, {{x} \over {1\, -x}}   \Bigr). 
\end{eqnarray}
Therefore,  we see that hypergeometric functions like
\begin{eqnarray}
\label{morehypergmodular}
  \hspace{-0.98in}&& \quad  \quad  \quad  \quad  \quad \quad \quad
 _2F_1\Bigl([-{{1} \over {6}}, \, {{5} \over {6 }}], \, [1], \, \, x  \Bigr), \, \, \,   \quad \quad  \quad 
 _2F_1\Bigl([{{5} \over {6}}, \, {{7} \over {6 }}], \, [1], \, \, x  \Bigr), 
\end{eqnarray}
are {\em not} classical modular forms, but are {\em related} to  classical modular forms, 
{\em being an order-one operator acting on a classical classical modular form}.

\vskip .2cm

{\bf Remark 1:}  The order-two linear differential operators  for
$\,  _2F_1([{{5} \over {6}}, \, {{7} \over {6 }}], \, [1], \, \, x)$,
and for  $\, _2F_1([{{1} \over {6}}, \, {{1} \over {6 }}], \, [1], \, \, x)$ which corresponds
to a classical modular form, are equivalent, up to a $\,\, x \, \rightarrow \,\, -x/(1-x)\,$
pullback.  The nome of the order-two linear differential operators  for
$\,  _2F_1([{{5} \over {6}}, \, {{7} \over {6 }}], \, [1], \, \, x)$ reads:
\begin{eqnarray}
\label{q5sur6.7sur6}
  \hspace{-0.98in}&& \,\, \, 
q \, \,= \, \, \, x \, \, \,+ {{x^2} \over {18}}  \, +{\frac {49\,{x}^{3}}{5184}}
\,  +{\frac {7201\,{x}^{4}}{2519424}} \,  +{\frac {6889571\,{x}^{5}}{5804752896}}
\, +{\frac {311739307\,{x}^{6}}{522427760640}} \, \,+ \, \cdots 
\end{eqnarray}        
which is {\em not} a globally bounded series. The equivalence of linear differential operators
{\em does not preserve the globally bounded character of the nome}. 

\vskip .1cm

{\bf Remark 2:} For any integers $\, n_1$,  $\, n_2$, $\, n_3$ 
the order-two linear differential  operator  annihilating $\,  _2F_1([a, \, b], \, [c], \, x)$
and the order-two linear differential  operator  annihilating $\,  _2F_1([a\, +n_1, \, b\, +n_2], \, [c \, +n_3], \, x)$,
are homomorphic\footnote[2]{Note that this is not the case for Heun functions. Recalling (\ref{alsoHeun})
  and changing $\, \gamma \, \rightarrow \, $
  $\gamma \, +1$, one sees easily that  the linear differential operators for $\, Heun(-1/27,2/27,1/3,2/3,1,1/2,-x) \, \, $
  and $\, Heun(-1/27,2/27,1/3,2/3,2,1/2,-x) \, \, $ are not equivalent.}. Denoting $\, N_2$ the order-two
linear differential  operator annihilating $\,  _2F_1([{{1} \over {6}}, \, {{5} \over {6 }}], \, [1], \, \, x)$,
and $\, P_2$ the order-two linear differential  operator  annihilating
$\,  _2F_1([{{7} \over {6 }}, \, {{5} \over {6}}], \, [1], \, \, x)$,
one has the homomorphism
\begin{eqnarray}
\label{N2P2}
  \hspace{-0.98in}&& \,\, \, \quad \quad \quad \quad \quad \quad \quad \quad \quad \quad 
P_2 \cdot \, (6 \, \theta \, + \, 1)  \, \, \,\,  = \, \, \,  \, (6 \, \theta \, + \, 7) \cdot \, N_2, 
\end{eqnarray}        
yielding the relation:
\begin{eqnarray}
\label{N2P2equal}
\hspace{-0.98in}&& \,\, \, \quad \quad \quad  \quad \quad 
  _2F_1\Bigl([{{7} \over {6}}, \, {{5} \over {6 }}], \, [1], \, \, x\Bigr)
 \, \, \,= \, \, \,  \,  (6 \, \theta \, + \, 1) \Bigl[ \,  _2F_1\Bigl([{{1} \over {6}}, \, {{5} \over {6 }}], \, [1], \, \, x\Bigr) \Bigr].
\end{eqnarray}        
From this last identity  (\ref{N2P2equal}) we see that $\, \, _2F_1([{{5} \over {6}}, \, {{7} \over {6 }}], \, [1], \, \, x)\,$
is just a {\em homogeneous derivative of a classical modular form}
$\,\,  _2F_1([{{1} \over {6}}, \, {{5} \over {6 }}], \, [1], \, \, x)\,$ (see (\ref{Specialliste2})).

\vskip .2cm

\subsubsection{A non minimal order telescoper associated with $\,  _2F_1([ -\, {{1} \over {6}}, \, {{5} \over {6 }}], \, [1], \, \,  \, z )$. \\}
\label{subnegative}
$\,$

\vskip .01cm

Using Koutschan's creative telescoping program we have obtained 
the telescoper of the rational function $\,\, 1/(1  + x\,y  + z\,x +3\,(x^2 \,+y^2)) \, $. It is
an order-four linear differential operator $\, W_4$ which factors\footnote[5]{But no direct-sum factorisation.} as
$\, W_4 \, \, = \, \, \, U_2 \cdot \, V_2 \, $ where $\,  \, V_2 \, $ is a linear differential
operator of order two with two {\em algebraic solutions}. A well-suited linear combination of these
two algebraic solutions gives the integer coefficient series expansion
\begin{eqnarray}
\label{mockclassimodularsolU2}
 \hspace{-0.98in}&& \, 
\,\, 1 \,  -9\,x^2 \,  +135 \, x^4  \, -2268 \, x^6 \,  +40095 \, x^8
\, -729729 \, x^{10} \, + 13533156 \, x^{12} \, + \, \, \cdots
\end{eqnarray}             
which is actually the diagonal of the rational function  $\, 1/(1  + x\,y  + z\,x  \, +3\,(x^2 \,+y^2))$.
Therefore the order-two linear differential operator $\,  \, V_2$ is the
{\em minimal order linear differential operator} for 
the diagonal (\ref{mockclassimodularsolU2}). The creative telescoping method, however,  provides
a {\em larger  order}\footnote[2]{Non minimal order linear differential 
operator as far as the diagonal of the rational function is concerned.}
 {\em telescoper} $\, W_4$ and, consequently, a ``companion''
to this minimal order operator  $\,  \, V_2$. The linear differential operator ``companion''
of order two, $\, U_2$, admits the solution:
\begin{eqnarray}
  \label{mockclassimodularsolU2bis}
  \hspace{-0.98in}&& \,  \, \,  
 \Bigl(1\, +\, {{81} \over {4}} \, x^2\Bigr)^{-1} \cdot \, {{d H(x)} \over { d x}}
 \quad \,   \hbox{where:}  \quad   \, H(x) \, \, = \, \, \,
  _2F_1\Bigl([{{5} \over {6}}, \, {{7} \over {6 }}], \, [1], \, \,  -\, {{ 315} \over {16}} \cdot \, x^2 \Bigr).
\end{eqnarray}
The series expansion of (\ref{mockclassimodularsolU2bis}) is a globally bounded series\footnote[1]{ The series
  (\ref{mockclassimodularsolU2bis}) becomes a series with integer coefficients
with $ \, x \,  \rightarrow \, \, 96 \, x$.}.
 From (\ref{furthermore}) this last hypergeometric function is related  to 
\begin{eqnarray}
  \label{mockclassimodularsolU2bis2}
  \hspace{-0.98in}&& \quad  \quad \quad \quad \quad \ \quad \quad  \quad 
 _2F_1\Bigl([ -\, {{1} \over {6}}, \, {{5} \over {6 }}], \, [1],
 \, \,  \, {{ 315 \, x^2} \over {16 \, +315 \, x^2}}  \Bigr),
\end{eqnarray}
which is related, using (\ref{relatedtomodforms}), to:
\begin{eqnarray}
  \label{mockclassimodularsolU2ter}
  \hspace{-0.98in}&& \quad  \quad  \quad  \, \,  \, \,  
 _2F_1\Bigl([{{1} \over {6}}, \, {{1} \over {6 }}], \, [1], \, \,  \, {{ 315 \, x^2} \over {16 \, +315 \, x^2}}  \Bigr)
  \\
 \hspace{-0.98in}&& \quad    \quad \,  \, \,      \quad  \quad     
 \, \, = \,  \, \,  \Bigl(1\, + \, {{315} \over {16}} \, x^2\Bigr)^{1/6} \cdot \,
 _2F_1\Bigl([{{1} \over {12}}, \, {{5} \over {12 }}], \, [1],
\, \,  \, -\, {{315 } \over { 64  }} \cdot  \, x^2 \cdot \,  (16 \, +315 \, x^2)  \Bigr).
 \nonumber
\end{eqnarray}
Hence, if a $\, _2F_1$ hypergeometric function appears as the diagonal of a rational function, or
a (globally bounded) solution of a factor of a (non minimal order) telescoper, it seems often to be still related to
a {\em classical modular form}: in this
case here, it is the {\em derivative of a classical modular form}.

\vskip .2cm

 \section{Diagonals of rational functions in three variables: derivatives of modular forms}
\label{subthree}

\vskip .1cm

Let us give a simple example of diagonal of rational function of three variables
yielding a derivative of a classical modular form (or a derivative of a Heun function). 
Let us consider the following rational function of three variables: 
\begin{eqnarray}
\label{Ratfonc3}
  \hspace{-0.98in}&& \quad  \quad  \quad  \quad  \quad  \quad  \quad  \quad 
\, \,  
R(x, \, y, \, z)  \, \, \,  = \, \,  \quad 
  {{3 \, \, x^3 \, y} \over { 1 \, \,\, + \, x \, +\, y \, \,+ \,z}}.
\end{eqnarray}
The diagonal of (\ref{Ratfonc3}) has the following series expansion with {\em integer coefficients}:
\begin{eqnarray}
\label{Ratfonc3ser}
\hspace{-0.98in}&& \quad  \, 
\,\,\, -30\, x^3 \,\,\,\, +840\, x^4 \,\,\,\, -20790 x^5 \,\,\, \,+504504 \, x^6
\,\, \,\,-12252240\, x^7 \,\,\,\, +299304720\, x^8
\nonumber \\
\hspace{-0.98in}&& \quad \quad   \quad \quad \quad \quad \quad
\, -7362064710\,\, x^9 \, \,\, +182298745200\,\, x^{10}
\,\,\,  \, + \,\,\, \cdots
\end{eqnarray}
The telescoper of this rational function  of three variables (\ref{Ratfonc3})
gives an order-three linear differential operator $\, L_3 \, = \, \, L_1 \oplus \, L_2 \, $
which is the direct sum (LCLM) of an
order-one linear differential operator  $\, L_1$ and an order-two linear differential
operator $\, L_2$, where:
\begin{eqnarray}
\label{Ratfonc3HeunL1L2}
  \hspace{-0.98in}&& \quad  \quad  \,  \quad \,  \, 
L_1 \, \, = \, \, \, \, \,  x \cdot \, D_x \, \, \, -1, \quad \quad
\\
\hspace{-0.98in}&& \quad  \quad  \,  \quad \,  \, 
L_2 \, \, = \, \, \,\, \, 
(1 \, +27\, x) \cdot \, (1 \, +30\, x) \cdot \, x  \, \cdot \,D_x^2
\, \,  \,\, \, -3 \cdot\,  x \, \cdot \,D_x \,\, \, \,  +180\,x \, +3.
 \nonumber 
\end{eqnarray}
The order-one linear differential operator $\, L_1$ has the simple solution
$\, y(x) \, =\, x$, and the order-two linear differential operator
has the following Heun solution: 
\begin{eqnarray}
\label{Ratfonc3Heun}
\hspace{-0.98in}&& \quad \quad 
x \cdot \,
Heun \Bigl({{9} \over {10}}, \, 0, \, {{1} \over {3}},
                   \, {{2} \over {3}}, \, 2, \,1, \, -27 \cdot \,x \Bigr)
\,\, = \, \, \, \, \, x \,\, \, \, \, -30\, x^3 \,\,\,\, +840\, x^4 \,\,\,\, -20790 x^5
 \nonumber \\
\hspace{-0.98in}&& \quad  \quad \quad \quad \quad \quad \quad \,\,\, \,   +504504\, x^6
\,\, \,\,-12252240\, x^7 \,\,\,\, +299304720\, x^8 \, \, \, \,  + \, \, \cdots
\nonumber \\
\hspace{-0.98in}&& \quad  \quad   \quad \quad 
\, \, =  \, \, \,  \,
-x \cdot \, _2F_1\Bigl([{{1} \over {3}}, {{2} \over {3}}], \, [1], \, \, -27 \cdot \, x\Bigr)
 \\
\hspace{-0.98in}&& \quad  \quad \, \quad \quad \quad  \quad  \quad \quad  \quad \quad  \quad
\, \, +\, 2 \cdot \, x \cdot \, (1\, + \, 27 \, x ) \cdot \,
  _2F_1\Bigl([{{4} \over {3}}, {{5} \over {3}}], \, [2], \, \, -27 \cdot \, x)
\nonumber 
\\
\hspace{-0.98in}&& \quad \quad  \quad \quad 
\, \, =  \, \, \,  \,
   {\cal L}_1 \Bigl( \,   _2F_1\Bigl([{{1} \over {3}}, {{2} \over {3}}], \, [1], \, \, -27 \cdot \, x)  \Bigr) 
\\
\hspace{-0.98in}&& \quad  \quad  
\hbox{where:} \quad  \quad \quad \quad \quad \quad \quad  \quad 
     {\cal L}_1  \, \,\, \,  =  \, \, \,  \,
     -x \,\,\,\,  \,  -\, {{1\, + 27 \, x} \over {3}} \cdot \, x \cdot \, {{d} \over {d x}}.
\end{eqnarray}
It can also be written alternatively as
\begin{eqnarray}
\label{Ratfonc3Heunbis}
\hspace{-0.98in}&& \, \, 
x \cdot \,
Heun \Bigl({{9} \over {10}}, \, 0, \, {{1} \over {3}},
 \, {{2} \over {3}}, \, 2, \,1, \, -27 \cdot \,x \Bigr)
\end{eqnarray}
\begin{eqnarray}                 
\hspace{-0.98in}&& \quad  \, \, \, 
\, \, =  \, \, \,  \,
x \cdot \, {{5 \, + \, 108\, x} \over { (1\, + \, 54 \, x )^{4/3}}}
\cdot \,  _2F_1\Bigl([{{1} \over {6}}, \, {{2} \over {3}}], \, [1], \, \,
      {{108 \cdot  \, x \cdot \, (1 \, + \, 27\, x) } \over { (1\, + \, 54 \, x )^2 }} \Bigr)
\nonumber \\
\hspace{-0.98in}&& \quad  \quad \quad  \quad   \, 
\, \, -\, \, 4 \, x  \cdot \, {{1 \, + \, 27\, x} \over { (1\, + \, 54 \, x )^{10/3}}} \cdot \,
_2F_1\Bigl([{{7} \over {6}}, \, {{5} \over {3}}], \, [2], \, \,
   {{108 \cdot \, x \cdot \, (1 \, + \, 27\, x) } \over { (1\, + \, 54 \, x )^2 }} \Bigr)
\nonumber
\end{eqnarray}
\begin{eqnarray}                        
\hspace{-0.98in}&& \quad  \, \, \, 
\, \, =  \, \, \,  \,
x \cdot \, {{(1 \, + \, 18\, x) \, (1 \, + \, 24\, x )} \over {
    (1\, + \, 36 \, x \, + \, 216\, x^2 )^{7/6}}}
\cdot \,  _2F_1\Bigl([{{1} \over {12}}, \, {{7} \over {12}}], \, [1], \, \,
      {{1728 \cdot  \, x^3 \cdot \, (1 \, + \, 27\, x) } \over {
          (1\, + \, 36 \, x \, + \, 216\, x^2 )^2 }} \Bigr)
\nonumber \\
\hspace{-0.98in}&& \quad  \quad  \, \,  \, 
      -84 \, x^3 \cdot \, {{(1 \, + \, 27\, x) \, (1 \, + \, 24\, x )^2} \over {
    (1\, + \, 36 \, x \, + \, 216\, x^2 )^{19/6}}}
\cdot \,  _2F_1\Bigl([{{13} \over {12}}, \, {{19} \over {12}}], \, [2], \, \,
      {{1728 \cdot  \, x^3 \cdot \, (1 \, + \, 27\, x) } \over {
 (1\, + \, 36 \, x \, + \, 216\, x^2 )^2 }} \Bigr)
\nonumber
 \end{eqnarray}
\begin{eqnarray}                       
  \hspace{-0.98in}&& \quad  \quad  \,\, =  \, \, \,  \, \,
 {\cal L}_1 \Bigl( \,  _2F_1\Bigl([{{1} \over {12}}, \, {{7} \over {12}}], \, [1], \, \,
      {{1728 \cdot  \, x^3 \cdot \, (1 \, + \, 27\, x) } \over {
                     (1\, + \, 36 \, x \, + \, 216\, x^2 )^2 }} \Bigr) \Bigr)              
 \nonumber \\
  \hspace{-0.98in}&& \quad  \quad  \, \,
\,\, =  \, \, \,  \, \,
{\cal M}_1 \Bigl( \,  _2F_1\Bigl([{{1} \over {12}}, \, {{5} \over {12}}], \, [1], \, \,
 -\,   {{1728 \, x^3 \, (1 \, + \, 27\, x) } \over { (1\, + \, 24 \, x)^3 }}  \Bigr) \Bigr),
\end{eqnarray}
where:
\begin{eqnarray}
 \hspace{-0.98in}&& \quad  \quad  \, \,
 {\cal L}_1   \, \, =  \, \, \,  \,
  x \cdot \, {{(1 \, + \, 18\, x) \cdot \, (1 \, + \, 24\, x )} \over {
  (1\, + \, 36 \, x \, + \, 216\, x^2 )^{7/6}}} \, \,   \, \,   \, \, 
  -\, {{1} \over {3}} \cdot \,   {{ (1 \, + \, 27\, x )} \over {
 (1\, + \, 36 \, x \, + \, 216\, x^2 )^{1/6}}} \, \cdot \, x \cdot \, D_x,
\nonumber 
\end{eqnarray}
\begin{eqnarray}
 \hspace{-0.98in}&& \quad  \quad  \, \,
{\cal M}_1   \, \, =  \, \, \,  \,
\, \, {{1 \, + \, 30 \, x } \over {
 (1\, + \, 24 \, x)^{5/4}}} \cdot \, x \, \,   \, \,   \,
- \,{{1} \over {3}} \cdot \,  \, {{(1 \, + \, 27\, x) } \over {
 (1\, + \, 24 \, x)^{1/4}}} \, \cdot \, x \cdot \, D_x.
\end{eqnarray}

\vskip .2cm

Again we see that the derivative of a  classical modular form, or more generally an order-one linear
differential operator like (\ref{Ratfonc3Heun}) acting on a  classical modular form, is
{\em no longer a  classical modular form}.
With this example we see, one more time (recall section (\ref{Derivclassforms})),
that a Heun function which has a series expansion
with integer coefficients, is not necessarily a  classical modular form but can be 
{\em an order-one linear differential operator acting on a classical modular form}.

\vskip .2cm

\section{Automorphic forms associated with a Shimura curve}
\label{Shimura}

\subsection{ The pullback in $\, _2F_1([{{ 1} \over { 24}}, \, {{ 7} \over { 24}}], \, [{{ 5} \over {6}} ], x) \, $
  and $\,  \,  _2F_1([{{ 5} \over { 24}}, \, {{ 11} \over { 24}}], \, [{{ 7} \over {6}} ], x)$ is special. }
\label{Kilianapp}

Like all the {\em Belyi coverings}~\cite{Belyi}, the  pullback $\,  {{27} \over { 4}} \cdot \, {{x^2} \over {(1\, -x)^3 }}\, $
in (\ref{Ratfonc4gg2hyp}) is ``special''. It is such that:
\begin{eqnarray}
\label{suchthatx}
 \hspace{-0.98in}  \quad   \quad   \quad  \quad   \, \,  \,  \,\,
  \Bigl( {{27} \over { 4}} \cdot \, {{x^2} \over {(1\, -x)^3 }}\Bigr)
  \,  \circ \, \Bigl( {\frac {480 \cdot \, (1 \, -x) \cdot \,  x}{ (17 \,x \, -\, 32)^{2} }} \Bigr)
  \nonumber \\
  \hspace{-0.98in}  \quad   \quad   \quad  \quad   \quad   \quad   \quad   \quad   \quad  
  \, \, = \, \, \,  \Bigl(  {{27} \over { 4}} \cdot \, {{x^2} \over {(1\, -x)^3 }}\Bigr)
  \,  \circ \, \Bigl(  {\frac { 15 \cdot \, (17 \,x \, -32) \cdot \, x}{1024 \cdot \, (1 \, -x)^{2}}}\Bigr).
\end{eqnarray}
It has already been seen
to occur in another framework~\cite{Kilian},
namely\footnote[1]{See equation
  $\, \, w(x) \, = \, \,  {{27} \over { 4}} \cdot \, {{x^2} \over {(1\, -x)^3 }}, \, $  page 3165
in~\cite{Kilian}.} the walk
in a Weyl chamber of the Lie algebra $\,   \mathfrak{sl}_3 (\mathbb{C})$.            
It actually occurs in the well-known ``Kernel equation'' for
that particular walk described in~\cite{Kilian}
\begin{eqnarray}
\label{suchthatxGGGG}
  \hspace{-0.98in} \quad \quad  \quad \quad  \quad  \quad \quad  \quad 
  G(x, \, y) \, \, \, + G(0, \, 0) \, \, \, \, = \, \, \, \, \, G(x, \, 0)  \, \, \, + G(0, \, y), 
  \\
  \hspace{-0.98in}    \quad  \quad  \quad  \quad  \hbox{where:}
  \quad \quad   \quad \quad   \quad \quad  \quad 
  G(x, \, y) \, \, \, = \, \, \,  \, L(x, y) \cdot \, H(x, \, y), 
\end{eqnarray}
and where the generating function  $\, H(x, \, y)$  of the walk, and
the Kernel of the walk  $\, L(x, \, y)$, read respectively:
\begin{eqnarray}
\label{suchthatxHL}
  \hspace{-0.98in} \quad \, \, 
  H(x, \, y) \, \,  = \,  \,\, {{ 1\, -x\, y} \over {(1\, -x)^3 \cdot \, (1\, -y)^3}},  \quad \quad
  L(x, \, y) \, \,  =  \,  \,\, {{27} \over {4}} \cdot \, (y \, + \, x\, y^2 \, +x^2 \, -3\, x\, y).
  \nonumber 
\end{eqnarray}
Noticeably, $\, G(x, \, y)$ is the sum  of the particular rational function  pullback
$\, w(x) \,  =  \, \,  {{27} \over { 4}} \cdot \, {{x^2} \over {(1\, -x)^3 }} \, $
and of another rational function of $\, y$:
\begin{eqnarray}
\label{suchthatxsplit}
  \hspace{-0.98in} \quad \,   \quad  \quad  \quad  \quad  \quad  \quad 
  G(x, \, y)  \, \, \, = \, \,  \,  \,  \, \,  {{27} \over { 4}} \cdot \, {{x^2} \over {(1\, -x)^3 }}
  \, \, \,\,  \,  +    {{27} \over { 4}} \cdot \, {{y} \over {(1\, -y)^3 }}.
\end{eqnarray}
Note that this additional rational function of $\, y$ corresponds to the duality $\, x \, \leftrightarrow \, \, 1/x$:
\begin{eqnarray}
\label{suchthatxsplitLH}
  \hspace{-0.98in} \quad \quad  \quad  \quad  \quad  \, \, \,
  G(x, \, y)  \, \, \, = \, \,  \,  L(x, y) \cdot \, H(x, \, y)
 \,  \,\,  = \, \, \, \,  w(x)  \, \, \,\,  \, - \, w\Bigl( {{1} \over {y}}\Bigr).
\end{eqnarray}
The genus-zero curve $\, L(x, \, y) \, \, = \, \, \, 0 \, $ has the rational parametrisation
\begin{eqnarray}
\label{ratioparam}
\hspace{-0.98in} \quad \quad  \quad  \quad  \quad \quad \quad 
  y \, \, = \, \, \, {\frac {225 \, \,{t}^{2}}{ 32 \, \cdot \, (1 \, -\,t)  \cdot \, ( 17\,t \, -32) }},
 \quad \quad \quad \quad \quad \hbox{and:}
 \nonumber  \\
 \hspace{-0.98in} \quad \quad  \quad 
  x \, = \, \, {\frac {  480 \cdot \, (1 \, -t) \cdot \,  t}{ (17 \,t \, -32)^{2}}}
 \quad \quad \quad \, \,  \hbox{or:} \quad \quad \quad \, \, 
  x' \, = \, \,  {\frac { 15 \cdot \, (17 \,t \, -32) \cdot \, t }{ 1024 \cdot \, (1 \, -t)^{2}}}, 
\end{eqnarray}
where one recovers, in the last two  rational parametrisations (\ref{ratioparam}) for $\, x$, 
the two  rational functions in the identity (\ref{suchthatx}). The two  rational functions
are the  rational parametrisation of the (symmetric) genus-zero curve
$\,  \, x^2 \, x'^2 \, \,  -3 \, x\, x' \, \,  +x \, +x' \, \,  = \,  \, \, 0$ (corresponding to the elimination of $\, y$
in  $\, L(x, \, y) \, \, = \, \, \, 0$ and  $\, L(x', \, y) \, \, = \, \, \, 0$). These
two rational functions (\ref{ratioparam})
are simply related by an involution:
\begin{eqnarray}
\label{ratioparaminvol}
 \hspace{-0.98in} \quad \quad \quad \quad \quad \, \, \,
  x'(t)  \, \, = \, \, \, x\Bigl( {{32\, t} \over { 49 \, t \, - \, 32}}   \Bigr),   \quad \quad  \quad
  x(t)  \, \, = \, \, \, x'\Bigl( {{32\, t} \over { 49 \, t \, - \, 32}}   \Bigr).
\end{eqnarray}

\vskip .1cm

\subsection{ $\, _2F_1([{{ 1} \over { 24}}, \, {{ 7} \over { 24}}], \, [{{ 5} \over {6}} ], x) $, 
  $\,  \,  _2F_1([{{ 5} \over { 24}}, \, {{ 11} \over { 24}}], \, [{{ 7} \over {6}} ], x)$
  and $\, _2F_1([{{ 1} \over { 24}}, \, {{ 5} \over { 24}}], \, [{{ 3} \over {4}} ], x)$,
 $\,  _2F_1\Bigl([{{ 7} \over { 24}}, \, {{ 11} \over { 24}}], \, [{{ 5} \over {4}} ], x) \, $
Shimura examples.}
\label{Shimuraexample1}
The order-two linear differential operator
\begin{eqnarray}
\label{L2theta}
  \hspace{-0.98in}&& \,  \quad  \quad   \quad   \quad   \quad   \quad   \quad  
L_2 \, \,  \, \, = \, \, \,  \, \,
\theta \cdot \, \Bigl(\theta \, - \, {{1} \over {6}}   \Bigr) \,\,\,
 \,  \, - x \cdot \, \Bigl(\theta \, + \, {{1} \over {24}}   \Bigr)  \cdot \, \Bigl(\theta \, + \, {{7} \over {24}}   \Bigr),  
\end{eqnarray}
has the two $\, _2F_1$ hypergeometric solutions:
\begin{eqnarray}
\label{L2thetahypergeom}
  \hspace{-0.98in}&& \quad  \, \,   \quad   \quad   \quad  
_2F_1\Bigl([{{ 1} \over { 24}}, \, {{ 7} \over { 24}}], \, [{{ 5} \over {6}} ],  \, \, x \Bigr),
 \quad \quad \quad   
 x^{1/6} \cdot \, _2F_1\Bigl([{{ 5} \over { 24}}, \, {{ 11} \over { 24}}], \, [{{ 7} \over {6}} ],\, \,  x\Bigr), 
\end{eqnarray}
A {\em modular equation of level five} has been given in~\cite{Voight} on
this example of $\, _2F_1$ hypergeometric function corresponding to automorphic forms associated with Shimura curves.
This {\em modular equation of level five}
corresponds to the elimination of $\, u$ between
\begin{eqnarray}
\label{ordertwogoodAB}
  \hspace{-0.98in}&&  \,  \quad     \quad   \quad  
 A(u) \, \, = \, \, - \,{\frac { 1350000 \cdot \, {u}^{6}}{225\,{u}^{2}+18\,u+1}}, \,
 \nonumber \\
  \hspace{-0.98in}&& \quad    \quad   \quad  
B(u)     \, \, = \, \,   A\Big({\frac {11\,u+2}{252\,u-11}}  \Bigr)   \, \, = \, \,\,
-\, {\frac {  2160 \cdot \left( 11\,u+2 \right) ^{6}}{ \left( 225\,{u}^{2}+18\,u+1 \right)  \left( 252\,u-11 \right)^{4}}},
\end{eqnarray}

Changing $\, x $ into $\, 1/x$, the order-two  linear differential operator (\ref{L2theta}) becomes
the order-two linear differential operator
\begin{eqnarray}
\label{L2thetaunsurx}
  \hspace{-0.98in}&& \,  \, \,   \quad     \quad   \quad  
 L_2\Bigl(x \, \rightarrow\, {{1}\over {x}}\Bigr) \,  \, \, = \, \, \,\, \, 
 \Bigl(\theta \, - \, {{1} \over {24}}   \Bigr)  \cdot \, \Bigl(\theta \, - \, {{7} \over {24}}   \Bigr)
  \, \,\, \,\, \, - x \cdot \,  \theta \cdot \, \Bigl(\theta \, + \, {{1} \over {6}}   \Bigr),  
\end{eqnarray}
which has the two $\, _2F_1$ hypergeometric solutions: 
\begin{eqnarray}
 \hspace{-0.98in}&&  \quad    \quad   \quad   \,  \,
 x^{1/24} \cdot \,  _2F_1\Bigl([{{ 1} \over { 24}}, \, {{ 5} \over { 24}}], \, [{{ 3} \over {4}} ], x\Bigr),
 \quad  \quad  \quad
  x^{7/24} \cdot \,  _2F_1\Bigl([{{ 7} \over { 24}}, \, {{ 11} \over { 24}}], \, [{{ 5} \over {4}} ], x\Bigr).
\end{eqnarray}
Instead of (\ref{L2thetaunsurx}) one can also introduce the conjugate of (\ref{L2thetaunsurx}) by $\, \,x^{1/24}$
\begin{eqnarray}
\label{L2thetaunsurxalso}
  \hspace{-0.98in}&& \,  \, \,   \quad  \quad   \quad    \quad     \quad   \quad    \quad  
\theta \cdot \, \Bigl(\theta \, - \, {{1} \over {4}}   \Bigr) \,  \, \,  \,  \,
- x \cdot \, \Bigl(\theta \, + \, {{1} \over {24}}   \Bigr)  \cdot \, \Bigl(\theta \, + \, {{5} \over {24}}   \Bigr),   
\end{eqnarray}
which has the two  $\, _2F_1$ hypergeometric solutions: 
\begin{eqnarray}
 \hspace{-0.98in}&&  \quad    \quad   \quad   \quad  \, \,  \, \, 
 _2F_1\Bigl([{{ 1} \over { 24}}, \, {{ 5} \over { 24}}], \, [{{ 3} \over {4}} ], x\Bigr), \quad  \quad  \quad
  x^{1/4} \cdot \,  _2F_1\Bigl([{{ 7} \over { 24}}, \, {{ 11} \over { 24}}], \, [{{ 5} \over {4}} ], x \Bigr).
\end{eqnarray}
One can also introduce the  order-two linear differential operator
\begin{eqnarray}
\label{L2thetacal}
\hspace{-0.98in}&& \,  \quad  \quad   \quad   \quad   \quad  \, \,   \quad
{\cal L}_2 \,  \,\, = \, \, \, \, \,
D_x^2 \,  \, \,\, + \, {\frac {135\,{x}^{2} \, -167\,x \, +140}{ 576  \cdot \, {x}^{2} \cdot \, (x-1) ^{2}}},                  
\end{eqnarray}
which has the two hypergeometric solutions: 
\begin{eqnarray}
  \label{set1}
 \hspace{-0.98in}&&  \, \, \quad \quad \quad \quad \quad  \quad  \quad
x^{5/12} \cdot \, (1\, -x)^{1/4} \cdot \,
_2F_1\Bigl([{{ 1} \over { 24}}, \, {{ 7} \over { 24}}], \, [{{ 5} \over {6}} ],  \, \, x \Bigr),
   \nonumber \\
  \hspace{-0.98in}&&  \, \, 
\,  \quad \quad \quad \quad \quad  \quad   \quad
x^{7/12} \cdot \,  (1\, -x)^{1/4} \cdot \,
_2F_1\Bigl([{{ 5} \over { 24}}, \, {{ 11} \over { 24}}], \, [{{ 7} \over {6}} ],\, \,  x\Bigr), 
\end{eqnarray}
and
\begin{eqnarray}
\label{L2thetacal2}
\hspace{-0.98in}&& \,  \quad  \quad  \quad   \quad   \quad   \quad   \quad \, \, \,
{\cal M}_2 \,\, \, \, = \, \, \, \,
D_x^2 \,\,\, \, +  {\frac {140 \,{x}^{2} \, -167\,x \, +135}{ 576 \cdot \, {x}^{2} \cdot \, (x-1) ^{2}}},                  
\end{eqnarray}
which has the two hypergeometric solutions: 
\begin{eqnarray}
  \label{set21}
   \hspace{-0.98in}&&  \, \, 
\,  \quad \quad \quad \quad \quad  \quad
S_1 \, \, = \, \, \,
x^{3/8} \cdot \, (1\, -x)^{1/4} \cdot \,
_2F_1([{{ 1} \over { 24}}, \, {{ 5} \over { 24}}], \, [{{ 3} \over {4}} ], x), 
 \\
 \label{set22}
 \hspace{-0.98in}&&  \, \, \quad \quad  \quad \quad \quad \quad
S_2 \, \, = \, \, \,
x^{5/8} \cdot \,  (1\, -x)^{1/4} \cdot \,
_2F_1([{{ 7} \over { 24}}, \, {{ 11} \over { 24}}], \, [{{ 5} \over {4}} ], x).
\end{eqnarray}

\subsection{Identities on Shimura $\, _2F_1$ hypergeometric functions and modular equations. }
\label{identitiesShimura}

There exists {\em an algebraic series} $\, y(x)$ such that the two hypergeometric (\ref{set21}), (\ref{set22})
{\em actually verify the two following identities}:
\begin{eqnarray}
  \label{ModularinftyL2inftysolident2}
  \hspace{-0.98in}&&  \quad   \quad   \quad  \quad \quad 
 w^{3/8}   \cdot \,     \rho \cdot \, y'(x)^{1/2}   \cdot \,     x^{3/8} \cdot \, (1-x)^{1/4} \cdot \,
 _2F_1\Bigl( [{{ 1} \over {24 }}, \,  {{5 } \over {24 }}], \, [{{ 3} \over {4 }} ], \, x\Bigr)
 \nonumber \\
  \hspace{-0.98in}&&  \quad   \quad   \quad  \quad  \quad \quad   \quad 
  \, \, = \, \,  \,  y(x)^{3/8} \cdot \, (1\, -y(x))^{1/4} \cdot \,
 _2F_1\Bigl( [{{ 1} \over {24 }}, \,  {{5 } \over {24 }}], \, [{{ 3} \over {4 }} ], \, y(x)\Big),
\end{eqnarray}
and (with the {\em same} $\, \rho$ and  $\, w$ )
\begin{eqnarray}
  \label{ModularinftyL2inftysolident}
  \hspace{-0.98in}&&  \quad   \quad   \quad  \quad \quad 
w^{5/8}   \cdot \,   \rho \cdot \, y'(x)^{1/2}   \cdot \,     x^{5/8} \cdot \, (1-x)^{1/4} \cdot \,
_2F_1\Bigl( [{{ 7} \over {24 }}, \,  {{11 } \over {24 }}], \, [{{ 5} \over {4 }} ], \, x\Bigr)
\nonumber \\
\hspace{-0.98in}&&  \quad   \quad   \quad   \quad \quad \quad   \quad 
  \, \, = \, \,  \,   y(x)^{5/8} \cdot \, (1\, -y(x))^{1/4} \cdot \,
_2F_1\Bigl( [{{ 7} \over {24 }}, \,  {{11 } \over {24 }}], \, [{{ 5} \over {4 }} ], \, y(x)\Bigr),
\end{eqnarray}
where  the two complex constants $\, \rho$ and  $\, w$  are given by:
\begin{eqnarray}
  \label{ModularinftyL2inftysoldefrho}
  \hspace{-0.98in}&& \quad \quad \quad  \quad  \quad \,\,\,\,\,\,
 \rho \, \, = \, \, {\frac{7}{25}} \, -{\frac {24\,i}{25}}, 
 \quad  \quad \quad \,\,
      w  \, \, = \, \,  {{1} \over {\rho^2}} \, \, = \, \, 
 - \, {{ 527} \over {625}} \, + \, {{ 336 \, i} \over {625}}.
\end{eqnarray}  
These two complex numbers  $\, w$ and $\, \rho$ {\em are on the unit circle}
$\, |w| \, = \, \,  |\rho| \, =\, \, 1$
but are not N-th root of unity.
The algebraic series $\, y(x)$ is given by the (symmetric genus-zero)
{\em modular equation of level\footnote[1]{The level of the (quaternionic) modular equation is the
reduced norm of $\, \alpha$ such that
$\, \Phi_L(j(\tau), \, j(\alpha \, \tau)) \, = \, \, 0$, see section 2.1 in~\cite{Baba2}. For the definition
of modular polynomials for these quaternionic cases see section 3.1 page 8 of~\cite{Baba}.
It depends only on the {\em integer index} $\, n:= [\Lambda: \beta \, \Lambda \beta^{-1} \, \bigcap \Lambda]$.} five}
$\,\, P(x, \, y) \, = \, \, 0\,\, $ where the symmetric polynomial $\, P(a, \, b)\,$ reads
\begin{eqnarray}
\label{Modularinfty}
  \hspace{-0.98in}&&
 P(a, \, b) \,  \,\,    = \,  \, \,   \, 6979147079584381377970176000000 \cdot \, {a}^{6}{b}^{6} 
 \nonumber \\
 \hspace{-0.98in}&&  \quad                   
\, +4434969287855682628249190400000 \cdot \, {a}^{5}{b}^{5} \cdot \, (a+b)
\nonumber \\
 \hspace{-0.98in}&& \, 
 +1752976676930715648  \cdot \, {a}^{4}{b}^{4} \cdot \,
 \Bigl(669871503125 \cdot \, ({a}^{2} \,+ \,{b}^{2})  -862324029349392 \, a b \Bigr)
\nonumber \\
  \hspace{-0.98in}&& \quad         
+25572696989368320 \cdot \,{a}^{3} {b}^{3} \cdot \, (a+b)  \cdot \,
 \Bigl(6484356150250 \cdot \, ({a}^{2} \,+ \,{b}^{2})
\nonumber \\
  \hspace{-0.98in}&& \quad  \quad         \quad \quad \quad \quad \quad \quad \quad \quad \quad 
                     \, +170835586907964203\,ab\Bigr)
\nonumber \\
  \hspace{-0.98in}&&  \quad
+57330892800 \, \cdot \, {a}^{2} \, {b}^{2} \cdot \,
  \Bigl(229748649317860805\, ({a}^{4}\, +b^4)
\nonumber \\
  \hspace{-0.98in}&& \quad  \quad     \quad  \quad 
\, -55981841121913535287822\,({a}^{3}b\, +a{b}^{3})
+ 170909095105030243444933\,{a}^{2}{b}^{2} \Bigr)
\nonumber 
 \end{eqnarray}
  \begin{eqnarray}
\label{Modularinfty2}
  \hspace{-0.98in}&&  \quad         
+829440 \cdot \,ab \cdot \, (a+b)
 \cdot \, \Bigl(672749994932560009201\,({a}^{4}+\,{b}^{4})
\nonumber \\
  \hspace{-0.98in}&& \quad \quad  \quad          \quad \quad  \quad \quad 
  +561918211777371330718156199 \cdot \,({a}^{3}b \, +a{b}^{3})
 \nonumber \\
  \hspace{-0.98in}&& \quad \quad \quad  \quad  \quad  \quad \quad  \quad \quad  \quad \quad 
 \, -52602925387952898241194053424 \,{a}^{2}{b}^{2} \Bigr)
 \nonumber
  \end{eqnarray}
\begin{eqnarray}
\label{Modularinfty2-2}
  \hspace{-0.98in}&& \quad         
 +9849732675807611094711841 \cdot \,({a}^{6}\, +b^6)
\nonumber \\
  \hspace{-0.98in}&& \quad  \quad  \quad         \quad 
 \, -10462859500072645481855465070150 \cdot \,({a}^{5}b +\,a{b}^{5})
 \nonumber \\
  \hspace{-0.98in}&& \quad \quad   \quad  \quad \quad   \quad 
 +26604718024918444951713833439099375 \cdot \,({a}^{4}{b}^{2}\, +{a}^{2}{b}^{4})
 \nonumber \\
  \hspace{-0.98in}&& \quad \quad \quad  \quad   \quad    \quad  \quad \quad 
+10825060619732076783684974180027500\,{a}^{3}{b}^{3}
  \nonumber \\
  \hspace{-0.98in}&&  \quad 
+1680315840\, \cdot \, (a+b)\cdot \, \Bigl(5559917313492231481\,({a}^{4} +\,{b}^{4})
\nonumber \\
 \hspace{-0.98in}&& \quad      \quad    \quad        \quad   \quad        \quad         
                    \, +2234383287388481537541244 \cdot (\,{a}^{3}b \, + \,a{b}^{3})
                    \nonumber \\
 \hspace{-0.98in}&& \quad       \quad   \quad      \quad   \quad      \quad   \quad      \quad 
\, +62713512471894189372026006\,{a}^{2}{b}^{2} \Bigr)
\nonumber
\end{eqnarray}
  \begin{eqnarray}
\label{Modularinfty3}
  \hspace{-0.98in}&&       \quad            
 +3322961675259430805968122201600 \cdot \,({a}^{4} \, +\, b^4)
 \nonumber \\
  \hspace{-0.98in}&&     \quad   \quad          
\, -70573817430654328766881842418944000 \cdot \, ({a}^{3}b \, + a{b}^{3})
 \nonumber \\
  \hspace{-0.98in}&&   \quad 
 -104957798363467459886332890685209600 \,{a}^{2}{b}^{2}
 \\
  \hspace{-0.98in}&&  \, 
+11860766958110577033461760 \, \cdot (a+b)
\left( 44289025\,{a}^{2}-3617524994\,ab+44289025\,{b}^{2} \right)
 \nonumber \\
  \hspace{-0.98in}&& \quad \quad \quad  \, 
+49824586485654547652165013405696 \cdot \, (625\, a^2 \, +1054\, a\, b \, +625\, b^2)
 \, \,\, = \,\, \, \, 0, \nonumber
\end{eqnarray}
which is parametrised by:
\begin{eqnarray}
\label{Level5inftyparam}
\hspace{-0.98in}&& \quad  \quad  \quad 
 a(v) \, \, = \, \, \,
 -{\frac {225\,{v}^{2}+18\,v+1}{1350000 \cdot \,{v}^{6}}}, 
\\
\hspace{-0.98in}&&   \quad  \quad  \quad 
 b(v) \, \, \, = \, \, \, a\Bigl( {\frac {11\,v+2}{252\,v-11}}  \Bigr)
\, \, \, = \, \, \,
 -{\frac { (225\,{v}^{2}+18\,v+1)  \cdot \, (252\,v-11)^{4}}{2160\, \cdot \, (11\,v+2)^{6}}}.
\nonumber
\end{eqnarray}
The algebraic series $\, y(x)$ in
(\ref{ModularinftyL2inftysolident}) or (\ref{ModularinftyL2inftysolident2}),
given by the {\em modular equation}
(\ref{Modularinfty}) of level five
$\, \,P(x, \, y) \, = \, \, 0\,\,$ reads
\begin{eqnarray}
  \label{Modularinftyexpans}
  \hspace{-0.98in}&& \, \, 
 y(x) \, \, = \, \, \, \,  w \cdot \, x \, \, \,\, + \Bigl({\frac {172937\,w}{168750}}
\,\,  +{\frac{103}{270}}\Bigr) \cdot \, {x}^{2} \, \,\, +  \Bigl({\frac {338124694601\,w}{398671875000}}
\, \, +{\frac{270081319}{637875000}}\Bigr) \cdot \,  {x}^{3}
\nonumber \\
  \hspace{-0.98in}&&\,\,\,\, \, \, \quad  \quad \quad \quad 
+ \Bigl({\frac {46359287214498287\,w}{67275878906250000}} \, \,
 +{\frac{42068837566753}{107641406250000}} \Bigr) \cdot \,  {x}^{4}
\\
  \hspace{-0.98in}&& \quad  \quad \,\,  
\, + \, \Bigl({\frac {25717676203788236624381\,w}{45411218261718750000000}}
\, \, +{\frac{25116272080576089139}{72657949218750000000}}\Bigr)  \cdot \, {x}^{5}
\, \, \, \,  \, \,  + \, \, \, \cdots     \nonumber         
\end{eqnarray}  
where $\, w$ was given previously in (\ref{ModularinftyL2inftysoldefrho}):
\begin{eqnarray}
  \label{Modularinftyexpansdefw}
  \hspace{-0.98in}&&
\quad \quad \quad  \quad  \,\,\,
w \, \, = \, \,  -{\frac{527}{625}} \, +{\frac {336\,i}{625}},
 \quad   \quad    \quad  \,\,
625\, w^2 \, +1054\, w \, +625\, \, = \, \, \, 0.
\end{eqnarray}

\subsection{Automorphic forms and Schwarzian equations: Schwarz map and Schwarz function. }
\label{automSchwarz}

It is straightforward to check that the Taylor expansion (\ref{Modularinftyexpans})
is such that the two identities (\ref{ModularinftyL2inftysolident})
or (\ref{ModularinftyL2inftysolident2}) are actually verified.
This algebraic series (\ref{Modularinftyexpans})
  {\em is actually a solution  of the Schwarzian equation:}
\begin{eqnarray}
  \label{ModularinftyexpansdefwSchwarz}
  \hspace{-0.98in}&& \quad  \quad  \quad   \quad  \quad   \quad  \quad  
\{y(x), \, x\} \, \,\, \,
+ {\frac {140\,{y(x)}^{2}-167\,y(x)+135}{288\, \cdot \, (y(x)\, -1)^{2}{y(x)}^{2}}} \cdot \, y'(x)^2
\nonumber \\
  \hspace{-0.98in}&&\,\,\,\, \quad  \quad  \quad \quad  \quad  \quad  \quad  \quad  \quad  \quad  \quad 
\, \,  -{\frac {140\,{x}^{2}-167\,x+135}{288\, \cdot \, (x\, -1)^{2} \cdot \,{x}^{2}}}
 \, \,\, = \, \, \,\, 0.  
\end{eqnarray}
This is a consequence of the fact that the
{\em algebraic transformation}\footnote[2]{Such a transformation is called
a {\em modular correspondence}.} $\, x \, \rightarrow \, y(x)$, given by the
modular equation of level five (\ref{Modularinfty}),
is a {\em symmetry} of the order-two linear differential operator  (\ref{L2thetacal2})
(see for instance~\cite{Youssef}): 
\begin{eqnarray}
 \label{ModularinftyexpansdefwSchwarz}
 \hspace{-0.98in}&& \quad  \quad \quad  \,    \quad \quad  \quad \quad
y'(x)^{1/2}  \cdot \, {\cal M}_2 \cdot y'(x)^{-1/2}
\, \, \, = \, \,\,  \, pullback\Bigl({\cal M}_2, \, \, y(x)\Bigr).
\end{eqnarray}
The (Shimura) {\em modular correspondences} $\, x \, \rightarrow \, y(x)$
are {\em algebraic} solutions of that Schwarzian equation (\ref{ModularinftyexpansdefwSchwarz})
{\em which thus ``encapsulates'' all the  modular correspondences}.

Let us introduce $\, \tau$ the ratio\footnote[1]{Such a ratio is called the Schwarz map.}
of the two hypergeometric (\ref{set21}), (\ref{set22})
solutions of (\ref{L2thetacal2} ). One  has the following identity:
\begin{eqnarray}  
\label{modidentitytaufirst}
 \hspace{-0.98in}&&  \,  \quad    \quad \,  \quad    \quad     \,   \,   \,  
\tau \, \, = \, \, \, {{S_2} \over { S_1}}  \, \, = \, \, \,\, 
x^{1/4} \cdot \,
{{   _2F_1\Bigl( [{{ 7} \over {24 }}, \,  {{11 } \over {24 }}], \, [{{ 5} \over {4 }} ], \, x\Bigr)
} \over {
 _2F_1\Bigl( [{{ 1} \over {24 }}, \,  {{5 } \over {24 }}], \, [{{ 3} \over {4 }} ], \, x\Bigr)
}} \, \, = \, \, \,
  \int {{ d x} \over { 4 \cdot \, S_1^2}}  
 \\
  \hspace{-0.98in}&&  \,  \quad    \quad  \quad    \quad
\quad \quad   \quad      \quad    \, \, = \, \, \,
{{1} \over {4}} \cdot \,  \int {{ d x} \over { x^{3/4} \cdot \, (1 \, -\, x)^{1/2}  \cdot \,
 _2F_1\Bigl( [{{ 1} \over {24 }}, \,  {{5 } \over {24 }}], \, [{{ 3} \over {4 }} ], \, x\Bigr)^2
}}.
\nonumber 
\end{eqnarray}
This identity corresponds to the (automorphy theory) relation that one of
the two solutions of (\ref{L2thetacal2}) can be written (up to an overall factor) as $\, f'(\tau)^{1/2}$ 
(the square root of the $\, \tau$-derivative of an automorphic function $\, x \, = \, \, f(\tau)$): 
\begin{eqnarray}  
\label{autommodidentitytaufirst}
\hspace{-0.98in}&&  
\Bigl({{d x} \over {d \tau}} \Bigr)^{1/2} \, = \, \, 2 \cdot \,  S_1, 
 \quad \hbox{i.e.} \quad        {{d x} \over {d \tau}}\, = \, \,  4 \cdot \,  S_1^2, \quad \hbox{i.e.} \quad   
 d \tau \, = \, \,         {{d x} \over { 4 \cdot \,  S_1^2}}, \quad \hbox{i.e.} \quad    
 \tau \, = \, \,    \int {{ d x} \over { 4 \cdot \, S_1^2}},      
\nonumber 
\end{eqnarray}
when the other solution  $\, S_2$ of (\ref{L2thetacal2}) can be written (up to an overall
factor) as $\, \tau \cdot \, f'(\tau)^{1/2}$:
\begin{eqnarray}  
\label{autommodidentitytauS2first}
  \hspace{-0.98in}&& \quad     \quad \quad \quad  \quad    \quad     \quad    
 \tau \cdot \, \Bigl({{d x} \over {d \tau}} \Bigr)^{1/2} \, \, = \, \, \,\, 
      {{S_2} \over { S_1}} \cdot \,  2 \cdot \,  S_1    \, \, = \,\,  \,\,  2 \cdot \,  S_2.        
\end{eqnarray}
The {\em Schwarz map} $\, \tau$,  given by (\ref{modidentitytaufirst}), seen as a function of $\, x$, 
is a {\em differentially algebraic function}.
It verifies the Schwarzian equation:    
\begin{eqnarray}  
\label{SchwarzEQU}
  \hspace{-0.98in}&&  \,  \quad   \quad  \quad   \quad  \quad \quad   \quad \quad     \,
 \{ \tau(x), \, x \} \,\,  \, \,   \,
 - \,   \,  {\frac {140\,{x}^{2} \, -167\,x \, +135}{288 \cdot \,{x}^{2} \cdot \, (x-1)^{2}}}
 \, \, \,  \, = \, \, \, \, \, 0.
\end{eqnarray}
Relation (\ref{modidentitytaufirst}) which can be rewritten
\begin{eqnarray}  
\label{rewritten}
  \hspace{-0.98in}&&  \,  \quad \quad   \,  \quad    \quad  \quad \quad   \quad \quad
   \tau^4 \, \, \, = \, \, \, \,
x \cdot \,
{{   _2F_1\Bigl( [{{ 7} \over {24 }}, \,  {{11 } \over {24 }}], \, [{{ 5} \over {4 }} ], \, x\Bigr)^4
} \over {
 _2F_1\Bigl( [{{ 1} \over {24 }}, \,  {{5 } \over {24 }}], \, [{{ 3} \over {4 }} ], \, x\Bigr)^4
}}, 
\end{eqnarray}
yields the following expansion of the {\em compositional inverse} of the Schwarz map
called the {\em Schwarz function} (here $\, z$ denotes $\,z \, = \, \,  \tau^4$)
\begin{eqnarray}  
  \label{Schwarzfunction}
 \hspace{-0.98in}&&  \,    \, \quad 
 x(\tau) \, \, = \, \, \, \, \, z \, \,  \, \,\,  -{\frac{103}{270}}{z}^{2} \, 
 \, \,  \, +{\frac{53399}{680400}}{z}^{3} \, 
\,\,  -{\frac{51411991}{4378374000}}{z}^{4} \,\, \,  +{\frac{4865571860153}{3376251758880000}}{z}^{5}
\nonumber \\
\hspace{-0.98in}&& \quad 
\, -{\frac{534844409962319}{3464034304610880000}}{z}^{6} \, 
\,\,  +{\frac{19352805342627235628117}{1291989881079911901888000000}}{z}^{7}
 \,\,  \,\,   + \, \, \, \cdots                 
\end{eqnarray}
which can be seen as an {\em automorphic function} of the variable $\, \tau$.
This automorphic function verifies the Schwarzian equation:
\begin{eqnarray}  
 \label{SchwarzSchwarzfunction}
  \hspace{-0.98in}&&  \,    \, \, \,  \quad \quad \quad \quad
\{x(\tau), \, \tau\} \,\,\, \,
+ \, {\frac {140\,{x(\tau)}^{2}-167\,x(\tau)+135}{288\, \cdot \, (x(\tau)\, -1)^{2} \cdot \,{x(\tau)}^{2}}}
\cdot \, \Bigl({{d x(\tau)} \over { d \tau}}\Bigr)^2 \, \, = \, \, \,\, 0.
\end{eqnarray}
Recalling the  modular equation (\ref{Modularinfty})  of level five 
$  \, P(x, \, y) \, \, = \, \, 0$,
{\em one easily verifies that this  modular equation of level five  is such that} 
\begin{eqnarray}
\label{Modularinftyxy}
\hspace{-0.98in}&& \quad \quad \quad  \quad 
P\Bigl(x(\tau), \, \,   x\Bigl(  {\cal E} \cdot \, \tau \Bigr) \Bigr)
\, \,   = \, \,  \, \,
P\Bigl(x(\tau), \, \, x\Bigl(\Bigl( -\, {{3} \over {5}} \, + \, i \cdot \,  {{4} \over {5}} \Bigr)    \cdot \, \tau \Bigr)\Bigr)
\, \, = \, \,\, 0,  
\end{eqnarray}
where $\,  {\cal E} \, = \, \, w^{1/4}$ (see (\ref{invtaucritbis3}) below)
but also
\begin{eqnarray}
\label{Modularinftyxy2}
\hspace{-0.98in}&& \quad \quad \quad  \quad 
P\Bigl(x(\tau), \, \,   x\Bigl(  {{\tau} \over { {\cal E}}}  \Bigr) \Bigr)
\, \,   = \, \,  \, \,
P\Bigl(x(\tau), \, \, x\Bigl(\Bigl( -\, {{3} \over {5}} \, -\, i \cdot \,  {{4} \over {5}} \Bigr)    \cdot \, \tau \Bigr)\Bigr)
\, \, = \, \,\, 0,  
\end{eqnarray}
corresponding to the $\, x \, \leftrightarrow \, y$ symmetry.
In other words, the   modular equation of level five is {\em also parametrised by the  automorphic Schwarz
function} (\ref{Schwarzfunction}).
We thus have two parametrisations of the  modular equation of level five: a rational
parametrisation (\ref{Level5inftyparam}), and a parametrisation (uniformisation) by
{\em automorphic functions}\footnote[1]{This is  the well-known Poincar\'e result~\cite{Whittaker}
  that, {\em whatever the genus of an algebraic curve},
this  algebraic curve is {\em uniformised by
  automorphic functions of a new variable} (here $\, \tau$). }.

Recalling the two $\, _2F_1$  hypergeometric identities 
(\ref{ModularinftyL2inftysolident}) or (\ref{ModularinftyL2inftysolident2}),
and taking their ratio, one finds
the {\em following simple covariance of the ratio}
$\, \tau(x) \, $ by  the algebraic series (\ref{Modularinftyexpans}): 
\begin{eqnarray}
 \label{invtauinfty}
 \hspace{-0.98in}&& \quad   \, \, \, \,  \quad  
\tau(y(x)) \, \, = \, \, \,   w^{1/4}   \cdot \, \tau(x),
\quad  \quad   \quad \hbox{where:} \quad   \quad  \, \,
 w \, \,   \, \, = \, \, \,  -{\frac{527}{625}} \, +{\frac {336\,i}{625}}.
\end{eqnarray}
Do note that  $\, w^{1/4} \, $ takes a simpler form\footnote[2]{See th 2.5 in J. Voight and J. Willis
  paper~\cite{Willis} for the simplicity of the complex numbers like
  $\,  {\cal E}$.}:
\begin{eqnarray}
\label{invtaucritbis3}
  \hspace{-0.98in}&& \quad   \, \, \,     \, \, \,   \quad
\tau(y(x)) \, \, = \, \, \,   {\cal E}     \cdot \, \tau(x),
\quad \quad \quad \, \hbox{where:} \quad \quad \quad \,
 {\cal E}     \, \, = \, \,
 -\, {{3} \over {5}} \,\, + \, i \cdot \, {{ 4} \over {5}}. 
\end{eqnarray}
Recalling the chain rule relation for the Schwarzian derivative of composition of functions
\begin{eqnarray}
\label{chainrule}
  \hspace{-0.98in}&&\quad   \quad \quad \quad \quad \,
 \{ \tau(y(x)), \, x \} \,\,   \,  =  \,   \,  \,  \,
 \{ \tau(y), \, y \}_{y=y(x)} \cdot \,  y'(x)^2 \, \, \,  \, + \{ y(x), \, x \},           
\end{eqnarray}
one deduces from the Schwarzian equation (\ref{SchwarzEQU}) that:
\begin{eqnarray}
\label{chainrule}
  \hspace{-0.98in}&& \quad \quad \quad \, 
\{ \tau(y(x)), \, x \} \,\,   \,  =  \,  \,  \,\{ {\cal E}     \cdot \, \tau(x), \, x \}
\,\, \,\,   \,  =  \,  \,  \, \{\tau(x), \, x \}
 \,\,   \,  =  \,  \,  \, \,
{\frac {140\,{x}^{2} \, -167\,x \, +135}{288 \cdot \,{x}^{2} \cdot \, (x-1)^{2}}}
 \nonumber \\
  \hspace{-0.98in}&&\quad   \quad \quad \quad \quad \quad
 \,\,   \,  =  \,  \,  \,\,
{\frac {140\,{y(x)}^{2} \, -167\,y(x) \, +135}{288 \cdot \,{y(x)}^{2} \cdot \, (y(x)-1)^{2}}} \cdot \,  y'(x)^2 
\, \,  \, + \, \{ y(x), \, x \}.           
\end{eqnarray}
One thus recovers the Schwarzian equation (\ref{ModularinftyexpansdefwSchwarz}) on
the algebraic correspondence $\, x \, \rightarrow \,  y(x)$.
Conversely, and more generally, if the Schwarz map $\, \tau(x)$ verifies 
the Schwarzian equation 
\begin{eqnarray}
\label{convers1}
  \hspace{-0.98in}&&\quad \quad \quad \quad \quad \quad \quad \quad \quad
\{ \tau(x), \, x \} \,\,  \, \,   \,\, - \,   \, W(x)
 \, \, \,  \, = \, \, \, \, \,\, 0, 
\end{eqnarray}
and the algebraic correspondence $\, x \, \longrightarrow \,  y(x)\, $
verifies the Schwarzian equation
\begin{eqnarray}
  \label{converse2}
  \hspace{-0.98in}&& \quad   \quad \quad  \quad   \quad  \quad  
\{y(x), \, x\} \, \,\, \,
+ W(y(x)) \cdot \, y'(x)^2 \, \, - \, W(x) \, \,\, = \, \, \,\, 0, 
\end{eqnarray}
with the same rational function $\, W(x)$, one deduces using the
the chain rule relation (\ref{chainrule}) that:
\begin{eqnarray}
  \label{converse3}
  \hspace{-0.98in}&& \quad  \quad \quad  \quad \quad  \quad  \quad  \quad   \quad
  \{\tau(y(x)), \, x\} \,  \, \,\, = \, \, \,\, \,  \{\tau(x), \, x\}.      
\end{eqnarray}
Consequently this means that $\, \tau(y(x)$ {\em can only be
a linear fractional transformation of} $\, \tau(x)$: 
\begin{eqnarray}
  \label{converse3}
  \hspace{-0.98in}&&  \quad  \quad \quad  \quad \quad  \quad  \quad \quad  \quad
 \tau(y(x)) \, \, \, = \, \, \, \,
 {{ a \cdot \, \tau(x)\,  + \, b} \over { c  \cdot \, \tau(x) \, + \, d}}.       
\end{eqnarray}
If one of the modular correspondence $\, \, x \, \rightarrow \, y(x)\, $ 
is such that (see (\ref{invtaucritbis3}))    
$\,  \tau(y(x)) \, \, = \, \, \,   {\cal E}     \cdot \, \tau(x)$,
the other {\em modular correspondences}
$\,\, x \, \rightarrow \, y(x) \, $ of the form (\ref{converse3})
commuting with that modular correspondence will  also be
such that $\, \,  \tau(y(x)) \, \, = \, \, \,  \alpha    \cdot \, \tau(x)$.

The action of the modular correspondence on the Schwarz map $\, \tau$, given by
this simple relation (\ref{invtaucritbis3}), is an {\em infinite order} transformation.
Such a simple relation (\ref{invtaucritbis3}) for {\em Shimura automorphic forms}
has to be compared with the action of the modular correspondence on the Schwarz map $\, \tau$ 
for {\em classical modular forms} (cusp forms\footnote[2]{This correspond to the fact that
  one of solutions has logarithmic terms in its formal series  solution
  expansion at $\, x= \, 0$. }  with a nome  $\, q$)
where we have transformations like
$\, \,\, q \,\, \rightarrow \,\,  \,  q^N \, \,$ where $\, N\, $ is an integer.

\vskip .2cm

\subsection{Derivative of  Shimura automorphic functions.}
\label{derivShim}
Let us introduce the order-two linear differential operator 
\begin{eqnarray}
\label{V2}
\hspace{-0.98in}&&  \quad    \quad   \quad \quad \quad  \quad   \quad   \quad 
  \theta \cdot \, \Bigl(\theta \, + \, {{3} \over {4}}   \Bigr)
  \,  \,  \, \,\, 
  - x \cdot \, \Bigl(\theta \, + \, {{29} \over {24}}   \Bigr)  \cdot \, \Bigl(\theta \, + \, {{25} \over {24}}   \Bigr),  
\end{eqnarray}
which has the two $\, \, _2F_1\, $  hypergeometric solutions: 
\begin{eqnarray}
\label{derivShimusolu}
 \hspace{-0.98in}&&  \quad  \, \,\,   \quad   \quad   \quad  
 _2F_1\Bigl([{{ 25} \over { 24}}, \, {{ 29} \over { 24}}], \, [{{ 7} \over {4}} ], x\Bigr), \quad  \quad  \quad
  x^{-3/4} \cdot \,  _2F_1\Bigl([{{ 7} \over { 24}}, \, {{ 11} \over { 24}}], \, [{{ 1} \over {4}} ], x \Bigr).
\end{eqnarray}
The linear differential operator (\ref{V2}) is homomorphic to (\ref{L2thetaunsurxalso}).  
One deduces from that homomorphism the following identity
\begin{eqnarray}
\label{identShimuderiv}
 \hspace{-0.98in}&&  \quad    \quad   \quad   \quad  
  (4 \, \theta \, + \, 1) \Bigl[\,  _2F_1\Bigl([{{ 7} \over { 24}}, \, {{ 11} \over { 24}}], \, [{{ 5} \over {4}} ], x \Bigr)\Bigr]
  \, \, = \, \, \, \, 
   _2F_1\Bigl([{{ 7} \over { 24}}, \, {{ 11} \over { 24}}], \, [{{ 1} \over {4}} ], x \Bigr).
\end{eqnarray}
Hypergeometric functions like (\ref{derivShimusolu}) are {\em not} Shimura $\, _2F_1$ hypergeometric functions:
they do not correspond to automorphic forms, but {\em derivatives of  automorphic forms}.

\vskip .2cm

Along this line, it is tempting to revisit the arguments of section (\ref{Derivclassforms}) 
for classical modular forms, to {\em (Shimura) automorphic forms}. 
Let us consider the rational function (\ref{Ratfonc4gg0}) of
section (\ref{vanHoeijVidunas}), which yields the Heun functions (\ref{HeunShimura})
and therefore the Shimura $\, _2F_1$ functions (\ref{Ratfonc4gg2hyp}). 
Recalling the identity (\ref{Ratfoncfour20deriv}) of section (\ref{Derivclassforms}) 
that the diagonal of the partial homogeneous derivative
of a rational function is the  homogeneous derivative of that diagonal
\begin{eqnarray}
\label{deriv}
  \hspace{-0.98in}&&  \quad  \quad \quad \quad 
 x \cdot \, {{d} \over { dx}} \Bigl( Diag(R(x, \,y, \, z, \, u)) \Bigr)
 \, \, = \, \,  \, Diag\Bigl( u \cdot \, {{\partial R(x, \,y, \, z, \, u)} \over {\partial u}}  \Bigr),  
\end{eqnarray}
it is tempting to calculate the telescoper of the homogeneous partial derivative with respect to $\, u$
of the rational function (\ref{Ratfonc4gg0}). One remarks that the  homogeneous partial derivative with
respect to $\, u \, $ of this rational function (\ref{Ratfonc4gg0}),
is very simple: it is equal to $\, u $ times the square of that rational function.
The telescoper of 
\begin{eqnarray}
\label{Ratfonc4gg0square}
  \hspace{-0.98in}&& \, \,    \quad  \quad \quad 
\, \, 
u \cdot \,  {{\partial R(x, \,y, \, z, \, u)} \over {\partial u}}   \\
\hspace{-0.98in}&&  \quad    \quad   \quad  \quad   \quad   \quad   \, \,   = \, \,  \,   \, 
  {{ u\cdot \, x^2 \, y^2 \, z^2} \over { (1 \, \,  \,  \, - x \,  y \,  z \, u \, \, 
 \, +\, x \,  y \,  z   \cdot \, (x \, +y\, +z) \,   \, \, + \, x \,y  + \, y \,z \, + \, x \, z)^2}},
 \nonumber 
\end{eqnarray}
is an order-three linear differential operator $\, M_3$ which is actually homomorphic
to the order-three linear differential operator $\, L_3$
given by (\ref{telesHeun}) and yielding
the Heun functions (\ref{HeunShimura}) or equivalently
squares of Shimura $\, _2F_1$ functions (\ref{Ratfonc4gg2hyp}). 
This homomorphism reads:
\begin{eqnarray}
\label{Ratfonc4gg0squarehomo}
\hspace{-0.98in}&& \quad \quad \quad \, \quad \quad \quad \quad \quad \quad \quad 
   M_3 \cdot \, \theta  \, \, \, \, = \, \,   \, \,  \, \,
                   x^2 \cdot \, \theta\,  \cdot \,  L_3.
\end{eqnarray}
From (\ref{Ratfonc4gg0squarehomo}) one finds that the solutions of  $ \, M_3$
(and thus the solutions of the telescoper of (\ref{Ratfonc4gg0square})) are
simply obtained from the action of the homogeneous derivative $\, \theta$  on the
Shimura solutions of the order-three linear differential operator $\, S_3$, i.e. 
on the  Heun functions (\ref{HeunShimura}) or the
(square of) Shimura $\, _2F_1$  hypergeometric functions (\ref{Ratfonc4gg2hyp}).
This example of rational function (\ref{Ratfonc4gg0square}) thus yields
a telescoper which solutions are {\em not} Shimura $\, _2F_1$  automorphic forms, but
{\em derivatives of Shimura $\, _2F_1$ automorphic forms}.

\vskip .2cm

{\bf Anecdotal remark:} recalling (\ref{3F2homomany})
where the parameters in (\ref{3F2homomany}) verify (\ref{Fuchscondi}) and (\ref{apparentcondi3}),
one can also write (\ref{identShimuderiv})  as $\, _3F_2([ 7/24, 11/24, 5/4], \, [5/4, 1/4], \,x)$
but also, since here $\, e \, = 1/4$,  as the Heun function 
\begin{eqnarray}
\label{anecdotalHeun}
\hspace{-0.98in}&& \quad \quad \quad \, \quad \quad \quad \quad \quad \quad \quad 
  Heun\Bigl(0, \, 0, \, {{7 } \over {24}}, \,{{11 } \over {24}}, \, \, {{5 } \over {4}}, \,  {{3 } \over {2}}, \, \,    x \Bigr). 
\end{eqnarray}
This simple form of the Heun function is a consequence of the fact that one has $\, \, \gamma \, = \, e \, +1$ in this case.
With $\, e \, = \, 1/2 \,\, $ the identities (\ref{3F2homomany}) become:
\begin{eqnarray}
\label{anecdotalHeun2}
\hspace{-0.98in}&& \quad \quad  \quad \,   
 Heun\Bigl({{72} \over {5}}, \, {{231} \over {40}}, \,
  {{7 } \over {24}}, \,{{11 } \over {24}}, \, \, {{5 } \over {4}}, \,  {{3 } \over {2}}, \, \,    x \Bigr)
\,  \, \, \, = \, \, \,   \,
(2\, \theta \, +1) \Bigl[  \,  _2F_1([{{7 } \over {24}}, \,{{11 } \over {24}}], \, [{{5 } \over {4}}], \, \, x\Bigr)   \Bigr]
\nonumber \\
 \hspace{-0.98in}&& \quad  \quad \quad \quad \quad  \quad  \quad  \quad                   \,  \, = \, \,   \, \,
_3F_2([{{7 } \over {24}}, \,{{11 } \over {24}}, \, \, {{3 } \over {2}}], \, [{{1 } \over {2}}, \, {{5 } \over {4}}], \, \, x\Bigr) 
\end{eqnarray}        

\vskip .2cm

\subsection{Identities linking hypergeometric functions that are related
with  Shimura curves.}
\label{identities}

In fact, several identities linking  $\, _2F_1$ hypergeometric functions  related
with {\em Shimura curves}, 
appear in the litterature. For example we see in~\cite{VidunasFilipuk} (equation (4.8) page 14):
\begin{eqnarray}
  \label{I50}
 \hspace{-0.98in}&&    \quad  \, \, 
 _2F_1\Bigl([{{5} \over {42}}, \, {{19} \over {42}}], \, [{{5} \over {7}}],  \, x  \Bigr)
\, \, \, \, = \, \, \, \,
\Bigl( {{  6561  \,-13851 \, x \,   -9261 \, x^2  \, + 16807 \, x^3 } \over {6561 }} \Bigr)^{-1/28} 
\\
\hspace{-0.98in}&& \quad \quad  \quad \quad  \quad  \, \, \times \, \, 
_2F_1\Bigl([{{1} \over {84}}, \, {{29} \over {84}}], \, [{{6} \over {7}}],
\, \,  {{ x^2 \cdot \, (1\, -x) \cdot \, (49\, x  - \, 81)^7} \over {
4 \cdot \, ( 6561  \,-13851 \, x \,   -9261 \, x^2  \, + 16807 \, x^3 )^3}}  \Bigr),
\nonumber 
\end{eqnarray}
where the $\,_2F_1$ hypergeometric function on the RHS of the identity  (\ref{I50}) corresponds
to a {\em Shimura curve} with {\em elliptic points} $ \, (2, \, 3,  \, 7)$. To see that  the hypergeometric function on the LHS also
corresponds to a Shimura curve is not obvious on the difference of exponants\footnote[1]{The difference of exponants
  of a $\, _2F_1([a,b],[c],x)$hypergeometric function is the triplet $\, (c\, -a\, -b, \, b-a, \, 1\, -c)$.
  These rational numbers $\, (c\, -a\, -b, \, b-a, \, 1\, -c)$ have to be (up to a sign)
reciprocals of integer, and furthermore in some table given by Takeuchi~\cite{Takeuchi}.} of the hypergeometric function.
However considering the order-two linear differential operator
$\, N_2$ annihilating $ \, _2F_1([{{5} \over {42}}, \, {{19} \over {42}}], \, [{{5} \over {7}}],  \, x)$,
one finds that the pullback of  $\, N_2$ by one of the  Euler's hypergeometric transformations, namely
$\, x \, \rightarrow \, 1/x$, has the solution $\, x^{5/42}\cdot \, _2F_1([11/42, 23/42],[2/3],x)$
which is a Shimura  hypergeometric function of the $\,(3, \, 3, \, 7)$ type.

Note that the set of Gauss hypergeometric functions
that are associated with Shimura curves {\em is a finite set}~\cite{Voight,Takeuchi}.

\vskip .2cm   

Similarly in~\cite{FangTingTuYifang} we find on page 2, equation (3), the identity:
\begin{eqnarray}
 \hspace{-0.98in}&& \quad \quad   \quad \,
_2F_1\Bigl([{{1} \over {20}}, \, {{1} \over {4}}], \, [{{4} \over {5}}],
 \,\,  {\frac { 64 \cdot \,x \cdot \, (1\, -x \, -{x}^{2})^{5}}{ (1\, -{x}^{2})  \, (1\, +4\,x \, -{x}^{2})^{5}}} \Bigr)
 \\
\hspace{-0.98in}&& \quad \quad \quad   \quad  \quad   \quad  \, \, 
\, \, = \, \, \, (1 \, - \, x^2)^{1/20} \, \cdot  (1 \, + 4\,  x\, - \, x^2)^{1/4} \, \cdot \, \, 
 _2F_1\Bigl([{{3} \over {10}}, \, {{2} \over {5}}], \, [{{9} \over {10}}],  \, \,  x^2 \Bigr),
\nonumber
\end{eqnarray}
where the  $\, _2F_1$ hypergeometric function on the RHS of the identity corresponds to a Shimura curve with
elliptic points $ \, (5, \, 2, \, 5)$.
The hypergeometric function on the LHS also corresponds
to a Shimura curve.  We also see the identity:
\begin{eqnarray}
\hspace{-0.98in}&& 
  \Bigl(1+ \, (4 +2\,b) \cdot \, x \,  \, - \, (1+2\,b) \cdot \, {x}^{2}\Bigr)^{-1/2} 
 \\
\hspace{-0.98in}&& \quad 
\times \,  _2F_1\Bigl([{{1} \over {8}}, \, {{3} \over {8}}], \, [{{3} \over {4}}],\, \, \,
- \,{\frac { 4 \cdot \,  (1 \, +b)^{4} \cdot \,  x \cdot \, (1 \, -\, (7 -4\,b)/3 \cdot \, {x}^{2})^{4}}{
(1 \, +x)  \cdot \, (1 \, -3\,x)  \cdot \,
 (1 \, \,\, + \, (4 \,+2\,b) \cdot \, x \, \, \, - \, ( 1 \, +2 \,b) \cdot \, {x}^{2})^{4}}}\Bigr)
\nonumber \\
\hspace{-0.98in}&& \, \,  \,  \,  
\, \, = \, \, \, {\frac { (1 \, +x)^{1/8} \cdot \, (1 \, -3\,x)^{1/8}}{ (1\, + a \cdot \,  x)^{5/4}}} \cdot \,
_2F_1\Bigl([{{5} \over {24}}, \, {{3} \over {8}}], \, [{{3} \over {4}}],\, \,          
                   \,{\frac {12 \cdot \, a \cdot \, x \cdot \, (1\,  -{x}^{2})
                   \cdot \, (1\,  -9\,{x}^{2}) }{(1\, +\, a  \cdot \,  x)^{6}}}\Bigr),
\nonumber
\end{eqnarray}
for $\, \, a^2 \, + \, 3 \, = \, \, 0\, $  and  $\,\,  b^2 \, + \, 2 \, = \, \, 0$, where
the $\, _2F_1$  hypergeometric function on the left of the identity, corresponds to an {\em automorphic form
associated with a Shimura curve}
with elliptic points $(4, \, 4, \, 4)$.

\vskip .2cm

\subsection{A level three modular equation for  Shimura Heun functions.}
\label{HeunTRUE}

Similar calculations can be performed on the Heun example  (\ref{trueHeun}).
This Heun example has a {\em level three modular equation}~\cite{Voight}. 
Introducing the following ratio (Schwarz map):
\begin{eqnarray}
\label{BabaHeununiftau}
  \hspace{-0.98in}&& \quad  \quad \quad  \quad \quad  \quad \quad
 \tau \, \, \,  = \, \, \,  x^{1/3} \cdot \,
{{Heun({{27} \over {2}}, \, {{47} \over {18}}, \, {{5} \over {12}},\, {{11} \over {12}}, \,\, {{4} \over {3}}, \,  {{1} \over {2}}, \, \, x)
} \over {
Heun({{27} \over {2}}, \, {{7} \over {36}}, \, {{1} \over {12}},\, {{7} \over {12}}, \,\, {{2} \over {3}}, \,  {{1} \over {2}}, \, \,  x) }}. 
\end{eqnarray}
Introducing
\begin{eqnarray}
\label{BabaHeununif}
  \hspace{-0.98in}&& \quad  \quad  \quad \quad  \quad 
\tau(z) \, = \, \, \eta(t(z)) \, \, = \, \, \,
 \,{\frac { 108 \cdot \,(t-1)^{3}}{ (t+1)^{2} \, (9\,{t}^{2}-10\,t+17) }},
\quad   \quad
\nonumber \\
   \hspace{-0.98in}&& \quad  \quad \quad \quad \quad 
\tilde{\tau}(z) \, = \, \,\,
\eta\Bigl({{10} \over {9}} \, -t(z)\Bigr) \,\, \, = \, \, \,
\,{\frac {12 \cdot \, (1 -9\,t)^{3}}{ (19\, -9\,t)^{2} \cdot \, (9\,{t}^{2}-10\,t+17) }},         
\end{eqnarray}
the elimination of the variable $\, t \, $ yields
the (genus-zero)  {\em level three modular equation}~\cite{Voight} which reads: 
 \begin{eqnarray}
\label{BabaHeununifMod}
  \hspace{-0.98in}&&   \quad \quad  \, \, \, \, 
 9604\, \cdot \, A^2\, \, B^2 \,\cdot  \, (9\, \, A^2 +14\, \, A\, \, B +9\, \, B^2)
\nonumber \\
\hspace{-0.98in}&&  \quad \quad\,  \, \, \, \, 
-2940 \, \cdot \, A\, \, B\,\cdot  \, (A+B)\, \, (25\, \, A^2 +518\, \, A\, \, B +25\, \, B^2)
\nonumber \\
\hspace{-0.98in}&&  \quad  \quad \, \, \, \, \, 
+ 5 \, \cdot \, \Bigl(3125\, \, (A^4+B^4) \,
+334944\, \, (A^3\, \, B \, +A\, \, B^3) +1605078\, \, A^2\, \, B^2 \Bigr)
 \nonumber \\
\hspace{-0.98in}&&  \quad \quad \quad \quad \quad \quad  \, \, \, \, \, 
+480 \, \cdot \, (A+B) \,\cdot  \, (375\, \, A^2-13513\, \, A\, \, B+375\, \, B^2)
\\
\hspace{-0.98in}&&  \quad   \quad  \, \, \, \, \, 
+ 5120 \,\cdot  \, (135\, \, A^2+142\, \, A\, \, B+135\, \, B^2)
\, \, + 884736\, \cdot \, (A+B)
  \, \, \, = \, \, \, \, 0.   \nonumber      
 \end{eqnarray}

 \vskip .2cm

 \section{Heun functions that are pullbacked $\, _2F_1$ hypergeometric functions
   but are not related to classical modular forms or  Shimura automorphic forms.}
\label{ReduceMaierHeun}

Let us recall Maiers's identity (see Theorem 3.8 in~\cite{ReduceMaier}) 
 \begin{eqnarray}
\label{ReduceMaier}
\hspace{-0.98in}&& \quad  \quad  \quad  \quad   \quad  
Heun\Bigl(\omega,  \, \, \alpha \, \beta \cdot \, \eta, \,  \,\alpha, \,  \,\beta, \, \, {{\alpha \, +\, \beta \, + \, 1} \over {3}},
\,  \, {{\alpha \, +\, \beta \, + \, 1} \over {3}}, \, \, \, \eta \cdot \, x \Bigr)
\nonumber \\
\hspace{-0.98in}&& 
 \quad \quad    \quad  \quad   \quad  \quad     \quad    \quad    \, \, = \, \, \, \,
   _2F_1\Bigl([{{\alpha} \over {3}}, \, {{\beta} \over {3}}], \, [{{\alpha \, +\, \beta \, + \, 1} \over {3}}],
  \, \,  \, x \cdot \, (3 \, -3\, x \,  +x^2) \Bigr),  
 \end{eqnarray}
 where  $\, \omega$ is a sixth root of unity and $\, \eta$
 read\footnote[1]{ Note that the {\em normalised accessory parameter} $\, q/\alpha/\beta \,$
   is a equal to the constant $\, \eta$.}: 
\begin{eqnarray}
\label{ReduceMaierwhere}
  \hspace{-0.98in}&& \quad  \quad  \quad  \quad  \quad  \quad  \, \, \,
 \omega \, \, = \, \, \,     {{1} \over {2}} \, +i \cdot \, {{3^{1/2}} \over {2}},
 \quad  \quad  \quad   \eta \, \, = \, \, \,    {{1} \over {2}} \, +i \cdot \, {{3^{1/2}} \over {6}}.
\end{eqnarray}                  
Such a Heun function is thus a pullbacked $\, _2F_1\,$ hypergeometric function 
{\em for all the values of the parameters} $\, \alpha$,  $\, \beta$. This two parameters ($\, \alpha$,  $\, \beta$)
space is large enough to encapsulate a set of interesting subcases.

$\bullet$ For selected values of the  two parameters $\, \alpha$,  $\, \beta$,
the Heun function (\ref{ReduceMaier}) actually reduces to
Shimura $\, _2F_1\,$ hypergeometric functions. For instance for $\, \alpha \, =\, 1$ and  $\, \beta \, = \, \, 1/4$, 
we recover the {\em Shimura}  $\, _2F_1\,$ hypergeometric function (\ref{Table344}) related to
the {\em telescoper of the rational function} (\ref{Ratfonc4gg0}). The relation (\ref{ReduceMaier})  becomes: 
\begin{eqnarray}
\label{ReduceMaierShimura}
\hspace{-0.98in}&& \quad  \quad  \quad \quad  \quad   \quad  \quad  
Heun\Bigl(\omega,  \, \,  \, {{\eta} \over {4}}, \,  \, 1, \,  \, {{1} \over {4}}, \, \, {{3} \over {4}}, \, \, {{3} \over {4}},
\, \, \, \eta \cdot \, x \Bigr)
\nonumber \\
\hspace{-0.98in}&& \quad 
 \quad \quad    \quad  \quad   \quad  \quad   \quad  \quad      \quad              \, \, = \, \, \,
   _2F_1\Bigl([{{1} \over {3}}, \, {{1} \over {12}}], \, [{{3} \over {4}}],
     \, x \cdot \, (3 \, -3\, x \,  +x^2) \Bigr).  
 \end{eqnarray}
 Many other values of  $\, (\alpha$,  $\, \beta) \, $ yield Shimura $\, _2F_1\,$ hypergeometric functions:
 \begin{eqnarray}
\label{ReduceMaierShimuramay}
   \hspace{-0.98in}&&\quad \quad
(\alpha,  \,  \, \beta)     \,\, =  \,\,\, \,  \Bigl({{1} \over {8}}, \, \,{{9} \over {8}}  \Bigr), \quad
\Bigl({{1} \over {3}}, \, \,{{4} \over {3}}  \Bigr), \quad  \Bigl({{3} \over {8}}, \, \,{{11} \over {8}}  \Bigr),
\quad  \Bigl({{2} \over {5}}, \, \,{{7} \over {5}}  \Bigr), \quad  \Bigl({{1} \over {4}}, \, \,{{5} \over {4}}  \Bigr), \quad
 \Bigl({{2} \over {3}}, \, \,{{7} \over {6}}  \Bigr),
 \nonumber \\
 \hspace{-0.98in}&& \quad \quad  \quad 
 \quad  \Bigl({{1} \over {12}}, \, \,{{19} \over {12}}  \Bigr), \quad
 \Bigl({{5} \over {32}}, \, \,{{53} \over {32}}  \Bigr), \quad  \Bigl({{3} \over {4}}, \, \,{{9} \over {8}}  \Bigr), \quad
 \Bigl({{1} \over {5}}, \, \,{{17} \over {10}}  \Bigr), \quad  \Bigl({{1} \over {2}}, \, \,{{5} \over {4}}  \Bigr), \quad
\Bigl({{1} \over {2}}, \, \,{{5} \over {4}}  \Bigr), 
\nonumber \\
 \hspace{-0.98in}&& \quad \quad  \quad \quad \quad
 \quad \quad  \Bigl({{1} \over {8}}, \, \,{{13} \over {8}}  \Bigr), \quad
 \Bigl({{1} \over {2}}, \, \, 1  \Bigr), \quad  \Bigl({{5} \over {8}}, \, \, 1  \Bigr), \quad
 \Bigl({{1} \over {6}}, \, \,{{5} \over {13}}  \Bigr), \quad  \Bigl({{1} \over {4}}, \, \,{{7} \over {4}}  \Bigr). \quad
\end{eqnarray}

\vskip .2cm 

$\bullet$ In \ref{notmodularsub} we have seen that one  can introduce a nome,
for that linear differential Heun operator (\ref{Heun}), when  $\, \gamma\, = \, \, 1$.
This condition is, of course,
{\em necessary but not sufficient} to reduce to a
classical modular form. In that $ \,\gamma \, = \, \, 1$ subcase where one can introduce a nome, the previous
Heun-to-$_2F_1$ reduction (\ref{ReduceMaier}) reads (since we have
$ \,\gamma \, = \, \, (\alpha+\beta+1)/3 \,\, =  \, 1$): 
\begin{eqnarray}
\label{ReduceMaiergamma1}
\hspace{-0.98in}&& \quad  \quad  \quad  \quad  \quad   \quad  
Heun\Bigl(\omega,  \, \, \alpha \, (2 \, - \, \alpha) \cdot \, \eta, \,  \,\alpha, \,  \, 2 \, - \, \alpha, \, \, 1,
\,  \, 1, \, \, \, \eta \cdot \, x \Bigr)
\nonumber \\
\hspace{-0.98in}&& 
 \quad \quad    \quad  \quad   \quad  \quad  \quad   \quad      \quad              \, \, = \, \, \,
   _2F_1\Bigl([{{\alpha} \over {3}}, \, {{2 \, - \, \alpha} \over {3}}], \, [1],
     \,\, \, x \cdot \, (3 \, -3\, x \,  +x^2) \Bigr),  
 \end{eqnarray}
 For selected values of $\, \alpha$ we actually get $\, _2F_1$ hypergeometric functions
 corresponding to {\em classical modular curves} (see (\ref{Specialliste2})). The selected values
 are $\, \alpha \, = \, \, 1, \, 1/2, \, 1/4$, corresponding respectively to:
 \begin{eqnarray}
\label{ReduceMaiergamma1respectiv}
\hspace{-0.98in}&& 
  \quad \quad    
   _2F_1\Bigl([{{1} \over {3}}, \, {{1} \over {3}}], \, [1],
 \,\, \, x \cdot \, (3 \, -3\, x \,  +x^2) \Bigr),  \quad  \quad  _2F_1\Bigl([{{1} \over {6}}, \, {{1} \over {2}}], \, [1],
\,\, \, x \cdot \, (3 \, -3\, x \,  +x^2) \Bigr),
\nonumber \\
\hspace{-0.98in}&& 
 \quad  \quad \quad \quad \quad  \quad  \quad  \quad   \quad  _2F_1\Bigl([{{1} \over {12}}, \, {{7} \over {12}}], \, [1],
 \,\, \, x \cdot \, (3 \, -3\, x \,  +x^2) \Bigr).  
 \end{eqnarray}
 Let us now, consider a rational value of $\, \alpha$, different from $\,  1, \, 1/2, \, 1/4$,
 for instance $\, \alpha \, = \, \, 5/8$.
 The identity (\ref{ReduceMaiergamma1}) becomes
 \begin{eqnarray}
\label{ReduceMaiergamma15sur8}
\hspace{-0.98in}&& \quad  \quad \quad   \quad  \quad   \quad  
Heun\Bigl(\omega,  \, \,  {{55} \over {64}} \cdot \, \eta, \,  \, {{5} \over {8}}, \,  \,  {{11} \over {8}}, \, \, 1,
\,  \, 1, \, \, \, \eta \cdot \, x \Bigr)
\nonumber \\
\hspace{-0.98in}&& 
 \quad \quad    \quad  \quad   \quad \quad   \quad    \quad     \quad              \, \, = \, \, \,
   _2F_1\Bigl([{{5} \over {24}}, \, {{11} \over {24}}], \, [1],
     \,\, \, x \cdot \, (3 \, -3\, x \,  +x^2) \Bigr).  
 \end{eqnarray}
 The series expansion of (\ref{ReduceMaiergamma15sur8}) gives a $\, _2F_1$ 
 {\em globally bounded series}\footnote[1]{This series
   can be recast into a series with integer coefficients after
   the $\, x \,  \, \rightarrow \,  \, 2304 \cdot \, x \,  $ rescaling. }.
 The Heun function (\ref{ReduceMaiergamma15sur8}) is annihilated by
 the order-two linear differential operator
 \begin{eqnarray}
\label{L2ReduceMaiergamma15sur8}
   \hspace{-0.98in}&& \,\,   \,    \,  
L_2 \, \,\,\,  = \, \,\, \,  64  \cdot \, x \, \cdot \, (x^2 \, -3 \,x \, +3)\cdot \, D_x^2
 \,\, \,\,  +192\cdot \, (x-1)^2 \cdot \, D_x \,\,  \, \, +55\cdot \, (x-1), 
 \end{eqnarray}
 which yields the following series expansion for the nome of $\, L_2$:
 \begin{eqnarray}
\label{nomeReduceMaiergamma15sur8}
   \hspace{-0.98in}&& \quad \quad \quad 
q \, \, = \, \,  \,\, 
 x \, \,\, \,  +{\frac {41\,{x}^{2}}{96}} \, \, \,
 +{\frac {24925\,{x}^{3}}{147456}} \,\,  \,  +{\frac {21593429\,{x}^{4}}{382205952}}
\,\,  +{\frac {57805761947\,{x}^{5}}{4696546738176}}
\nonumber \\
 \hspace{-0.98in}&& \quad  \quad \quad \quad \quad  \quad \quad 
\, -{\frac {4091796188773\,{x}^{6}}{2254342434324480}} \, \,  \,
-{\frac {400657105197062971\,{x}^{7}}{93492089436304834560}} \,  \,  \, \,  \,\,  + \, \, \, \cdots 
\end{eqnarray}
This series is {\em not} globally bounded. Therefore (\ref{ReduceMaiergamma15sur8}) {\em does not
correspond to a classical modular form}, as expected since the $\, _2F_1$ hypergeometric
function in the RHS of (\ref{ReduceMaiergamma15sur8}) is not in the list of the twenty-eight
$\, _2F_1$ hypergeometric functions (\ref{Specialliste2}).
With this example (\ref{ReduceMaiergamma15sur8}) we see that a Heun function can actually correspond
to a globally bounded series, being reducible to a pullbacked $\, _2F_1$ hypergeometric function,
without necessarily corresponding to a classical modular form\footnote[2]{We cannot totally exclude the fact
  that a nome series like (\ref{nomeReduceMaiergamma15sur8}) could correspond to
 an order-one linear differential operator
  acting on a classical modular form (see the second $ \, _2F_1$ hypergeometric function
  of (\ref{morehypergmodular}) and equation (\ref{q5sur6.7sur6}) in \ref{negative}).}.
We have not been able to find such Heun functions as diagonal
of rational functions or even as solutions of telescopers of rational functions.

\vskip .2cm

\section{The $\, x \, \leftrightarrow \, 1/x$ and $\, x \, \leftrightarrow \, A/x$ dualities.}
\label{Duality}

Using the (multi-Taylor) definition
of the diagonal  of a  rational function,
it is straightforward to show, for any {\em positive} integer $\, n$,  that 
the diagonal of a  rational function  $\, R(x, \, y, \, z, \, w)$ and  the 
diagonal of a  rational function  $\, R(x^n, \, y^n, \, z^n, \, w^n)$ are simply related.  Denoting these two diagonals
respectively $\, D_1(x) \, \, = \, \,  \,  Diag( R(x, \, y, \, z, \, w))$  and
$ \,D_n(x) \, \, = \, \,  \,  Diag( R(x^n, \, y^n, \, z^n, \, w^n))$,
one has the simple relation: $\, D_n(x) \, \, = \, \,  \,   D_1(x^n)$. Of course this demonstration
cannot be extended to negative integers $\, n$, in particular $\, n= \, -1$.

Let us revisit example 5  where the diagonal of the rational function (\ref{Ratfoncfour10})
is the Heun series expansion (\ref{H10}). Let us consider, instead of the rational function (\ref{Ratfoncfour10}),
the rational function where the four variables $(x, \, y, \, z, \, w)$  have been changed into their reciprocal 
$(1/x, \, 1/y, \, 1/z, \, 1/w)$ in the rational function $\, R(x, \, y, \, z, \, w)$ given by  (\ref{Ratfoncfour10}):
\begin{eqnarray}
\label{Ratfoncfour10recip}
  \hspace{-0.98in}&& \quad  \quad   \quad  \quad   
\, \,  \,  \,  
{\cal R}(x, \, y, \, z, \, w)  \, \, \,  = \, \,  \,    \, \,
 R\Bigl({{1 } \over {x}}, \, {{1 } \over {y}}, \, {{1 } \over {z}}, \, {{1 } \over {w}}\Bigr)
\nonumber \\
  \hspace{-0.98in}&& \quad  \quad    \quad  \quad \quad  \quad
\, \, \,  = \, \,  \,   \,  \,  
 {\frac {xyzw}{xyzw \, \, \, - xyw \,\, \,  -xzw \,\,\,   -y w \, \, \, -z w\, \,\,  -xy \, \,\,  -z \, \, \, -1}}. 
\end{eqnarray}
The diagonal of this ``reciprocal''  rational function (\ref{Ratfoncfour10recip})  reads:
\begin{eqnarray}
\label{SS10recip}
  \hspace{-0.98in}&& \,  
Diag\Bigl({\cal R}(x, \, y, \, z, \, w)\Bigr)  \, \, 
                     \, \, = \, \, \,\,
 -x \, \, \, -5\,x^2 \, \, \,-73\, x^3 \,\,
-1445\, x^4 \,\,  -33001\, x^5 \, \,\,  + \, \, \, \cdots 
\end{eqnarray}
The telescoper of the diagonal of this  rational function (\ref{Ratfoncfour10recip}) of four variables
reads:
\begin{eqnarray}
\label{S10recip}
  \hspace{-0.98in}&& \quad  \quad  \quad  
M_3    \, \,\,  = \, \, \,\,  \, 5\,x \, \, -1 \, \, \, \,
 +(1 \,-10\,x \,+x^2) \cdot \, x \, \cdot \, D_x \, \, \,  +3 \,  \cdot \, (x-17) \cdot \, x^3 \cdot \, D_x^2
\nonumber \\
 \hspace{-0.98in}&& \quad  \quad  \quad   \quad  \quad   \quad  \quad  \quad
\, \,  +(1 \,-34\,x \,+x^2)\cdot \, x^3 \cdot \, D_x^3.
\end{eqnarray}             
This order-three linear differential operator $\, M_3$ given by  (\ref{S10recip}) {\em actually corresponds to 
the telescoper $\, L_3$ given by} (\ref{S10}) {\em pullbacked by} $\, x \, \rightarrow \, 1/x$.
Recalling the solution (\ref{H10}) of the telescoper $\, L_3$
given by (\ref{S10}), one finds easily that the diagonal series expansion (\ref{SS10recip}),
solution of the order-three linear differential operator $\, M_3$,
can be written as
\begin{eqnarray}
\label{H10recip}
\hspace{-0.98in}&& \quad
-x \cdot \, (1 \, -34\,\, x \, +x^2)  \,\,  \times
  \nonumber \\
 \hspace{-0.98in}&& \quad \quad \,\,
 Heun\Bigl(577 \, +408 \cdot \, 2^{1/2}, \,  \,  {{663} \over {2}} \, +234 \cdot \, 2^{1/2}, \, \, 
 {{3} \over {2}}, \,  \, {{3} \over {2}},\, \,   1,\, {{3} \over {2}}, \, 
 \,\,  \, (17\,+12\cdot  \,2^{1/2}) \cdot \, x \Bigr)^2
  \nonumber \\
  \hspace{-0.98in}&&
\, = \, \, \,\, \, -x \cdot \, (1 \, -34\,\, x \, +x^2)  \,\,  \times
  \nonumber \\
 \hspace{-0.98in}&& \quad \quad \,\,
 Heun\Bigl(577 \, -408 \cdot \, 2^{1/2}, \,  \,  {{663} \over {2}} \, -234 \cdot \, 2^{1/2}, \, \, 
 {{3} \over {2}}, \,  \, {{3} \over {2}},\, \,   1,\, {{3} \over {2}}, \, 
 \,\,  \, (17\, -12\cdot  \,2^{1/2}) \cdot \, x \Bigr)^2
  \nonumber \\
  \hspace{-0.98in}&& \quad  \,\,    \quad
\, \, \, = \, \, \,\, \, -x \, \, \, -5\,x^2 \, \, \,-73\, x^3 \,\,\,  -1445\, x^4 \,\, \, -33001\, x^5
 \,\,\,  -819005\, x^6 \, \, \,\,  \, + \, \, \, \cdots 
\end{eqnarray}
In other words there is a simple relation between the diagonal of  (\ref{Ratfoncfour10})
and the diagonal of its ``reciprocal'' (\ref{Ratfoncfour10recip}):
\begin{eqnarray}
\label{H10recip}
  \hspace{-0.98in}&& \quad \quad \quad \quad  \quad \quad  \quad 
Diag\Bigl({\cal R}(x, \, y, \, z, \, w)\Bigr)  \,  \, \, = \, \, \,\,
 Diag\Bigl( R\Bigl({{1 } \over {x}}, \, {{1 } \over {y}}, \, {{1 } \over {z}}, \, {{1 } \over {w}}\Bigr)\Bigr)
 \nonumber \\
 \hspace{-0.98in}&& \quad \quad \quad \quad \quad \quad \quad \quad  \quad  \quad \quad 
 \, \, = \, \, \,\,  -x \cdot \,   Diag\Bigl( R(x, \, y, \, z, \, w)\Bigr). 
\end{eqnarray}
This is confirmed by the homomorphism between the two telescopers $\, L_3$ and $\, M_3$:
\begin{eqnarray}
\label{H10recip}
\hspace{-0.98in}&& \quad \quad  \quad  \quad \quad \quad \quad \quad  \quad 
                     M_3 \cdot \, x \, \, \, = \, \, \, x^2 \cdot \, L_3.
\end{eqnarray}
Recalling the fact that the order-three linear differential operator $\, M_3$  corresponds to 
the telescoper $\, L_3$ pullbacked by $\, x \, \rightarrow \, 1/x$,
this homomorphism means a slightly puzzling homomorphism between the telescoper $\, L_3$
and itself pullbacked by $\, x \, \rightarrow \, 1/x$.

\vskip .2cm 

\subsection{The $\, x \, \longleftrightarrow \, A/x$ dualities:  dualities on the telescopers.}
\label{Duality1}

Let us revisit  example 1, considering, similarly,  the diagonal of the rational function
of four variables (\ref{Ratfoncfour_7}), its telescoper  $\, L_3$ given by (\ref{L3SS7}),
and the corresponding Heun solutions  (\ref{H1solhyp}). We introduce, as previously, 
a new  rational function (for the rational function (\ref{Ratfoncfour_7})
of four variables $\, R(x, \, y, \, z, \, w)$) corresponding to simple involutions
of the form $\, t \, \,  \rightarrow \,  \, A/t$ on the four variables:  
\begin{eqnarray}
\label{Ratfoncfour7recip}
  \hspace{-0.98in}&& \quad  \quad   \quad  \quad   \quad 
\, \,  
{\cal R}(x, \, y, \, z, \, w)  \, \, \,  = \, \,  \,    \, \,
 R\Bigl(-{{ 1 } \over {4\, x}}, \, -{{1 } \over {4\, y}}, \, -{{1 } \over {4\, z}}, \, \, {{1 } \over {w}}\Bigr)
\nonumber \\
  \hspace{-0.98in}&& \quad  \quad \quad  \quad \quad  \quad \quad  \quad
\, \, \, \,  = \, \,\,   \,    \, \,{\frac { 64\, xyzw}{64\, xyzw \, -4\,xw \, +16\,yz \, +w \, -4\,x \, -4\,y \, -4\,z}}.
\end{eqnarray}
The telescoper of this  rational function (\ref{Ratfoncfour7recip}) reads:
\begin{eqnarray}
\label{M3Ratfoncfour7recip}
  \hspace{-0.98in}&& 
 M_3 \, \, = \, \, \,  \,   -15  \, +32\, x \,  \, \,\,
 + (3 \, - \, 32 \, x) \cdot \, (5 \, + \, 32 \, x) \cdot \, x \, \cdot \,  D_x
  \, \, \,  \, - 96 \cdot \, (3 \, + \, 32 \, x) \cdot \,  x^3 \cdot \, D_x^2
\nonumber \\
 \hspace{-0.98in}&& \quad \quad \quad \quad \quad \quad    \quad \quad   \,
  + (1 \,+ 4\, x) \cdot \, (1\, -\, 16\, x) \cdot \, x^3 \cdot \, D_x^3.
\end {eqnarray}
This order-three linear differential operator (\ref{M3Ratfoncfour7recip})
is {\em exactly the telescoper}  $\, L_3$ given by (\ref{L3SS7}),
{\em pullbacked by} $\,\, x \, \rightarrow \, \, -1/64/x$.
The telescoper  $\, M_3$ has the following Heun solutions (to be compared with (\ref{H1solhyp}))
\begin{eqnarray}
\label{M3Ratfoncfour7recipsolA} \quad  \quad \quad  \quad \quad  \quad 
  \hspace{-0.98in}&& x^{3/4} \cdot  Heun\Bigl(- \, {{1} \over {4}}, \, {{7} \over {256}}, \, {{3} \over {8}},
  \,  {{3} \over {8}}, \,   {{3} \over {4}}, \,  {{1} \over {2}}, \, -4 \, x\Bigr)^2, \, \, 
\end{eqnarray}
\begin{eqnarray}
\label{M3Ratfoncfour7recipsolB}\quad  \quad \quad  \quad \quad  \quad 
  \hspace{-0.98in}&& 
 x^{5/4} \cdot  Heun\Bigl(- \, {{1} \over {4}}, \, {{31} \over {256}}, \, {{5} \over {8}},
  \,  {{5} \over {8}}, \,   {{5} \over {4}}, \,  {{1} \over {2}}, \, -4 \, x\Bigr)^2, 
\end{eqnarray}
and:
\begin{eqnarray}
\label{M3Ratfoncfour7recipsol2}
  \hspace{-0.98in}&& \, \,\, \,
    x \cdot      Heun\Bigl(- \, {{1} \over {4}},  {{7} \over {256}}, {{3} \over {8}},
 \,  {{3} \over {8}},   {{3} \over {4}},   {{1} \over {2}},  -4 \, x\Bigr)
\cdot    Heun\Bigl(- \, {{1} \over {4}},  {{31} \over {256}},  {{5} \over {8}},
    {{5} \over {8}},    {{5} \over {4}},  {{1} \over {2}}, \, -4 \, x\Bigr).            
\end{eqnarray}
These (non globally bounded) solutions (\ref{M3Ratfoncfour7recipsolA}), (\ref{M3Ratfoncfour7recipsolB}),
and (\ref{M3Ratfoncfour7recipsol2})
are different from the diagonal of the 
rational function (\ref{Ratfoncfour7recip}) which is trivial in that case. We are in the situation where
the solutions of the
telescoper are different from  the diagonal of the rational function: they are 
``Periods''~\cite{ KontZagier} over {\em non-evanescent cycles} of the algebraic variety corresponding
to the rational function. Example 1 was seen to correspond to a classical modular form
with an integer series and a formal series with a logarithm at $\, x= \, 0$. The point at infinity
$\, x= \infty$ is an elliptic point with no logarithm, and series that are {\em not} globally bounded. 
Let us note that this  order-three linear differential operator (\ref{M3Ratfoncfour7recip})
has the {\em same singularities} as the order-three operator  $\, L_3$ given by (\ref{L3SS7}).
Consequently the first parameter of these  Heun functions solutions and of the Heun function (\ref{H1solhyp})
solution of $\, L_3$ is the same.

More generally the telescoper $\, T_R$ of a rational function $\, R(x, \, y, \, z, \, w)$
and the  telescoper $\, T_{\cal R}$ of the rational function
\begin{eqnarray}
\label{Ratfoncfour7recipnew}
  \hspace{-0.98in}&& \quad  \quad   \quad  \quad   \quad    \quad  \quad \, \,  
{\cal R}(x, \, y, \, z, \, w)  \, \, \,  = \, \,  \,    \, \,
 R\Bigl({{ A_1 } \over {x}}, \, {{A_2} \over { y}}, \, {{A_3 } \over {z}}, \, \, {{A_4 } \over {w}}\Bigr), 
\end{eqnarray}
are simply related. The  telescoper $\, T_{\cal R}$ is the  telescoper $\, T_R$ pullbacked by
the involution $\, \, x \, \rightarrow \,\,  A/x \, \, $
where $\,\,  A \, = \,\,  A_1 \, A_2 \, A_3 \, A_4$.

One can easily perform the same calculations for examples 2, 3, 4, 6 of section (\ref{subfour}),
and find exactly similar results\footnote[1]{With a set of new (non globally bounded)
  Heun functions like (\ref{M3Ratfoncfour7recipsol2}).}, namely
the fact that the telescoper   $\, T_{\cal R}$  of the new rational function (\ref{Ratfoncfour7recipnew})
is exactly the telescoper $\, T_R$  of the rational function, pullbacked by $\, x \, \rightarrow \, \, A/x$.
Note that, for these examples, there exists a choice of the constant $\, A$ such that the
singularities of $\, T_R$ and  $\, T_{\cal R}$ {\em remain the same}. For examples
2, 3, 4, 6  one must take respectively $\, A\, = -1/27, \,-1/128, \, -1/16, \, 1/144$. For the
first lattice Green example of section (\ref{subIntroduction}) take $\, A \, = \, 1/4$.

\vskip .1cm 

{\bf Remark:} Do note that the other Shimura example of section (\ref{othervanHoeijVidunas}) has been build that way:
the rational function (\ref{Ratfonc4gg0OTHER}) is actually the
rational function of  (\ref{Ratfonc4gg0}) where $\, (x, \, y, \, z, \, w)$ has
been changed into $\, (1/x,\, 1/y, \, 1/z, \, 1/w)$ . Again we have that
the telescoper  $\, M_3$ of the ``recipocal'' rational function is exactly
the  telescoper  $\, L_3$ of the rational function, pullbacked by $ \,\, x \, \rightarrow \, \, 1/x$.

\vskip .2cm 

\subsection{Periods of extremal rational surfaces: the $\, x \, \leftrightarrow \, 1/x$ duality on the telescoper.}
\label{Duality2}

Let us recall the rational function (\ref{Ratfonc4}) which diagonal was the sum of two classical modular forms.
Let us consider the rational function where the three variables $(x, \, y, \, z)$  have been changed into their reciprocal 
$(1/x, \, 1/y, \, 1/z)$ in the rational function $\, R(x, \, y, \, z)$ given by  (\ref{Ratfonc4}):
\begin{eqnarray}
\label{Ratfoncfour4recip}
  \hspace{-0.98in}&& \quad  \quad   \quad  \quad   
\, \,  
{\cal R}(x, \, y, \, z)  \, \, \,  = \, \,  \,    \, \,
 R\Bigl({{1 } \over {x}}, \, {{1 } \over {y}}, \, {{1 } \over {z}}\Bigr)
\nonumber \\
  \hspace{-0.98in}&& \quad  \quad \,  \quad \quad  \quad \quad  \quad
\, \, \,  = \, \,  \,   \,  \,  
 {\frac {{x}^{3}yz}{{x}^{3}yz \, \, \, +{x}^{3}y \, \,\,  +{x}^{3}z \, \,\,  +{x}^{2}yz \,  \,\, +{x}^{3} \,  \,\, +{x}^{2}z \,\, \, -1}}.
\end{eqnarray}
The telescoper of this rational function (\ref{Ratfoncfour4recip}) is an order-four linear differential operator
which is the direct sum of two order-two operators $\,  L_2^{(p)} \oplus \, M_2^{(p)}$ which are
precisely the two order-two linear differential  operators occurring of section (\ref{subthree2})
pullbacked by $\, x \, \rightarrow \, 1/x$. The solutions (\ref{Ratfonc4HeunSol2.3}) and (\ref{spurious})
where one changes  $\, x \, \rightarrow \, 1/x$
are solutions of these new linear  pullbacked differential  operators $\,  L_2^{(p)}$ and $\, M_2^{(p)}$.
Such (classical modular forms) solutions are actually expandable at $\, x \, = \, 0$ and correspond to {\em globally bounded}
series expansions. The solution of $\,  L_2^{(p)}$, analytic at $\, x \, = \, 0$, can be written as
a Heun function
\begin{eqnarray}
  \label{Ratfoncfour4recipHeunv}
  \hspace{-0.98in}&& \quad  \quad \quad  
  x\cdot \,  Heun\Bigl({{1} \over {2}} \, - {{i \, \sqrt{3}} \over {2}}, \,{{1} \over {2}} \, - {{i \, \sqrt{3}} \over {6}},
 \, 1, \, 1, \, 1, \, 1, \, \Bigl({{-3 \, +i \sqrt{3}} \over {18}}\Bigr) \cdot \, x    \Bigr)
 \nonumber \\
 \hspace{-0.98in}&& \quad  \quad  \quad \quad 
 \, \,  = \, \,  \,
  x  \, \,-{\frac{1}{9}}{x}^{2} \, +{\frac{1}{81}}{x}^{3}
 \, \, -{\frac{7}{6561}}{x}^{4} \,\, +{\frac{1}{59049}}{x}^{5} \,\, +{\frac{11}{531441}}{x}^{6}   \,\,  \, +  \,  \, \, \cdots  
\end{eqnarray} 
to be compared with  (\ref{Ratfonc4HeunSol2.3}). Again these nice classical modular forms globally bounded
series expansions are different from the diagonal of (\ref{Ratfoncfour4recip}) which is equal to zero.

\end{document}